\documentclass{report}
\usepackage{amsfonts,amsmath,amssymb}
\usepackage[dvips]{graphics}
\usepackage{epsfig}
\usepackage{amscd}
\usepackage{verbatim}
\newcommand{\Rl}{\mathbb{R}}
\newcommand{\Nl}{\mathbb{N}}
\newcommand{\Ir}{\mathbb{Z}}
\newcommand{\Cx}{\mathbb{C}}
\newcommand{\A}{\mathcal{A}}
\newcommand{\HH}{\mathcal{H}}

\newtheorem{theorem}{Theorem}[section]
          
\newtheorem{lemma}[theorem]{Lemma}              
\newtheorem{claim}[theorem]{Claim}
\newtheorem{proposition}[theorem]{Proposition}
\newtheorem{corollary}[theorem]{Corollary}
\newtheorem{definition}[theorem]{Definition}
\newtheorem{conjecture}[theorem]{Conjecture}
\newcommand{\rem}[1]{{\bf Remark:}}

\newcommand{\Section}[1]{\setcounter{equation}{0}\section{#1}}

\newcommand{\eq}[1]{(\ref{#1})}
\newenvironment{proof}{\noindent {\bf Proof: }}{\QED\medskip}
\def\QED{{\hspace*{\fill}{\vrule height .5ex width 1ex }\quad} 
    \vskip 0pt plus20pt}
\newcommand{\be}{$$}
\newcommand{\ee}{$$}
\newcommand{\bea}{\begin{eqnarray}}
\newcommand{\eea}{\end{eqnarray}}
\newcommand{\beann}{\begin{eqnarray*}}
\newcommand{\eeann}{\end{eqnarray*}}

\newcommand{\ket}[1]{\vert{#1}\rangle}
\newcommand{\bra}[1]{\langle{#1}\vert}
\newcommand{\wslim}{{\rm w}^*\mbox{-}\lim}
\newcommand{\Prob}{{\rm Prob}}

\newcommand{\unity}{{1\hskip -3pt \rm{I}}}

\newcommand{\GS}{\mathcal{G}}
\newcommand{\ip}[2]{\langle{#1}|{#2}\rangle}

\newcommand{\tvec}{\mathbf{t}}
\newcommand{\avec}{\mathbf{a}}

\newcommand{\nvec}{\mathbf{n}}
\newcommand{\mvec}{\mathbf{m}}

\newcommand{\Sym}{\operatorname{Sym}}
\newcommand{\Hil}{\mathcal{H}}
\newcommand{\Par}{\mathbb{P}}
\newcommand{\ps}{\textit{ps}}

\newcommand{\qbinom}[3]{\left[{\begin{matrix}
	#1 \\ #2\end{matrix}}\right]_{#3}}
\newcommand{\Span}{\operatorname{span}}

\newcommand{\Spin}{\textrm{S}}
\newcommand{\Proj}{\operatorname{Proj}}
\newcommand{\Obs}{\mathcal{A}}
\newcommand{\M}{\mathbb{M}}

\newcommand{\Rayleigh}[2]{\frac{\ip{#1}{#2 #1}}{\ip{#1}{#1}}}
\newcommand{\calK}{\mathcal{K}}
\newcommand{\Hpp}{H^{++}}
\newcommand{\HXXZ}{H^{\rm{XXZ}}}

\newcommand{\allup}{\ket{\uparrow\dots\uparrow}}
\newcommand{\alldown}{\ket{\downarrow\dots\downarrow}}
\newcommand{\sbinom}[2]{\scriptstyle{\left(\hspace{-3pt}
	\begin{array}{c}\scriptstyle #1 \\ 
        \scriptstyle #2\end{array}\hspace{-3pt}\right)}}
\newcommand{\sqbinom}[3]{\left[\hspace{-3pt}
	\begin{array}{c}\scriptstyle #1 \\ 
        \scriptstyle #2\end{array}\hspace{-3pt}\right]_{#3}}
\newcommand{\pn}[1]{f_q({#1})}

\newcommand{\floor}[1]{\left\lfloor{#1}\right\rfloor}
\newcommand{\ceil}[1]{\left\lceil{#1}\right\rceil}
\newcommand{\EXP}[2]{{\langle{#1}\rangle}_{#2}}
\newcommand{\ran}{\operatorname{ran}}

\DeclareMathAlphabet{\mathol}{OT1}{cmr}{l}{ol}

\newcommand{\up}{\ket{\uparrow}}
\newcommand{\down}{\ket{\downarrow}}
\newcommand{\uu}{\ket{\uparrow \uparrow}}
\newcommand{\ud}{\ket{\uparrow \downarrow}}
\newcommand{\du}{\ket{\downarrow \uparrow}}
\newcommand{\dd}{\ket{\downarrow \downarrow}}
\newcommand{\C}{\mathbb{C}}
\newcommand{\alphavec}{\mathbf{\alpha}}

\newcommand{\Z}{\mathbb{Z}}
\newcommand{\ZGC}[2]{Z^{GC}(#1,#2)}
\newcommand{\op}[3]{\bra{#1} #2 \ket{#3}}

\newcommand{\avg}[2]{\langle{#1}\rangle_{#2}}
\newcommand{\avgGC}[2]{\langle{#1}\rangle^{GC}_{#2}}
\newcommand{\supp}{{\rm supp}}
\newcommand{\Num}{\mathsf{N}}
\newcommand{\navg}{\avgGC{\Num}{\Sigma,\mu}}
\newcommand{\Sc}{\mathcal{S}}

\newcommand{\sech}{\, {\rm sech}}
\newcommand{\real}{\, {\rm Re}}

\newcommand{\RawWedge}{\wedge}
\newcommand{\CapitalNu}{N}
\newcommand{\inlineTFinmath}[1]{$#1$}
\newcommand{\Mvariable}{M}
\newcommand{\SkeletonIndicator}{\%}

\begin{document}
\pagenumbering{roman}
\begin{center}
{\Large \bf Some properties for the low-lying spectrum 
of the ferromagnetic, quantum XXZ spin system
}
\vskip 0.2in
\centerline{By}
\vskip 0.1in
\centerline{\large  Shannon L. Starr}
\centerline{\large B.A. University of California, Berkeley 1996}
\vskip 0.3in
\centerline{\large  DISSERTATION}
\vskip 0.1in
\centerline{\large Submitted in partial satisfaction of the requirements for the
degree of}
\vskip 0.1in
\centerline{\large  DOCTOR OF PHILOSOPHY}
\vskip 0.1in
\centerline{\large in}
\vskip 0.1in
\centerline{\large MATHEMATICS}
\vskip 0.2in
\centerline{\large in the}
\vskip 0.1in
\centerline{\large  OFFICE OF GRADUATE STUDIES}
\vskip 0.1in
\centerline{\large of the}
\vskip 0.1in
\centerline{\large  UNIVERSITY OF CALIFORNIA}
\vskip 0.1in
\centerline{\large  DAVIS}
\end{center}
\vskip 0.1in
{\large Approved:}
\begin{center}
\centerline{\underbar{\hskip 2.5in}}
\vskip 0.15in
\centerline{\underbar{\hskip 2.5in}}
\vskip 0.15in
\centerline{\underbar{\hskip 2.5in}}
\vskip 0.2in
\centerline{\large Committee in Charge}
\vskip 0.2in
\centerline{\large 2001}
\end{center}
\newpage
\large
\tableofcontents


\newpage
\large
{\Large \bf ACKNOWLEDGEMENTS} \\

I wish to thank my advisor, Bruno Nachtergaele, who has been exceptionally patient
and generous with his time.
He has shared many ideas with me, as well as introduced me to the fascinating subject
of quantum spin systems.
Without him, none of this research would have been done.
I would like to thank my other first collaborators, Pierluigi Contucci and Oscar Bolina,
who also helped me to understand the XXZ model.
I would like to thank Professors Janko Gravner, Albert Schwartz and Craig Tracy.
All of them have been my teachers, and all have helped me get out of some tight spots
in my research.
I would like to thank Professors Kuperberg and Khovanov for taking time to teach me
many things that I should know. 
I would like to thank Wolfgang Spitzer, for being a fellow researcher,
as well as a teacher on many occasions.
I would like to thank my friends in the math department for creating a stimulating
environment, which seems to get better all the time.
I would like to thank my family, which is too large to mention by name.
They have always supported my education, and given me much needed reality checks.
I especially thank: my sister Amy, for being my connection to the \textit{real world};
my brother Jason, for being my connection to \textit{his world}; my mother Susan, for loving
me with no strings attached; and Carmeliza for so much, including putting up with
my problems, without ever burdening me with hers.
I would also like to thank Robyn, Heath and the Navasca family, who I think of as part of my
own family.
Finally I would like to thank the members of the band ``Tool'', whose music has
been a source of inspiration during many late nights of study.

This material is based on work supported by the National Science Foundation
under Grant No. DMS0070774.


\newpage 
\begin{center}
\underline{\bf \Large Abstract} 
\end{center}

\medskip

\large

We consider several aspects of the XXZ quantum spin system.
These aspects are: existence of a spectral gap above infinite-volume ground states for arbitrary spin-$\Spin$ XXZ chains;
description of droplet ground states for the XXZ Hamiltonian with up-spin boundary conditions; and a
constructive proof of nonexistence of spectral gap above interface ground states in dimensions two and higher.

In Chapter 1 we introduce the XXZ spin system, list some background.
We also give a complete summary of the rest of the paper.

In Chapter 2, we review some known results, including : the existence of interface ground states of the quantum XXZ model 
in all dimensions and for all values of spin \cite{ASW}, \cite{GW}; completeness of this list for ground states in one
dimension \cite{Mat}, \cite{KN3}; 
quantum group $\textrm{SU}_q(2)$ symmetry for the spin-$\frac{1}{2}$, spin chain \cite{PS};
existence and exact calculation of nonvanishing spectral gap above infinite-volume ground states
for the infinite spin-$\frac{1}{2}$, spin chain \cite{KN1}.

In Chapter 3, we extend the result of \cite{KN1} by proving the existence of a nonzero spectral gap in one-dimension for 
spins higher than $\frac{1}{2}$.
We show that the gap depends on the magnetization, $M$, of the ground state through an even, $2\Spin$-periodic function, which
is the filling factor for one-dimensional interfaces.
We give numerical estimates for the gap, and observe that
the gap is maximized at some value of anisotropy strictly
the totally anisotropic Ising model and the isotropic model.
Based on our numerical evidence we observe that the gap of a spin-$\Spin$ chain grows like linearly with $\Spin$, 
and we conjecture that the rescaled gap converges to a fixed function of $\Delta$, in the classical $\Spin \to \infty$ limit.

In Chapter 4, we introduce two models a one-dimensional quantum spin droplet.
The first is a linear XXZ spin chain with up-spin boundary fields,
and a pinning field, which is an external, negative magnetic field at a single site in the interior of the spin chain.
For special choices of the boundary field  and pinning field strengths the model admits frustration free ground states,
which describe a droplet of down-spins centered at the pinned site, on a background sea of up-spins.
We analyze the interfaces between the domains of up-spins and down-spins.
We show that in the limit that the droplet grows to ifninity, the right-interface is a convex combination of kink ground states
with discrete Gaussian weights.
After this, we consider the droplet Hamiltonian without pinning fields, which is a more realistic model.
We show that it also admits droplet ground states, which may be expressed as a mixture of states obtained
by tensoring an antikink state on the left half of the spin chain to a kink state on the right half.
We also prove that there is a spectral gap above these droplet ground states.
Our proof uses a method of ``cutting'' and ``pasting'' spin chains which is based on finding intervals where the spin are all
up or all down.
Finally, we prove that the amplitude for dispersion relation of droplet state energy versus momentum is decaying exponentially
with the droplets' size.
I.e., the droplets have a mass which grows exponentially with their size.

In Chapter 5, we reprove a known result, that there can be no spectral gap above infinite-volume interface ground states of the 
XXZ Hamiltonian  in dimensions two and higher \cite{KN2}, \cite{Mat2}.
The original proof was based on the Goldstone theorem, which applies because the interface states break $\textrm{U}(1)$-symmetry.
Using similar ideas, we construct explicit excitations, and develop rigorous upper-bounds for the spectral gap.
We show that in the thermodynamic limit the decay of the spectral gap is at least as fast as  
$1/R^2$, where $R$ is the linear size of the interface..
Our technique shows that excitations which minimize energy are spin waves, constrained to the quantum interface.
In the course of our proof we develop an equivalence of ensembles (EOE) result 
to show that canonical and grand canonical ground states of the XXZ model are equivalent on small subvolume.


\newpage
\pagenumbering{arabic}
\markright{  \rm \normalsize CHAPTER 1. \hspace{0.5cm}
Introduction}
\large 
\chapter{Introduction}

The main subject of this thesis is the quantum XXZ spin system, which
is a quantum spin system with Hamiltonian
\begin{equation}
\label{IntroXXZDef}
H_{\Lambda} 
  = - \sum_{\substack{\{x,y\} \in \Lambda \\ |x-y| = 1}} 
  \left(\frac{1}{\Delta} S_x^1 S_y^1
  + \frac{1}{\Delta} S_x^2 S_y^2 + S_x^3 S_y^3\right)\, ,
\end{equation}
defined on the domain $\Lambda \subset \Ir^d$, where $|\cdot|$ is the
usual $l^1$-norm, and $S_x^\alpha$ is the spin matrix for spin
in the coordinate direction $\boldsymbol{e}_\alpha$ at site $x$.
$\Delta$ is a number between 1 and $\infty$. 
This is a deformation of the usual isotropic Heisenberg model, for which
$\Delta = 1$.
At $\Delta = \infty$ one has the Ising spin system, instead.
This is a ferromagnetic model because of the overall factor $-1$ in front of 
the sum.

Spin is the most fundamental purely quantum phenomena. It is connected to the Pauli exclusion
principle via the 
celebrated spin-statistics theorem, which says that all integer-spin particles are bosons and
all integer-plus-a-half spin particles are fermions (c.f.\ \cite{SW}).
Since its discovery, spin has been the key to understanding magnetism, 
and it plays a part in many
modern theories such as the BCS theory for superconductivity.
There are several different types of magnetism, the most well-known being
ferromagnetism, antiferromagnetism, and ferrimagnetism.
The first quantum model proposed for ferromagnetism was Heisenberg's model,
in which electrons are tightly bound to atoms of some regular crystal.
For each atom, the unpaired valence electrons' spins add by the usual quantum rules 
for addition of angular momentum, i.e.\ one takes the tensor product of $\textrm{SU}(2)$
for each electron.
Due to the short range of magnetic fields, one makes the approximation that
each atom interacts only with its nearest neighbors.
The interaction term is $-J(x,y) \boldsymbol{S}_x \cdot \boldsymbol{S}_y$.
Since the interaction is $\textrm{SU}(2)$-invariant, one can decompose the spin
of each atom into its irreducible components,
and the Hamiltonian will block diagonalize in such a way that the different
irreducibles, for a single atom, are not coupled.  
Consequently, the lowest energy state will be a state such that at each atom
the spin falls in only one of the irreducible representations,
in this way one usually considers only the model where
each atom of the crystal is equipped with a single irreducible representation
of $\textrm{SU}(2)$.
(One should also consider the orbital angular momentum of the electrons,
but as can be calculated by the Einstein-de Haas experiment, for most atoms
this is negligible in comparison to spin angular momentum.
Also, for many atoms, iron being one, the interaction is not a direct exchange
as we have indicated, 
but the product of two antiferromagnetic exchanges with some other atom, such
as oxygen, which results in an effective direct exchange.)
The Heisenberg model is thus
$$
H^{\textrm{Heis}}_\Lambda = - \sum_{\substack{\{x,y\} \subset \Lambda \\
  |x-y|=1}} J(x,y) \boldsymbol{S}_x \cdot \boldsymbol{S}_y\, ,
$$
where $\Lambda$ represents the sites for the various atoms.
The most basic model is a homogeneous magnet, i.e.\ all 
coupling constants $J(x,y)$ are equal to a constant.
This way, one arrives at the XXX model, which is a special case of the XXZ model,
obtained by setting $\Delta = 1$. 
This model is often simply called the Heisenberg model.
Although it is apparently quite simple, it does describe some magnets
surprisingly well.
It also has subtle features, for example it is a longstanding open
problem to prove that the isotropic Heisenberg ferromagnet model has a phase 
transition in dimensions three and higher, and to prove
long range order for the ground state of the spin-$\frac{1}{2}$ model in dimension 
two.
It is important to note the difference between the isotropic and the anisotropic
model, here.
For the XXZ model Tom Kennedy proved, in 1985, \cite{Ken}
that the low-temperature equilibrium
states possess long range order in dimension two and higher if \textit{any} anisotropy
is present, i.e.\ for all $\Delta > 0$.

A seemingly more complicated model of magnetism would allow, instead of the
$\textrm{SU}(2)$-symmetric exchange $-\boldsymbol{S}_x \cdot \boldsymbol{S}_y$,
an anisotropic exchange of the form
$$
J_1 S_x^1 S_y^1 + J_2 S_x^2 S_y^2 + J_3 S_x^3 S_y^3\, .
$$
It turns out that for certain purposes this is actually an easier model to study.
It is certainly physically relevant, because some materials do have anisotropic exchanges.
Also, by introducing the more general Hamiltonian, one can gain insight
into the original Heisenberg model, for example by observing the
dependence of ground states, or low excitations on $(J_1,J_2,J_3)$.
For example, in the model we will study in this paper, $J_1=J_2 = \Delta^{-1}$,
while $J_3=1$.
This is called the XXZ model.
It turns out that in all dimensions one can write down some of the ground states
of such a model, and in one-dimension it has been proved that the known list
is complete \cite{Mat}, \cite{KN3}.
For one dimension, one also knows the complete list of ground states
of the isotropic model \cite{KN3}.
But the two models differ in that for the XXZ model a spectral gap exists
above the one-dimensional ground states \cite{KN1}; whereas, for the XXX model,
one can obtain arbitrarily low-energy excitations of the
ground state by introducing spin-waves, which continuously
rotate a spin through $180^\circ$ over a long interval.
But for the XXZ model, the anisotropy damps these spin waves, because there is a 
high energy cost for deviation of the spin from alignment with the 
$\boldsymbol{e}_3$-axis.
Therefore the anisotropic exchange results in more stable ground states,
and in fact one can obtain a theory of domain walls on the basis of this alone.
For the XXX model, one must assume crystal defects and impurities or surface
effects via boundary conditions to account
for domain walls.
For the XXZ model, in the thermodynamic limit, domain walls exist as stable ground states.
Also, as mentioned before, some properties, such as long range
order in three dimensions, are known to exist for the XXZ model with $\Delta > 1$,
and expected for the XXX model, but still have not been proved.
It is a tantalizing thought that one might be able to prove this by taking $\Delta \to 1$,
although the non-commutativity of the thermodynamic limit with the limit $\Delta \to 1$
precludes any obvious implementation of this idea.

The XXZ chain has been much studied in the field of exactly solvable models.
The Heisenberg model was proposed by Heisenberg in 1929.
In 1931, Hans Bethe proposed a method of solution which, in principle, solves
not only of the XXX model, but also the XXZ model.
Later, with the advent of the variational Yang-Yang equation, the Bethe ansatz was put
on solid mathematical ground; however the proof is only valid for
the antiferromagnet.
Since its inception, the Bethe ansatz has become an industry of its own;
c.f.\ the excellent text by Korepin, Bogoliubov, Izergin \cite{KBI}.
In general the XYZ chain
$$
H^{\textrm{XYZ}} = -\frac{1}{2} \sum_{x=1}^L \left[
  J_1 S_x^1 S_{x+1}^1 + J_2 S_x^2 S_{x+1}^2 + J_3 S_x^3 S_{x+1}^3 \right]\, ,
$$
has been shown to be equivalent to a zero field, eight-vertex model by Baxter;
c.f.\ \cite{Bax}.
The XXZ model is equivalent to a six-vertex model, also known as an ice-type model.
Such models were solved by Lieb \cite{L1,L2,L3}, who used properties of the
XXZ model in his solution.
Alcaraz et.al.\ \cite{ABBBQ} used the XXZ chain with imaginary boundary fields 
to determine critical exponents of the Ashkin-Teller and Potts models.
More recently, Greg Kuperberg used the six-vertex model, and in particular
a result of Korepin and Izergin, in his proof of the alternating
sign matrix conjecture \cite{Kup}.
Even more recently, the number of ASM's has shown up in the coefficients
of eigenstates of the XXZ model for special values of $\Delta$,
and has led to new conjectures \cite{RS}, \cite{BdGN}.

Having mentioned only a few of the many interesting results associated to the
Bethe ansatz and the exact solvability of the XXZ model, we make clear that
our results are entirely independent of the Bethe ansatz.
There are two basic reasons for this.
The first is that the solutions of the Bethe ansatz have been shown to be complete
only for $\Delta=1$, i.e.\ th XXX model, by Babbit and Thomas
\cite{Thom}, \cite{BT}.
Despite efforts by Babbit and Gutkin \cite{BG}, \cite{Gut} 
to prove the analog for the XXZ model, 
their results are not complete, and only address excitations which are
perturbations of the translation-invariant state $\omega_\uparrow$.
The subject of our work are non-translation invariant kink ground states.
It seems that a rigorous result regarding the Bethe ansatz for excitations of kink 
ground states would be
extremely difficult, due to the fact that a kink ground state has both an infinite
number of up-spins and an infinite number of down-spins.
The second reason for choosing not to use the Bethe ansatz is that the results
of this thesis are not tenable by the Bethe ansatz, in any case.
Even putting questions of rigor aside, the Bethe ansatz ceases to be a viable 
method for the XXZ chain with spin $>1/2$, or spin systems on lattices in dimensions
$>1$.
There are some integrable spin chains with spin greater than $1/2$,
for example those worked out by Babujian and Tsvelik \cite{BabTsv},
but the XXZ model for higher spin is not one of these.
This rules out two of the results of this paper: a proof of non-vanishing
spectral gap for spin chains with spin $\Spin>1/2$, and an upper-bound on the 
rate of vanishing for the spectral gap above interface states in dimensions two
and higher.
Our third result, is a quantum model spin droplets, as ground states of the XXZ model 
with positive boundary fields.
We give approximate formulas for the entire list of ground states, in the 
limit that the number of down-spins becomes infinite.
Our results remain valid for states with infinitely 
many down-spins and infinitely many up-spins, and
therefore extracting useful information from the Bethe ansatz is 
almost certainly impossible.

For our purposes, more useful than the supposed complete solvability of the
XXZ model, is the existence of a quantum group symmetry.
In comparison to many of the results surrounding the Bethe ansatz, the quantum 
group symmetry was discovered fairly recently, it was first set down
concretely for the XXZ model by Pasquier and Saleur in 1989 \cite{PS}.
Originally discovered by Woronowicz \cite{Wo1,Wo2,Wo3,Wo4} 
a quantum group is a smooth deformation of a Lie group in the category
of Hopf algebras,
or in the version of Drinfel'd \cite{Dri} and Jimbo \cite{Jim} a smooth
deformation of the universal enveloping algebra of a Lie algebra in the
category of Hopf algebras.
A quantum group possesses module structures which are very similar to those of 
Lie groups and Lie algebras.
In particular there is a product, a coproduct, an identity, a coidentity,
and an antipode.
However, for quantum groups, neither the product or coproduct is generally
commutative.
The coproduct allows one to make a 
representation on the tensor product of any two representations.
This is important for quantum spin chains, because the Hilbert space is naturally
expressed as a large tensor product of two-dimensional irreps of $\textrm{SU}(2)$.
It turns out that the two-dimensional representations of the quantum group
$\textrm{SU}_q(2)$, for real values of $q$, are equivalent to the two-dimensional
irreps of $\textrm{SU}(2)$ (c.f.\ \cite{Kas}).
Thus, there is a natural $\textrm{SU}_q(2)$ representation on any ordered 
spin-$\frac{1}{2}$, spin chain.
The fundamental result of Pasquier and Saleur, and later Alcaraz, Salinas and 
Wreszinski, \cite{ASW}, was that for a particular value of $q$, namely
$q$ a solution of $\Delta = \frac{q + q^{-1}}{2}$, the XXZ chain with 
special ``kink'' boundary 
conditions
$$
H^{\textrm{kink}}_{[1,L]} = H^{\textrm{XXZ}}_{[1,L]} + A(\Delta) (S_L^3 - S_1^3)\, ,\qquad
A(\Delta) = \frac{1}{2} \sqrt{1 - \Delta^{-2}}\, ,
$$
actually commutes with the entire $\textrm{SU}_q(2)$ representation.
This gives a system of commuting variables, and it stands that one can compute
the eigenvalues of $H^{\textrm{kink}}_{[1,L]}$, by first finding the invariant 
subspaces of $\textrm{SU}_q(2)$.
At this level, the tool of the quantum group symmetry is exactly the same as
the $\textrm{SU}(2)$ symmetry is for the XXX model.
Moreover, there are analogous results, such as the ordering-of-energy levels
result by Lieb and Mattis \cite{LM}, which is apparently also true in the 
framework of $\textrm{SU}_q(2)$-symmetric Hamiltonians.
A notable use of the quantum group symmetry is Koma and Nachtergaele's proof of
the existence of a spectral gap above the infinite-volume ground state \cite{KN1}.
One drawback of the quantum group approach is that it cannot be extended to 
the infinite-volume Hamiltonian.
I.e.\ at the present time there is no rigorously defined representation of 
$\textrm{SU}_q(2)$
on the algebra of quasilocal observables which commutes with the kink Hamiltonian.
In fact by a work of Fannes, Nachtergaele and Werner \cite{FNW2}, no such 
representation can exist.
Still, there is enticing algebraic (as opposed to rigorous analytic), work by Jimbo and Miwa 
\cite{JM}, which indicates
that in the thermodynamic limit the antiferromagnet possesses a symmetry of the
quantum affine algebra $\textrm{U}_q(\widehat{\mathfrak{sl}}_2)$.
For rigorous analytic results, the $\textrm{SU}_q(2)$ symmetry of the finite-volume
spin chains is the most that is available, but is often quite useful.

There are several reasons for studying the XXZ ferromagnet,
as opposed to say studying the antiferromagnet.
The first, is that ferromagnets are abundant in nature, so there is hope of
developing a theory with real physical significance.
For some efforts in the direction of physical applications of the XXZ ferromagnet, 
see \cite{KY1}, \cite{KY2}.
We note that in terms of the most obvious physical features of the theory, more is
already proved for the antiferromagnet than the ferromagnet, such as long range order in 
dimensions two and higher \cite{KLS}, owing to reflection positivity.
For our type of analysis, the ferromagnet is more natural to study, since
the kink states have a simple asymptotic structure, and the interface is exponentially
localized.
By avoiding the Bethe ansatz, we seek to provide direct physical arguments to explain
the low-lying spectrum.
Along these lines, more rigorous results have been proved for the ferromagnet
(such as the complete list of ground states and existence of a spectral gap in one-dimension).
Finally, for finite volumes, more is known about the ground states of the ferromagnet
than the antiferromagnet.
Even for the XXX model, there is no simple formula for th ground state of the antiferromagnet,
the best that is available is the ordering of energy-levels result of Lieb and Mattis \cite{LM}.
In contrast, for any choice of anisotropy $1 \leq \Delta \leq \infty$, any spin 
$\Spin \in\frac{1}{2} \Nl$, and in any dimension, one can write down the ground state of
the ferromagnetic XXZ model.
Thus the ferromagnetic model is more amenable to rigorous research, at least for a beginner.
It would be an interesting point of further research to see what one can prove
along the lines of spectral gaps, etc., for the antiferromagnet, but one which we leave
open for now.

The outline of the thesis is as follows:
In Chapter 2, we introduce some preliminaries about the 
XXZ spin chain.
In chapter 3, we generalize a result on the existence of a spectral
gap for the infinite-volume 1-d spin system, which had been proved
for spin-$\frac{1}{2}$, but which we show to be true for all spin $S>\frac{1}{2}$,
as well.
In chapter 4, we present two models for spin droplets, with the property that
the ground states have a domain of down spins surrounded on the left and the 
right by domains of up-spins (contrasted to the kink state which has just down-spins
to the left and up-spins to the right).
We show that in the limit that the size of the droplet goes to infinity, the
ground states have a very simple form, and also possess a spectral gap above.
In chapter 5, we reconsider an old theorem that the spectral gap above interface
ground states must vanish in the infinite-volume limit in all dimensions greater
than 1.
We reprove the theorem, constructively, giving an upper-bound for the size
of the gap which is of the order $1/R^2$, where $R$ is the linear size 
of a plane parallel to the interface.
We briefly summarize each of these chapters, now.

We begin in chapter 2 by stating some important facts, 
which are all already in the literature.
First, we state a theorem due to Alcaraz, Salinas and Wreszinski
\cite{ASW} on the ground states for the ferromagnetic XXZ model in
any dimension, and with any spin-$\Spin$.
They showed that with the kink boundary conditions one can write down
a simple formula for the ground states.
For spin-$1/2$ and in one-dimension, the formula is 
\begin{equation}
\label{IntroGSFormula}
\psi_0(n) = \sum_{\{x_1,x_2,\dots,x_n\} \subset [1,L]}
  \prod_{k=1}^n q^{x_k} S_{x_k}^-\ \ket{\textrm{all down}}_{[1,L]}\, .
\end{equation}
To motivate the kink boundary conditions, we consider the classical
(spin goes to infinity) model.
After this detour, we return to the Alcaraz, Salinas \& Wreszinski theorem, which
we state in slightly more generality than its authors did,
because we want to allow for graphs which are not subsets of $\Ir^d$.
As far as proof goes, the generalization is trivial, but it is useful
in Chapter 4, where one model for a droplet Hamiltonian is shown to be
equivalent to a kink Hamiltonian on a different graph.
The discovery of Alcaraz, Salinas and Wreszinski, and particularly the choice
of kink boundary conditions, was motivated by the earlier discovery
that the finite-volume XXZ Hamiltonian is actually invariant under the action of 
the quantum group $\textrm{SU}_q(2)$.
This was probably known at some level for some time in connection with the 
Bethe-ansatz solvability of the model, but was first stated concretely for the 
XXZ Hamiltonian,
including the choice of boundary conditions for the quantum group symmetry,
by Pasquier and Saleur in \cite{PS}.
We give a brief introduction to quantum groups and in particular $\textrm{SU}_q(2)$
in chapter 2, immediately following the main theorem of Alcaraz, Salinas \& Wreszinski.
We demonstrate the main result of Pasquier and Saleur that the representation of 
$\textrm{SU}_q(2)$
commutes with the kink Hamiltonian.

After this, we give a brief review of the work of Gottstein and Werner, 
\cite{GW}, which was done
contemporaneously with and independently of \cite{ASW}.
Gottstein and Werner, like Alcaraz, Salinas \& Wreszinski, determined the finite-volume ground states,
though only for dimension-one and spin-$\frac{1}{2}$.
However, they did considerably more because they also discovered infinite-volume
ground states.
In fact they noted a key feature of both the finite- and infinite-volume 
ground states, which is that they are 
``frustration-free'' meaning they minimize each of the translation invariant
nearest-neighbor interactions, instead of simply minimizing the sum.
This fact, which allows the definition of infinite volume ground states, 
also gives an explanation of why the same ground states work
for dimensions greater than one and spins greater than $1/2$
(although Gottstein \& Werner did not explicitly note the latter).
In Gottstein and Werner's work, in order to treat infinite-volume 
zero-energy (zero-energy = frustration-free) ground states, they constructed
an entire theory of infinite-volume zero-energy states as states on the
approximate inductive limit of finite-volume zero-energy observables.
This theory is somewhat more complicated than the usual definition of states
on the algebra of quasilocal observables, because whereas the local observable
algebra $\Obs_\Lambda$ always decomposes as 
$\Obs_{\Lambda'} \otimes \Obs_{\Lambda\setminus \Lambda'}$ for any 
$\Lambda' \subset \Lambda$,
the same is not true of the local zero-energy observable 
algebra $\mathcal{B}_\Lambda$.

The Gottstein \& Werner theory is a natural extension, and is most useful for the 
zero-energy observables with respect to the kink Hamiltonian.
For this case, the 
approximate-inductive limit $\mathcal{B}_\infty$ is shown to be a $C^*$ algebra,
with a Hilbert-space representation for which the representations of all operators
$B \in \mathcal{B}_\infty$ are normal.
In fact even more is true, because the Hilbert space representation 
coincides with the GNS representation of all the infinite-volume kink ground states,
and is also the Guichardet Hilbert space (synonymous with incomplete tensor product)
of all quasilocal perturbations of the vector
\begin{equation}
\label{IntroOmegaDef}
\ket{\Omega} = \bigotimes_{x \in \Ir} \ket{\Omega(x)}\, ,\quad
\ket{\Omega(x)} = \begin{cases} \ket{\uparrow}\, ,& x>0\, ; \\
\ket{\downarrow}\, ,& x\leq 0\, .\end{cases}
\end{equation}
Using their theory, Gottstein \& Werner show that the complete list of infinite-volume,
zero-energy ground states, w.r.t\ the kink interaction, 
consists of two translation invariant ground states defined by vectors
$\ket{\textrm{all up}}$ and $\ket{\textrm{all down}}$,
as well as the infinite-volume kink states which they define
\begin{equation}
\label{IntroInfVolGSDef}
\psi(n) = \sum_{\substack{(k,l) \in \Nl^2 \\ k-l=n}}
  \sum_{\substack{\{y_1,\dots y_l\} \subset (-\infty,0]\\
  \{x_1,\dots,x_k\} \subset [1,\infty)}}
  \prod_{i=1}^l q^{-y_i} S_{y_i}^+ \prod_{j=1}^k q^{x_j} S_{x_j}^-\ \ket{\Omega}\, .
\end{equation}
In later work, \cite{Mat}, Matsui showed that for the case $\Delta>1$,
the zero energy ground states with respect to the kink and antikink interactions
are the complete list of pure, infinite-volume ground-states.
After this, Koma and Nachtergaele \cite{KN3} gave a new proof of this fact;
one which did not explicitly use the existence of a spectral gap above the
ground states.
In particular, their argument extends to the case $\Delta = 1$, to show that
the complete list of pure, infinite-volume ground states for the XXX model
is the sphere of translation-invariant states.

In addition to giving the finishing argument for the complete list of
ground states of the XXZ model, Koma and Nacthergaele also showed that 
there is a spectral gap above all the ground states \cite{KN1}
for the spin chain with spin equal to $1/2$.
In fact this work predates the complete list of ground states, they show that
there is a spectral gap above all the zero-energy ground states, and the 
existence of the spectral gap was an implicit part of Matsui's argument,
though not Koma \& Nachtergaele's argument, for completeness of the ground states.
Koma and Nachtergaele's proof for the spectral gap is based upon showing that
for finite-volumes the spectral gap is bounded in each sector of
fixed magnetization, $S_{\textrm{tot}}^3 = M$, uniformly in $M$. 
In fact not only is the bound uniform, the spectral gap itself is a constant
independent of $M$.
This remarkable fact owes to the quantum group symmetry.
It is known that the ground states form the unique $(L+1)$-dimensional irrep of
$\textrm{SU}_q(2)$ in the tensor product 
$\Hil_\Lambda = \bigotimes_{x \in \Lambda} \Cx_x^2$.
Koma and Nachtergaele prove a lemma which shows that the next lowest energy
band is an $(L-1)$-dimensional irrep. 
Therefore, the next lowest energy band intersects every sector, 
except $M=\pm L/2$, which are
one-dimensional, hence exhausted by the ground state band.
This shows that the gap is constant, independent of $M$ (for 
$-L/2<M<L/2$), and it also shows that the band intersects the $L$-dimensional
sector for which $L/2-M=1$; i.e.\ the subspace of vectors with just one down-spin.
It is a fact that in this sector, the Hamiltonian can be easily diagonalized
by transfer matrix methods.
In this way, Koma \& Nachtergaele prove that for a finite-volume spin chain of length $L$, the
spectral gap is exactly equal to
$$
\gamma_L = 1 - \Delta^{-1} \cos(\pi/L)\, .
$$
They parlay the result for finite spin-chains into an equivalent result for
infinite-volumes.
For infinite-volumes the spectral gap above a particular ground state 
$\omega$ is defined in terms of the GNS representation 
$(\Hil^{\textrm{GNS}},H^{\textrm{GNS}},\Omega^{\textrm{GNS}})$ as the smallest 
number $\gamma$ for which
$$
\ip{\Omega^{\textrm{GNS}}}{\pi(X)^* (H^{\textrm{GNS}})^3 \pi(X) \Omega}
  \geq \gamma 
  \ip{\Omega^{\textrm{GNS}}}{\pi(X)^* (H^{\textrm{GNS}})^2 \pi(X) \Omega}\, ,
$$
is valid for all strictly local observables $X$.
Koma \& Nachtergaele show that the spectral gap above all the infinite-volume ground states
is
$$
\gamma = 1 - \Delta^{-1}\, .
$$
Note that this does agrees with the known result that there is no spectral
gap above the infinite-volume ground states for the XXX model.

In chapter 3 we generalize the result of Koma \& Nachtergaele to show that for spins greater
than $1/2$, the spectral gap above any of the infinite-volume ground states 
is still nonvanishing.
Unlike the original paper of Koma \& Nachtergaele, we cannot exploit the quantum group symmetry,
because it is nonexistent for spin-1 and higher. 
A related fact is that for spin-$\Spin$, $\Spin>1/2$, 
the spectral gap ceases to be  independent of $M$ for finite volumes. 
Instead one finds that for $L \gg |M|$, the gap is approximately an even,
$2\Spin$-periodic function of $M$.
This is due to the dependence of the gap on the filling factor $M \mod 2\Spin$,
which in the Ising limit, is the magnetization at the site
$x$ separating all $\ket{-\Spin}$'s to the left and all $\ket{+\Spin}$'s to the 
right.
One can show that the spectral gap for sectors with $|M| \ll 2\Spin$,
is bounded below, uniformly in $L$, and therefore that the gap above all
the kink ground states is nonzero.
Unlike Koma \& Nachtergaele, the proof is existential, instead of constructive.
Essentially, we use exponential localization of the interface to reduce the
problem of low-energy perturbations to an equivalent problem in finite dimensions,
where we know a spectral gap exists.
The spectral gap exists in finite dimensions because the spectrum
is discrete, but in the proof we do not demonstrate explicit bounds for the
finite-dimensional spectral gap.
This is the main difference between our result and the original result of
Koma \& Nachtergaele, since they give very clear lower bounds on the spectral gap in 
finite-dimensions.
One can also bound from below the spectral gap above the translation-invariant
ground states, by other, easier techniques.
True to the lack of quantum-group symmetry, the gap above the 
translation-invariant ground states is quite different than the gap above
the kink ground states.
The gap above the translation-invariant ground states is $2\Spin(1-\Delta^{-1})$,
for all spins $\Spin$.
However, at least for the Ising limit, the gap of the kink ground states is exactly
1.
For $1<\Delta<\infty$, the kink state gap is always lower than the 
translation-invariant gap (this is obvious because the kink states have large
domains which look essentially translation-invariant, so that any low-energy
perturbation to the translation-invariant ground states would also work for the 
kink ground state), but both grow linearly with $\Spin$.

Although our proof of the spectral gap is not constructive, there is an important
inequality used, which lends itself to a numerical approximation of the 
gap for finite volumes.
The inequality gives lower bounds for the spectral gap of the full kink Hamiltonian,
which is a $(2\Spin+1)^L \times (2\Spin+1)^L$ matrix, in terms of the spectral
gap of a Hamiltonian of much reduced size.
Specifically the reduced matrix has shape
 $p(L,2\Spin,n) \times p(L,2\Spin,n)$, where $p(J,K,N)$ is the number of
partitions of $N$ with at most $L$ parts of size each at most $K$.
Here $n=\Spin L - M$, which is typically on the order of $\frac{1}{2} \Spin L$.
The partition number can be bounded by $(L+1)^{2\Spin}$.
If one fixes $\Spin$ and lets $L$ grow, to approximate the thermodynamic limit,
then the new system becomes much smaller than the original one as $L$ becomes 
large.
The determination of the lower-bound matrix uses the theory of symmetric 
functions. 
Empirically, the lower bounds seem to converge quickly as $L \to \infty$,
much faster than the rate of convergence for the exact value of the spectral gap,
as can be determined by Lancz\"os iteration.
From these numerical lower bounds one can deduce a startling feature of the 
spectral gap, which is that for $\Spin\geq 5/2$ the spectral gap is not maximized
at $\Delta=\infty$, the Ising limit, but at some nontrivial value of
$1<\Delta<\infty$.
In other words, in the vicinity of the Ising limit, decreasing the anisotropy
leads to more stable ground states. 
In fact, comparison with Lancz\"os iteration for the full Hamiltonian, as well
as perturbation theory for the full Hamiltonian, shows this feature to be 
true for $\Spin\geq 3/2$.
This is entirely different than the situation for $\Spin=1/2$.
From this, one may guess that there is a semi-classical behavior (in the limit
$\Spin \to \infty$) for the low-lying spectrum, which is possibly quite
different than the low-lying spectrum for the spin-$1/2$ model.
We give additional numerical evidence for this, although we provide
no analytical proof of this fact.
A reasonable argument is currently in the works, but will be published
later as a separate result.

In chapter 4, we present two models for droplet states in a quantum spin
system, and derive the ground states for both. For the second model,
which is the more realistic of the two, we also demonstrate a spectral gap above
the ground states.
For classical spin systems, such as the Ising spin system, droplet states
have an important role.
For the Ising model, a droplet state can be defined to be the equilibrium
state for the model with all $+$-boundary conditions, in the canonical
ensemble with a fixed density $\frac{1}{2} < \rho < 1$, and temperature in the 
range $\beta_0 < \beta$,
in the thermodynamic limit.
The need for an upper bound on the temperature is apparent: if the temperature is
too high then one will have a disordered phase where every spin is a Bernoulli
random variable, independent of every other spin, and with identical mean values 
$\rho$.
Dobrushin, Koteck\'y and Shlosman showed \cite{DKS}, that if $\beta$ is in the 
correct range, then the equilibrium state is a sum of translates 
of states characterized by two domains, one nearly all up-spins, one 
nearly all down-spins, 
separated by a contour whose shape is given by the Wulff construction.
Thus a droplet in the sense we defined before matches the physical meaning of a 
droplet, as observed in condensation of crystals, and described by the 
phenomenological Wulff construction.
Other notable work on this area of research for classical spin systems is
\cite{BIV}, \cite{BIV2}, \cite{Pfi}, \cite{SS}.
In particular, in \cite{SS}, it is shown how the droplet states help to understand
dynamical properties of non-equilibrium states, which is quite interesting since
non-equilibrium statistical mechanics is a relatively open field.

While the progress for the classical model is impressive, almost nothing is known
about the same problem for the quantum model. 
I.e., can one derive states described by Wulff droplets in dimensions two and higher
starting from a quantum interaction instead of a classical interaction?
This question will probably be open for some time, since currently,
it is not even
known what are all the interface ground states of the quantum XXZ model, in dimensions
two and higher.
In fact, even for the interface ground states which have been determined,
one does not have a proof of the stability properties (see \cite{MN} for more
on the current state of the problem).
What can be done is to obtain a model for a one-dimensional droplet.
A one-dimensional droplet state should have a domain of down-spins on a background of
up-spins.
Unlike, for the classical model, even at zero-temperature, there are quantum
fluctuations.
Therefore, the interesting aspect of the problem is in determining the nature
of the quantum interfaces between the three domains, of all up-spins to
the far left, all down-spins in the middle, and all up-spins to the far right.
In analogy with the Ising model, one expects to obtain droplet states for the
canonical ensemble with a particular choice of density of down-spins, with
either periodic boundary conditions (i.e.\ spin ring) or with a
boundary field favoring up-spins.
We consider a non-periodic spin chain with boundary field 
$-A(\Delta) (S_1^3 S_L^3)$, which does favor up-spins at the edge,
and we are later able to deduce the results for a periodic spin chain.
The choice of magnitude for the boundary field is important since we know
$A(\Delta)$ is the energy of a kink interface.

As with the classical case, even without periodic boundary conditions,
there is a recovered translation symmetry in the thermodynamic limit.
To break this symmetry, we consider a toy model of a droplet with a pinning
field at some site in the center of the spin chain, which has the effect of pinning
the droplet of down-spins to be centered at that site.
This is a toy model because one should not require anything as unphysical
as a pinning field (although one may interpret such a field as the effect
of an impurity) to demonstrate droplet states.
Also, with the correct choice of amplitude for the pinning field, namely 
the field equals $2 A(\Delta) S_x^3$, the model becomes nothing more than
an antikink Hamiltonian glued to a kink Hamiltonian.
For such a simple model, we can give an explicit formula for the ground state.
Then the main challenge is to extract from this formula the behavior of the quantum 
interfaces at the two edges of the down-spin droplet.
This is done straightforwardly using explicit formulae obtained
in Section \ref{ExactCalculationsSection}.

In chapter 4, following the toy model, we present a real model for the spin
droplet, without the unphysical pinning field.
For this model we are able to show that the translation symmetry is recovered,
and that for large droplets, the ground states are approximately equal to the sum of
states which are
a tensor product of an antikink on the left with a kink on the right.
This shows the quantum interfaces for a spin droplet are the same as for a kink
or antikink state.
We also demonstrate a spectral gap above these states.
For our results, we need the size of the droplet to become large, but there
is no requirement on the density of down-spins.
It can range from zero to one, as long as the absolute number of down-spins is
large enough.
The main technique for the proof is to make use of the spectral gap result
of \cite{KN1}.
If we can find a region with nearly all down-spins, then we can
decompose the droplet spin chain
$$
H^{++}_{[1,L]} = H^{XXZ}_{[1,L]} - A(\Delta) (S_1^3 + S_L^3)\, ,
$$
into a sum of an antikink and a kink spin chain, and a two-site droplet
spin chain
$$
H^{++}_{[1,L]} = H^{+-}_{[1,x]} + H^{-+}_{[x+1,L]} + H^{++}_{[x,x+1]}\, .
$$
Then, using the spectral gap result of Koma \& Nachtergaele, we can show that any low-energy
state must be close to an antikink on sites $[1,x]$ and a kink on sites
$[x+1,L]$.
In order to find an interval of nearly all down-spins, we introduce a lemma,
which we call Corollary \ref{IHS:Cor}, which shows that for the droplet Hamiltonian,
and in fact for any Hamiltonian which is a finite perturbation of the XXZ Hamiltonian,
in any long-enough finite energy state, there exists an interval, such that
the restriction of the state to that interval is close to a convex combination
of the two states $\omega_{\uparrow}$ and $\omega_{\downarrow}$.
To obtain an interval  where the restriction of the state is just
$\omega_\downarrow$, we use an induction argument on the number of down-spins.
Since we do not assume the number of down-spins is a positive fraction of the 
total number of sites, the induction argument is a little technical.
It mainly relies upon exact calculations which can be performed for the kink 
and antikink states, to determine the probabilities of finding intervals of
all up-spins or all down-spins.
A corollary of our result, is that for periodic spin chains, one has the same form for 
ground states: namely a kink state tensored to an antikink states.
Also, if one considers the GNS representation for the translation invariant
all-up state, then one can decompose the Hilbert space into a direct sum
of sectors of given finite magnetization.
Our results imply that in sectors where $M$ is large but finite, the lowest-energy
states also look like a linear combination of states which can be described as
an semi-infinite antikink-state tensored to a semi-infinite kink state.
A surprising feature is that droplets of a given size, at different locations interact
very weakly.
In fact two droplets with their centers displaced just one unit, have an interaction
on the order of $q^n$, where $n$ is the number of down-spins in the droplet.
This weak interaction leads to a very flat dispersion relation for droplet
energy versus momentum.
In other words, the mass of the individual droplets grows exponentially with the
number of down-spins comprising the droplet.

In chapter 5 we consider the spectral gap above the $(1,1,\dots,1)$ interface
ground state in dimensions two and higher.
These ground states are the natural extension of the kink ground states given by Alcaraz, Salinas \& Wreszinski.
However, the behavior of the excitation spectrum in dimensions greater than one
is much different than in one dimension.
Specifically, there is a continuous $\textrm{U}(1)$ symmetry of the XXZ Hamiltonian,
realized as global rotations about the $\boldsymbol{e}_3$-axis.
This symmetry, as well as translation symmetry, is broken in the ground state due 
to phase-locking.
Thus by the Goldstone theorem, as implemented by \cite{LPW}, there are gapless
excitations in dimensions two and higher.
This is the argument for the vanishing of the spectral gap in dimensions two and higher
given by Koma and Nachtergaele \cite{KN4}, and later by Matsui \cite{Mat2}.
One point of interest in this result is that one does not know whether the
interface ground states are stable to small thermal fluctuations in dimensions
two and higher.
In contrast, in one-dimensional quantum spin systems the KMS state is unique
\cite{Araki}.
Hence the translation symmetry of the Hamiltonian is not broken.
This means the unique KMS state in each sector is the translation-invariant state,
which shows that kink states are not stable to thermal fluctuations.
The nonexistence of a spectral gap, proved by such general schemes as the Goldstone 
theorem, does not completely illuminate the nature of the low-energy excitations.
Are such excitations localized in space?
Is the gap vanishing with a rate that depends on the dimension?
Is the shape of the interface significantly changed by these low-lying excitations?
How do the excitations depend on the anisotropy and the sector number?
These are some of the questions which we answer.

Chapter 5 is broken into two parts, corresponding to two papers.
The first paper shows what type of variational states one would use to obtain
lowest-energy excitations of an interface ground state in the grand-canonical ensemble.
In the grand-canonical ensemble, one does not separate out subspaces corresponding
to different numbers of down-spins, as is done in the canonical ensemble.
By allowing states with a fluctuating number of down-spins, one can obtain simpler
expressions for the ground states.
Specifically, the ground states can be parametrized by a continuous, real parameter, $\mu$,
which we call the chemical potential.
Defining $l(x) = x_1 + \dots + x_d$ (in $d$-dimensions), the ground state is
\begin{equation}
\label{IntroGCGSDef}
\psi^{\textrm{GC}}(\mu) 
  = \bigotimes_{x \in \Lambda} \frac{q^{(\mu - l(x))/2} \ket{\uparrow}_x 
  + q^{(l(x) - \mu)/2} \ket{\downarrow}_c}{\sqrt{q^{\mu - l(x)} + q^{l(x)-\mu}}} \, .
\end{equation}
Since this is a simple-tensor product state, it has many properties resembling a classical
state, including ``independence'' of spins at different sites
(i.e. statistical independence of the spin observbles at different sites with respect to this
state). 
By perturbing a single site of the ground state to yield a spin orthogonal to the 
the original spin of that site, one can produce an orthogonal state.
The number of such excitations equals the number of sites in the domain $\Lambda$.
We explicitly consider the perturbation
$$
X(f) = \sum_{x \in \Lambda} f(x) (S_x^+ - S_x^-)\, ,
$$
which does produce an orthogonal state due to the fact that the coefficients
$q^{\pm (\mu - l(x))/2}$ of the up and down spin at each site are real.
By considering the Hamiltonian acting on such a perturbation, one discovers that
$X(f) \psi^{\textrm{GC}}(\mu)$ can have a low energy, but only if $f$ obeys strict 
relations in the $(1,1,\dots,1)$ direction.
Imposing these restrictions still leaves freedom in the definition of $f$ in one plane
perpendicular to the $(1,1,\dots,1)$-direction.
Thus $f$ has one degree of freedom reduced, but in dimensions two and higher, this still leaves
at least one additional degree of freedom.
The plane perpendicular to the $(1,1,\dots,1)$-direction is the plane of the interface, 
which we will call $\pi$.
If the portion of $\Lambda$ intersecting the plane has diameter $R$, and 
if we assume $f$ varies continuously over this region
like $f(x) = \phi(x/R)$ for some smooth function $\phi$ on a continuous domain, then 
the leading-order contribution to the energy of $X(f) \psi^{\textrm{GC}}(\mu)$ is given by
$$
 \frac{1}{R^2} \int_{R^{-1} \pi} \|\nabla \phi(x)\|^2\, dx
$$
times a constant depending on $\mu$.
($\mu$ also enters the definition of $f$ through the relations for $f$ in the $(1,1,\dots,1)$
direction. The interface of the grand-canonical ground state is
localized about the ``plane'' $\{x :  |l(x)-\mu|<1\}$,  and the restriction of $f$ in the
$(1,1,\dots,1)$ direction has the effect that the perturbation $X(f)$ is also localized about 
this plane.)
The norm of the state $X(f) \psi^{\textrm{GC}}(\mu)$ is equal to
$$
\int_{R^{-1} \pi} |\phi(x)|^2\, dx
$$
times a constant depending on $\mu$.
Thus, taking the Rayleigh quotient, one has the variational formula for the eigenvalue problem
of the Laplace operator on the plane $\pi$.
This is not surprising since it tells us that the lowest excitation is a multidimensional
analog of a spin-wave on the plane of the interface.

The second part of chapter 5 is devoted to the same problem in the canonical ensemble.
In fact this is the more rigorous approach, since in this framework one can produce excitations
which are not only orthogonal to one ground state, but to them all.
(More precisely there is only one ground state per sector of total spin, so an excitation which
is entirely in one sector and orthogonal to the one ground state in its sector is 
automatically orthogonal to all other ground states in other sectors.
The difficulty with the canonical ensemble is that the ground states do not have the simple form
of \eq{IntroGCGSDef}, and so it is not so trivial to construct a class of perturbations
which are genuinely orthogonal.
There are two steps which allow one to bypass this obstacle.
The first step, which is trivial but important, is to use the variational inequality.
If $\psi_0(n)$ is the unique ground state in the sector with $n$ down-spins, 
then the energy of the spectral gap in this sector is
\begin{equation}
\label{IntroVarIneq}
\gamma = \inf_{\phi \not\in \Cx \psi_0(n)} \frac{\ip{\phi}{H^{\textrm{kink}} \phi}}
  {\|\phi\|^2} \cdot \frac{\|\psi_0(n)\|^2 \|\phi\|^2}
  {\|\psi_0(n)\|^2 \|\phi\|^2 - |\ip{\psi_0(n)}{\phi}|^2}\, .
\end{equation}
This allows us to bound the spectral gap by considering non-orthogonal perturbations
of the ground state.
The second step is to reduce the calculation of certain local observables with respect
to the canonical ensemble to the calculation of the same observables in the grand-canonical
ensemble.
This is done by proving an equivalence-of-ensembles result.
More specifically, for small subdomains $\Lambda_0 \subset \Lambda$, the restriction of the state
$\omega_0(\dots) = \ip{\psi_0(n)}{\dots \psi_0(n)}$ to the observable algebra $\Obs_{\Lambda_0}$,
is close to a grand-canonical ground state.
This can be understood easily enough: although the number of total down-spins in the volume
$\Lambda$ is a fixed number, the number of down-spins in any subvolume is varying.
If the subvolume is small enough, then the distribution of down-spins becomes exponential,
with Gaussian corrections of lower-order, by a usual central limit theorem argument.
In fact, the relevant quantity to estimate turns out to be $\|\psi_0(m)\|^2$ for all values of
$m$ near $n$.
One can evaluate $\|\psi^{\textrm{GC}}(\mu)\|^2$ exactly for each $\mu$, due to the 
factorization property of the state.
Moreover the states $\psi^{\textrm{GC}}(\mu)$ form an exponential generating function for
the canonical ground states.
So, alternatively to the central limit theorem, one can use Hayman's method (c.f.\ \cite{Wilf})
to determine the distribution of down-spins.
This allows for a more refined version of equivalence of ensembles which is proved in
the last two sections.
As well as demonstrating explicit upper-bounds for the spectral gap, which shows that the
gap vanishes at least (and probably at most) as fast as $1/R^2$, we show how the spectral
gap depends on the sector number 
$M = \frac{1}{2} |\Lambda| -n$.
It turns out that the most fundamental effect is due to the partial filling factor of 
the interface plane, due to having particle number $n$ which is not a perfect multiple of
the size of a plane.
This is elucidated in the paper.
Also, we show how the gap depends on the anisotropy parameter $\Delta$ : it depends through
two functions of $\Delta$ which measure the first and second moments for particle number
in the one-dimensional grand-canonical ensemble.
This is the subject of the appendix.
Our estimates are only valid for $1<\Delta<\infty$.
However, we point out that for $\Delta=1$, even in one-dimension there are gapless excitations
due to spin-waves, and for $\Delta = \infty$, the 11-interface is degenerate
with all the other interfaces with the same global geometry and equal perimeter.
Our arguments rely on nondegeneracy of ground states for finite volumes, as well 
the Koma \& Nachtergaele spectral gap result above the kink states in one dimension, and so it is natural
to only consider $\Delta$ in the range $(1,\infty)$.


\newpage
\pagestyle{myheadings}
\chapter{Preliminaries}
\thispagestyle{myheadings} 
\markright{  \rm \normalsize CHAPTER 2. \hspace{0.5cm}
Preliminaries}

\Section{Definitions}

The quantum XXZ spin chain is a mathematical model for magnetism at the 
level of atoms or ions in a crystal.
This is purely a spin system, consequently, no issues of charge of the atoms 
enter into the XXZ model.
There is no kinetic energy for the atoms; in fact, the atoms are
fixed at sites of a subset of a discrete lattice.
For our purposes, the only lattices we consider in detail are the 
$d$-dimensional integer lattices $\Ir^d$.
Physicists call these simple cubic lattices.
The most sensible parameter range for $d$ is $d=1,2,3$.
Our results, however, are of two types, those true for $d=1$,
and those which are true for all $d>1$.

By $\Lambda$ we will usually mean a subset of $\Ir^d$.
It is the collection of sites for our spin system.
Unless otherwise specified, $\Lambda$ will be a finite set; i.e.\
$|\Lambda|<\infty$.
The configuration space for the spin of a single site 
$x \in \Lambda$ is  the Hilbert space
$\Hil_{x} = \Cx_x^{2\Spin + 1}$, and the Hilbert space for the
entire spin system is $\Hil_\Lambda = \bigotimes_{x \in \Lambda} \Hil_x$.
The subscript $x$ simply indicates which spin the Hilbert space describes.
$\Hil_x$ carries an action of $SU(2)$. 
More precisely, it is a $(2\Spin+1)$-dimensional irreducible representation. 
This is natural in the quantum theory of spin.
A good reference for quantization of angular momentum is \cite{Edmonds},
a good mathematical treatment of the representation theory of $SU(2)$ 
is \cite{Simon}.
What is most important for us is that there is a set of three operators 
$S_x^1$, $S_x^2$ and $S_x^3$ defined with respect to a basis
$\{\ket{+\Spin}_x,\ket{\Spin-1}_x,\dots,\ket{-\Spin}\}$ by
$$
S_x^3 \ket{m}_x = m_x \ket{m}_x\, ,
$$
and
$$
S_x^{\pm} \ket{m}_x = \sqrt{\Spin(\Spin-1) - m(m\pm 1)} \ket{m\pm 1}_x,
$$
where 
$$
S_x^{\pm} = S_x^1 \pm i S_x^2\, .
$$
For us this is the most convenient way to specify the representation of 
$\textrm{SU}(2)$.
We also define the spin-vector $\boldsymbol{S}_x = (S_x^1,S_x^2,S_x^3)$.

With this notation, the one-dimensional
spin-$\Spin$ XXZ Hamiltonian is defined as
$$
H^{\textrm{XXZ}}_L 
  = \sum_{x=1}^{L-1} \Delta^{-1}(\Spin^2 - \boldsymbol{S}_x \cdot
    \boldsymbol{S}_{x+1}) 
    + (1 - \Delta^{-1}) (\Spin^2 - S_x^3 S_{x+1}^3)\, .
$$
We assume $1\leq \Delta \leq +\infty$.
The Hamiltonian is obviously Hermitian.
It is easy to see that it is also nonnegative:
Indeed, 
\begin{align*}
\boldsymbol{S}_x \cdot \boldsymbol{S}_{x+1} 
  &= \frac{1}{2}(\boldsymbol{S}_x + \boldsymbol{S}_{x+1})
  \cdot (\boldsymbol{S}_x + \boldsymbol{S}_{x+1})
    - \frac{1}{2} \boldsymbol{S}_x  \cdot \boldsymbol{S}_x 
    - \frac{1}{2} \boldsymbol{S}_{x+1} \cdot \boldsymbol{S}_{x+1}\\
  &= \frac{1}{2} \mathcal{C}_{x,x+1} - 
	\frac{1}{2}(\mathcal{C}_x + \mathcal{C}_{x+1}) \\
  &= \frac{1}{2} \mathcal{C}_{x,x+1} - 
	\Spin(\Spin+1)\, ,
\end{align*}
where $\mathcal{C}_{\Lambda} = \boldsymbol{S}_\Lambda \cdot \boldsymbol{S}_\Lambda$ 
is the Casimir operator for the representation of $\textrm{SU}(2)$ on $\Hil_\Lambda$,
and in particular $\mathcal{C}_{x,x+1}$ has eigenvalues
$J(J+1)$ for each $J=2\Spin,2\Spin-1,\dots,0$.
The Casimir operators $\mathcal{C}_x$ and $\mathcal{C}_{x+1}$ are
identically $\Spin(\Spin+1)$ because $\Hil_x$ and $\Hil_{x+1}$ are
spin-$\Spin$ irreducible representations of $\textrm{SU}(2)$.
This means that $\Spin^2 - \boldsymbol{S}_x \cdot \boldsymbol{S}_{x+1}$
has minimum eigenvalue equal to 0, when $J=2\Spin$.
The minimum eigenvalue of $\Spin^2 - S_x^3 S_{x+1}^3$ is bounded
below by $\Spin^2 - \|S_x^3\| \|S_{x+1}^3\|$, and 
$\|S_x^3\|=\|S_{x+1}^3\|=\Spin$.
So, in fact, each of the summands in $H^{\textrm{XXZ}}_L$ is nonnegative.

There are vectors which minimize each pair interaction simultaneously, namely
$\ket{+\Spin,+\Spin,\dots,+\Spin}$ and $\ket{-\Spin,-\Spin,\dots,-\Spin}$.
Here we have adopted the notation 
$$
\ket{m_1,m_2,\dots,m_L} = 
\ket{m_1}_1\otimes\ket{m_2}_2\otimes\cdots\otimes\ket{m_L}_L\, .
$$
For $\Delta>1$, these two vectors are the only absolute ground states 
because they are the only vectors which maximize every pair $S_x^3 S_{x+1}^3$, 
simultaneously, and 
also maximize each $\mathcal{C}_{x,x+1}$.
They are the vacuum states for $S_{\textrm{tot}}^-$ and $S_{\textrm{tot}}^+$,
respectively, 
in the $(2 L \Spin +1)$-dimensional irreducible subrepresentation of
$\Hil_L = \bigotimes_{x=1}^L \Hil_x$.
These two states are ground states in the thermodynamic limit, 
$L \to \infty$.
Or, more accurately, there are states $\omega_\uparrow$, $\omega_\downarrow$
in $\Obs_0^*$, where $\Obs_0$ is the set of quasi-local observables and
$\Obs_0^*$ is the space of continuous linear functionals on $\Obs_0$, 
given by $\omega_{\uparrow,\downarrow}(A_\Lambda) 
= \bra{\pm \Spin}_{\Lambda} A_\Lambda \ket{\pm \Spin}_{\Lambda}$
for any $A \in \Obs_\Lambda$, and these two states are ground states.
For the general framework of states in a quantum spin system see \cite{BR}
or \cite{Simon2}.
By definition, a state $\omega$ is a ground state iff 
$$
\omega(A^* \delta(A)) 
  := \lim_{\Lambda \nearrow \Ir} \omega(A^* [H_\Lambda,A])
  \geq 0\, ,
$$
for all $A \in \Obs_0$.
Since $H^{\textrm{XXZ}}_\Lambda \geq 0$ and 
$H^{\textrm{XXZ}}_\Lambda \ket{\pm S}_\Lambda = 0$, it is trivial to check
that $\omega_{\uparrow,\downarrow}(A^*_\Lambda [H_\Lambda,A_\Lambda]) \geq 0$ 
for any
$A_\Lambda \in \Obs_\Lambda$, with $|\Lambda|<\infty$.
Since 
$$
\Obs_0 
  = \textrm{cl}(\bigcup_{\Lambda\subset\Ir, |\Lambda|<\infty} \Obs_\Lambda)\, ,
$$
it is clear that $\omega_{\uparrow,\downarrow}$ are infinite-volume ground 
states.
The states $\omega_{\uparrow,\downarrow}$ are also pure states, which is to 
say extremal elements of the convex set of all normalized states.
This is apparent from the definitions.
However, they are not the only pure, infinite-volume ground states.
The first natural question which arises about this Hamiltonian is 
\begin{description}
\item{\textbf{1.}} What is the complete list of pure, infinite-volume
ground states?
\end{description}

A complete solution to this problem was eventually found by Koma and Nachtergaele
in reference \cite{KN3}, although to place the problem in its proper perspective
one should consider a number of other results, most notably \cite{ASW} and \cite{GW}. 
In order to motivate the treatment of the quantum problem it is instructive to 
consider the classical $XXZ$ model.
It is important to note, however, that this was not the original
motivation behind exact results for the ferromagnetic model.
Historically, the Bethe ansatz and the quantum group symmetry played a much 
more active role in leading to the formula for the ground state of the XXZ model.

\section{The Classical XXZ Model}
The purpose in examining the classical XXZ model is to give an intuition for 
the quantum model,
the ultimate goal being to find all the infinite-volume ground states.
It is typically the case that one can obtain infinite-volume ground states by 
considering a sequence of finite-volume Hamiltonians with the ``correct'' boundary 
terms added.
This is the case for the XXZ model, as will be shown in the next sections.
In this section we try to ``guess'' the correct boundary conditions, of course
with the caveat that having read the original papers \cite{PS}, \cite{ASW}, 
\cite{GW} and \cite{KN3}, we know beforehand what the correct boundary 
conditions are.
We treat the case of finite anisotropy, $1 \leq \Delta < \infty$, separate
from the infinitely anisotropic, or so-called Ising limit.

\subsection{Finite Anistropy}
To obtain the classical $XXZ$ model, one formally takes $\Spin \to \infty$.
Thus, one scales
$$
\Spin^{-2} H^{\textrm{XXZ}} = 
  \sum_{x=1}^{L-1} \Delta^{-1}(1 - \frac{1}{\Spin^2} \boldsymbol{S}_x \cdot
    \boldsymbol{S}_{x+1}) 
    + (1 - \Delta^{-1}) (1 - \frac{1}{\Spin^2} S_x^3 S_{x+1}^3)\, 
$$
and in the limit $\Spin \to \infty$, obtains
$$
H^{\textrm{cl}}(\{\boldsymbol{\sigma}_x\}_{x=1}^L)
  = \sum_{x=1}^{L-1} \Delta^{-1}(1 - \boldsymbol{\sigma}_x \cdot
    \boldsymbol{\sigma}_{x+1}) 
    + (1 - \Delta^{-1}) (1 - \sigma_x^3 \sigma_{x+1}^3)\, ,
$$
where $\boldsymbol{\sigma}_x =(\sigma_x^1,\sigma_x^2,\sigma_x^3)$ 
is a unit vector for each $x = 1,\dots,L$.
Actually, it is not the Hamiltonian itself which is well-defined in the limit,
but rather the partition function associated to the Hamiltonian:
c.f.\ \cite{L4}.
Still, one may define a nonnegative functional on the space of all sequences
$(\Rl^3)^{\Ir}$ by
$$
H^{\textrm{cl}}(\{\boldsymbol{\sigma}_x\}_{x\in \Ir})
  = \sum_{x=-\infty}^{\infty} \Delta^{-1}(1 - \boldsymbol{\sigma}_x \cdot
    \boldsymbol{\sigma}_{x+1}) 
    + (1 - \Delta^{-1}) (1 - \sigma_x^3 \sigma_{x+1}^3)\,  .
$$
Although this may take the value infinity, it is always well-defined,
since all the summands are nonnegative.
In spherical coordinates 
$$
\boldsymbol{\sigma}_x = (\cos \theta_x, \sin \theta_x \cos\phi_x, 
\sin \theta_x \sin\phi_x)\, ,
$$
where $\theta_x \in [0,\pi]$ and $\phi \in [0,2\pi)$.
Then
$$
H^{\textrm{cl}} 
  = \sum_{x=-\infty}^{\infty} [1 - \cos \theta_x \cos \theta_{x+1}
    - \Delta^{-1} \sin \theta_x \sin \theta_{x+1} \cos(\phi_x - \phi_{x+1})]\,  .
$$
It is clear that the choice of $\{\phi_x\}_{x \in\Ir}$ which minimizes $H^{\textrm{cl}}$
is $\phi_x \equiv \phi$ for some $\phi \in [0,2\pi)$.
Thus every minimum-energy state of $H^{\textrm{cl}}$ is a planar rotation.
With this assumption, we have
$$
H^{\textrm{cl}} 
  = \sum_{x=-\infty}^{\infty} [1 - \cos \theta_x \cos \theta_{x+1}
    - \Delta^{-1} \sin \theta_x \sin \theta_{x+1}]\,  .
$$
In order to have a minimum-energy state, $H^{\textrm{cl}}$ should be extremal with
respect to every $\theta_x$.
Hence
\begin{align*}
0 &= \frac{\partial H^{\textrm{cl}}}{\partial \theta_x} 
  = \sin \theta_x (\cos \theta_{x-1} + \cos \theta_{x+1}) 
  - \Delta^{-1} \cos \theta_x (\sin \theta_{x-1} + \sin \theta_{x+1}) \\
\Leftrightarrow
\tan \theta_x &= \Delta^{-1} \frac{\sin \theta_{x-1} + \sin \theta_{x+1}}
  {\cos \theta_{x-1} + \cos \theta_{x+1}} \\
  &= \Delta^{-1} \tan \frac{\theta_{x-1} + \theta_{x+1}}{2}\, .
\end{align*}
Defining $t_x = \tan \frac{1}{2} \theta_x$, and using the angle addition law
for tangent, gives
\begin{equation}
\label{Classical equations}
\frac{2 t_x}{1-t_x^2} = \frac{1}{\Delta} \frac{t_{x-1} + t_{x+1}}{1 - t_{x-1} t_{x+1}}\, .
\end{equation}
If we make the ansatz that $t_x = q^x \tau$ for some $q,\tau \in \Rl^+$,
then we observe that $1-t_x^2 = 1 - t_{x-1} t_{x+1}$, so $\tau$ drops from the
equations, and we are left just with the condition
$$
2 = \Delta^{-1} (q^{-1} + q)
\Leftrightarrow \Delta = \frac{q + q^{-1}}{2}\, .
$$
\begin{definition}
For $0<\Delta<\infty$ let $q \in (0,1)$ be the unique solution of the quadratic
equation $\Delta = \frac{1}{2}(q + q^{-1})$, i.e.\ $q = \Delta - \sqrt{\Delta^2 - 1}$.
For $\Delta = 1$, $q=1$ and for $\Delta = +\infty$, $q=0$.
\end{definition}

Hence there are two one-parameter classes of solutions to \eq{Classical equations}:
the so-called \textit{kink} solution $t_x = q^x \tau$ or 
$\theta_x = 2 \tan^{-1} (q^x \tan \frac{1}{2} \theta_0)$, and the so-called 
\textit{antikink} solution $t_x = q^{-x} \tau$ or
$\theta_x = 2 \tan^{-1} (q^{-x} \tan \frac{1}{2} \theta_0)$.
It may be preferable to write the solutions in terms of the original coordinates
$\{\boldsymbol{\sigma}_x\}$.
The fact that $\phi$ is fixed means that the kink and antikink solutions are 
planar waves:
\begin{equation}
\label{classical solution 1}
\boldsymbol{\sigma}_x = \sigma_x^3 \boldsymbol{e}_3 + \sqrt{1 - (\sigma_x^3)^2} 
(\cos(\phi) \boldsymbol{e}_1 + \sin(\phi) \boldsymbol{e}_2)\, .
\end{equation}
Then the wave is specified just by the sequence $\sigma_x^3$.
While there is no simple linear recurrence for the $\sigma_x^3$, 
there is the next best thing, a linear-fractional recurrence relation:
\begin{equation}
\label{classical solution 2}
\begin{gathered}
\textrm{ kink : } \quad 
  \sigma^3_{x+1}
 = \frac{ \sigma_x^3 + A(\Delta)}{A(\Delta) \sigma^3_x + 1}\, ,\\
\textrm{ antikink : } \quad 
  \sigma^3_{x+1}
 = \frac{A(\Delta) \sigma_x^3 - 1}{-\sigma^3_x + A(\Delta)}\, ,
\end{gathered}
\end{equation}
where $A(\Delta)$ is a special value defined by
\begin{equation}
\label{definition of A}
A(\Delta) = \sqrt{1 - \Delta^{-2}} = \frac{1 - q^2}{1 + q^2}\, .
\end{equation}

For these solutions to be valid, we must require $1 \leq \Delta < \infty$, i.e.\
we exclude the case $q=0$, which corresponds to the Ising limit.
In the isotropic limit, $\Delta =1$, $q=1$, all the solutions are translation-invariant
with $\boldsymbol{\sigma}_x \equiv \boldsymbol{\sigma}$ for any vector
$\boldsymbol{\sigma}$ on the two-sphere.
For $0<q<1$ there are still two translation-invariant ground states corresponding to
$\theta_0=0$, all up-spins, or $\theta_0 = \pi$, all down-spins.
For any other choice of $\theta_0$, we observe the behavior,
\begin{gather*}
\textrm{kink :} \qquad
\lim_{x \to -\infty} \theta_x = \pi\, , 
\quad \lim_{x \to +\infty} \theta_x = 0\, ;\\
\textrm{antikink :} \qquad
\lim_{x \to -\infty} \theta_x = 0\, , 
\quad \lim_{x \to +\infty} \theta_x = \pi\, .
\end{gather*}
\begin{figure}
\begin{center}
\resizebox{6truecm}{1truecm}{\includegraphics{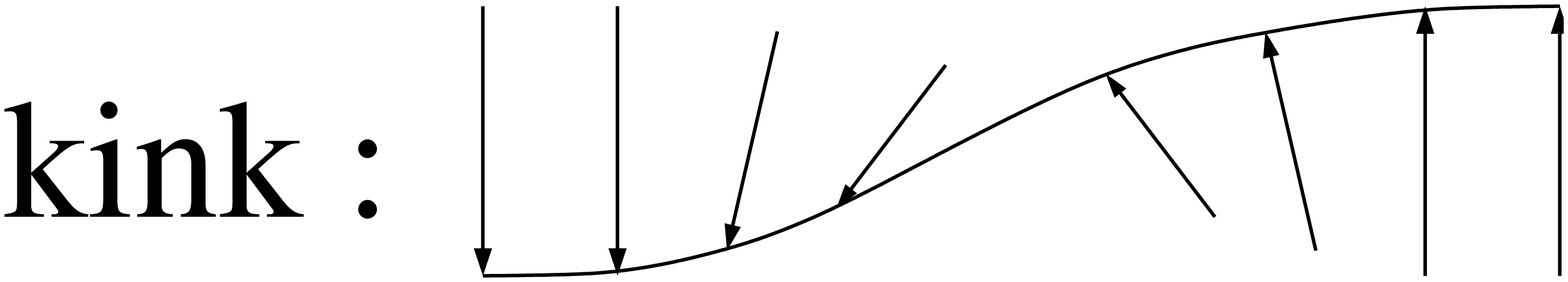}}\qquad
\resizebox{6truecm}{1truecm}{\includegraphics{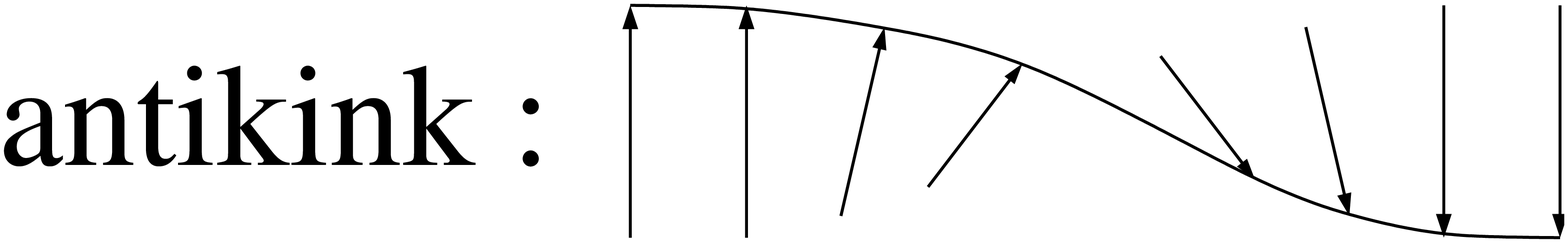}}
\end{center}
\end{figure}
So the profiles for $\sigma_x^3$ are schematically as in the figure above.

We can calculate the energy stored by the kink and antikink states.
The energy of the interaction between site $x$ and $x+1$ is
$$
h(\theta_x,\theta_{x+1}) = 
  1 - \cos \theta_x \cos \theta_{x+1}
    - \Delta^{-1} \sin \theta_x \sin \theta_{x+1} \cos(\phi_x - \phi_{x+1})\,  .
$$
Defining 
$$
\alpha_x = \frac{1}{2}(\theta_x - \theta_{x+1})\, ,\quad
\beta_x = \frac{1}{2}(\theta_x + \theta_{x+1})\, ,
$$
we have
\begin{align*}
h_{x,x+1} 
  &= \sin^2 \alpha_x + \sin^2 \beta_x 
  + \Delta^{-1} (\sin^2 \alpha_x - \sin^2 \beta_x) \\
  &= (1 + \Delta^{-1}) \sin^2 \alpha_x + (1-\Delta^{-1}) \sin^2 \beta_x\, .
\end{align*}
Completing the square, we have
$$
h_{x,x+1} 
  = (\sqrt{1+\Delta^{-1}} \sin \alpha_x - \sqrt{1-\Delta^{-1}} \sin \beta_x)^2
  + 2 \sqrt{1 - \Delta^{-2}} \sin \alpha_x \sin \beta_x\, .
$$
It is a straightforward calculation to see that for the kink state
$$
\sqrt{1+\Delta^{-1}} \sin \alpha_x - \sqrt{1-\Delta^{-1}} \sin \beta_x = 0\, .
$$
So the energy of each interaction is
\begin{align*}
h_{x,x+1} &= 2 \sqrt{1 - \Delta^{-2}} \sin \alpha_x \sin \beta_x \\
  &= A(\Delta) (\cos \theta_{x+1} - \cos \theta_x)\, .
\end{align*}
There is no need to calculate each interaction energy as a function
of $q$ and $\tau$ because, since $h_{x,x+1} = \eta_{x+1} - \eta_x$, the sum
$H^{\textrm{cl}}(\{\boldsymbol{\sigma}_x\}_x)$ is a telescoping sum:
\begin{align*}
H^{\textrm{cl}}(\{\boldsymbol{\sigma}_x\}_x)
  &= \lim_{L \to \infty} A(\Delta) (\cos \theta_L - \cos \theta_{-L}) \\
  &= A(\Delta) (\cos 0 - \cos \pi) \\
  &= 2 A(\Delta)\, .
\end{align*}
Thus we see that all kink states have the same ground state energy, independent of the
choice of $\theta_0 \in (0,\pi)$.
Similarly, the energy of the antikink states is also $2 A(\Delta)$.

Even though the energy of the kink states is higher than the energy of the 
translation-invariant states, it should be noted that both are ground states.
The reason is that the translation-invariant ground states,
$\omega_{\uparrow,\downarrow}$, and the kink states are mutually singular (have
no absolutely continuous part w.r.t.\ one another).
To see this it is sufficient to observe that they differ at infinity.
Since we will be interested in analyzing kink states for finite chains and their 
approach to the thermodynamic limit, it is useful to define a new Hamiltonian
which differs from $H^{\textrm{cl}}$ by only boundary terms, such that the kink
states are absolute ground states.
Thanks to our previous analysis, we know exactly how to do this, namely,
add to each interaction the term 
$A(\Delta) (\cos\theta_{x} - \cos\theta_{x+1})$.
Or, in terms of the original variable $\boldsymbol{\sigma}_x$,
\begin{equation}
\label{classical kink}
H^{-+,\textrm{cl}}_L(\{\boldsymbol{\sigma}_x\}_{x=1}^L)
  = \sum_{x=1}^{L-1} [1 - \sigma_x^3 \sigma_{x+1}^3 
  - \Delta^{-1} (\sigma_x^1 \sigma_{x+1}^1 + \sigma_x^2 \sigma_{x+1}^2) 
  + A(\Delta) (\sigma_{x}^3 - \sigma_{x+1}^3)]\, .
\end{equation}
Note that because of the telescoping sum, this is only a boundary field,
i.e.
$$
H^{-+,\textrm{cl}}_L = H^{\textrm{cl}}_L + A(\Delta)(\sigma_1^3 - \sigma_L^3)\, .
$$
The significance of this change is not only that the kink states now have
energy equal to the all up- and all down-spin states:
It is that the kink states now minimize each pair interaction separately, 
instead of simply minimizing the sum of all pair interactions together.
Such states are called frustration-free ground states signifying that they 
are not ``frustrated'' on any bond.
They play a very significant role in the classification of ground states for the
quantum XXZ model, as we will see shortly.

\subsection{``Ising-type'' states}
We can treat the Ising limit $q \to 0$ in entirely the same manner as above, 
except that we must be careful whenever we use $q^{-1}$.
It is simpler and more instructive to treat the Ising limit directly,
since in most circumstances one gains intuition for the XXZ model from the
Ising model, and not the other way around.
The Ising Hamiltonian is 
$$
H^{\textrm{Ising}}(\{\boldsymbol{\sigma}_x\}_{x=1}^L)
  = \sum_{x=1}^{L-1} (1 - \sigma_x^3 \sigma_{x+1}^3)\, ,
$$ 
and we can replace the unit vector $\boldsymbol{\sigma}_x$ with just its
third component which we rename simply $\sigma_x$.
We point out that this is not the Ising model, since, among other things 
$\sigma_x$ is allowed to vary continuously between $+1$ and $-1$.
In the quantum model with $\Spin=1/2$, the Ising limit really does correspond 
to the Ising system.
For higher $\Spin$, the Ising limit is not an Ising model because the configuration
space has cardinality $2\Spin+1$.
It is also not obviously a clock model or Potts model.
The simplest way to think of it is as a Ising spin-ladder model.
The Hamiltonian above can be rewritten
$$
H^{\textrm{Ising}}(\{\boldsymbol{\sigma}_x\}_{x=1}^L)
  = \sum_{x=1}^{L-1} (1 - \sigma_x \sigma_{x+1})\, .
$$ 
It is clear that this model is ferromagnetic in the strictest sense, i.e.
all spins prefer to be aligned as much as possible in the $\boldsymbol{e}_3$
direction.
One might then, precipitately, guess that the only ground states are 
$\omega_{\uparrow,\downarrow}$.
This is incorrect, it is possible to change the boundary conditions at infinity
to force a domain wall.
Namely, for finite volumes $\Lambda = [-L,L]$, by adding a boundary field 
which forces an up-spin on one boundary and a down-spin on the other.
In fact our kink Hamiltonian 
$H^{-+,\textrm{Ising}} = H^{\textrm{Ising}} + \sigma_{-L} - \sigma_L$ is sufficient.
The ground states are then kink states centered at any point $k \in \Ir$,
$$
\omega^{-+}(k) = \{\boldsymbol{\sigma}_x\}_{x\in\Ir} \, :\qquad 
  \sigma_x = \begin{cases} -1 & \textrm{for $x\leq k$,}\\
+1 & \textrm{for $x>k$;}\end{cases}
$$
as well as the translation-invariant states $\omega_{\uparrow,\downarrow}$.

\begin{figure}
\begin{center}
\resizebox{10truecm}{6truecm}{\includegraphics{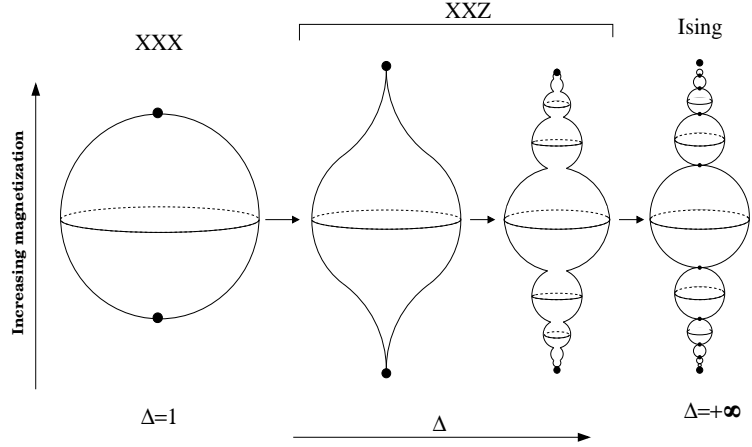}}
\end{center}
\caption{One interpretation of the geometry of the ground state space for the classical 
XXZ model}
\label{Interpretation}
\end{figure}
In fact there are more ground states than these.
Observing that $H^{-+,\textrm{Ising}}$ can be written as a sum of nearest neighbor
interactions
\begin{gather*}
H^{-+,\textrm{Ising}}(\{\sigma_x\}_x) 
  = \sum_{x=-\infty}^{\infty} h(\sigma_x,\sigma_{x+1}) \\
h(\sigma_x,\sigma_{x+1}) 
  = 1 - \sigma_x \sigma_{x+1} + \sigma_{x+1} - \sigma_x
  = (1 - \sigma_x)(1 + \sigma_{x+1})\, ,
\end{gather*}
we see that it is possible to have a frustration-free ground state with one 
spin, say at site $k$, in any orientation as long as all the spins to the
left of $k$ are aligned along the south-pole and all the spins to the right 
are aligned along the north-pole.
Thus we have a family of solutions $\omega^{-+} : \Ir \times S^2(\Rl^3) \to (S^3(\Rl^3))^{\Ir}$,
given by 
$$
\omega^{-+}(k,\boldsymbol{\varsigma}) = \{\boldsymbol{\sigma}_x\}_{x \in \Ir}\, :
\qquad \boldsymbol{\sigma}_x 
  = \begin{cases} - \boldsymbol{e}_3 & \textrm{for $x < k$,}\\
  \boldsymbol{\varsigma} & \textrm{for $x=k$,}\\
  +\boldsymbol{e}_3 & \textrm{for $x>k$.}\end{cases}
$$
Of course $\omega^{-+}(k,-\boldsymbol{e}_3) = \omega^{-+}(k+1,+\boldsymbol{e}_3)$.
Also, note that 
$$
\omega_{\downarrow,\uparrow} = \wslim_{k \to \pm \infty} 
	\omega^{-+}_k(\boldsymbol{\sigma}_k)\, ;
$$
for any sequence $\{\boldsymbol{\sigma}_k\}_{k \in \Ir}$.
However, neither $\omega_{\downarrow}$ nor $\omega_{\uparrow}$ is a
quasilocal perturbation of any state in the span of 
$\{\omega^{-+}_k(\boldsymbol{\sigma}) : k \in \Ir, \boldsymbol{\sigma} \in S^2(\Rl^3)\}$,
just as in the case $1<\Delta<\infty$.
A pictorial interpretation of the ground state space for the different kink models is 
shown in Figure \ref{Interpretation}:
For the isotropic (XXX) model, all ground states are translation-invariant, with
any choice of unit vector for all sites;
for the XXZ model with $1<\Delta<\infty$, there is a continuous family of points obtained
by specifying the spin at, say, the origin, and then choosing every other spin to have
the same angle $\phi$ and azimuthal angle 
$\theta_x = 2 \tan^{-1} (q^x \tan \frac{1}{2} \theta_0)$.
For the Ising model, the kink states are just what we have described above.

Also, with respect to the pair interaction 
$h^{-+}_{x,x+1} = 1 - \sigma_x \sigma_{x+1} + \sigma_{x+1} - \sigma_x$, the states
$\omega^{+-}_k$ are all frustration-free.
Of course, one can also define Ising limit analogues of antikink states, and the
results are parallel (or rather antiparallel) to those for kink states.

\subsection{Higher dimensions}
For statistical models one can not often solve exactly 
for physical properties in one dimension; it is even more unlikely
to solve a model in two dimensions; and it is almost never the
case that one can find exact data rigorously in three dimensions and higher.
However, for the XXZ model with certain domain wall-boundary fields, 
one can solve for the ground states in all dimensions.
The reason this is possible is that the one-dimensional ground states
defined by equation \eq{classical solution 1} and \eq{classical solution 2}
are frustration-free ground states.
As long as one can minimize the energy of each bond separately, it is possible
to add extra bonds and sites to the one-dimensional model, as long as one makes
sure that the state defined on the new site minimizes the energy of the new bond.
Of course there are consistency conditions which must be satisfied, and for 
example  if there is a non-simply-connected loop of oriented bonds, as occurs
in a spin ring, then the ground state will not have a simple closed form. 
(One can retain something reminiscent of frustration-free ground states for a 
spin ring by choosing $q$ to be an $L$th root of unity. 
Then it is possible to have a state such that its energy with repect to each bond is
zero.
However in this case $\Delta < 1$,
so that the XY-plane is the easy plane, and the Hamiltonian is no longer
nonnegative.
The ``frustration-free'' state one obtains is then not a ground state, it lies
somewhere in the middle of the spectrum.)

\begin{figure}
\begin{center}
\resizebox{10truecm}{5truecm}{\includegraphics{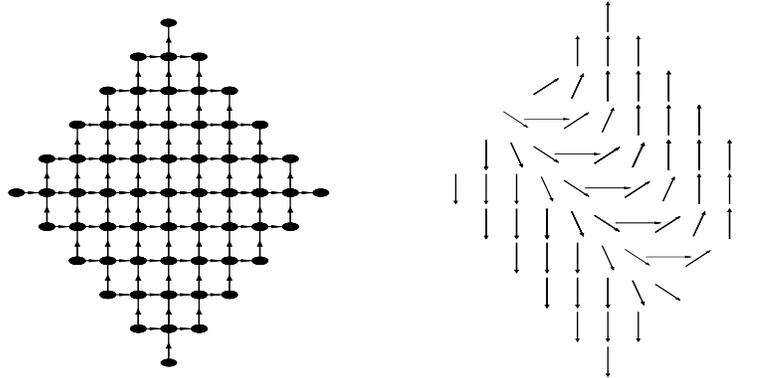}}
\end{center}
\caption{Sketch of a directed graph which supports a height function,
and a ground state}
\label{2d interface state}
\end{figure}

The specific condition for a frustration-free ground state of the classical system 
is the following.
Suppose $(\Lambda,\mathcal{B})$ is a directed graph; so that
$\Lambda$ a finite set, and 
$\mathcal{B}$ a collection of ordered pairs of points from $\Lambda$,
which are usually called directed edges,
but which we will call \textit{oriented bonds}.
We can then define the kink Hamiltonian on sequences
$\{\boldsymbol{\sigma}_x\} \in (S^2(\Rl^3))^{\Lambda}$ 
by
$$
H^{-+}_{(\Lambda,\mathcal{B})}(\{\boldsymbol{\sigma}_x\})
  = \sum_{(x,y) \in \mathcal{B}} \Big[1 - \sigma_x^3 \sigma_y^3 
  - \Delta^{-1}(\sigma_x^1 \sigma_y^1 + \sigma_x^2 \sigma_y^2)
  + A(\Delta) (\sigma_x^3 - \sigma_y^3) \Big]\, .
$$
One generally says that $(\Lambda,\mathcal{B})$ is \textit{connected} if, 
for any pair $x,y \in \Lambda$, there is a finite sequence of points 
$x_0,x_1,\dots,x_n$ such that $x_0 = x$, $x_n = y$ and for all $k=0,\dots,n-1$
either $(x_k,x+{k+1}) \in \mathcal{B}$ or $(x_{k+1},x_i) \in \mathcal{B}$.
We call such a sequence an \textit{unoriented path}
to distiguish it from the more common use of the word ``path'' wherein 
each $(x_k,x_{k+1})$ must all be oriented bonds.
We call the latter an \textit{oriented path}.
An oriented path is always also an unoriented path, but usually not vice-versa.
For any two points $x$,$y$ in a connected graph $(\Lambda,\mathcal{B})$, 
there are typically many choices of 
unoriented paths.
To construct a ground state for $H^{+-}_{(\Lambda,\mathcal{B})}$ requires 
some consistency condition among all these unoriented paths.
We define a \textit{height function} to be a function $l : \Lambda \to \Ir$
with the property that
$$
\forall (x,y) \in \mathcal{B}\, ,\
l(y)-l(x) = 1\, .
$$
The consistency condition we require is that $(\Lambda,\mathcal{B})$ supports
height functions, i.e.\ there exists some height function for 
$(\Lambda,\mathcal{B})$.
It is easy to see that if $(\Lambda,\mathcal{B})$ is connected and supports
height functions, then any two height functions differ by a constant.
It is also easy to see that if $(\Lambda,\mathcal{B})$ supports height functions
then there is no closed, oriented path, i.e.\ no unoriented path
$x_0,x_1,\dots,x_{n-1},x_n$ such that $x_0 = x_n$ and such that
$(x_k,x_{k+1}) \in \mathcal{B}$ for each $k=0,\dots,n-1$.
In fact $(\Lambda,\mathcal{B})$ supports height functions iff for any 
closed, unoriented path, the path crosses the same
number of oriented bonds going forward , $(x_k,x_{k+1}) \in \mathcal{B}$, as
it crosses going backward ,$(x_{k+1},x_k) \in \mathcal{B}$.
With this characterization, it is obvious that if a directed graph supports a 
height function, then so does any subgraph.
Similarly, the digraph $(\Lambda,\mathcal{B})$ supports height functions iff
each of its maximal, connected subgraphs supports height functions.

With this terminology we have the following lemma
\begin{lemma} (Classical XXZ model)
\label{classical general lemma}
If $(\Lambda,\mathcal{B})$ is connected and supports height functions, then 
for each real number $M$ satisfying $-|\Lambda|<M<|\Lambda|$ there is a sequence
$\{\boldsymbol{\sigma}_x\}$ with the property that
$$
H^{-+}_{(\Lambda,\mathcal{B})}(\{\boldsymbol{\sigma}_x\}) = 0\, ,\quad
\textrm{and}\quad
\sum_{x \in \Lambda} \sigma_x^3 = M\, .
$$
Moreover this sequence is uniquely determined by the number $M$,
up to scalar multiplication and
a uniform rotation of every $\boldsymbol{\sigma}_x$ about the 
$\boldsymbol{e}_3$-axis by some angle $\phi$.
For $M=\pm |\Lambda|$, one has $\boldsymbol{\sigma}_x = \pm \boldsymbol{e}_3$.
For $-|\Lambda|<M<|\Lambda|$, any $\phi \in \Rl/2\pi\Ir$, and any 
height funtion $l$, one can define the zero-energy state
by equation \eq{classical solution 1} and 
\begin{equation}
\label{general classical solution}
\sigma_x^3 = \frac{1 - q^{2 l(x)} \tau(l,M)}{1 + q^{2 l(x)} \tau(l,M)}\, ,
\end{equation}
where $\tau(l,M)$ is chosen to satisfy
$$
\sum_{x \in \Lambda} \frac{1 - q^{2 l(x)} \tau(l,M)}{1 + q^{2 l(x)} \tau(l,M)}
  = M\, .
$$
\end{lemma}

\begin{proof}
To be a zero-energy state means that for every $(x,y) \in \mathcal{B}$,
$$
0 = h(x,y) := 1 - \sigma_x^3 \sigma_y^3 
  - \Delta^{-1}(\sigma_x^1 \sigma_y^1 + \sigma_x^2 \sigma_y^2)
  + A(\Delta) (\sigma_x^3 - \sigma_y^3)\, .
$$
This necessitates $\phi_y = \phi_x$ and
$$
0 =  \Big[\sqrt{1+\Delta^{-1}} \sin \frac{1}{2} (\theta_y - \theta_x) 
  - \sqrt{1-\Delta^{-1}} \sin \frac{1}{2} (\theta_x + \theta_y)\Big]^2\, ,
$$
which in turn means
$$
t_y = q\, t_x\, ,\qquad  t_z := \tan \frac{1}{2} \theta_z\quad
\text{for all } z \in \Lambda\, .
$$
Clearly, if there exist $\phi$ and $\tau$ such that $\phi_x = \phi$
and $t_x = q^{l(x)} \tau$ for all $x$, then $h(x,y) = 0$ for all oriented
bonds $(x,y)$, since $l(y) = l(x)+1$.
Conversely if $h(x,y)=0$ for every oriented bond, then for any $x,y \in \Lambda$
not necessarily an oriented bond, consideration of a connecting sequence
$x = x_0,x_1,\dots,x_n=y$ leads us to the conclusion that $\phi_y = \phi_x$
and $t_y = q^{l(y) - l(x)} t_x$.
Taking $\phi = \phi_x$ and $\tau = q^{-l(x)} t_x$ for any $x \in \Lambda$, we have
have the same condition as before.
It can easily be seen that 
$$
\tan\frac{1}{2} \theta_x = q^{l(x)} \tau\quad 
  \Leftrightarrow\quad  \sigma_x^3 = \frac{1 - q^{2 x} \tau}{1 + q^{2 x} \tau}\, .
$$
So the lemma is proved once we observe that with $l$ fixed, the map
$$
M(\tau) = \sum_{x \in \Lambda} \frac{1 - q^{2 l(x)} \tau}{1 + q^{2 l(x)} \tau}
$$
is a strictly decreasing, continuous map from $(0,\infty)$ onto 
$(-|\Lambda|,|\Lambda|)$.
\end{proof}

This result is useful for example because one can make $\Lambda = \Ir^d$ into a 
directed graph by defining
$$
\mathcal{B} = \{(\boldsymbol{x},\boldsymbol{y}) \in \Ir^d\times \Ir^d : 
  \boldsymbol{y} - \boldsymbol{x} \in \{\boldsymbol{e}_1,\dots,\boldsymbol{e}_d\}\}\, .
$$
Then for any connected, finite subset $\Lambda \subset \Ir^d$, one can solve for 
the frustration-free ground states of $\Hil^{-+}_\Lambda$.
These ground states will have a $(1,1,\dots,1)$-interface between down spins
and up spins, because, up to adding a constant, the only choice for a height 
function is $l(\boldsymbol{x}) = \boldsymbol{x}\cdot (1,1,\dots,1)$.
A sketch of such an interface state for $d=2$ is shown in Figure \ref{2d interface
state}.

There is a generalization of this technique which allows for different  
anistropies on bonds going in different directions.
We delay this generalization to the spin-$1/2$ quantum model.
We derive it there mostly because the formula for ground states of the
XXZ model in dimensions higher than one is stated in \cite{ASW} 
in just this generality.

\section{The Quantum Kink Hamiltonian in  Finite Volumes}
\noindent
For each choice of $\Spin$, define the kink Hamiltonian 
\begin{align*}
H^{-+,(\Spin)}_L &:= H^{\textrm{XXZ}}_L + \Spin \sqrt{1 - \Delta^{-2}} (S_1^3 - S_L^3) 
  = \sum_{x=1}^{L-1} h^{(\Spin)}_{x,x+1} \\
h^{(\Spin)}_{x,x+1} &= \Spin^2  - S_x^3 S_{x+1}^3 
  - \frac{1}{2 \Delta} (S_x^+ S_{x+1}^- + S_x^- S_{x+1}^+) 
  - \Spin \sqrt{1 - \Delta^{-2}} (S_{x+1}^3 - S_x^3)\, .
\end{align*}
Note that we have placed the same boundary field as in the classical model, but 
scaled by $\Spin$, so that every term in the Hamiltonian is homogeneous of 
degree $\Spin^2$.
For $\Spin=1/2$, it is a simple calculation to verify
$$
h^{(1/2)}_{x,x+1} = \ket{\xi}\bra{\xi}\, ,\qquad
\xi = \frac{1}{\sqrt{1+q^2}}
  (q \ket{\downarrow \uparrow} - \ket{\uparrow \downarrow})\, .
$$
If $q=1$ then $\xi$ is the spin singlet, which is 
also the (unique up to scalar multiplication) antisymmetric tensor in 
$\Hil_x \otimes \Hil_{x+1}$.
Thus, for $q=1$, 
$$
h^{(1/2)}_{x,x+1} = \unity - T_{(x\  x+1)}\, ,
$$
where $(x\  x+1) \in \mathfrak{S}_L$ is the transposition, and 
$T : \mathfrak{S}_L \to \textrm{GL}(\Hil_L)$ is the standard action defined by its
image on simple tensors:
$$
T_\pi \bigotimes_{x=1}^L \ket{\psi_x}_x 
  = \bigotimes_{x=1}^L \ket{\psi_{\pi^{-1}(x)}}_x\, .
$$
Thus, for the isotropic model, any ground state is invariant under the action
of every nearest-neighbor transposition.
Since the nearest-neighbor transpositions generate the entire symmetric group,
this means the ground states of the isotropic model are exactly the symmetric 
tensors in the $L$-fold tensor product.
This is a standard result, which is related to the $\textrm{SU}(2)$-invariance
of the XXX Hamiltonian, because the subspace of symmetric tensors coincide
with the heighest-weight, $(L+1)$-dimensional, 
irreducible representation of $\textrm{SU}(2)$
in the tensor product $\bigotimes_{x=1}^L \Hil_x$.

\subsection{Ground States}

We will now state an important result of \cite{ASW}, which gives an analogue
of the last paragraph for $\Delta > 1$.
The proof we provide is not exactly the same as the original proof of Alcaraz, 
Salinas and Wreszinski, or Gottstein and Werner \cite{GW}.
This is mostly for pedagogical reasons.
The discovery of ASW was originally related to the quantum group 
symmetry of the model,
which was first explicitly pointed out in the paper \cite{PS}.
Gottstein and Werner use generating functions to calculate the finite-volume ground states,
which is fine, but one still needs a separate argument to show that all 
the ground states can be obtained in this way.
We will take a different, but more direct, approach.

The result is analogous to Lemma \ref{classical general lemma},
i.e. one can solve the model in all dimensions,
provided there exists a height function for the directed graph of 
oriented bonds between nearest-neighbor pairs.
In fact the formula from \cite{ASW} is even more general, 
because it allows for different anisotropies in different directions.
To put this in the framework of directed graphs that we already introduced, 
we need
to expand our definitions slightly.
Suppose $\Lambda$ is a finite set and 
$\mathcal{B}_1,\mathcal{B}_2,\dots,\mathcal{B}_d$ are disjoint collections 
of ordered
pairs from $\Lambda$, which we will interpret as oriented bonds in $d$ 
different directions.
Then, given numbers $1 \leq \Delta_i \leq +\infty$, we define the XXZ 
Hamiltonian with anisotropies $(\Delta_1,\dots,\Delta_d)$ by
\begin{gather*}
H^{-+,(\Spin)}_{(\Lambda,\mathcal{B}_1,\dots,\mathcal{B}_d)}
  (\Delta_1,\dots,\Delta_d)
  = \sum_{i=1}^d \sum_{(x,y) \in \mathcal{B}_i} h_{(x,y)}(\Delta_i)\, ,\\
h_{(x,y)}(\Delta_i) = \Spin^2  - S_x^3 S_y^3 
  - \frac{1}{2 \Delta_i} (S_x^+ S_y^- + S_x^- S_y^+) 
  + \Spin A(\Delta_i) (S_x^3 - S_y^3)\, .
\end{gather*}
We will say that $(\Lambda,\mathcal{B}_1,\dots,\mathcal{B}_d)$ is connected if
$(\Lambda,\bigcup_{i=1}^d \mathcal{B}_i)$ is.
We will call an unoriented path any sequence $x_0,\dots,x_n$ which is an unoriented
path w.r.t.\ $(\Lambda,\bigcup_{i=1}^d \mathcal{B}_i)$, and
we define a \textit{generalized height function} to be any function
$\boldsymbol{l} : \Lambda \to \Ir^d$ such that 
$\boldsymbol{l}(y) = \boldsymbol{l}(x) + \boldsymbol{e}_i$ whenever
$(x,y) \in \mathcal{B}_i$.
The existence of a generalized height function is equivalent to the property that
any closed unoriented path
traverses the same number of bonds of each type $i=1,\dots,d$ going forwards 
as going backwards:
$$
\# \{j : (x_j,x_{j+1}) \in \mathcal{B}_i\} 
  = \# \{j : (x_{j+1},x_j) \in \mathcal{B}_i\}\quad
\textrm{for every } i=1,\dots,d\, .
$$
One simple example of $(\Lambda,\mathcal{B}_1,\dots,\mathcal{B}_d)$ which supports
generalized height functions is any finite subset of $\Ir^d$ with the definitions
$\mathcal{B}_i = \{(x,y) : y-x=\boldsymbol{e}_i\}$.
In this case, if $\Lambda$ is connected, then every
generalized height function is defined by 
$\boldsymbol{l}(x)=x-x_0$ for some $x_0 \in \Ir^d$.
Just as for height function, if $(\Lambda,\mathcal{B}_1,\dots,\mathcal{B}_d)$ is 
connected and supports generalized height functions, then all 
generalized height functions differ by a constant.

\begin{theorem}
{(Alcaraz, Salinas, Wreszinski)}
\label{ASWTheorem}
If $(\Lambda,\mathcal{B}_1,\dots,\mathcal{B}_d)$ is a connected finite graph 
which supports a 
generalized height function $\boldsymbol{l}$, then for any choice 
$\Delta_1 ,\dots, \Delta_d \in [1,+\infty)$, and $\Spin$, the Hamiltonian
$$
H^{-+,(\Spin)}_{(\Lambda,\mathcal{B}_1,\dots,\mathcal{B}_d)}
(\Delta_1,\dots,\Delta_d)
$$
has a unique zero-energy state in every sector of total magnetization
$$
M=- |\Lambda| \Spin ,- |\Lambda| \Spin +1,\dots, |\Lambda| \Spin\, .
$$
Moreover, for a specified magnetization, the ground state is given by the 
simple formula
\begin{align*}
\Psi_0(M) = \sum_{\substack{\{m_x\} \in \{-\Spin,\dots,+\Spin\}^\Lambda \\
  \sum_x m_x = M}}\,
  \prod_{x=1}^L (q_1^{l_1(x)} q_2^{l_2(x)} \dots q_d^{l_d(x)})^{m_x} 
  \binom{2 \Spin}{\Spin + m_x}^{1/2}\ 
  \ket{\{m_x\}}\, ,
\end{align*}
where $l_i(x) = \boldsymbol{l}(x) \cdot \boldsymbol{e}_i$, and each
$q_i \in (0,1]$ is the solution of $\Delta_i = \frac{1}{2}(q_i + q_i^{-1})$.
\end{theorem}

\begin{proof}
We will first prove it for $\Spin = \frac{1}{2}$.
In this case, we are guided by the argument at the beginning of the section
which proves that the ground states of the XXX model are symmetric tensors.
The key is to realize the ground states of the XXZ model as symmetric tensors, as well,
but with a different choice of action of the symmetric group.

Let $\mathfrak{S}_\Lambda$ be the group of permutations of $\Lambda$.
The standard basis for $\Hil_\Lambda$ is the set of all $2^{|\Lambda|}$
simple tensors of the form 
$$
\ket{\{\sigma_x\}} = \bigotimes_{x \in \Lambda} \ket{\sigma_x}_x\, ,
$$
where $\{\sigma_x\}$ is a sequence of $+1/2$'s and $-1/2$'s.
Let $T : \mathfrak{S}_\Lambda \to \textrm{GL}(\Hil_\Lambda)$ be the usual action
defined by
$$
T_\pi \ket{\{\sigma_x\}} = \ket{\{\sigma_{\pi^{-1}(x)}\}}\, .
$$
We define a weight $\boldsymbol{W} : \{\pm 1/2\}^{\Lambda} \to \Ir^d$ by
$$
\boldsymbol{W}(\{\sigma_x\}) 
  = \sum_{x \in \Lambda} \left(\frac{1}{2} - \sigma_x\right) \boldsymbol{l}(x)\, .
$$
We define a new basis
$$
\ket{\{\sigma_x\}}' = \prod_{i=1}^d q_i^{W_i(\{\sigma_x\})}\, \ket{\{\sigma_x\}}\, ,
$$
and a new action $U : \mathfrak{S}_\Lambda \to \Hil_\Lambda$ by
$$
U_\pi \ket{\{\sigma_x\}}' = \ket{\{\sigma_{\pi^{-1}(x)}\}}'\, .
$$
With respect to the original basis,
$$
U_\pi \ket{\{\sigma_x\}} = \prod_{i=1}^d q_i^{W_i(\{\sigma_{\pi^{-1}(x)}\}) - 
W_i(\{\sigma_{x}\})}\ \ket{\{\sigma_{\pi^{-1}(x)}\}}\, .
$$

Hence, if $(x,y) \in \mathcal{B}_i$ and if $\tau = (x\ y)$ is the corresponding 
transpositon,
then $U_tau = \unity_{\Lambda \setminus \{x,y\}} \otimes U'_\tau$
where $U'_\tau$ acts on $\Hil_x \otimes \Hil_y$ by
\begin{alignat*}{2}
U'_\tau \ket{\uparrow\uparrow} &= \ket{\uparrow \uparrow}\, ,&\quad
U'_\tau \ket{\downarrow\uparrow} &= q_i \ket{\uparrow\downarrow}\, ,\\
U'_\tau \ket{\downarrow\downarrow} &= \ket{\downarrow \downarrow}\, ,&\quad
U'_\tau \ket{\uparrow\downarrow} &= q_i^{-1} \ket{\downarrow \uparrow}\, .
\end{alignat*}
with $\ket{\uparrow} \equiv \ket{+\frac{1}{2}}$ and 
$\ket{\downarrow} \equiv \ket{-\frac{1}{2}}$.
Note that, restricting attention to $\Hil_x \otimes \Hil_y$ and the representation of
the two-element group $\mathfrak{S}_{\{x,y\}} = \{\tau,e\}$,
the symmetric tensors are 
$$
\ket{\uparrow \uparrow}\, ,\ \ket{\downarrow \downarrow}\, ,\
\frac
{\ket{\downarrow \uparrow} + q_i \ket{\uparrow \downarrow}}
{\sqrt{1 + q_i^2}} \, ,
$$
and the antisymmetric tensor is
$$
\xi(q_i) = \frac
{q_i \ket{\downarrow \uparrow} - \ket{\uparrow \downarrow}}
{\sqrt{1 + q_i^2}} \, .
$$
Now we already know
$$
h_{(x,y)}(\Delta_i) = \ket{\xi(q_i)}_{(x,y)}\bra{\xi(q_i)}_{(x,y)}\, ;
$$
from which it obviously follows that
$$
h_{(x,y)}(\Delta_i) = \unity - U_\tau\, .
$$
Thus a state minimizes the interaction $h_{(x,y)}(\Delta_i)$ if and only if it is 
invariant under the action of $\tau = (x\ y)$.
Hence a state is frustration-free, i.e.\ minimizes every interaction, if and only if
it is invariant under every nearest-neighbor transposition.
But it is well known that for a connected graph, $\Lambda$, 
the nearest-neighbor transpositions generate the entire symmetric group 
$\mathfrak{S}_\Lambda$.
Hence, any state is frustration free iff it is invariant under the entire action of 
$U$ of $(\mathfrak{S}_\Lambda)$.
I.e. the frustration free states exactly coincide with the tensors which are symmetric 
with repsect to $U$.

We know a formula for the symmetric states using the basis $\ket{\{\sigma_x\}}'$
on which $U$ has the standard action:
$$
\Psi'_0(M) = 
 \sum_{\substack{\{\sigma_x\} \in \{\pm 1/2\}^\Lambda \\
  \sum_x \sigma_x = M}} \ket{\{m_x\}}'\, ,\qquad
M = -\frac{1}{2} |\Lambda|, -\frac{1}{2} |\Lambda|+1,\dots,\frac{1}{2}|\Lambda|\, .
$$
This means that $\Psi_0(M)$ is defined in terms of the usual basis by
$$
\Psi'_0(M) = 
 \sum_{\substack{\{\sigma_x\} \in \{\pm 1/2\}^\Lambda \\
  \sum_x \sigma_x = M}}\,
  \prod_{i=1}^d q_i^{W_i(\{\sigma_x\})}\,  \ket{\{\sigma_x\}}'\, ,\qquad
M = -\frac{1}{2} |\Lambda|, -\frac{1}{2} |\Lambda|+1,\dots,\frac{1}{2}|\Lambda|\, .
$$
Mutliplying by a constant 
$$
C(M) = \prod_{x \in \Lambda} \prod_{i=1}^d q_i^{-\sum_x l_i(x)/2}\, ,
$$
one obtains the formula
$$
\Psi_0(M) = C(M) \Psi'_0(M) 
 = \sum_{\substack{\{\sigma_x\} \in \{\pm 1/2\}^\Lambda \\ \sum_x \sigma_x = M}}\,
  \prod_{x\in \Lambda} \prod_{i=1}^d q_i^{l_i(x) \sigma_x}\,  \ket{\{\sigma_x\}}\, .
$$
This is the desired result for $\Spin = \frac{1}{2}$.

\begin{figure}
\begin{center}
\resizebox{10truecm}{5truecm}{\includegraphics{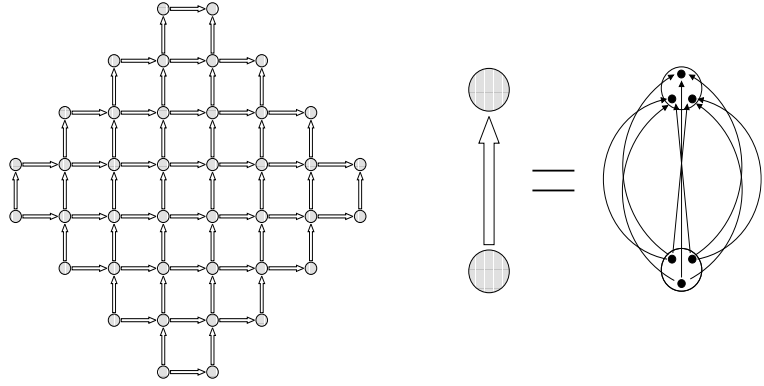}}
\end{center}
\caption{Schematic of two-dimensional spin system with $\Spin = \frac{3}{2}$ in terms of 
a spin system with $\Spin = \frac{1}{2}$.}
\label{Higher spin}
\end{figure}

The result for higher spin systems follows by embedding any spin-$\Spin$ system with
$|\Lambda|$ sites into a spin-$1/2$ spin system with $2 \Spin |\Lambda|$ sites.
Specifically, define $\Lambda_1,\dots,\Lambda_{2\Spin}$ for be disjoint copies of
$\Lambda$.
Let 
$$
\mathcal{B}_i' = \{(x_r,y_s) \in \Lambda_r \times \Lambda_s : (x,y) \in \mathcal{B}\, ,
r,s =1,\dots,2\Spin\}\, .
$$
For a single site $x \in \Lambda$, there is a natural way to identify
$\Hil^{(\Spin)}_x$ with a subspace of $\bigotimes_{s=1}^{2\Spin} \Hil^{(1/2)}_{x_s}$,
namely as the symmetric tensors.
We define an action of $\mathfrak{S}_{2\Spin}^{\times \Lambda}$ on
$\bigotimes_{s=1}^{2 \Spin} \Hil^{(1/2)}_{\Lambda_s}$, wherein
each copy $\mathfrak{S}_{2\Spin}^{(x)}$ acts by the usual permutation action
on $\bigotimes_{s=1}^{2\Spin} \Hil^{(1/2)}_{x_s}$.
Let $P^{\textrm{sym}}$ be the projection onto the set of all vectors which are
fixed by the entire action.
Then by our previous consideration, this is precisely the projection onto
$\Hil_\Lambda^{(\Spin)}$.
Then we see that
$$
H^{-+,(\Spin)}_{(\Lambda,\mathcal{B}_1,\dots,\mathcal{B}_d)}(\Delta_1,\dots,\Delta_d)
  = P^{\textrm{sym}} 
  H^{-+,(1/2)}_{(\bigcup_s \Lambda_s,\mathcal{B}_1',\dots,\mathcal{B}_d')}
  (\Delta_1,\dots,\Delta_d) P^{\textrm{sym}}\, .
$$
But also, the entire action of $\mathfrak{S}_{2\Spin}^{\times \Lambda}$ 
commutes with 
$H^{-+,(1/2)}_{(\bigcup_s \Lambda_s,\mathcal{B}_1',\dots,\mathcal{B}_d')}$
by the definitions of $\mathcal{B}_i'$.
(All bonds are allowed between any $(x_r,y_s)$ with $(x,y) \in \mathcal{B}_i$
for all $r$ and $s$.
Therefore, permuting the indices $r \to \pi_x(r)$ and $s \to \pi_y(s)$ does not 
change the sum.)
Hence, $H^{-+,(1/2)}_{(\bigcup_s \Lambda_s,\mathcal{B}_1',\dots,\mathcal{B}_d')}
(\Delta_1,\dots,\Delta_d)$
also commutes with $P^{\textrm{sym}}$, and we see that the subspace of ground states of
$H^{-+,(\Spin)}_{(\Lambda,\mathcal{B}_1,\dots,\mathcal{B}_d)}(\Delta_1,\dots,\Delta_d)
(\Delta_1,\dots,\Delta_d)$
in $\Hil_\Lambda^{(\Spin)}$  is exactly equal to
$$
\ran(P^{\textrm{sym}}) \cap 
\ker(H^{-+,(1/2)}_{(\bigcup_s \Lambda_s,\mathcal{B}_1',\dots,\mathcal{B}_d')}
(\Delta_1,\dots,\Delta_d))\, .
$$
But by its very definition as the symmetric tensors under the action 
$U$ of $\mathfrak{S}_{\bigcup_s \Lambda_s}$, one sees that the entire kernel of
$\ker(H^{-+,(1/2)}_{(\bigcup_s \Lambda_s,\mathcal{B}_1',\dots,\mathcal{B}_d')}
(\Delta_1,\dots,\Delta_d))$ is contained in the range of $P^{\textrm{sym}}$.
Hence the ground states of the spin-$\Spin$ model coincide with the ground states
of $\ker(H^{-+,(1/2)}_{(\bigcup_s \Lambda_s,\mathcal{B}_1',\dots,\mathcal{B}_d')}
(\Delta_1,\dots,\Delta_d))$.
Writing these in terms of the natural basis for the spin-$\Spin$ representation
gives
\begin{align*}
\Psi^{(\Spin)}_0(M) &= 
  \sum_{\substack{\{\sigma_{x,s}\} \in \{\pm 1/2\}^{\bigcup_s \Lambda_s} \\ 
  \sum_{x,s} \sigma_{x,s} = M}}\,
  \prod_{x\in \Lambda} \prod_{i=1}^d \prod_{s=1}^{2\Spin}
  q_i^{l_i(x) \sigma_{x,s}}\,  \ket{\{\sigma_{x,s}\}}\, \\
  &= \sum_{\substack{\{m_x\} \in \{-\Spin,\dots,+\Spin\}^\Lambda \\
  \sum_x m_x = M}}\,
  \bigotimes_{x\in \Lambda} \prod_{i=1}^d q_i^{l_i(x) m_x}
  \sum_{\substack{\sigma_{x,1},\dots,\sigma_{x,2\Spin} = \pm 1/2 \\
  \sum_s \sigma_{x,s} = m_x}} \,  \ket{\{\sigma_{x,s}\}}_x\, \\
  &= \sum_{\substack{\{m_x\} \in \{-\Spin,\dots,+\Spin\}^\Lambda \\
  \sum_x m_x = M}}\, \bigotimes_{x\in \Lambda} \prod_{i=1}^d q_i^{l_i(x) m_x}
  \binom{2\Spin}{\Spin + m_x}^{1/2} \ket{m_x}_x \\
  &= \sum_{\substack{\{m_x\} \in \{-\Spin,\dots,+\Spin\}^\Lambda \\
  \sum_x m_x = M}}\, \prod_{x\in \Lambda} \prod_{i=1}^d q_i^{l_i(x) m_x}
  \binom{2\Spin}{\Spin + m_x}^{1/2} \ket{\{m_x\}} \, ,
\end{align*}
and this last formula is exactly what we want.
\end{proof}
This formula is due to \cite{ASW}, although \cite{PS} had already calculated the 
partition function for the case $q$ a root of unity, and periodic boundary conditions.
For spin-$\frac{1}{2}$ it was arrived at independently in \cite{GW}.

\subsection{Quantum group symmetry}
For this section we will consider only $\Spin=\frac{1}{2}$, and we will
denote the representation of $\textrm{SU}(2)$, instead
as a representation of $\mathfrak{sl}(2)$.
The generators $S_x^\alpha$,
$\alpha=1,2,3$, of the representation
on $\Hil_x = \Cx_x^{2}$ actually define a representation of $\mathfrak{su}(2)$,
not $\textrm{SU}(2)$.
But since $\Hil_x$ is complex, we may as well consider the representations of 
$\mathfrak{sl}(2)$.
Indeed, it is most often more useful to work with the three operators
$(S^3,S^+,S^-)$ which are the generators of the representation of $\mathfrak{sl}(2)$,
rather than the operators $(S^1,S^2,S^3)$.
Of course there is an obvious way to go back and forth, 
between complex representations of $\textrm{SU}(2)$,
$\mathfrak{su}(2)$ and $\mathfrak{sl}(2)$.
But for now, we prefer $\mathfrak{sl}(2)$.

A main feature of the isotropic (XXX) ferromagnet is that 
$\Hil_\Lambda$ possesses a representation of $\mathfrak{sl}(2)$, and 
moreover that this representation commutes with $H^{\textrm{XXX}}$.
By definition of our Hilbert space, each tensor factor 
$\Hil_x = \Cx_x^{2}$ is equipped
with a representation of $\mathfrak{sl}(2)$.
There is then a canonical representation of $\mathfrak{sl}(2)$ on the 
tensor product $\Hil_\Lambda = \otimes_{x \in \Lambda} \Hil_x$, given
by the generators
$$
S_{\textrm{tot}}^\alpha 
  = \sum_{x \in \Lambda} S_x^\alpha
$$
for $\alpha=3,+,-$.
We reiterate that what we actually mean when we write $S_x^\alpha$ is the
operator
$$
S_x^\alpha \otimes \bigotimes_{x \neq y \in \Lambda} \unity_y\, .
$$
The reason we bring this up now is that we wish to examine for a moment the
very basics of Lie algebra representations.

Defining $U(\mathfrak{sl}(2))$ to be the universal enveloping algebra,
each representation $\Hil_x$ of $\mathfrak{sl}(2)$ extends to a
representation of $U(\mathfrak{sl}(2))$.
Moreover, there is a natural (unital) algebra structure on the tensor product
$U(\mathfrak{sl}(2))^{\otimes \Lambda}$, and the tensor product
$\Hil_\Lambda$ is most naturally a representation of this algebra.
The way one passes from a module of
$U(\mathfrak{sl}(2))^{\otimes \Lambda}$ to a module of 
$U(\mathfrak{sl}(2))$ (which is the same as a representation of $\mathfrak{sl}(2)$) is through
the coproduct 
$$
\Delta : U(\mathfrak{sl}(2)) \to U(\mathfrak{sl}(2)) \otimes U(\mathfrak{sl}(2))\, ,\qquad
\Delta(x) = x \otimes 1 + 1 \otimes x\, \forall x \in \mathfrak{sl}(2)\, .
$$
(We are abusing notation since we use the symbol $\Delta$ here for the coproduct, while elsewhere
it refers to the anisotropy.
Since the two uses of $\Delta$ are so different, we trust the reader can tell which we mean from
the context. In particular, for the rest of this section $\Delta$ means coproduct: we use $q$
to parametrize anisotropy, here.)
This is a homomorphism, as is checked through the calculation
$$
[x\otimes 1+1\otimes x,y\otimes 1 + 1\otimes y]
  = [x \otimes 1,y \otimes 1] + [1\otimes x,1\otimes y]
  = [x,y]\otimes 1 + 1 \otimes [x,y]\, .
$$
The coproduct is coassociative, so that
$$
\begin{CD}
U(\mathfrak{sl}(2)) @>\Delta>> U(\mathfrak{sl}(2)) \otimes U(\mathfrak{sl}(2)) \\
@VV{\Delta}V @VV{\textrm{id}\otimes\Delta}V\\
U(\mathfrak{sl}(2)) \otimes U(\mathfrak{sl}(2)) @>{\Delta\otimes\textrm{id}}>>
  U(\mathfrak{sl}(2)) \otimes U(\mathfrak{sl}(2)) \otimes U(\mathfrak{sl}(2))
\end{CD}
$$ 
is a commutative diagram, and the coproduct is even cocommutative
so that $\Delta = \tau_{1,2} \Delta$, where 
$\tau_{1,2} : U(\mathfrak{sl}(2))_1 \otimes U(\mathfrak{sl}(2))_2 \to
U(\mathfrak{sl}(2))_2 \otimes U(\mathfrak{sl}(2))_1 $
is the switch-flip.
By coassociativity, composing enough $\Delta$ maps gives a well-defined 
algebra homomorphism
$U(\mathfrak{sl}(2)) \to U(\mathfrak{sl}(2))^{\Lambda}$
as long as $\Lambda$ is an ordered set, and by cocommutativity,
there is a well-defined homomorphism even if $\Lambda$ is not ordered.
(This is a fundamental reason that one can analyze $H^{\textrm{XXX}}$ 
more easily than $H^{\textrm{XXZ}}$ in dimensions higher than one.)
The action of $\Delta$ then gives the rule
$$
S_{\textrm{tot}}^\alpha 
  = \sum_{x \in \Lambda} S_x^\alpha\, ,
$$
as previously claimed.

The cocommutativity has the important consequence that since 
$S_{\textrm{tot}}^\alpha$ is invariant under the action of the permutation group 
generated by all the $\tau_{x,y}$'s, then for any $\psi \in \Hil_\Lambda$,
the subrepresentation
$$
U(\mathfrak{sl}(2))\cdot \psi 
  :=\{x\cdot \psi : x \in U(\mathfrak{su}(2))\}
$$ 
is a subspace of $\textrm{Sym}(\Hil_\Lambda)$ iff $\psi \in \Sym(\Hil_\Lambda)$.
Since the highest dimensional irrep equals
$U(\mathfrak{sl}(2))\cdot \psi$ for 
$\psi = \ket{\textrm{all up}}_\Lambda$, we have a simple proof
that the symmetric tensors comprise the highest dimensional irrep of 
$\mathfrak{sl}(2)$ in $\Hil_\Lambda$.
This has the important consequence that the ground state space of 
$H^{\textrm{XXX}}$ is actually the highest dimmensional irrep of 
$\mathfrak{sl}(2)$ in $\Hil_\Lambda$.
Another tell-tale sign of the coproduct is that by ``forgetting'' the coproduct
we can obtain simple decomposition formulas.
For example, the ground states of $H^{\textrm{XXX}}_{[1,L]}$ can be given by the formula
$$
\psi_{[1,L]}(n) = \left(S_{\Lambda}^-\right)^n\, \ket{\textrm{all up}}_{[1,L]}\, ,
$$
in which case one also has the identity
$$
\psi_{[1,L]}(n) = \sum_{k=0}^n \psi_{[1,L_0]}(k) \otimes \psi_{[L_0+1,L]}(n-k)\, ,
$$
for any $1\leq L_0\leq L$ (where one interprets
$\psi_{[a,b]}(n) = 0$ whenever $n$ is negative or greater than $b-a+1$).

\begin{figure}
\begin{center}
\resizebox{7truecm}{7truecm}{\includegraphics{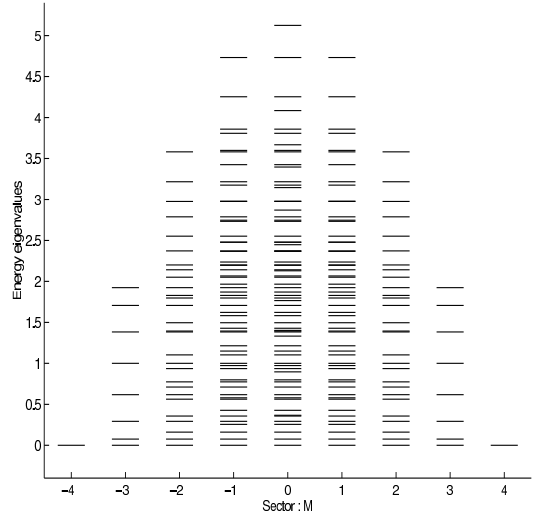}}
\end{center}
\caption{Spectrum of XXX model for $\Spin=1/2$, spin chain with 8 sites, 
decomposed according to sector. Note the upper envelope of the spectrum decreases
monotonically with $|M|$, as proved by Lieb and Mattis in \cite{LM}.}
\label{XXXspec}
\end{figure}

Of course, since the XXX Hamiltonian commutes with the entire representation,
not only is the ground state space equal to the highest dimensional irrep
(as we have proved),
but also every eigenspace for $H^{\textrm{XXX}}_\Lambda$ is a subrepresentation
of $\mathfrak{sl}(2)$, so that one can decompose $\Hil_\Lambda$ into
irreps of $\mathfrak{sl}(2)$, such that each irrep is also an eigenspace of 
$\Hil_\Lambda$.
We will now be able to interpret the $\mathfrak{sl}(2)$ symmetry
in terms of computational evidence.
We recall that a sector labelled by $M$ is the generalized eigenspace
of the action of $S_{\textrm{tot}}^3$ on $\Hil_\Lambda$ with eigenvalue $M$.
We will call the sector $\Hil_\Lambda^{(M)}$.
In terms of the sectors and eigenvectors of $H^{\textrm{XXX}}$,
we then have the following:
In the sector $\Hil_{\Lambda}^{(M)}$,
each eigenvector $\psi_M$ of $H^{\textrm{XXZ}}$ is also an eigenvector of 
$\mathcal{C}_{\textrm{tot}}$ (the image of the Casimir operator of 
$\mathfrak{sl}(2)$ in $\mathfrak{gl}(\Hil_\Lambda)$)
with eigenvalue $J(J+1)$ satisfying $M \in \{-J,-J+1,\dots,J\}$. 
If $J \neq M$, then the raising operator $S^+_{\textrm{tot}}$ maps the 
eigenvector $\psi_M$ to an eigenvector $\psi_{M+1}$
of $H^{\textrm{XXX}}$ with the same eigenvalue,
but such that $\psi_{M+1} \in \Hil_{\Lambda}^{(M+1)}$.
Similarly if $J \neq -M$, then the lowering operator $S^-_{\textrm{tot}}$
maps $\psi_M$ into an eigenvector $\psi_{M-1}$ with equal energy,
and $\psi_{M-1} \in \Hil_\Lambda^{(M-1)}$.
What this means in terms of the spectrum, is that there are constant-energy
bands sweeping through sectors with $M=-J,-J+1,\dots,+J$, corresponding
to the irreducible representations in the Clebsch-Gordon 
decomposition of $\Hil_\Lambda$ w.r.t.\ $\mathfrak{sl}(2)$.
This is clearly demonstrated in Figure \ref{XXXspec}.

\begin{figure}
\begin{center}
\resizebox{15truecm}{15truecm}{\includegraphics{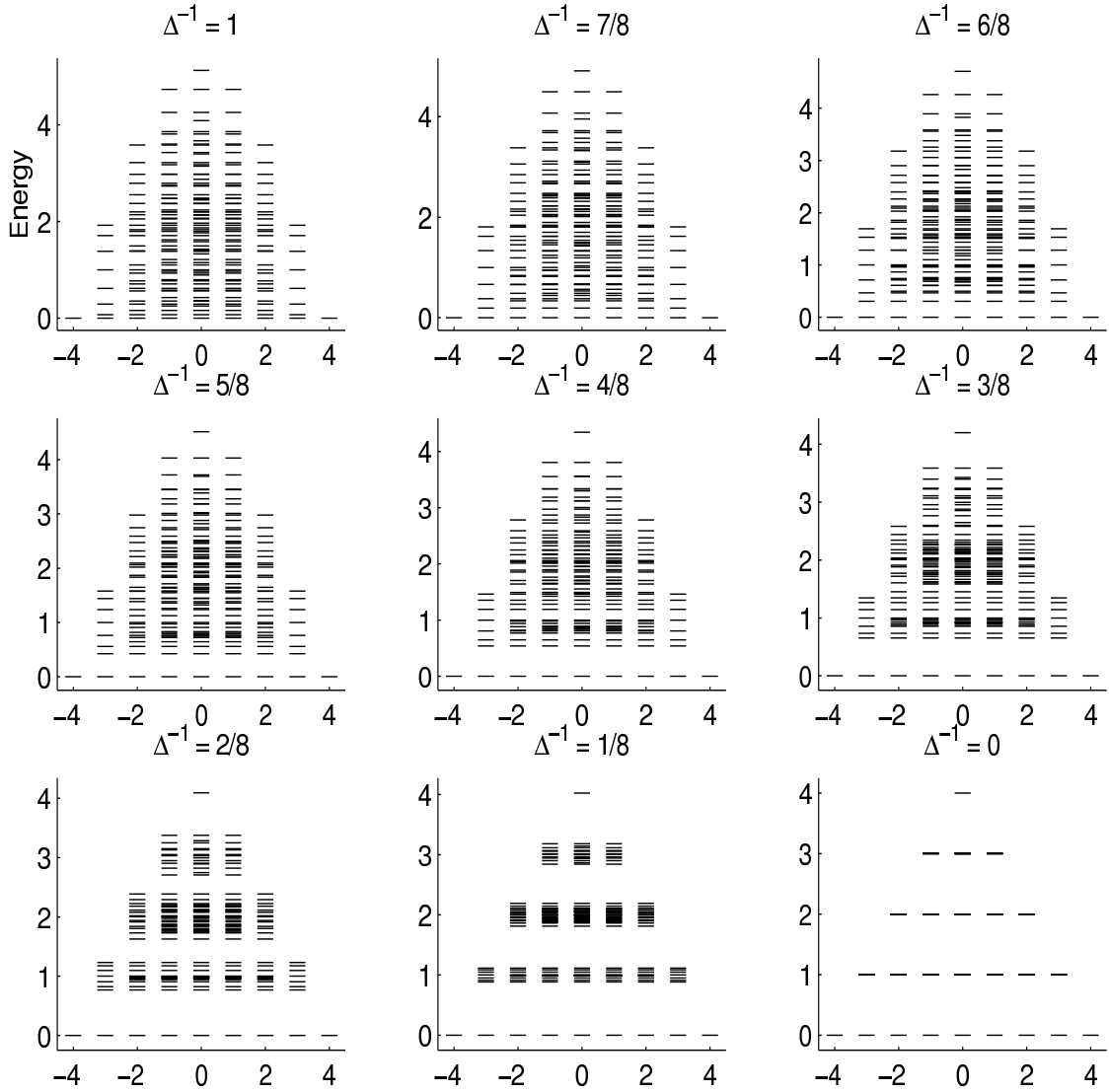}}
\end{center}
\caption{Spectrum of XXZ model for $\Spin=1/2$, $L=8$ sites, 
and anisotropy varying between isotropic $\Delta^{-1}=1$ and Ising-limit
$\Delta^{-1} = 0$.}
\label{XXZSpecArray}
\end{figure}

The kink Hamiltonian for $\Spin=1/2$ and 
$\Delta>1$ is clearly not $\mathfrak{sl}(2)$ symmetric,
which the interested reader may easily verify for himself.
Therefore, it is something of a surprise that the distinctive structure of the 
spectrum which was just noted for the XXX model, 
whose explanation relied entirely on the $\mathfrak{sl}(2)$ symmetry,
is still present in the kink Hamiltonian on a finite interval 
$\Lambda = \{1,2,\dots,L\}$ and $\Spin=1/2$ for all $\Delta \geq 1$.
This is clearly demonstrated in Figure \ref{XXZSpecArray}, where we have plotted
the spectrum for a range of $\Delta$ values between, and including, 
1 and $\infty$.
We would obviously like an explanation for this.
The answer is that while, for $\Delta>1$, the kink Hamiltonian
$H^{+-}_\Lambda$ is not invariant under the action of 
the universal enveloping algebra, $U(\mathfrak{sl}(2))$,
it is invariant under the action of the quantum universal enveloping
algebra $U_q(\mathfrak{sl}(2))$.
In order for the last statement to make sense, we will now define
$U_q(\mathfrak{sl}(2))$, observing what are for us the most important features.
This is not a comprehensive introduction to quantum groups.
For that the reader is referred to the very clear reference \cite{Kas}.
This is the reference which we are copying.

A quantum group is defined as an abstraction of the structure of a Lie algebra.
Given a Lie algebra $\mathfrak{g}$ one may form the universal enveloping algebra
$$
U(\mathfrak{g}) = \left(\Cx \oplus 
  \bigoplus_{n=1}^\infty \mathfrak{g}^{\otimes n}\right)
  \Big/\langle{xy-yx-[x,y] : x,y \in \mathfrak{g}}\rangle\, .
$$
Then $U(\mathfrak{g})$ is obviously a unital algebra, since it is the quotient
of a unital algebra $T(\mathfrak{g})$ by a two-sided ideal
$I(\mathfrak{g}) = \langle{xy-yx-[x,y] : x,y \in \mathfrak{g}}\rangle$.
One defines a map $\mu : U(\mathfrak{g}) \otimes U(\mathfrak{g}) \to U(\mathfrak{g})$,
given by $\mu(x,y) = xy$.
Since the product on $U(\mathfrak{g})$ is associative, one is guaranteed that
the following diagram commutes: 
\begin{equation}
\label{CD1}
\begin{array}{ccc}
U(\mathfrak{g})\otimes U(\mathfrak{g})\otimes U(\mathfrak{g})
  & \stackrel{\mu\otimes\textrm{id}}{\longrightarrow} 
  & U(\mathfrak{g})\otimes U(\mathfrak{g})\\
  \downarrow\vcenter{\rlap{$\scriptstyle\textrm{id}\otimes\mu$}} & 
  & \downarrow\vcenter{\rlap{$\scriptstyle\mu$}}\\
  U(\mathfrak{g})\otimes U(\mathfrak{g}) & \stackrel{\mu}{\longrightarrow}
  & U(\mathfrak{g})
\end{array}
\end{equation}
One also speaks of a map $\eta : \Cx \to U(\mathfrak{g})$ given by 
$\eta(\lambda)$ is mapped to the image of 
$$
\lambda + 0 + 0 + \dots \in \Cx \oplus \mathfrak{g} \oplus \mathfrak{g}^{\otimes 2} \oplus
  \dots
$$
which expresses the fact that the algebra is unital.
Then
\begin{equation}
\label{CD2}
\begin{array}{ccccc}
\Cx \otimes U(\mathfrak{g}) & \stackrel{\eta\otimes\textrm{id}}{\longrightarrow} &
  U(\mathfrak{g})\otimes U(\mathfrak{g}) &
  \stackrel{\textrm{id}\otimes \eta}{\longleftarrow} & U(\mathfrak{g}) \otimes \Cx\\
  & \searrow\vcenter{\rlap{$\scriptstyle\cong$}} 
  & \downarrow\vcenter{\rlap{$\scriptstyle\mu$}}
  & \swarrow\vcenter{\rlap{$\scriptstyle\cong$}}\\
  && U(\mathfrak{g})
\end{array}
\end{equation}
since the image of $1$ in the quotient is the left and right identity element.
One can define a one-sided inverse of $\eta$ by taking the map
$\tilde{\epsilon} : T(\mathfrak{g}) \to \Cx$
given by $\lambda \mapsto \lambda$ for $\lambda \in \mathfrak{g}^{\otimes 0}$,
and $x \mapsto 0$ for any $x \in \mathfrak{g}$.
This map is identically zero on $I(\mathfrak{g})$ so it factors through a homomorphism
$\epsilon : U(\mathfrak{g}) \to \Cx$.
Also, there is a map $\Delta : U(\mathfrak{g}) \to U(\mathfrak{g})\otimes U(\mathfrak{g})$,
which we have already define for $U(\mathfrak{su}(2))$, generated by
$\Delta : x \mapsto x\otimes 1 + 1\otimes x$, for any $x \in \mathfrak{g}$.
One can check (in fact is encouraged to check) that the following diagrams
commute
\begin{equation}
\label{CD3}
\begin{array}{ccc}
U(\mathfrak{g}) & \stackrel{\Delta}{\longrightarrow}
  & U(\mathfrak{g})\otimes U(\mathfrak{g}) \\
\downarrow\vcenter{\rlap{$\scriptstyle\Delta$}} & 
  &\downarrow\vcenter{\rlap{$\scriptstyle\textrm{id}\otimes\Delta$}}\\
U(\mathfrak{g})\otimes U(\mathfrak{g}) 
  & \stackrel{\Delta\otimes\textrm{id}}{\longrightarrow} 
  & U(\mathfrak{g})\otimes U(\mathfrak{g})\otimes U(\mathfrak{g})
\end{array} 
\end{equation}
and
\begin{equation}
\label{CD4}
\begin{array}{ccccc}
\Cx \otimes U(\mathfrak{g}) & \stackrel{\epsilon\otimes\textrm{id}}{\longleftarrow} &
  U(\mathfrak{g})\otimes U(\mathfrak{g}) &
  \stackrel{\textrm{id}\otimes \epsilon}{\longrightarrow} & U(\mathfrak{g}) \otimes \Cx\\
  & \nwarrow\vcenter{\rlap{$\scriptstyle\cong$}} 
  & \uparrow\vcenter{\rlap{$\scriptstyle\Delta$}}
  & \nearrow\vcenter{\rlap{$\scriptstyle\cong$}}\\
  && U(\mathfrak{g})
\end{array}
\end{equation}
(Note that defining $\Delta$ on $\mathfrak{g}$ determines $\Delta$ uniquely
on all of $U(\mathfrak{g})$ since $\Delta$ is an algebra morphism.
So, for example, from the fact that $1 x = x$ for any $x \in \mathfrak{g}$,
we have that $\Delta(1) \Delta(x) = \Delta(x)$, for any $x$, which forces
$\Delta(1) = 1\otimes 1$, 
while as a second example
$$
\Delta(xy) = \Delta(x) \Delta(y)
  = xy \otimes 1 + x\otimes y + y\otimes x + 1 \otimes xy
$$
for any $x,y \in \mathfrak{g}$.)
One can also check commutativity of the diagrams
\begin{gather}
\label{CD5}
\begin{array}{ccc}
U(\mathfrak{g})\otimes U(\mathfrak{g}) 
  & \stackrel{\mu}{\longrightarrow} & U(\mathfrak{g})\\
  \downarrow\vcenter{\rlap{$\scriptstyle(\textrm{id}\otimes\tau\otimes\textrm{id})
  (\Delta\otimes \Delta)$}}
  & & \downarrow\vcenter{\rlap{$\scriptstyle\Delta$}}\\
  (U(\mathfrak{g})\otimes U(\mathfrak{g}))\otimes(U(\mathfrak{g})\otimes U(\mathfrak{g}))
  & \stackrel{\mu\otimes\mu}{\longrightarrow} 
  & U(\mathfrak{g})\otimes U(\mathfrak{g})
\end{array}\, ,\\
\label{CD6}
\begin{array}{ccc}
U(\mathfrak{g})\otimes U(\mathfrak{g}) 
  & \stackrel{\epsilon\otimes\epsilon}{\longrightarrow} & \Cx\otimes\Cx \\
  \downarrow\vcenter{\rlap{$\scriptstyle\mu$}}
  & & \downarrow\vcenter{\rlap{$\scriptstyle\textrm{id}$}}\\
  U(\mathfrak{g})
  & \stackrel{\epsilon}{\longrightarrow} 
  & \Cx
\end{array}\, ,\\
\label{CD7}
\begin{array}{ccc}
\Cx
  & \stackrel{\eta}{\longrightarrow} & U(\mathfrak{g}) \\
  \downarrow\vcenter{\rlap{$\scriptstyle\textrm{id}$}}
  & & \downarrow\vcenter{\rlap{$\scriptstyle\Delta$}}\\
  \Cx\otimes\Cx
  & \stackrel{\eta\otimes\eta}{\longrightarrow} 
  & U(\mathfrak{g})\otimes U(\mathfrak{g}) 
\end{array}\, ,\\
\label{CD8}
\begin{array}{ccc}
\Cx & \stackrel{\eta}{\longrightarrow} & U(\mathfrak{g})\\
  \searrow\vcenter{\rlap{$\scriptstyle\textrm{id}$}} & 
  & \swarrow\vcenter{\rlap{$\scriptstyle\epsilon$}}\\
  & \Cx
\end{array}\, ,
\end{gather}
which express the fact that the algebra structure given by $\mu$ and $\eta$
is compatible with the coalgebra structure given by $\eta$ and $\Delta$.
Any linear space $H$ with linear maps $\mu : H\otimes H \to H$,
$\eta : \Cx \to H$, $\eta : H \to \Cx$ and $\Delta : H \to H \otimes H$,
satisfying equations \eq{CD1}--\eq{CD8} is called a bialgebra.
There is one last piece of the puzzle before defining a Hopf algebra.
Given two endomorphisms $f,g \in \operatorname{End}_{\Cx}(H)$, we define the
convolution $f\star g \in \operatorname{End}_{\Cx}(H)$, as the composition
$$
H \stackrel{\Delta}{\longrightarrow} H \otimes H 
  \stackrel{f\otimes g}{\longrightarrow} H \otimes H
  \stackrel{\mu}{\longrightarrow} H\, .
$$
(So for example, for $H = U(\mathfrak{g})$, one has 
$$
f\star g(x) = \mu\circ(f\otimes g)(x\otimes1 + 1\otimes x)
  = f(x) g(1) + f(1) g(x)\, ,
$$
while $f\star g(1) = f(1) g(1)$, and 
$$
f\star g(xy) = f(xy) g(1) + f(x) g(y) + f(y) g(x) + f(1) g(xy)\, ,
$$
for any $x,y \in \mathfrak{g}$.)
An antipode of $H$, if it exists, is by definition an endomorphism $S$ of $H$,
 such that
$S \star \textrm{id}_H = \textrm{id}_H\star S = \eta\circ \epsilon$.
(For $H = U(\mathfrak{g})$, this means that 
$S \star \textrm{id}_H$ and $\textrm{id}_H\star S$ act as the identity
on $\mathfrak{g}^{\otimes 0}$ and annihilates $\mathfrak{g}^{\otimes n}$
for each $n=1,2,3,\dots$.)
One can check that the map $S : U(\mathfrak{g}) \to U(\mathfrak{g})^{\textrm{op}}$
which is an antihomorphism $S(ab) = S(b)S(a)$ for any $a,b \in U(\mathfrak{g})$,
and which is further determined by the rule $S(x) = -x$ for any $x \in \mathfrak{g}$,
is an antipode.
And the reader who is new to Hopf algebras is strongly encouraged to do so,
at least by checking the value of the convolution $S \star \textrm{id}_H$
on the elements $1$, $x$ and $xy$.
A Hopf algebra is by definition a bialgebra $H$ with an antipode $S$.

Now our main example of a Lie algebra is $\mathfrak{sl}(2)$.
The Lie algebra of $\mathfrak{sl}(2)$ is a three-dimensional Lie algebra generated
by $X$, $Y$ and $H$,  with 
$$
[X,Y] = H\, ,\qquad
[H,X] = 2X\, ,\qquad
[H,Y] = -2Y\, .
$$
One definition of a quantum group is as a smooth deformation of $U(\mathfrak{g})$ --
for some Lie algebra $\mathfrak{g}$ -- in the category of Hopf algebras.
The example which interests us is $U_q(\mathfrak{sl}(2))$.
We define $U_q'(\mathfrak{sl}(2))$ to be the algebra generated by five elements 
$E$, $F$, $K$, $K^{-1}$, $L$ and the relations
\begin{gather*}
K K^{-1} = K^{-1} K = 1\, ,\\
K E K^{-1} = q^2 E\, ,\quad
K F K^{-1} = q^{-2} F\, ,\\
[E,F] = L\, ,\quad (q-q^{-1}) L = K - K^{-1}\, ,\\
[L,E] = q(E K + K^{-1} E)\, ,\quad
[L,F] = q^{-1}(F K + K^{-1} F)\, .
\end{gather*}
The parameter $q$ is allowed to be any nonzero complex number.
For $q$ different from $0$, $1$ and $-1$, one may reduce the number of generators by
one by defining 
$$
L = \frac{K - K^{-1}}{q - q^{-1}}\, ,
$$
and this is literally what one means by $U_q(\mathfrak{sl}(2))$.
However, it is important that $U'_q(\mathfrak{sl}(2))$ is well-defined even for 
$q=1$ because then one finds
$$
U(\mathfrak{sl}(2)) \cong U_1'(\mathfrak{sl}(2))/(K-1)\, ,
$$
so that $U'_q(\mathfrak{sl}(2))$ is really a deformation of $U(\mathfrak{sl}(2))$.
Of course, for $q=1$, one already has a Hopf algebra structure on $U(\mathfrak{sl}(2))$.
For $q \neq \pm1, 0$, one may define a Hopf algebra structure on $U_q(\mathfrak{sl}(2))$
by the definitions
\begin{gather*}
\Delta(E) = 1\otimes E + E\otimes K\, ,\quad
\Delta(F) = K^{-1} \otimes F + F \otimes 1\, ,\\
\Delta(K) = K \otimes K\, ,\quad
\Delta(K^{-1}) = K^{-1} \otimes K^{-1}\, ,\\
\epsilon(E) = \epsilon(F) = 0\, ,\quad
\epsilon(K) = \epsilon(K^{-1}) = 1\, ,
\end{gather*}
and
$$
S(E) = -E K^{-1}\, ,\quad
S(F) = - K F\, ,\quad
S(K) = K^{-1}\, ,\quad
S(K^{-1}) = K\, ,
$$
although we will not verify this here.

It is a well-known result that for $q$ any complex number 
other than a root of unity
the representation theory of $U_q(\mathfrak{sl}(2))$ is equivalent to the representation
theory of $U(\mathfrak{sl}(2))$.
We paraphrase some definitions and theorems from Kassel:
For any representation $V$, and any nonzero $\lambda \in \Cx$ one defines 
$V^\lambda$ to be the eigenspace of $K$ with eigenvalue $\lambda$.
If $V^\lambda \neq \{0\}$, then $\lambda$ is called a weight, and $V^\lambda$ is called
a weight space.
For any weight $\lambda$, a nonzero $v$ such that $K v = \lambda v$ and $E v = 0$
is called a highest weight vector.
For $\varepsilon = \pm 1$, and any $n \in \Nl$, one may define an $(n+1)$-dimensional
module $V_{\varepsilon,n}$ which is spanned by vectors $v_0,v_1,\dots,v_n$ with
\begin{align*}
K v_p &= \varepsilon q^{n-2p} v_p\, ,\\
E v_p &= \varepsilon [n-p+1] v_{p-1}\, ,\\
F v_{p-1} &= [p] v_p\, ,
\end{align*}
where $v_{-1} = 0$ and
$$
[n] := \frac{q^n - q^{-n}}{q-q^{-1}}\, .
$$
This is a simple module with highest weight vector $v_0$ of weight $\varepsilon q^n$.
Then, as proved in Kassel, one has the following results
\begin{theorem}
(Representation theory of $\textrm{U}_q(\mathfrak{sl}(2))$)
Any finite-dimensional simple $\textrm{U}_q(\mathfrak{sl}(2))$-module is isomorphic to
$V_{\varepsilon,n}$ for some choice of $\varepsilon$ and $n$.
Any finite-dimensional $\textrm{U}_q(\mathfrak{sl}(2))$-module is semisimple.
One has
$$
V_{\epsilon,n} \cong V_{\epsilon,0} \otimes V_{1,n} \cong V_{1,n} \otimes V_{\epsilon,0}\, ,
\quad V_{-1,0} \otimes V_{-1,0} = V_{1,0}\, .
$$
Denoting $V_{1,n}$ by $V_n$, then for $n\geq m$,
$$
V_n \otimes V_m \cong V_{n+m} \oplus V_{n+m-1} \oplus \dots \oplus V_{n-m}\, .
$$
\end{theorem}

We will now give some explicit representations of $U_q(\mathfrak{sl}(2))$.
Specifically, we will give a representation of $U_q(\mathfrak{sl}(2))$ on the vector space
$\Hil_\Lambda$ and show that is commutes with $H^{+-}$.
For the purpose of the representation, we will not distinguish between the elements
$E$, $F$, $K$ and $K^{-1}$ and their images in $\mathfrak{gl}(\Hil_\Lambda)$.
For concreteness let $\Lambda = [1,L]$.
Then
\begin{gather*}
K = \prod_{x=1}^L q^{2 S_x^3}\, ,\quad
K^{-1} = \prod_{x=1}^L q^{-2 S_x^3}\, ,\\
K^{-1} E = \frac{1}{\sqrt{2}} \sum_{x=1}^L \prod_{y=1}^{x-1} q^{-2 S_y^3}\, S_x^+\, ,\\
F K = \frac{1}{\sqrt{2}} \sum_{x=1}^L \prod_{y=x+1}^{L} q^{2 S_y^3}\, S_x^-\,
\end{gather*}
defines a representation.
We define $K^{-1} E$ and $F K$ instead of $E$ and $F$ as a matter of convention
(because these are the operators which are defined in \cite{KN1}).
To prove that this is a representation we should check the relations.
Evidently $K K^{-1} = K^{-1} K = \unity$.
Also,
$$
S_x^+ q^{2 S_x^3} = q^{-1} S_x^+ = q^{-2} q^{2 S_x^3} S_x^+\quad \textrm{and}\quad
S_x^- q^{-2 S_x^3} = q^{-1} S_x^- = q^{-2} q^{-2 S_x^3} S_x^-\, .
$$
Hence
$$
K S_x^+ K^{-1} = q^2 S_x^+\quad \textrm{and}
K S_x^- K^{-1} = q^{-2} S_x^-\, ,
$$
which implies that
$$
K E K^{-1} = q^2 E\, ,\quad
K F K^{-1} = q^{-2} F\, .
$$
Finally, we calculate $[E,F]$.
Note that owing to the relations above $K^{-1} E F K  = E F$.
Let $\theta$ be the Heaviside function
$$
\theta(x) = \begin{cases} 0 & x\leq 0\, ,\\ 1& x>0\, .\end{cases}\
$$
Then
\begin{align*}
EF &= K^{-1} E F K\\  
  &= \frac{1}{2} \sum_{x=1}^L \sum_{x'=1}^L
  \left(\prod_{y=1}^{x-1} q^{-2S_y^3}\right) S_x^+
  \left(\prod_{y'=x'+1}^{L} q^{2S_{y'}^3}\right) S_{x'}^- \\
  &= \frac{1}{2} \sum_{x=1}^L \sum_{x'=1}^L
  \left(\prod_{y=1}^{x-1} q^{-2S_y^3}\right) 
  \left(\prod_{y'=x'+1}^{L} q^{2S_{y'}^3}\right) 
  (1 + \theta(x-x') (q^{-2} - 1)) S_{x}^+ S_{x'}^- \, ,
\end{align*}
and
\begin{align*}
FE &= F K K^{-1} E \\
  &= \frac{1}{2} \sum_{x=1}^L \sum_{x'=1}^L
  \left(\prod_{y'=x'+1}^{L} q^{2S_{y'}^3}\right) S_{x'}^- 
  \left(\prod_{y=1}^{x-1} q^{-2S_y^3}\right) S_x^+ \\
  &= \frac{1}{2} \sum_{x=1}^L \sum_{x'=1}^L
  \left(\prod_{y'=x'+1}^{L} q^{2S_{y'}^3}\right) 
  \left(\prod_{y=1}^{x-1} q^{-2S_y^3}\right) 
  (1 + \theta(x'-x) (q^{-2} - 1)) S_{x'}^- S_{x'}^+ \, .
\end{align*}
Note that if $\sigma = \pm 1$ then 
$$
\frac{q^\sigma - q^{-\sigma}}{q - q^{-1}} = \sigma\, ,
$$
Thus
\begin{align*}
[E,F] &= EF-FE \\
  &= \frac{1}{2} \sum_{x=1}^L \left(\prod_{y=1}^{x-1} q^{-2S_y^3}\right) 
  \left(\prod_{y=x+1}^{L} q^{2S_{y'}^3}\right) 2 S_x^3 \\
  &= \sum_{x=1}^L \left(\prod_{y=1}^{x-1} q^{-2S_y^3}\right) 
  \frac{q^{2 S_x^3} - q^{-2 S_x^3}}{q-q^{-1}} 
  \left(\prod_{y=x+1}^{L} q^{2S_{y'}^3}\right) 2 S_x^3 \\
  &= \frac{1}{q-q^{-1}}
  \sum_{x=1}^L \left(q^{-2 \sum_{y=1}^{x-1} S_y^3 + 2 \sum_{y=x}^L S_y^3}
  - q^{-2\sum_{y=1}^x S_y^3 + 2 \sum_{y=x+1}^L S_y^3}\right) \\
  &= \frac{1}{q - q^{-1}} \left(q^{2 \sum_{y=1}^L S_y^3} - q^{-2 \sum_{y=1}^L S_y^3}\right)\\
  &= \frac{K - K^{-1}}{q - q^{-1}}\, ,
\end{align*}
as required.
One usually then defines the Casimir operator
$$
C = E F + \frac{q^{-1} K + q K^{-1}}{(q-q^{-1})^2}\, ,
$$
with the result that on a $(2J+1)$ irrep, the Casimir operator takes a constant value
$$
\frac{q^{-J-1} + q^{J+1}}{(q^{-1} - q)^2}\, .
$$
Also, denoting the representation on $\Hil_\Lambda$ by $K_{\Lambda}$,
$K^{-1}_\Lambda$, $E_\Lambda$ and $F_\Lambda$, one can check by definition that
for $\Lambda_1 = [a,b]$ and $\Lambda_2 = [b+1,c]$,
\begin{align*}
K_{\Lambda_1\cup \Lambda_2} = K_{\Lambda_1} K_{\Lambda_2}\, ,\quad
K^{-1}_{\Lambda_1\cup \Lambda_2} = K^{-1}_{\Lambda_1} K^{-1}_{\Lambda_2}\, ,\\
E_{\Lambda_1\cup \Lambda_2} = E_{\Lambda_2} + E_{\Lambda_1} K_{\Lambda_2}\, ,\quad
F_{\Lambda_1\cup \Lambda_2} = F_{\Lambda_1} + F_{\Lambda_2} K^{-1}_{\Lambda_1}\, .
\end{align*}
Note that in terms of the representation we have given it is easier to check that
$$
(K^{-1} E)_{\Lambda_1\cup \Lambda_2} 
  = (K^{-1} E)_{\Lambda_1} + K^{-1}_{\Lambda_1} (K^{-1} E)_{\Lambda_2}\, ,\quad
(F K)_{\Lambda_1\cup \Lambda_2} 
  = (F K)_{\Lambda_2}  + (F K)_{\Lambda_1} K_{\Lambda_2}\, .
$$
In fact, this is obvious since
$$
(K^{-1} E)_{[1,L]} = \sum_{x=1}^L K^{-1}_{[1,x-1]} E_{\{x\}}\, ,\quad
(F K)_{[1,L]} = \sum_{x=1}^L F_{\{x\}} K_{[x+1,L]}\, .
$$

Next, we will show that the entire representation of $U_q(\mathfrak{sl}(2))$
commutes with $H^{+-}$.
We already know that $\sum_{x=1}^L S_x^3$ does.
Hence $K = q^{\sum_{x=1}^L S_x^3}$ and $K^{-1} = q^{-2 \sum_{x=1}^L S_x^3}$ commute.
Suppose $L=2$, then 
$$
K^{-1} E = \frac{1}{\sqrt{2}} \left(S_1^+ + q^{-2 S_1^3} S_2^+\right)\, ,\quad
F K = \frac{1}{\sqrt{2}} \left(S_2^- + q^{2 S_2^3} S_1^- \right)\, .
$$
Hence
\begin{align*}
\sqrt{2} K^{-1} E \ket{\downarrow \downarrow} 
  &= \ket{\uparrow \downarrow} + q \ket{\downarrow \uparrow}\, \\
\sqrt{2} K^{-1} E \left(\ket{\uparrow \downarrow} + q \ket{\uparrow \downarrow}\right)
  &= (q + q^{-1}) \ket{\uparrow \uparrow}\, ,\\
\sqrt{2} F K \ket{\uparrow \uparrow}
  &= \ket{\uparrow \downarrow} + q \ket{\downarrow \uparrow}\, ,\\
\sqrt{2} F K \left(\ket{\uparrow \downarrow} + q \ket{\downarrow \uparrow}\right)
  &= (q + q^{-1}) \ket{\downarrow \downarrow}\, .
\end{align*}
This is the ground state space.
So the three-dimensional irrep corresponds to the ground state space.
For the one-dimensional irrep, $E$ and $F$ must both annihilate everything.
Thus we see that $E$ and $F$ do commute with $H^{+-}$ for the
case $L=2$.
Now we can use the coproduct $\Delta$ and the fact that $H^{+-}$ is a sum of
translation-invariant, nearest-neighbor interactions to show that
the entire representation of $U_q(\mathfrak{sl}(2))$ commutes with 
$H^{+-}$ for every $L$.
Specifically, for any $L>2$, and any $x,x+1 \in [1,L]$, we have
$$
E_{[1,L]} = E_{[x+2,L]} + E_{[x,x+1]} K_{[x+2,L]} + E_{[1,x-1]} K_{[x,x+1]} K_{[x+2,L]}\, .
$$
Then obviously $E_{[x+2,L]}$, $K_{[x+2,L]}$ and $E_{[1,x-1]}$ all commute with
$H^{+-}_{x,x+1}$.
And we have just shown that $E_{[x,x+1]}$ and $K_{[x,x+1]}$ commute with 
$H^{+-}_{x,x+1}$, because this is the case $L=2$.
So $E$ commutes with $H^{+-}_{x,x+1}$ for every $x,x+1 \in [1,L]$.
A similar argument shows that $F$ commutes with every $H^{+-}_{x,x+1}$, as well.
This symmetry explains the pictures of Figure \ref{XXZSpecArray}, and gives a 
useful tool for comprehending the kink quantum spin system.
We observe that we have not used the antipode, here, although it is an
important feature of a Hopf algebra, since it allows one to do much more
with the representation theory than one could do just with a bialgebra.
(The antipode at the level of a Lie algebra $U(\mathfrak{g})$ is the same as 
the inverse at the level of the Lie Group $G$.
This is used, for example, to define a representation on $V^*$ given any representation 
on $V$, by taking $(x\cdot f)(v) = f(S(x)\cdot v)$.
In our case we did not need this piece of plethysm, just the fact that one could
obtain a representation on $V \otimes W$, given representations on $V$ and $W$.)
Something else which will be useful later on is the Cartan automorphism $\omega$ on 
$U_q(\mathfrak{sl}(2))$, which is an involution such that
$$
E \stackrel{\omega}{\longleftrightarrow} F\, ,\quad
K \stackrel{\omega}{\longleftrightarrow} K^{-1}\, .
$$
This corresponds to the simulateneous reflection and spin-flip symmetry of the 
kink Hamiltonian.
Either the reflection or the spin-flip symmetry alone takes the kink Hamiltonian
to the antikink Hamiltonian.
The fact that these two Hamiltonians are not equal is related to the fact
that $\Delta$ is not cocommutative.

\section{The Infinite Volume Kink Hamiltonian}

\subsection{Zero-Energy Ground States}
We return now to our original question, which is ``What is the complete set of
infinite volume ground states for the quantum XXZ model?'' 
In \cite{GW}, Gottstein and Werner realized
that there are both finite and infinite-volume ground states which are actually
frustration free.
In fact, in finite volumes all ground states of the Hamiltonian with kink boundary 
conditions are frustration free.

For a finite volume ground state $\psi$, being frustration free means that
$H^{+-}_{x,x+1} \psi = 0$ for all $x,x+1 \in [1,L]$.
Recall that for the infinite system a \textit{state} is a bounded linear functional,
$\omega$, on the closed algebra of quasilocal observables $\Obs_\infty$.
One observes that for any local observable $A \in \Obs_\Lambda$, the derivation
$$
\delta(A) = \lim_{\Lambda' \nearrow \Ir} [H_{\Lambda'},A]\, ,
$$
is well-defined since the limit stabilizes (for $\Lambda' \supset \Lambda + \{-1,0,1\}$).
A ground state is determined by the inequality
$$
\omega(A^* \delta(A)) \geq 0\, ,
$$
which must hold for all strictly local observables.
This expresses the stability of a ground state to local perturbations.
To be a zero energy ground state, $\omega$ must satisfy a more stringent condition.
Specifically, if $H_\Lambda = \sum_{\{x,x+1\} \subset \Lambda} H_{x,x+1}$, where
$H_{x,x+1} = \tau^{-x} H_{0,1} \tau^x$ is a translation invariant pair interaction,
($\tau$ is left translation, one unit,)
then a zero-energy state is a state $\omega$ such that
$$
\omega(H_{x,x+1}) = \min_{\substack{\omega' \in \Obs^* \\ \omega'(\unity) = 1}}
  \omega'(H_{x,x+1})\, ,
$$
for all $\{x,x+1\} \subset \Ir$. 
Of course being a zero energy state implies that $\omega$ is a ground state,
because for any strictly local observable $A \in \Obs_\Lambda$,
$\omega$ restricted to the finite-dimensional algebra $\Obs_{\Lambda + \{-1,0,1\}}$
is a density matrix, in which case
$$
\omega(A^* A H_{x,x+1}) = \mathcal{E}_0 \omega(A^* A)\, ,
$$
where $\mathcal{E}_0$ is the minimum eigenvalue of $H_{\{x,x+1\}}$.
Then
\begin{align*}
\omega(A^* \delta(A)) 
  &= \sum_{\{x,x+1\} \subset \Lambda+\{-1,0,1\}} \omega(A^*(H_{x,x+1} - \mathcal{E}_0)A) \\ 
  &= \sum_{\{x,x+1\} \subset \Lambda+\{-1,0,1\}} \tilde\omega(H_{x,x+1} - \mathcal{E}_0)
  &\geq 0\, .
\end{align*}
And, of course, the reverse is not necessarily true; i.e.\ it is generally false that
all infinite-volume ground states are zero-energy.
However in the case of the kink Hamiltonian, one may hope that it is true that the
zero-energy ground states are the complete list of infinite-volume ground states,
because all the finite-volume ground states are zero-energy.
This turned out to be a correct prediction, on the part of Gottstein and Werner,
although the proof, eventually given by \cite{Mat} and \cite{KN3} involved
some real work beyond just the concept of zero-energy states.

The zero energy ground states give a nice intermediate step between
the finite-volume ground states of the previous section, and the general
infinite-volume ground states which one usually hopes to determine.
It may be hoped that for some remaining open problems, such as all infinite-volume
ground states of the XXZ chain in dimensions greater than one, that the
zero-energy ground states would be a useful starting point.
Gottstein and Werner found all the zero energy ground states for two spin chains,
the XXZ and XXX, as well as connecting the notion of zero-energy ground states to
some known results about Valence Bond Solid states.
They did this by constructing a theory parallel to the usual theory
of states on the quasilocal observable algebra, except now considering states on 
an algebra of zero-energy observables.
For zero energy observables, one starts from the Hilbert subspace 
$\mathcal{G}_\Lambda \subset \Hil_\Lambda$ corresponding to the zero-energy
vectors w.r.t.\ $H_\Lambda$.
The zero energy observables are then $\mathcal{B}_\Lambda$,
the algebra of operators on $\mathcal{G}_\Lambda$.
Note that the usual quasilocal observable algebra $\mathcal{A_\infty}$ is defined
as the inductive limit of local observable algebras $\mathcal{A}_\Lambda$,
with the property that for $\Lambda_1 \subset \Lambda$, 
$\mathcal{A}_{\Lambda_1} \subset \mathcal{A}_{\Lambda}$.
This property is entirely due to the fact that 
$\Hil_\Lambda = \Hil_{\Lambda_1} \otimes \Hil_{\Lambda\setminus \Lambda_1}$,
so that for $A \in \Obs_{\Lambda_1}$ one simply defines 
$A_{\Lambda_1} \otimes \unity_{\Lambda\setminus \Lambda_1} \in \Obs_\Lambda$.
But, for the zero-energy observables, it is not true that 
$\mathcal{G}_\Lambda = \mathcal{G}_{\Lambda_1} \otimes 
\mathcal{G}_{\Lambda\setminus \Lambda_1}$.
E.g., although $\ket{\uparrow\uparrow}$ and $\ket{\downarrow\downarrow}$
are zero energy vectors of $H^{+-}_{[1,2]}$, 
$\ket{\uparrow\uparrow\downarrow\downarrow}$ is not a zero energy vector in $H^{+-}_{[1,4]}$.
So, in order to define a zero-energy analog of the algebra of quasilocal observables,
one needs a new inductive limit.

A first step in this process is the observation that the projection 
$g_\Lambda : \Hil_\Lambda \to \mathcal{G}_\Lambda$ satsifies certain consistency 
conditions
\begin{alignat*}{2}
g_\Lambda &\neq 0 &\quad \textrm{for all $\Lambda$, and}\\
g_\Lambda &\leq i_{\Lambda \Lambda'}(g_{\Lambda'}) 
  &\quad \textrm{for all $\Lambda\supset \Lambda'$.}
\end{alignat*}
This encodes the fact that for every finite volume the set of zero-energy states 
is nonempty, and that the restriction of a zero-energy state is a zero-energy
state.
For the case of $H$ a sum of translation-invariant, nearest-neighbor interactions,
this is obvious because any restriction of a state on $\Lambda$ will clearly minimize
every nearest neighbor interaction in $\Lambda'$ (because it minimizes every
nearest neighbor interaction in $\Lambda$).
However these two simple assumptions are enough to guarantee that the set 
$K_z(\Obs_\infty)$ of infinite-volume zero-energy ground states is nonempty,
which one can prove by weak-$*$ compactness.
Gottstein and Werner use the notion of approximate inductive limits, which 
they review, in order to construct the Banach space of $j$-convergent limits
of local zero-energy observables, where
$$
j_{\Lambda \Lambda'} : \mathcal{B}_{\Lambda'} \to \mathcal{B}_\Lambda\, ,\quad
j_{\Lambda \Lambda'}(A) = g_\Lambda i_{\Lambda \Lambda'}(A) g_\Lambda\, .
$$
They then prove a theorem to show that $K_z(\Obs_\infty)$, the space of 
zero-energy ground states on $\Obs_\infty$ is isomorphic, via a direct construction
to the space $K(\mathcal{B}_\infty)$ of ground states on the zero-energy observables.
This means that a complete knowledge of the Hilbert spaces $\mathcal{G}_\Lambda$
is sufficient to construct the zero-energy ground states, which are a priori defined
in terms also of observables on the complementary Hilbert space 
$\mathcal{G}_\Lambda^\perp$. 
Unfortunately, the approximate inductive limit does not guarantee that $\mathcal{B}_\infty$
is in fact a $\textrm{C}^*$ algebra, only that it is an order unit space.
However, with the condition that for every $j$-convergent nets $A_\Lambda$, $B_\Lambda$,
the products $A_\Lambda B_\Lambda$ is also $j$-convergent, 
Gottstein and Werner prove that $\mathcal{B}_\infty$ is a $\textrm{C}^*$ algebra.

Gottstein and Werner go on to consider Hilbert space representations of 
$\mathcal{B}_\Lambda$, analogous to the GNS construction.
We will not repeat their results for the general case, but rather we state now
the major application of their work, which is a Hilbert space representation
of $\mathcal{B}_\infty$ and 
the classification of $K(\mathcal{B}_\infty)$ for 
the special case of $H_\Lambda = H^{+-}_\Lambda$.
\begin{theorem}
\textbf{(Gottstein \& Werner)}
As a convex set, the set of zero-energy states of the interaction $H^{+-}$ is 
isomorphic to the convex hull of the three quasi-equivalence classes:\\
(1) the set consisting only of the ``all spins up'' state $\omega_\uparrow$\\
(2) the set consisting only of the ``all spins down'' state $\omega_\downarrow$\\
(3) a set of ``kink states'', which is isomorphic to the set of density matrices
on a separable Hilbert space. Each of these states converges in the 
$\textrm{w}^*$ topology to $\omega_\uparrow$ (resp.\ $\omega_\downarrow$),
when shifted along the chain to right (resp.\ left) infinity.
\end{theorem}
We will not reproduce their proof.
It is an analytic result, and it relies upon two important steps.
The first is to realize that there is a simple formula for $j_{\Lambda \Lambda'}$,
a fact which can be traced straight back to the representation of $U_q(\mathfrak{sl}(2))$
on each $\Hil_\Lambda$, and particularly to the existence of the coproduct $\Delta$.
The second is to examine the left and right asymptotics of any net of ground states,
whose magnetic moments form a bounded net.
This is the second statement of (3).
In fact once this is known it is trivial to construct a Hilbert space representation
of all kink states: it is simply the GNS Hilbert space of all quasilocal perturbations
to the fiducial vector
$$
\Omega^{+-} = \bigotimes_{x \in \Lambda} \Omega^{+-}(x)\, ,\qquad
\Omega^{+-}(x) = \begin{cases} \ket{\uparrow} & x\geq 1\, ,\\ 
\ket{\downarrow} & x\leq 0\, .\end{cases}
$$
Note that $\Omega^{+-}$ is not a ground state, nor a zero energy state.
But there are many ground states, which can be labelled as the coefficients
of the Laurent series
$$
\Psi^{\textrm{GC}}(z) 
  = \bigotimes_{x=-\infty}^0 (\ket{\downarrow}_x + q^{-x} z^{-1} \ket{\uparrow}_x)
  \otimes \bigotimes_{x=1}^\infty (\ket{\uparrow}_x + q^x z \ket{\downarrow}_x)\, ,
$$
or directly by the definition
$$
\Psi_0(n) = \sum_{k=0}^\infty 
  \sum_{\substack{\{x_1,\dots,x_k\} \subset \Ir_{\leq 0} \\
  \{y_1,\dots,y_{n+k}\} \subset \Ir_{\geq 1}}}
  q^{-(x_1+\dots+x_k) + (y_1 + \dots + y_{n+k})} \prod_{j=1}^k S_{x_j}^+
  \prod_{j=1}^{n+1} S_{y_j}^-\, \Omega^{+-}\, .
$$

\subsection{Complete List of Infinite-Volume Ground States}

In \cite{GW}, it was proved that all the zero-energy ground states are of the form 
given in the last section. 
But this does not asnwer the main question posed.
It turns out, as was proved by Matsui in \cite{Mat}, and generalized
(in particular, generalized to the XXX model, which was not possible with Matsui's methods)
by Koma and Nachtergaele in \cite{KN3}, that the zero-energy ground states are the
complete list, not only for spin-$\frac{1}{2}$, but for all spin.
Of course, one must consider not just the kink interaction, but also the antikink interaction
since these give the same Heisenberg dynamics on $\Obs_\infty$.
Thus the translation invariant all up spins state $\omega_\uparrow$,
the translation invariant all down spins state $\omega_\downarrow$, the kink states,
and the antikink states, are the complete list of ground states for the infinite-volume
XXZ model.
We do not reproduce their proofs; however, we observe that a proof can easily be 
inferred from the methods and arguments of Chapter 4.
This is not surprising, since the main ideas used in Chapter 4 are adaptations
of techniques introduced in \cite{KN3}.

\section{Spectral Gap}

The next natural question after determining all the infinite-volume ground states,
is to determine what the low-lying excitation looks like.
The most fundamental question along these lines, is whether or not there exists 
a nonvanishing spectral gap in the thermodynamic limit, or if the spectral gap
does vanish, what is its rate of vanishing?
\begin{figure}
\begin{center}
\resizebox{6truecm}{6truecm}{\includegraphics{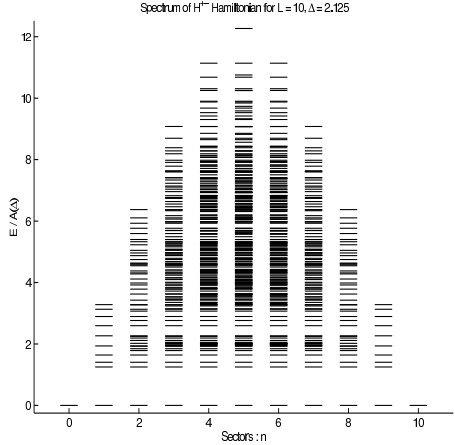}}
\end{center}
\caption{Spectrum of kink Hamiltonian: 
Note the energy gap between the first and second eigenvalues}
\end{figure}
In \cite{KN1}, Koma and Nachtergaele also provided a definite answer to that
question by showing that for the one-dimensional spin-$\frac{1}{2}$, there is a 
gap for any $q<1$, which persists in the thermodynamic limit.
It is the subject of Chapter 3 to verify that the same is true if one replaces 
spin-$\frac{1}{2}$ by spin-$\Spin$ for all $\Spin \in \frac{1}{2} \Nl_{\geq 2}$.
For dimensions greater than one, it is known, on the basis of general principles,
that the spectral gap must vanish in the thermodynamic limit.
In Chapter 5 we derive upper bounds for the spectral gap in dimensions two and higher,
which shows that the spectral gap generally vanishes at least as rapidly as $R^{-2}$,
where $R$ is the diameter of $\Lambda$ for the spin system.
However, before any extensions of the spectral gap can be understood, one must 
know the basic result of \cite{KN1}.
We include a full outline of \cite{KN1} here for two reasons:
First, their result is concrete, precise, and important;
Second, the method of proof is easy, and illustrates several tools
of quantum statistical mechanics well, including the transfer matrix method,
and quantum group symmetry.
For this reason, it is especially good for students new to the XXZ model to 
see that proof.

It is known generally that for $q=1$, there is no spectral gap in the thermodynamic limit.
To prove this, it suffices to consider excitations of the all-up-spin state, since the 
ground states of the isotropic model are all unitarily equivalent
(c.f.\ \cite{KN3}).
In this case, one can restrict attention to the ``spin-wave'' excitations.
Although literally we should consider a linear spin chain of length $L$,
it is mathematically simpler to consider a periodic spin chain.
Then the spin waves are 
$$
\psi^{\textrm{sw}}(k) = \sum_{x=1}^L e^{i k x} S_x^- \ket{\textrm{all up}}_{[1,L]}\, ,
$$
where $k$ is in the reciprocal lattice $\frac{2\pi}{L} \Ir/L$.
Then, defining the periodic spin chain
$$
H^{\textrm{XXX}}_{\Ir / L} 
  = \sum_{x=1}^L \left(\frac{1}{4} - \boldsymbol{S}_x \cdot \boldsymbol{S}_{x+1}\right)\, ,
$$
where $\boldsymbol{S}_{L+1}$ is identified with $\boldsymbol{S}_1$, we have
$$
H^{\textrm{XXX}}_{\Ir / L} \psi^{\textrm{sw}}(k) 
  = 2 \sin^2 \frac{k}{2}\ \psi^{\textrm{sw}}(k)\, .
$$
The actual ground state in this sector corresponds to $k=0$.
But an orthogonal excitation obtained by taking $k = 2 \pi /L$ gives an energy
of roughly $2 \pi^2/L^2$.
So the spectral gap decays at least as fast as $2 \pi/L^2$ for this spin chain.
We should actually be asking what is the spectral gap for the XXX model with free boundary
conditions.
Koma and Nachtergaele calculated this, as part of their exact calculation of the spectral
gap for the XXZ chain (since XXX is a special case of XXZ).
It turns out that the spectral gap is exactly the same as for the periodic chain, except 
that one can take $k=\pi/L$ instead of $2\pi/L$.
(For the free boundary conditions one can put a spin wave with wavelength equal to 
$2L$ instead of $L$.)

For $q<1$ one can ask whether the same procedure will work to produce arbitrarily
low-energy excitations of the unique ground state in the sector of one overturned
spin.
The answer is no.
Physically speaking, the anisotropy damps any such spin wave, because there is an
energy cost for not having the spin aligned nearly fully along the spin-up
or spin-down position.
At a quantitative level, one may actually diagonalize the kink Hamiltonian restricted
to this sector by using the transfer matrix method.
Koma and Nachtergaele begin their analysis of the spectral gap by 
doing just this.
We reproduce their analysis, now.
One defines for each $x \in [1,L]$, the vector
$D_x = S_x^-\, \ket{\textrm{all up}}_{[1,L]}$.
Then an arbitrary vector in the sector $\Hil_L^{(1)}$ of one overturned spin 
is given by the formula
$$
\psi = \sum_{x=1}^L a_x D_x\, ,
$$
for some sequence $a_1,\dots,a_L \in \Cx$.
It is easy to compute
\begin{align*}
H^{+-}_{[1,L]} D_1 
  &= \left(\frac{1}{2}  - A(\Delta)\right) D_1 - \frac{1}{2\Delta} D_2\, ,\\
H^{+-}_{[1,L]} D_x 
  &= D_x - \frac{1}{2\Delta} (D_{x+1} + D_{x-1})
  &\textrm{for} \quad x=2,3,\dots,L-1\, ,\\
H^{+-}_{[1,L]} D_L
  &= \left(\frac{1}{2} + A(\Delta)\right) D_L - \frac{1}{2\Delta} D_{L-1}\, .
\end{align*}
So if $\psi$ is an eigenvector of $H^{+-}_{[1,L]}$, say 
$H^{+-}_{[1,L]} \psi = \mathcal{E} \psi$, then 
\begin{gather*}
a_{y+1} = 2 \Delta (1 - \mathcal{E}) a_y - a_{y-1}\quad 
  \textrm{for}\quad 2\leq y\leq L-1\, ,\\
  a_2 = 2 \Delta [1/2 + A(\Delta) - \mathcal{E}] a_1\, ,\quad
  a_{L-1} = 2 \Delta [1/2 - A(\Delta) - \mathcal{E}] a_L\, .
\end{gather*}
These conditions can be rewritten
$$
\begin{pmatrix}a_{y+1} \\ a_y\end{pmatrix} 
  = T \begin{pmatrix} a_y \\ a_{y-1} \end{pmatrix}\, \quad \textrm{with}\quad
T = \begin{pmatrix}2 \Delta (1 - \mathcal{E}) & -1 \\ 1 & 0\end{pmatrix}\, 
$$
for $2 \leq y \leq L-1$, and
$$
a_2 = 2 \Delta(\frac{1}{2} + A(\Delta) - \mathcal{E})a_1\, ,\qquad
a_{L-1} = 2 \Delta (\frac{1}{2} - A(\Delta) - \mathcal{E})a_L\, .
$$
Combining these conditions, one has
\begin{equation}
\label{Eigenvalue condition}
a_L \begin{pmatrix}1 \\ 2 \Delta(\frac{1}{2} - A(\Delta) - \mathcal{E})\end{pmatrix}
  = a_1 T^{L-2} 
  \begin{pmatrix}2 \Delta(\frac{1}{2} + A(\Delta) - \mathcal{E}) \\ 1\end{pmatrix}\, .
\end{equation}
The matrix $T$ is called the transfer matrix for obvious reasons.
It has eigenvalues
$$
\lambda_{\pm} = \Delta(1 - \mathcal{E}) \pm \sqrt{\Delta^{2} (1 - \mathcal{E})^2-1}\, ,
$$
and eigenvectors 
$$
u_{\pm} = \begin{bmatrix}\lambda_{\pm} \\ 1\end{bmatrix}\, .
$$
In terms of these, equation \eq{Eigenvalue condition} can be rewritten
$$
a_L [ \beta_+ \lambda_- u_+ + \beta_- \lambda_+ u_-]
  = a_1 (\alpha_+ \lambda_+^{L-2} u_+ + \alpha_- \lambda_- u_-)\, ,
$$
where
\begin{align*}
\alpha_{\pm} &= \frac{1}{2} \left[1 \pm \frac{1}{\sqrt{\Delta^2(1 - \mathcal{E})^2 - 1}}
(\sqrt{\Delta^2-1}-\Delta \mathcal{E})\right]\, ,\\
\beta_{\pm} &= \frac{1}{2} \left[1 \pm \frac{1}{\sqrt{\Delta^2(1 - \mathcal{E})^2 - 1}}
(\sqrt{\Delta^2-1}+\Delta \mathcal{E})\right]\, .
\end{align*}
Or, since $u_+$ and $u_-$ are independent vectors
$$
a_L \beta_+ = a_1 \alpha_+ \lambda_+^{L-1}\, ,\quad
a_L \beta_- = a_1 \alpha_- \lambda_-^{L-1}\, .
$$
In case none of $\alpha_\pm$ and $\beta_\pm$ this means
$$
\lambda_+^{2L-2} = \frac{\alpha_-}{\alpha_+} \times \frac{\beta_+}{\beta_-}\, ,
$$
and taking into account the value of $\alpha_\pm$ and $\beta_\pm$, this gives
$\lambda_+^{2L} = 1$.
Hence $\lambda_+ = e^{i \pi l / L}$ where $l \in \Nl$.
For $l=1,\dots,L-1$, this leads to 
$$
\mathcal{E}_L(l) = 1 - \Delta^{-1} \cos(\pi l/L)\, .
$$
One can show that then the algebraic equations expressing the eigenvector are nonsingular,
so that these are actually eigenvalues.
Along with the ground state energy, which is zero, this gives all eigenvalues
for the sector $\Hil_{[1,L]}^{(1)}$.
Note the lowest excited energy in this sector is greater than $1 - \Delta^{-1}$.

Koma and Nachtergaele do not include this analysis just to show that spin waves
fail to have a lower energy than $1 - \Delta^{-1}$.
It is , in fact the basis of the calculation of the spectral gap for all sectors,
using the quantum group symmetry to extend the results. 
We observe that any $L-1$-dimensional irreducible representation of 
$SU_q(2)$ in the tensor product
$\Hil_L$ must intersect the sector with one downspin.
Thus, to extend the spectral gap in the sector $\Hil_{[1,L]}^{(1)}$ to the entire
Hilbert space $\Hil_{[1,L]}$, 
it would suffice to show that for every choice of magnetization, $1<m<L$, the
lowest excited state actually lies in an $L-1$-dimensional irreducible representation.
In that case the eigenvector corresponding to the lowest excited energy
can be raised by $E$ (i.e.\ the quantum group raising operator $S_q^+$)
to a state in the sector with one downspin.
Since $E$ commutes with $H^{+-}$, this means that the raised vector has the same energy
as the lowest excited energy in the sector of magentization $m$.
But, as we have shown the raised vector, which is an eigenvector in $\Hil_{[1,L]}^{(1)}$
other than the ground state, must have energy at least equal to 
$1 - \Delta^{-1} \cos(\pi/L)$.
The argument is reminiscent of the ordering-of-eigenvalues theorem of Lieb and Mattis
\cite{LM},
except that here it is applied to the ferromagnet instead of the antiferromagnet.
Koma and Nachtergaele prove that the main hypothesis is true, i.e.\ for the XXZ spin chain,
as well as for any other spin chain with a quantum group symmetry, satisfying certain 
conditions, the lowest excitations occur in the $L-1$-dimensional irreps of $SU_q(2)$.

Define
$$
\epsilon_n^{(m)} = \min_{\substack{0 \neq \psi \perp \ker H_n \\
\psi \in \Hil_{S^3 \geq \Spin n - m} }}
  \frac{\ip{\psi}{H_n \psi}}{\|\psi\|^2}
$$
for each $m=1,2,\dots,L-1$.
Then they prove the following lemma
\begin{lemma}
\label{KN1Lemma}
Consider an $\textrm{SU}_q(2)$ invariant spin-$\Spin$ ferromagnetic spin chain
of $L$ sites with a nearest neighbor Hamiltonian $H_L = \sum_{x=1}^{L-1} h_{x,x+1}$,
and for which the space of all ground states of a finite chain of $n$ sites is
the irreducible representation of maximal spin ($=n\Spin$), for
$2 \leq n\leq L$.
Let $\gamma_n$ denote the spectral gap of $H_n$ and let $\epsilon_n^{(m)}$ be as
defined.
If
$$
\epsilon_n^{(2\Spin)} \geq \epsilon_{n+1}^{(2\Spin)}\, ,
$$
for all $n$, $2 \leq n \leq L-1$, then
$$
\gamma_L = \epsilon_L^{(2\Spin)}\, .
$$
\end{lemma}
Note that if $\Spin = 1/2$, then this gives exactly what we want.
An induction proof on $L$ then shows that for every $L$, the spectral gap
in each sector $m=L-1,\dots,-L+1$ is $1 - \Delta^{-1} \cos(\pi/L)$.

One should note that Lemma \ref{KN1Lemma} leads to a natural conjectured
generalization.
One knows, by the classification of the finite-volume ground states and the
quantum group symmetry, that the ground states of the kink Hamiltonian comprise
the unique highest-dimensional ($L+1$-dimensional) irrep of $\textrm{SU}_q(2)$
in $\Hil_{[1,L]}$.
By the lemma, one also knows that the second lowest energy levels correspond
to a $L-1$-dimensional irrep.
Suppose one defines for each $j=\frac{L}{2},\frac{L-2}{2},\dots,\frac{1}{2}
\textrm{ or } 0$, $E(j)$ to be the lowest energy of any $2j+1$ irrep of 
$\textrm{SU}_q(2)$ in $\Hil_{[1,L]}$.
(Since $H^{+-}$ acts as a multiple of the Casimir operator on each irrep, we
can speak of the energy of an irrep.)
We know that $E\left(\frac{L}{2}\right) < E\left(\frac{L}{2} -1\right) < E(j)$
for any other $j$.
Based on numerical evidence (some of which is included in Figure \ref{XXZSpecArray}),
we conjecture that $E(j) < E(j')$ whenever $j>j'$.
To my knowledge, this conjecture was first formulated by Wolfgang Spitzer.

Since the spectral gap $\gamma_L = 1 - \Delta^{-1} \cos \pi/L$ is nonvanishing
in the limit $L \to \infty$, it is natural to suppose that there is a nonzero
spectral gap above the infinite-volume ground states, as well.
Koma and Nachtergaele prove that this is true, as well.
To understand the slight subtlety of this statement it helps to know 
the actual definition of the spectral gap for infinite-volume ground
states.
This definition is closely related to the definition of a ground state,
which we recall is that
$$
\lim_{\Lambda \nearrow \Ir}  \omega(A^* [H_\Lambda,A]) \geq 0
$$
for all strictly local observables $A$.
This means that no local perturbation of $\omega$ can lower the energy.
The locally perturbed state is obtained here as $\omega(A^* \cdot A)$.
If one has a GNS representation of a ground state as a vector $\Omega$
with zero energy, then for any strictly local observable $A$,
one has $\pi(A) \Omega \in \operatorname{Dom}(H^{\textrm{GNS}})$.
The reason for this is that
$$
H^{\textrm{GNS}} \pi(A) \Omega = [H^{\textrm{GNS}},\pi(A)] \Omega
  = \pi(\delta(A)) \Omega\, ,
$$
where 
$$
\delta(A) = \lim_{\Lambda \nearrow \Ir} [H_\Lambda,A]\, ,
$$
is also a strictly local observable (for $H$ a finite range interaction).
In fact the subspace of vectors $\pi(A) \Omega$ is a core for all powers of 
$H^{\textrm{GNS}}$.
This means that the orthogonal complement to the ground state space is spanned
(in the sense that the closure of the span equals the desired subspace)
by the vectors $H^{\textrm{GNS}} \pi(A) \Omega$.
The reason for this is that $H^{\textrm{GNS}}$ is self-adjoint and annihilates
all ground states.
Thus, the infinite volume analog of the definition
$$
\gamma = \inf_{\psi \perp \textrm{g.s. space}} \frac{\ip{\psi}{H \psi}}{\|\psi\|^2}\, ,
$$
is the definition: $\gamma$ is the spectral gap above the ground state $\omega$
iff $\gamma$ is the smallest number
making the following inequality true for all strictly local observables $A$,
$$
\ip{\Omega}{\pi(A^*) (H^{\textrm{GNS}})^3 \pi(A) \Omega}
  \geq \gamma \ip{\Omega}{\pi(A^*) (H^{\textrm{GNS}})^2 \pi(A) \Omega}\, .
$$
Koma and Nachtergaele prove that for any of the infinite volume ground states
($\omega_\uparrow$, $\omega_\downarrow$, kink states and antikink states),
the spectral gap $\gamma$ is equal to $1 - \Delta^{-1}$.
We omit their proof; however, see Section \ref{ProofSectionHigherSpinSpectralGap},
the proof of Theorem \ref{Our theorem}, for essentially the same argument.

\section{Exact Calculations for Ground States Properties in One Dimension}
\label{ExactCalculationsSection}
Several calculations for the XXZ model can be carried out exactly due to 
known algebraic identities.
We put these in two categories: those which 
use basic hypergeometric functions, which are the $q$-analogues
of hypergeometric functions; and those which use Gaussian polynomials,
which are the $q$-analogues of binomial coefficients.
The two cases are distinguished by the choice of the lattice $\Lambda$.
If $\Lambda$ is infinite, i.e.\ $\Lambda = \Ir$ or $\Ir_{\geq 1}$, then basic
hypergeomtric functions may apply.
If $\Lambda = \{1,\dots,L\}$ fo some $L < \infty$, then the best results
are obtained with Gaussian polynomials.
It seems that the hypergeometric functions give better, i.e.\ more algebraically
succinct, results than the Gaussian polynomials, which makes sense because
for $\Lambda$ infinite there are no boundary effects, so that one can take advantage
of the discrete translation symmetry of the Hamiltonian.
We will time, and again, use exact formulas for finite volumes in the subsequent
chapters, but here we would like to state and prove some simple results
for the infinit-volume ground states of the XXZ model.
It is hoped that this type of calculation, which uses standard identities from
the theory of basic hypergeometric functions, may serve as a model for future
exact calculations.

We recall the formula for the ground states bi-infinite lattice, $\Lambda = \Ir$.
These are vectors in the incomplete tensor product
$\bigotimes_{x \in \Ir} (\Cx_x^2,\Omega_x)$, where
$$
\Omega_x = \begin{cases}\ket{\downarrow} & x \leq 0\, ,\\ \ket{\uparrow} & x \geq 1\, .
\end{cases}
$$
We define $\Omega = \bigotimes \Omega_x$.
Then the ground states themselves are defined by
\begin{equation}
\label{bi-infinite canonical gs}
\Psi_0(n) = \sum_{k=0}^\infty 
  \sum_{\substack{\{x_1,\dots,x_k\} \subset \Ir_{\leq 0} \\
  \{y_1,\dots,y_{n+k}\} \subset \Ir_{\geq 1}}}
  q^{-(x_1+\dots+x_k) + (y_1 + \dots + y_{n+k})} \prod_{j=1}^k S_{x_j}^+
  \prod_{j=1}^{n+1} S_{y_j}^-\, \Omega\, ,
\end{equation}
for $n \geq 0$, and $\Psi_0(-n) = \mathcal{R} \mathcal{F} \Psi_0(n-1)$,
where $\mathcal{F}$ is the spin flip which sends each $\ket{\sigma_x}_x$
to $\ket{-\sigma_x}$, and $\mathcal{R}$ is reflection which interchanges 
$\sigma_{x}$ and $\sigma_{-x}$ for every $x=0,-1,-2,\dots$.
It is easier to handle the generating function for the ground states,
rather than the ground states themselves.
We define
$$
\Psi^{GC}_0(z) = \sum_{n\in \Ir} \Psi_0(n) z^n\, ,
$$
with the observation that
\begin{equation}
\label{bi-infinite gc gs}
\Psi^{GC}_0(z) = \prod_{x \in \Ir_{\leq 0}} (1 + q^{-x} z^{-1} S_x^-)
  \prod_{y \in \Ir_{\geq 1}} (1 + q^y z S_y^+)\ \Omega\, .
\end{equation}
We now begin calculations based on such states.

\subsection{Application of basic hypergeometric functions}
The hypergeometric functions are defined by
$$
_rF_s(a_1,\dots,a_r ; b_1,\dots,b_s;z)
  = \sum_{n=0}^\infty \frac{(a_1)_n \cdots (a_r)_n}{(b_1)_n \cdots (b_s)_n}\,
  \frac{z^n}{n!}\, ,
$$
where $r$ and $s$ are natural numbers, and $(a)_n$ is the shifted factorial
$$
(a)_0 = 1\, ,\quad
(a)_n = \prod_{k=1}^{n} (a+k-1)\, .
$$
They are ubiquitous in mathematical physics.
Basic hypergeometric functions are functions of $z$ whose coefficients,
instead of being rational numbers, are rational functions of another 
indeterminate $q$.
One defines the $q$-shifted factorial by
$$
(a;q)_0 = 1\, ,\quad
(a;q)_n = \prod_{k=1}^n (1 - a q^{k-1})\, .
$$
For $|q|<1$, one may also define the $q$-shifted factorial for $n=\infty$:
$$
(a;q)_\infty = \prod_{k=1}^\infty (1 - a q^{k-1})
$$
from which one has the alternative definition
$$
(a;q)_n = \frac{(a;q)_\infty}{(a q^n;q)_\infty}\, .
$$
The basic hypergeometric functions are
$$
_r \phi_s(a_1,\dots,a_r ; b_1,\dots,b_s;z)
  = \sum_{n=0}^\infty \frac{(a_1;q)_n \cdots (a_r;q)_n}
  {(b_1;q)_n \cdots (b_s;q)_n}\,
  \frac{[(-1) q^{\binom{n}{2}}]^{1+s-r}}{(q;q)_n} z^n\, .
$$
The standard reference for basic hypergeometric functions is \cite{GR}.
It contains the proofs of all the identities stated in this section.

The first thing we would like to do is find a formula for 
$$
Z(n) = \|\Psi_0(n)\|^2\,
$$
where $\Psi_0(n)$ is defined by \eq{bi-infinite canonical gs}.
It is certainly easier to calculate
$$
Z^{\textrm{GC}}(z) = \|\Psi_0^{\textrm{GC}}(z)\|^2\, .
$$
In fact, one immediately has the formula
\begin{align*}
Z^{\textrm{GC}}(z) 
  &= \prod_{x \in \Ir_{\leq 0}} (1 + q^{-2x} |z|^{-2})
  \prod_{y \in \Ir_{\geq 1}} (1 + q^{2y} |z|^2) \\
  &= (1 + |z|^{-2}) \prod_{x=1}^\infty (1 + q^{2x} |z|^{-2})
  \prod_{y=1}^\infty (1 + q^{2y} |z|^2) \\
  &= (1 + |z|^{-2}) (- q^2 |z|^{-2};q^2)_\infty (-q^2 |z|^2;q^2)_\infty\, .
\end{align*}
We wish to extract the formulae for the norm of the ground states from the 
formula for the norm of their generating function.
We recall that $\Psi_0(-n) = \mathcal{R} \mathcal{F} \Psi_0(n-1)$,
where $\mathcal{R} \mathcal{F}$ is an isometry of the incomplete tensor
product.
Thus $Z(-n)=Z(n-1)$ for all $n$.
This can also be seen from the generating function since
\begin{equation}
\label{Z relation 1}
Z^{\textrm{GC}}(z^{-1}) = (1 + |z|^2) (-q^2 |z|^2;q^2)_\infty
 	(-q^2 |z|^{-2};q^2)_\infty
	= z^2 Z^{\textrm{GC}}(z)\, .
\end{equation}
Since 
$$
Z^{\textrm{GC}}(z) = \sum_{n=-\infty}^\infty Z(n) |z|^{2n}\, ,
$$
this does indeed show that $Z(-n) = Z(1-n)$.
In a similar vein, 
\begin{align*}
Z^{\textrm{GC}}(q z) &= (1 + q^{-2} |z|^{-2})
	(-|z|^{-2};q^2)_\infty (q^4 |z|^2;q^2)_\infty \\
  &= (1 + q^{-2} |z|^{-2}) (1 + |z|^{-2}) (-q^2 |z|^{-2};q^2)_\infty
  \frac{(q^2 |z|^2;q^2)_\infty}{(1 + q^2 |z|^2)} \\
  &= \frac{1 + q^{-2} |z|^{-2}}{1 + q^2 |z|^2} Z^{\textrm{GC}}(z) \\
  &= q^{-2} |z|^{-2} Z^{\textrm{GC}}(z)\, .
\end{align*}
Thus 
\begin{equation}
\label{Z relation 2}
Z^{\textrm{GC}}(z) = q^2 |z|^2 Z^{\textrm{GC}}(q z)\, ,
\end{equation} 
which implies
$$
Z(n) = q^{2n} Z(n-1)\, .
$$
This implies
$$
Z(n) = q^{n(n+1)} Z(0)\, .
$$
This is remarkable because it gives us the formula for the normalizations
modulo one constant $Z(0)$.
To actually calculate $Z(0)$ requires the first identity from the theory of
basic hypergeometric functions.

The binomial theorem
$$
(1 - z)^{-a} = \sum_{n=0}^\infty \frac{(a)_n}{n!} z^n =:\, _1F_0(a;-;z)
$$
is the most basic identity in the theory of hypergeometric series.
It has the $q$-analogue
$$
_1\phi_0(a;-;z) := \sum_{n=0}^\infty \frac{(a;q)_n}{(q;q)_n} z^n
  = \frac{(a z;q)_\infty}{(z;q)_\infty}\, ,
$$
which was proved independently by Cauchy, Heine, and others.
If we let $z = t/a$, then
$$
\frac{(t;q)_\infty}{(t/a;q)_\infty} 
  = \sum_{n=0}^\infty \frac{(a;q)_n}{a^n} \frac{t^n}{(q;q)_n}\, .
$$
We observe that $\lim_{a \to \infty} (t/a;q)_\infty = (0;q)_\infty = 1$, while
$$
\lim_{a \to \infty} \frac{(a;q)_n}{a^n}
  = \lim_{a \to \infty} (-1)^n (1-a^{-1})(q-a^{-1})\cdots(q^{n-1} - a^{-1})
  = (-1)^n q^{\binom{n}{2}}\, .
$$
Thus we have
$$
(t;q)_\infty = \sum_{n=0}^\infty \frac{(-1)^n q^{\binom{n}{2}}}
  {(q;q)_n} t^n\, .
$$

Replacing $q$ by $q^2$ and $t$ by $-q^2 |z|^{-2}$ and $q^2 |z|^2$, we obtain
$$
Z^{\textrm{GC}}(z) = (1 + |z|^{-2}) 
  \sum_{n=0}^\infty \frac{q^{n(n+1)}}{(q^2;q^2)_n} |z|^{-2n}
  \sum_{m=0}^\infty \frac{q^{m(m+1)}}{(q^2;q^2)_m} |z|^{2m}\, .
$$
Since $Z(0)$ is the constant term in this series we have
\begin{align*}
Z(0) &= \sum_{n=0}^\infty \left[\frac{q^{2n(n+1)}}{(q^2;q^2)_n^2}
  + \frac{q^{n(n+1) + (n+1)(n+2)}}{(q^2;q^2)_n (q^2;q^2)_{n+1}}\right] \\
  &= \sum_{n=0}^\infty \frac{q^{2n(n+1)}}{(q^2;q^2)_n^2}
  \left[1 + \frac{q^{(n+1)(n+2) - n(n+1)}}{(q^2;q^2)_{n+1}/(q^2;q^2)_n}\right] \\
  &= \sum_{n=0}^\infty \frac{q^{2n(n+1)}}{(q^2;q^2)_n^2}
  \left[1 + \frac{q^{2(n+1)}}{1-q^{2(n+1)}}\right] \\
  &= \sum_{n=0}^\infty \frac{q^{2n(n+1)}}{(q^2;q^2)_n (q^2;q^2)_{n+1}}\, .
\end{align*}
This series may be evaluated with the second main identity.

One may recall the Gauss summation formula
$$
_2F_1(a,b;c;1) := \sum_{n=0}^\infty \frac{(a)_n (b)_n}{n! (c)_n}
  = \frac{\Gamma(c) \Gamma(c-a-b)}{\Gamma(c-a) \Gamma(c-b)}\, ,\qquad
\real(c-a-b) > 0\, .
$$
Heine's $q$-analogue is the following
$$
_2\phi_1(a,b;c;q,\frac{c}{ab}) 
  := \sum_{n=0}^\infty \frac{(a;q)_n (b;q)_n}{(q;q)_n (c;q)_n} 
  \left(\frac{c}{ab}\right)^n
  = \frac{(\frac{c}{a};q)_\infty (\frac{c}{b};q)_\infty}
  {(c;q)_\infty (\frac{c}{ab};q)_\infty}\, .
$$
If we set $c=q^2$ and take $a,b, \to \infty$ then the right hand side of the
equation above becomes $(q^2;q)_\infty^{-1}$, and since
$$
\lim_{\substack{a\to \infty \\ b \to \infty}} \frac{(a;q)_n}{a^n} 
\frac{(b;q)_n}{b^n} = q^{n(n-1)}\, ,
$$
left hand side becomes
$$
\sum_{n=0}^\infty \frac{q^{n(n-1)}}{(q;q)_n (q^2;q)_n} q^{2n}
 = \sum_{n=0}^\infty \frac{q^{n(n+1)}}{(q;q)_n (q;q)_{n+1}} (1-q)\, .
$$
This means
$$
\sum_{n=0}^\infty \frac{q^{n(n+1)}}{(q;q)_n (q;q)_{n+1}} 
  = \frac{1}{(1-q)(q^2;q)_\infty} = \frac{1}{(q;q)_\infty}\, .
$$
Replacing $q$ by $q^2$, again, we find
$$
Z(0) = \sum_{n=0}^\infty \frac{q^{2n(n+1)}}{(q^2;q^2)_n (q^2;q^2)_{n+1}}
  = \frac{1}{(q^2;q^2)_\infty}\, ,
$$
and hence
\begin{equation}
\label{PartitionFunctionIdentity}
\boxed{
Z(n) = \frac{q^{n(n+1)}}{(q^2;q^2)_\infty}\, .}
\end{equation}

The next natural question one would ask is what is the magnetic profile
of $\Psi_0(n)$?
In other words, for each $x \in \Ir$, what is the expectation
$$
\EXP{\boldsymbol{S}_x}{n} := \frac{\ip{\Psi_0(n)}{\boldsymbol{S}_x \Psi_0(n)}}
  {\|\Psi_0(n)\|^2}\, ?
$$
By the rotational symmetry about the $\boldsymbol{e}_3$-axis, we know a priori
that $\EXP{S_x^1}{n} = \EXP{S_x^2}{n} = 0$.
What remains is to calculate $\EXP{S_x^3}{n}$.
Once again the formula for the generating function is quite trivial.
Since $\Psi^{\textrm{GC}}_0(z)$ is a simple tensor product,
it is easy to calculate
$$
\EXP{S_x^3}{z}^{\textrm{GC}}
  = \frac{(\bra{\uparrow} + q^x z \bra{\downarrow}) S_x^3 
  (\ket{\uparrow} + q^x z \ket{\downarrow})}{1 + q^{2x} |z|^2}
  = \frac{1}{2} \cdot \frac{1 - q^{2x} |z|^2}{1 + q^{2x} |z|^2}\, .
$$
We observe that 
\begin{align*}
\EXP{S_x^3}{z}^{\textrm{GC}}
  &= Z^{\textrm{GC}}(z)^{-1} 
  \sum_{n=-\infty}^\infty \EXP{S_x^3}{n} Z(n) |z|^{2n} \\
  &= \frac{Z(0)}{Z^{\textrm{GC}}(z)} 
  \sum_{n=-\infty}^\infty \EXP{S_x^3}{n} q^{n(n+1)} |z|^{2n}\, .
\end{align*}
Now from its formula, we see that
$$
\EXP{S_{x+1}^3}{z}^{\textrm{GC}} = \EXP{S_x^3}{qz}^{\textrm{GC}}\, ,
$$
and
$$
\EXP{S_{-x}^3}{z}^{\textrm{GC}} = \EXP{S_x^3}{z^{-1}}^{\textrm{GC}}\, .
$$
Using these relations with \eq{Z relation 1} and \eq{Z relation 2},
we can derive
$$
\EXP{S_x^3}{n} = \EXP{S_0^3}{n-x}
$$
and
$$
\EXP{S_x^3}{-n} = - \EXP{S_{-x}^3}{n-1}\, .
$$
From this we determine 
$$
\EXP{S_x^3}{n} = \operatorname{sign}(n-x) q^{2nx - x(x-1)} 
  \EXP{S_0^3}{|n+\frac{1}{2}-x|-\frac{1}{2}}\, .
$$
This means that it suffices to calculate $\EXP{S_0^3}{n}$
for $n=0,1,2,\dots$.

We observe
\begin{align*}
Z^{\textrm{GC}}(z) \EXP{S_0^3}{z}^{\textrm{GC}} & := 
\sum_{n = -\infty}^\infty \EXP{S_0^3}{n} Z(0) q^{n(n+1)} |z|^{2n} \\
  &= \frac{1}{2} \cdot \frac{1 - |z|^2}{1 + |z|^2} 
  (1 + |z|^{-2}) 
  \sum_{n=0}^\infty \frac{q^{n(n+1)}}{(q^2;q^2)_n} |z|^{-2n} 
  \sum_{m=0}^\infty \frac{q^{m(m+1)}}{(q^2;q^2)_m} |z|^{2m} \\
  &= -\frac{1}{2} (1 - |z|^{-2}) 
  \sum_{n=0}^\infty \frac{q^{n(n+1)}}{(q^2;q^2)_n} 
  |z|^{-2n} 
  \sum_{m=0}^\infty \frac{q^{m(m+1)}}{(q^2;q^2)_m} |z|^{2m} \, .
\end{align*}
This implies
\begin{align*}
\EXP{S_0^3}{n} &= - \frac{1}{2 Z(0)}
  \sum_{k=0}^\infty \frac{q^k(k+1)}{(q^2;q^2)_k}
  \left[ \frac{q^{(n+k)(n+k+1)}}{(q^2;q^2)_{n+k}} 
  - \frac{q^{(n+k+1)(n+k+2)}}{(q^2;q^2)_{n+k+1}} \right]\\
  &= - \frac{1}{2} q^{-n(n+1)} (q^2;q^2)_\infty
  \sum_{k=0}^\infty \frac{q^{k(k+1) + (n+k)(n+k+1)}}{(q^2;q^2)_k (q^2;q^2)_{n+k}}
  \left[1 - \frac{q^{n+k+1}}{1 - q^{n+k+1}}\right] \\
  &= - \frac{1}{2} q^{-n(n+1)} (q^2;q^2)_\infty
  \sum_{k=0}^\infty \frac{q^{k(k+1) + (n+k)(n+k+1)}(1-2 q^{n+k+1})}
  {(q^2;q^2)_k (q^2;q^2)_{n+k+1}}\, .
\end{align*}
For future reference we repeat
\begin{equation}
\label{profile -- Heine}
\EXP{S_0^3}{n} =
- \frac{1}{2} q^{-n(n+1)} (q^2;q^2)_\infty
  \sum_{k=0}^\infty \frac{q^{k(k+1) + (n+k)(n+k+1)}(1-2 q^{n+k+1})}
  {(q^2;q^2)_k (q^2;q^2)_{n+k+1}}\, .
\end{equation}
From this formidable looking series we can extract a simple question:
What is
$$
(z;q)_\infty \sum_{k=0}^\infty \frac{q^{k^2} z^k}{(q;q)_k (z;q)_k}\, ?
$$
By this we mean, is there a simple product formula which equals the
sum above?
As far as we know, this question is not answered in \cite{GR}.
This is related to \eq{profile -- Heine} by the fact that
\begin{align*}
& q^{-n(n+1)} (q^2;q^2)_\infty
  \sum_{k=0}^\infty \frac{q^{k(k+1) + (n+k)(n+k+1)}}
  {(q^2;q^2)_k (q^2;q^2)_{n+k}} \\
&\qquad \qquad =(q^2;q^2)_\infty q^{-n(n+1)}
  \sum_{k=0}^\infty \frac{q^{2k^2 + (2n+2)k + n(n+1)}}
  {(q^2;q^2)_k (q^{2n+2};q^2)_k (q^2;q^2)_n} \\
&\qquad \qquad =(q^2;q^2)_\infty q^{-n(n+1)}
  \sum_{k=0}^\infty \frac{q^{2k^2 + (2n+2)k + n(n+1)}}
  {(q^2;q^2)_k (q^{2n+2};q^2)_k (q^2;q^2)_n} \\
&\qquad \qquad = \frac{(q^2;q^2)_\infty}{(q^2;q^2)_n} 
  \sum_{k=0}^\infty \frac{q^{2k^2} (q^{2n+2})k}
  {(q^2;q^2)_k (q^{2n+2};q^2)_k} \, .
\end{align*}
So the calculation of this term is the same as our simplified question,
with $q \to q^2$ and $z \to q^{2n+2}$.
The second term in \eq{profile -- Heine} is also obtained by setting 
$z \to q^{2n+4}$.

The answer to our question is not given by Heine's $q$-analogue of the Gauss
summation formula, which states
$$
(q z;q)_\infty \sum_{k=0}^\infty \frac{q^{k^2} z^k}{(q;q)_k (qz;q)_k} = 1\, .
$$
But, if we define 
$$
f(z) = (z;q)_\infty \sum_{k=0}^\infty \frac{q^{k^2} z^k}{(q;q)_k (z;q)_k}\, ,
$$
then by Heine's formula,
\begin{align*}
f(z) &= (z;q)_\infty \sum_{k=0}^\infty \frac{q^{k^2} z^k}{(q;q)_k (qz;q)_k} \cdot 
  \frac{1 - q^k z}{1-z} \\
  &= \frac{(z;q)_\infty}{1-z} \left[\frac{1}{(qz;q)_\infty}
  - z \sum_{k=0}^\infty \frac{q^{k^2} (q z)^k}{(q;q)_k (qz;q)_k}\right] \\
  &= 1 - z f(q z)\, .
\end{align*}
This allows the continued-product formula
\begin{align*}
f(-z) &= 1 + z f(-qz) \\
  &= 1 + z(1 + qz f(-q^2 z)) \\
  &= 1 + z(1 + qz( 1 + q^2 z f(-q^3 z))) \\
  &= \dots \\
  &= 1 + z(1 + qz (1 + q^2 z(1 + q^3 z( \cdots (1 + q^n z( \cdots \\
  &= 1 + z + q z^2 + q^3 z^3 + q^6 z^4 + \dots + q^{\binom{n}{2}} z^n + \dots
\end{align*}
I.e.\ 
$$
f(z) = \sum_{k=0}^\infty (-1)^k q^{\binom{k}{2}} z^k\, .
$$
(By a similar argument, if we let 
$$
f_n(z) = (q^n z;q)_\infty \sum_{k=0}^\infty 
  \frac{q^{k^2} z^k}{(q;q)_k (q^n z;q)_k}\, ,
$$
then we have 
$$
f_n(z) = \sum_{k=0}^n (-1)^k \frac{(q^{1-n};q)_k}{(q;q)_k} q^{\binom{k}{2}}
  (q^n z)^k\, ,
$$
for all $n \in \Ir$.
For the special case that $n$ is a positive integer, this leads to a terminating
series
$$
f_{n}(z) = \sum_{k=0}^{n-1} \qbinom{n}{k}{q} q^{k^2} z^k\, ,
$$
where
$$
\qbinom{n}{k}{q} = \frac{(q;q)_n}{(q;q)_k (q;q)_{n-k}}
$$
is the Gaussian polynomial also called $q$-binomial coefficient.
Of course we are interested in the case $n=0$, which is not terminating.)

We now observe that
$$
f(z) - z f(qz) = 1 - 2 z f(q z)
  = 1 + 2 \sum_{n=1}^\infty (-1)^n z^n q^{\binom{n}{2}}\, .
$$
But by a very simple calculation one may verify
$$
\EXP{S_0^3}{n} = -\frac{1}{2} \left[ f_(q^{2n+2};q^2) - q^{2n+2} f(q^{2n+4};q^2)
  \right]\, ,
$$
where we have put a second $q^2$ to remind ourselves that we have replaced
$q$ by $q^2$ in the definition of $f$.
This means
$$
\EXP{S_0^3}{n} = -\frac{1}{2} \left[ 1 - 2 q^{2n+2} \sum_{k=0}^\infty
  (-1)^k q^{k(k+3+2n)}\right]\, .
$$
From this one may determine
$$
\EXP{S_x^3}{n} = 
  \begin{cases}
  -\frac{1}{2}  + q^{2(n+1-x)} \sum_{k=0}^\infty
  (-1)^k q^{k(k+1+2(n+1-x))} &\quad \textrm{if } n\geq x\, , \\
  +\frac{1}{2} - q^{2(x-n)} \sum_{k=0}^\infty
  (-1)^k q^{k(k+1+2(x-n))} &\quad \textrm{if } n\leq x-1\, . 
  \end{cases}
$$
From this formula, one can see the exact behavior of the magnetic profile,
and particularly that the interface is exponentially localized.
I.e.\ the third component of spin approaches $+1/2$ and $-1/2$ as
one moves to the right or left of the interface, exponentially fast
and with rate $\ln\frac{1}{q^2}$.


\newpage
\pagestyle{myheadings} 
\markright{  \rm \normalsize CHAPTER 3. \hspace{0.5cm} 
The spectral gap for the 1d, Spin-$\Spin$ model : $\Spin > \frac{1}{2}$}
\chapter{The spectral gap for the 1d, Spin-$\Spin$ model : $\Spin > \frac{1}{2}$}
\thispagestyle{myheadings}

\section{Acknowledgements}
The results of this section are due to Bruno Nachtergaele, Tohru Koma,
and myself.
More specifically, the proof of the existence of a nontrivial gap was originally
done entirely by Nachtergaele and Koma for the case of $\Spin = 1$, which I then
generalized to $\Spin > 1$.
The numerical analysis of the spectral gap, using the lower bounds obtained by
Koma and Nachtergaele's proof, was carried out by me under 
the direction of Bruno Nachtergaele.
I will also present other numerical results in support of conjectures made
by Bruno Nachtergaele and myself.

\newpage

\section{Introduction}
The purpose of this chapter is to prove the existence of a 
spectral gap above the infinite-volume ground states for the ferromagnetic
XXZ chain for every spin $\Spin = 1,\frac{3}{2},2,\frac{5}{2},\dots$.
If $\Lambda = [a+1,b]$, the spin-$\Spin$ XXZ Hamiltonian on the finite chain
$\Lambda$ is
$$
H^{\textrm{XXZ}}_\Lambda = - \sum_{x=a+1}^{b-1} (\Delta^{-1} S_x^1 S_{x+1}^1
  + \Delta^{-1} S_x^2 S_{x+1}^2 + S_x^3 S_{x+1}^3- \Spin^2)\, .
$$
The operators $S_x^\alpha$ are the spin-$\Spin$ matrices acting on the
site $x$.
For the case we are considering, which is a ferromagnetic model with
spin alignment preferred along the third axis, $\Delta >1$.
We define another parameter $q$ to be the solution to the equation
$\Delta = \frac{1}{2} (q + q^{-1})$ with the restriction that 
$0\leq q <1$.
It has been shown that the correct finite volume Hamiltonians for
determining behavior above the infinite-volume ground states 
have a boundary field, with opposite direction on the opposite endpoints of
$\Lambda$.
Precisely, the finite volume Hamiltonian we will consider is
\begin{equation}
\label{Finite Hamiltonian}
H^{\Spin}_\Lambda = H^{\textrm{XXZ}}_\Lambda + 
  \Spin \sqrt{1 - \Delta^{-2}} (S_{a+1}^3 - S_b^3)\, .
\end{equation}
It is important to note that the boundary terms can be included in a nearest
neighbor interaction
$$
h^{\Spin}_{x,x+1} 
  = - \Delta^{-1} S_x^1 S_{x+1}^1
    - \Delta^{-1} S_x^2 S_{x+1}^2 - S_x^3 S_{x+1}^3
    - \Spin \sqrt{1 - \Delta^{-2}} (S_b^3 - S_{a+1}^3)\, .
$$

We gather here a few facts from Chapter 2, which we will
need for our current arguments.
We will denote the finite-volume ground states by
\begin{equation}
\label{ASW formula}
\Psi^{\Spin}_0([1,L],N) 
  = \sum_{\substack{\{n_x\}_{x=1}^L \\ \sum_x n_x = N}} q^{\sum_x x n_x}
  \prod_{x=1}^L \binom{2 \Spin}{n_x}^{1/2} \ket{\Spin-n_1,\Spin-n_2,\dots,
  \Spin - n_L}\, .
\end{equation}
This is the unique ground state of the kink Hamiltonian $H_{[1,L]}^{\Spin}$
which is simultaneously an eigenvector of
$S^3_{\textrm{tot}}$ with eigenvalue $M = \Spin L - N$.
For spin-$\frac{1}{2}$, we denote
the GNS vectors for the infinite-volume kink ground states as
\begin{equation}
\begin{split}
\label{Bi-infinite spin-1/2 ground state}
&\Psi^{1/2}_0(\Ir,N) = \sum_{k = 0 \vee N}^\infty
  \sum_{x_1 < x_2 < \dots < x_k \leq 0 < y_1 < y_2 < \dots < y_{k-N}}
  q^{\sum_{i=1}^{k-N} y_i - \sum_{i=1}^k x_i} \times \\
& \hspace{85 pt} \times \prod_{i=1}^k S_{x_i}^+ 
  \prod_{i=1}^{k-N} S_{y_i}^-\, \ket{\Omega}\, ,
\end{split}
\end{equation}
where
$$
\ket{\Omega} = \bigotimes_{x \in \Ir} \ket{\Omega(x)}_x\, ,\qquad
\ket{\Omega_x} = \begin{cases} \ket{\uparrow} & x>0\\ 
\ket{\downarrow} & x\leq 0\end{cases}\, .
$$
In \cite{KN3}, it was shown that the zero-energy states introduced by
Gottstein \& Werner are the complete list of ground states for 
$\Spin=\frac{1}{2}$. More generally, the following theorem is proved
for arbitrary $\Spin \in \frac{1}{2} \Nl$:
\begin{theorem}{(Koma \& Nachtergaele)}
\label{Complete ground states} 
For the spin-$\Spin$ XXZ ferromagnetic chain with the anisotropic
coupling $\Delta > 1$, the following statements are valid:
There are two translationally invariant pure ground states,
namely $\omega_\uparrow$ and $\omega_\downarrow$.
Any pure infinite-volume ground state that is not translation invariant
is either a kink, or an antikink ground state, belonging to the set
described in \cite{GW}.
\end{theorem}

In \cite{KN1} the spectral gap for the finite-volume kink ground states
in the case $\Spin = \frac{1}{2}$ was calculated exactly.
This was then used to obtain the spectral gap above the infinite-volume 
ground states.
\begin{proposition}{(Koma \& Nachtergaele)}
\label{Spin-1/2 spectral gap}
For the $\textrm{SU}_q(2)$ invariant spin-$\frac{1}{2}$ ferromagnetic XXZ chain
with $L \geq 2$ and $\Delta \geq 1$, one has
$$
\gamma_L = 1 - \Delta^{-1} \cos(\pi/L)\, .
$$
Above any of the infinite-volume ground states the spectral gap is
$$
\gamma = 1 - \Delta^{-1}\, .
$$
\end{proposition}
We recall that this formula  
is specific to $\Spin=\frac{1}{2}$, because it relies on 
the $\textrm{SU}_q(2)$-symmetry of the spin-$\frac{1}{2}$ model.
The quantum group symmetry is absent for all other choices of $\Spin$.
More specifically, for spin-$\frac{1}{2}$, it is proved that the spectral gap is
a constant independent of the sector, except that the spectral gap doesn't exist
in the all up-spin or all down-spin sector because these are each one-dimensional.
In contrast, for $\Spin>\frac{1}{2}$, the spectral
gap will not be the same in all sectors; instead it depends
on the ``filling factor'' of $N$. 
This means the spectral gap is an even, $2\Spin$-periodic
function of the sector, for sectors with $|M|\ll L$, in the limit that $L \to \infty$.
There are some general techniques for estimating the spectral gap
for quantum spin systems in \cite{FNW}, \cite{Na1}, \cite{Na2}, none of which  
depend on the quantum group symmetry.
These techniques do not seem to be directy applicable to the XXZ model for 
$\Spin>\frac{1}{2}$, or at least not more directly applicable than the argument
which we present here.

The main theorem for this chapter is the following:
\begin{theorem}
\label{Our theorem}
For any half-integer $\Spin > \frac{1}{2}$, 
and any $\Delta > 1$, there exists a nonvanishing spectral gap 
$\gamma > 0$ above all the infinite-volume ground states of the spin-$\Spin$
ferromagnetic XXZ model.
Moreover, above the translation-invariant ground states the gap is exactly
$2 \Spin (1 - \Delta^{-1})$.
\end{theorem}

The proof relies upon the existence of a gap for $\Spin=\frac{1}{2}$,
and the explicit formulas for the infinite-volume kink states.
Part of the proof is a result which says that the gap 
$\gamma$ for the XXZ Hamiltonian is bounded below by the gap 
$\tilde{\gamma}$ of a Hamiltonian on a much reduced state space.
In Section \ref{Numerics}, we use this technique to develop numerical
recipes for estimating the spectral gap, and present the results
for $\Spin = 1,3/2,2,5/2$.

\section{Spin Ladder Representation}
We wish to reduce the problem of calculating the spectral gap for 
$\Spin>1/2$, to a form where we can make use of Proposition
\ref{Spin-1/2 spectral gap}.
The way we do this is by replacing the spin-$\Spin$ chain with a spin-$1/2$
ladder with $2\Spin$ legs.
This will have the advantage that we obtain lower bounds by disregarding some
of the interactions in the spin ladder.
By neglecting these bonds, the resulting spin system 
ceases to be equivalent to a spin-$\Spin$ spin chain.
So the spin ladder representation is a key part of the proof.

The state space for the spin-$\Spin$ XXZ Hamiltonian is 
$\bigotimes_{x \in \Lambda} \Hil_x$, where
$\Hil_x = \Cx_x^{2\Spin+1}$ is equipped with an irreducible
representation of $SU(2)$.
The representation $\Hil_x$ 
can be rewritten as the heighest-weight irreducible
representation in the tensor product 
of two-dimensional representations $\bigotimes_{m=1}^{2\Spin} 
\Cx^2_{(x,m)}$.
The subscript $(x,m)$ is just a placeholder. 
We denote 
\begin{gather*}
\Hil^{(m)}_x := \Cx^2_{(x,m)}\, ,\quad
\Hil^{\Spin}_x := \bigotimes_{m=1}^{2\Spin} \Hil^{(m)}_x\, ,\\
\Hil^{(m)}_\Lambda := \bigotimes_{x \in \Lambda} \Hil_x^{(m)}\, ,\quad 
\Hil^{\Spin}_\Lambda := \bigotimes_{x \in \Lambda}
\bigotimes_{m=1}^{2\Spin} \Hil_x^{(m)}\, .
\end{gather*}
The following diagram should help with the definition:
\begin{equation*}
\begin{array}{ccccccccccc}
\Hil_1^{(1)} & \otimes & \Hil_2^{(1)} & \otimes 
  & \dots & \otimes & \Hil_L^{(1)} & = & \Hil_{[1,L]}^{(1)}\\
\otimes & & \otimes & & & & \otimes && \otimes\\
\Hil_1^{(2)} & \otimes & \Hil_2^{(2)} & \otimes 
  & \dots & \otimes & \Hil_L^{(2)} & = & \Hil_{[1,L]}^{(2)}\\
\otimes & & \otimes & & & & \otimes && \otimes\\
\vdots & & \vdots & & \ddots & & \vdots && \vdots\\
\otimes & & \otimes & & & & \otimes && \otimes\\
\Hil_1^{(2\Spin)} & \otimes & \Hil_2^{(2\Spin)} & \otimes 
  & \dots & \otimes & \Hil_L^{(2\Spin)} & = & \Hil_{[1,L]}^{(2\Spin)}\\
\shortparallel & & \shortparallel & & & 
  & \shortparallel && \shortparallel  \\
 \Hil_1^{\Spin} & \otimes & \Hil_2^{\Spin}  
  & \otimes & \dots & \otimes & \Hil_L^{\Spin} & = & \Hil_{[1,L]}^{\Spin} 
\end{array}
\end{equation*}
We denote the projection of $\Hil_x^{\Spin}$ onto $\Hil_x$
by $P^{\textrm{Sym}}_x$.
Alternatively, this is the symmetrization projection defined
by its action on simple tensors
$$
P^{\textrm{Sym}}_x \bigotimes_{m=1}^{2\Spin} \ket{v_m}_{(x,m)}
  = \frac{1}{2\Spin!} \sum_{\pi \in \mathfrak{S}_{2\Spin}} 
  \bigotimes_{m=1}^{2\Spin} \ket{v_{\pi^{-1}(m)}}_{(x,m)}\, .
$$
This simply indicates the well-known fact that the highest-weight irreducible 
representation in a tensor product of two-dimensional representations of 
$\textrm{SU}(2)$ is the subspace of symmetric tensors.
The subscript refers to the order of the tensor factors.
We define $P^{\textrm{Sym}}_\Lambda = \prod_{x \in \Lambda} 
P^{\textrm{Sym}}_x$.
(Since $\{P^{\textrm{Sym}}_x\}_{x \in \Lambda}$ is a commuting family,
the order of the product does not matter.)
Then the Hamiltonian $H^{\Spin}_\Lambda$ can be recovered as 
\begin{equation}
\label{Equivalent Hamiltonians}
P^{\textrm{Sym}}_\Lambda H^{\Spin}_\Lambda P^{\textrm{Sym}}_\Lambda
  = 2 \Spin P^{\textrm{Sym}}_\Lambda 
  \sum_{m=1}^{2\Spin} \sum_{x,x+1 \in \Lambda} h^{1/2}_{(x,m),(x+1,m)}
  P^{\textrm{Sym}}_\Lambda\, .
\end{equation}
This is not technically the same operator since the displayed operator
has a larger domain, but it is identically zero on 
$\ker P^{\textrm{Sym}}_\Lambda$, which is the orthogonal complement of
$\mathcal{D}^{\Spin}_\Lambda$ in $\Hil^{\Spin}_\Lambda$.
The state space $\Hil^{\Spin}_\Lambda$ should be thought of as a spin
ladder with $2 \Spin$ legs defined by the state space $\Hil^{(m)}_\Lambda$,
$m=1,\dots,2\Spin$.
The Hamiltonian $\tilde{\Hil}^{\Spin}_\Lambda$ acts on each of the legs 
separately, but in $\Hil^{\Spin}_\Lambda$ the legs are coupled through the
conjugation by $P^{\text{Sym}}_\Lambda$.
Each $P^{\text{Sym}}_x$ acts on the rung $\Hil^{\Spin}_x$, 
and connects the states on different legs.

We define the operators
$$
\tilde{H}^{(m)}_\Lambda = \sum_{x,x+1 \in \Lambda} h^{1/2}_{(x,m),(x+1,m)}\, ,
\quad m=1,2,\dots,2\Spin\, ,
$$
and $\tilde{H}^{\Spin}_\Lambda = \sum_{m=1}^{2\Spin} \tilde{H}^{(m)}_\Lambda$.
By equation \eq{Equivalent Hamiltonians}, the ground states
of $H^{\Spin}_\Lambda$ are the subset of ground states of 
$\tilde{H}^{\Spin}_\Lambda$ in the range of 
$P^{\textrm{Sym}}_\Lambda$.
We define this subspace as $\GS^{\Spin}_\Lambda$.
Then the spectral gap for $H^{\Spin}_\Lambda$ is defined as
$$
\gamma^{(\Spin)}_\Lambda = \inf_{\psi \in \mathcal{D}^{\Spin}_\Lambda\, ,\, 
 \psi \perp \GS^{\Spin}_\Lambda} \frac{\ip{\psi}{H^{\Spin}_\Lambda \psi}}
  {\ip{\psi}{\psi}}\, ,
$$
which, in view of \eq{Equivalent Hamiltonians},
is equivalent to the formula
$$
\gamma^{(\Spin)}_\Lambda = 2 \Spin 
  \inf_{\psi \not\in \ker(P^{\textrm{Sym}}_\Lambda)\, ,\,
  \psi \perp \GS^{\Spin}_\Lambda}
  \frac{\ip{P^{\textrm{Sym}}_\Lambda \psi}
  {\tilde{H}^{\Spin}_\Lambda P^{\textrm{Sym}}_\Lambda \psi}}
  {\ip{P^{\textrm{Sym}}_\Lambda \psi}{P^{\textrm{Sym}}_\Lambda \psi}}\, .
$$
(Note that since $\GS^{\Spin}_\Lambda \subset \mathcal{D}^{\Spin}_\Lambda$,
$\psi \perp \GS^{\Spin}_\Lambda$ iff $P^{\textrm{Sym}}_\Lambda \psi \perp
\GS^{\Spin}_\Lambda$.)

We define the subspace $\Hil^{\Spin}_0(\Lambda)$ to be the kernel of
$\tilde{H}^{\Spin}_\Lambda$.
Since each $\tilde{H}^{(m)}_\Lambda$ is actually a spin-$\frac{1}{2}$
XXZ Hamiltonian acting on the sites $\Lambda \times \{m\}$, we can use
\eq{ASW formula} to obtain
$$
\Hil^{\Spin}_0(\Lambda) 
  = \Span\{ \bigotimes_{m=1}^{2\Spin} \Psi_0^{(m)}(\Lambda,n_m)
  : \nvec = (n_1,\dots,n_{2\Spin}) \in [0,L]^{2\Spin}\}\, .
$$
We define $\Hil^{\Spin}_{0,\perp}(\Lambda) = \Hil^{\Spin}_0(\Lambda) \cap 
(\GS^{\Spin}_\Lambda)^\perp$.
We also define $\Hil^{\Spin}_{\textrm{exc}}(\Lambda) = 
\Hil^{\Spin}_0(\Lambda)^\perp$.
By Proposition \ref{Spin-1/2 spectral gap}, it is clear that
$\tilde{H}^{\Spin}_\Lambda \geq (1 - \Delta^{-1}) 
\Proj(\Hil^{\Spin}_{\textrm{exc}})$.
Then we have the following estimate for $\gamma^{(\Spin)}$.
\begin{lemma}
\label{Key lemma}
For any state $\phi \perp \GS^{\Spin}_\Lambda$, there exist two states
$\psi \in \Hil^{\Spin}_{0,\perp}(\Lambda)$, and 
$\psi' \in \Hil^{\Spin}_{\textrm{exc}}(\Lambda)$ such that
$$
P^{\textrm{Sym}}_\Lambda \phi = \psi + \psi'\, .
$$
Moreover, if $\phi \not \in \ker(P^{\textrm{sym}})$, we have
$$
\frac{\ip{P^{\textrm{Sym}}_\Lambda \phi}{\tilde{H}^{\Spin}_\Lambda 
P^{\textrm{Sym}}_\Lambda \phi}}{\|P^{\textrm{Sym}}_\Lambda \phi\|^2}
  \geq (1 - \Delta^{-1}) \left(1 - \frac{\|P^{\textrm{Sym}}_\Lambda \psi\|^2}
  {\|\psi\|^2}\right)\, .
$$
\end{lemma}

\begin{proof}
If $\phi \perp \GS^{\Spin}_\Lambda$, then also 
$P^{\textrm{Sym}}_\Lambda \phi \perp \GS^{\Spin}_\Lambda$ because
$$
\ip{P^{\textrm{Sym}}_\Lambda \phi}{\phi'} 
  = \ip{\phi}{P^{\textrm{Sym}}_\Lambda \phi'} 
  = \ip{\phi}{\phi'}
  = 0
$$
for any $\phi' \in \GS^{\Spin}_\Lambda \subset 
P^{\textrm{Sym}}_\Lambda(\Hil^{\Spin}_\Lambda)$.
Since $\Hil^{\Spin}_\Lambda = \GS^{\Spin}_\Lambda \oplus 
\Hil^{\Spin}_{0,\perp}(\Lambda) \oplus \Hil^{\Spin}_{\textrm{exc}}(\Lambda)$,
it is clear that there exist $\psi \in \Hil^{\Spin}_{0,\perp}(\Lambda)$,
$\psi' \in \Hil^{\Spin}_{\textrm{exc}}(\Lambda)$ with
$P^{\textrm{Sym}}_\Lambda \phi = \psi + \psi'$.
By Proposition \ref{Spin-1/2 spectral gap},
\begin{align*}
\ip{P^{\textrm{Sym}}_\Lambda \phi}{\tilde{H}^{\Spin}_\Lambda 
  P^{\textrm{Sym}}_\Lambda \phi}
  &= \ip{\psi + \psi'}{\tilde{H}^{\Spin}_\Lambda(\psi+\psi')} \\
  &= \ip{\psi'}{\tilde{H}^{\Spin}_\Lambda \psi'} \\
  &\geq (1-\Delta^{-1}) \|\psi'\|^2 \\
  &= (1-\Delta^{-1}) \|P^{\textrm{Sym}}_\Lambda \phi\|^2
  (1 - \frac{\|\psi\|^2}{\|P^{\textrm{Sym}}_\Lambda \phi\|^2})\, .
\end{align*}
Using Cauchy-Schwarz, we estimate
$$
\|\psi\|^2 
  = \ip{\psi}{P^{\textrm{Sym}}_\Lambda \phi} 
  = \ip{P^{\textrm{Sym}}_\Lambda \psi}{P^{\textrm{Sym}}_\Lambda \phi} 
  \leq \|P^{\textrm{Sym}}_\Lambda \psi\|\, 
  \|P^{\textrm{Sym}}_\Lambda \phi\|\, ,
$$
and so
$$
\frac{\|\psi\|}{\|P^{\textrm{Sym}}_\Lambda \phi\|}
  \leq \frac{\|P^{\textrm{Sym}}_\Lambda \psi\|}{\|\psi\|}\, ,
$$
which proves the lemma.
\end{proof}

We define 
$$
\delta^{(\Spin)}_\Lambda 
  = \sup_{\psi \in \Hil^{\Spin}_{0,\perp}(\Lambda)}
  \frac{\ip{\psi}{P^{\textrm{Sym}}_\Lambda \psi}}{\ip{\psi}{\psi}}\, .
$$
Then an immediate corollary is
\begin{corollary}
\label{Useful corollary}
$\gamma^{(\Spin)}_\Lambda \geq 2 \Spin (1 - \Delta^{-1}) 
  (1 - \delta^{(\Spin)}_\Lambda)$.
\end{corollary}
We can define a reduced Hamiltonian
$$
H^{\Spin,\textrm{Red}}_\Lambda = 2\Spin(1-\Delta^{-1})
\left[\unity - \Proj(\Hil^{\Spin}_0(\Lambda)) P^{\textrm{Sym}}_\Lambda 
\Proj(\Hil^{\Spin}_0(\Lambda))\right]\, ,
$$
on the Hilbert space $\Hil_0(\Lambda)$.
Then the corollary says 
$\gamma^{(\Spin)}_\Lambda \geq \gamma^{(\Spin),\textrm{Red}}_\Lambda$.
What is important to note is that while the original state space,
$\mathcal{D}^{\Spin}_\Lambda$ is $(2\Spin + 1)^{|\Lambda|}$-dimensional,
and the spin ladder state space $\Hil_\Lambda^{\Spin}$ is
$2^{(2\Spin+1)|\Lambda|}$-dimensional,
the reduced state space is only $(|\Lambda|+1)^{2\Spin}$-dimensional.
In addition, the reduced Hamiltonian is not more difficult to calculate
(as is often the case), in fact it is easier because the basis states
for $\Hil^{\Spin}_0(\Lambda)$ have a simple formula, and are close to
classical configurations of the Ising model.
These benefits are the motivation for the approximations of
Section \ref{Numerics}, as well as for Theorem \ref{Our theorem}.

An important fact for each of the operators introduced in this section,
is that they all commute with the total third component of spin
$$
S^3_{\textrm{tot}} = \sum_{x \in \Lambda} \sum_{m=1}^{2\Spin} S^3_{(x,m)}\, .
$$
We define $\Hil^{\Spin}(\Lambda,N)$ to be the $N$th sector, i.e.\ the
eigenspace of $S^3_{\textrm{tot}}$ with eigenvalue $2 \Spin |\Lambda| - N$.
Then we define $\Hil^{\Spin}_0(\Lambda,N)$, $\GS^{\Spin}(\Lambda,N)$ and
$\Hil_{0,\perp}(\Lambda,N)$ as the intersection of $\Hil^{\Spin}(\Lambda,N)$
with $\Hil^{\Spin}_0(\Lambda)$, $\GS^{\Spin}_\Lambda$ and 
$\Hil^{\Spin}_{0,\perp}(\Lambda)$, respectively.
Note that $\GS^{\Spin}(\Lambda,N)$ is the one-dimensional space spanned
by $\Psi^{\Spin}_0(\Lambda,N)$.
In terms of the embedding $\bigotimes_{x \in \Lambda} \mathcal{D}^{\Spin}_x
\hookrightarrow \Hil^{\Spin}_\Lambda$, equation \eq{ASW formula} becomes
$$
\Psi^{\Spin}_0(\Lambda,N) = \sum_{\{(x_i,m_i)\}_{i=1}^N}
  q^{\sum_{i=1}^N x_i} \prod_{i=1}^N S_{(x_i,n_i)}^- 
  \ket{\uparrow}_{\Lambda \times [1,2\Spin]}\, .
$$
There is one more piece of notation before we proceed.
Given any $\nvec \in [0,|\Lambda|]^{2\Spin}$, we define
\begin{equation}
\label{Ladder state definition}
\Psi^{\Spin}_0(\Lambda,\nvec) = \bigotimes_{m=1}^{2\Spin} 
\Psi^{1/2}_0(\Lambda\times\{m\},n_m)\, .
\end{equation}
Then
$$
\Hil^{\Spin}_0(\Lambda,N) = \Span\{ \Psi^{\Spin}_0(\Lambda,\nvec) : 
  \nvec \in [0,|\Lambda|]^{2\Spin}, 
  |\nvec| = N\}.
$$
where $|\nvec| := n_1 + n_2 + \dots + n_{2\Spin}$.

Since $P^{\textrm{Sym}}_\Lambda$ commutes with $S^3_{\textrm{tot}}$,
we see that
$$
\delta^{\Spin}_\Lambda = \sup_{1\leq N\leq 2 \Spin |\Lambda|}\ 
  \sup_{\psi \in \Hil_{0,\perp}(\Lambda,N)} 
  \frac{\ip{\psi}{P^{\textrm{Sym}}_\Lambda \psi}}{\|\psi\|^2}\, .
$$
The main element of the proof of Theorem \ref{Our theorem} is
the following,
\begin{proposition}
\label{Finite proposition}
Given any sequence of triples $(\Lambda_\alpha,N_\alpha,\psi_\alpha)$,
such that:
\begin{itemize}
\item $\Lambda_\alpha$ is a finite interval,
\item $0 \leq N_\alpha \leq 2 \Spin |\Lambda_\alpha|$, and
\item $\psi_\alpha \in \Hil^{\Spin}_{0,\perp}(\Lambda_\alpha,N_\alpha)$,
\end{itemize}
the following is true
$$
\limsup_{\alpha \to \infty} \frac{\ip{\psi_\alpha}
  {P^{\textrm{Sym}}_{\Lambda_\alpha} \psi_\alpha}}
  {\ip{\psi_\alpha}{\psi_\alpha}} < 1\, .
$$
\end{proposition}

\section{Proof}
\label{ProofSectionHigherSpinSpectralGap}
In this section we will prove Proposition \ref{Finite proposition}
first, and then Theorem \ref{Our theorem}.

\begin{proof}\hspace{-10pt} \textbf{(of Proposition \ref{Finite proposition})}
The proof is by contradiction.
Thus we assume the existence of a sequence 
$(\Lambda_\alpha,N_\alpha,\psi_\alpha)$ satisfying the hypotheses of 
the proposition, and also such that
\begin{equation}
\label{Final hypothesis}
\limsup_{\alpha \to \infty} \frac{\ip{\psi_\alpha}
  {P^{\textrm{Sym}}_{\Lambda_\alpha} \psi_\alpha}}
  {\ip{\psi_\alpha}{\psi_\alpha}} = 1\, .  
\end{equation}
We also assume, for convenience that each $\psi_\alpha$ is normalized.
By taking an appropriate subsequence, we can replace the $\limsup$
in the formula above with a $\lim$, and we assume this is done.
Our method of proof will be to show that under the hypotheses given, and if
$|\Lambda_\alpha|, N_\alpha \to \infty$ in such a way that 
$|\Lambda_\alpha| - (2\Spin)^{-1} N_\alpha \to \infty$, then
we can construct a limit state $\omega$ from the states 
$\omega_\alpha = \|\psi_\alpha\|^{-2} \ip{\psi_\alpha}{\dots \psi_\alpha}$
with the property that it is an infinite-volume ground state and also is
orthogonal to every infinite-volume ground state, clearly a contradiction.
But in order to prove this we must first show that 
$|\Lambda_\alpha|,M_\alpha,|\Lambda_\alpha|-(2\Spin)^{-1}M_\alpha \to \infty$.
\begin{lemma}
If $(\Lambda_\alpha,M_\alpha,\psi_\alpha)$ is a sequence satisfying the
hypotheses of the proposition, and also equation \eq{Final hypothesis}
(with $\limsup$ replaced by $\lim$), then $|\Lambda_\alpha| \to \infty$.
\end{lemma}
\begin{proof}
Let us suppose first that 
\begin{equation}
\label{Finite constant length}
|\Lambda_\alpha| \equiv L
\end{equation}
independent of $\alpha$.
Then we can map $\psi_\alpha$ to a state 
$\psi'_\alpha \in \Hil^{\Spin}_{0,\perp}(\Lambda:=[1,L],N_\alpha)$ for every $\alpha$.
But $\Hil^{\Spin}_{0,\perp}(\Lambda)$ is a finite-dimensional Hilbert space,
and the intersection of the range of $P^{\textrm{Sym}}_\Lambda$
with $\Hil^{\Spin}_0(\Lambda)$ is exactly $\GS^{\Spin}_\Lambda$
by equation \eq{Equivalent Hamiltonians}. 
So it is clear that
$\Proj(\Hil^{\Spin}_{0,\perp}(\Lambda)) P^{\textrm{Sym}}_\Lambda
\Proj(\Hil^{\Spin}_{0,\perp}(\Lambda))$ is strictly smaller than the identity
operator on $\Hil^{\Spin}_{0,\perp}(\Lambda)$.
Thus the conditions of the proposition along with \eq{Finite constant length}
contradict \eq{Final hypothesis}.
Now, in the general case, if $|\Lambda_\alpha|$ does not converge to
$+\infty$, then there is some $L$ such that $|\Lambda_\alpha| = L$ infinitely
often, and by taking the appropriate subsequence, we again have a 
contradiction.
Therefore, it must be that $|\Lambda_\alpha| \to \infty$.
\end{proof}

By taking an appropriate subsequence, we may assume that 
$|\Lambda_\alpha| \nearrow \infty$.
We assume this is done.
\begin{lemma}
\label{Infinite magnetization lemma}
If $(\Lambda_\alpha,N_\alpha,\psi_\alpha)$ is a sequence satisfying the
hypotheses of the proposition, and also equation \eq{Final hypothesis}
(with $\limsup$ replaced by $\lim$), then $N_\alpha \to \infty$.
\end{lemma}

\begin{proof}
We first assume 
\begin{equation}
\label{Finite constant magnetization}
N_\alpha \equiv N\, ,
\end{equation}
independent of $\alpha$ in order to prove a contradiction.
The proof is similar to the previous lemma, and essentially follows
from the fact that for a finite-dimensional vector space the spectral 
gap is always positive.
But this time the finiteness comes from $N<\infty$, not $|\Lambda|<\infty$,
and we need to demonstrate that the subspaces 
$\Hil^{\Spin}_0(\Lambda_\alpha,N)$ actually converge to a single 
finite-dimensional space $\Hil^{\Spin}_0(\Ir_+,N)$.

By the previous lemma, we know that $|\Lambda_\alpha| \nearrow \infty$.
By taking a unitary transformation, if necessary, we assume that 
$\Lambda_\alpha = [1,L_\alpha]$, with $L_\alpha \nearrow \infty$.
Then we can write
$$
\psi_\alpha = \sum_{\substack{\nvec \in \Nl^{2\Spin} \\ |\nvec|=N}}
  C_\alpha(\nvec) \Psi^{\Spin}_0([1,L_\alpha],\nvec)\, .
$$
We define two new vectors
\begin{gather*}
\psi'_\alpha = \psi_\alpha \otimes \ket{\uparrow}_{[L_\alpha+1,\infty)
  \times[1,2\Spin]} \\
\psi''_\alpha = \sum_{\substack{\nvec \in \Nl^{2\Spin} \\ |\nvec|=N}}
  C_\alpha(\nvec) \Psi^{\Spin}_0(\Ir_+,\nvec)\, ,
\end{gather*}
where $\Psi^{\Spin}_0(\Ir_+,\nvec)$ is defined as the tensor product of
$\Psi^{1/2}_0(\Ir_+\times\{m\},n_m)$ over $m=1,\dots,2\Spin$, and
$$
\Psi^{1/2}_0(\Ir_+,N) = 
  \sum_{1\leq x_1 < \dots < x_N < \infty}
  q^{\sum_{i=1}^N x_i} \prod_{i=1}^N S_{x_i}^- 
  \ket{\uparrow}_{\Ir_+}\, .
$$
It is trivial to check that 
$\Psi^{1/2}_0([1,L],N) \otimes \ket{\uparrow}_{[L+1,\infty)}$
converges in norm to $\Psi^{1/2}_0(\Ir_+,N)$, as $L \to \infty$.
Indeed, from our definition
\begin{align*}
\ip{\Psi^{1/2}_0(\Ir_+,N)} 
{\Psi^{1/2}_0([1,L],N) \otimes \ket{\uparrow}_{[L+1,\infty)}}
  &= \|\Psi^{1/2}_0([1,L],N) \otimes \ket{\uparrow}_{[L=1,\infty)}\|^2 \\
  &= \frac{q^{N(N+1)} \prod_{k=1}^L (1 - q^{2k})}
  {\prod_{k=1}^N (1 - q^{2k}) \prod_{k=1}^{L-N} (1-q^{2k})}\\
  &\to \frac{q^{N(N+1)}}{(q^2;q^2)_\infty} \\
  &= \|\Psi^{1/2}_0(\Ir_+,N)\|^2\, .
\end{align*}
From this it follows that 
$$
\Psi^{\Spin}_0([1,L_\alpha],\nvec)\otimes \ket{\uparrow}_{[L_\alpha+1,\infty)
  \times [1,2\Spin]}  \stackrel{\|.\|}\longrightarrow
  \Psi^{\Spin}_0(\Ir_+,\nvec)
$$
for each $\nvec \in \Ir^{2\Spin}$ such that $|\nvec| = N$.
Note that this is a finite set of multiindices $\nvec$, specifically,
the cardinality is $\binom{2\Spin+N-1}{N}$.
Thus 
$$
\lim_{\alpha \to \infty} \|\psi_\alpha' - \psi_\alpha''\| = 0,
$$
which implies that $(\unity - P^{\textrm{Sym}}_\Lambda) \psi_\alpha'' \to 0$
as $\alpha \to \infty$ for every finite $\Lambda \subset \Ir_+$
(because the same is true for $\psi'_\alpha$ by hypothesis).
But also, 
$$
\Psi^{\Spin}_0([1,L_\alpha],N) \otimes \ket{S}_{[L_\alpha+1,\infty)}
  \stackrel{\|.\|}\longrightarrow \Psi^{\Spin}_0(\Ir_+,N)
$$
as $\alpha \to \infty$.
So, since $\psi'_\alpha \perp \Psi^{\Spin}_0([1,L_\alpha],N) 
\otimes \ket{S}_{[L_\alpha+1,\infty)}$, we have
$$
\lim_{\alpha \to \infty} \ip{\psi''_\alpha}{\Psi^{\Spin}_0(\Ir_+,N)} = 0\, .
$$
Since the subspace $\Hil^{2\Spin}_0(\Ir_+,N)$ is finite-dimensional, and all
the $\psi'_\alpha$ have norm 1, there is a limit point $\psi''$ of the sequence
$\{\psi''_\alpha\}$.
This vector satisfies $\|\psi''\| = 1$, 
$P^{\textrm{Sym}}_\Lambda \psi'' = \psi''$ 
for every finite $\Lambda \subset \Ir_+$, and
$\ip{\psi''}{\Psi^{\Spin}_0(\Ir_+,N)} = 0$.
But, since the intersection of $\Hil^{\Spin}_0(\Ir_+,N)$ 
with $P^{\textrm{Sym}}_{\Ir_+} = \lim_{\Lambda \nearrow \Ir_+} 
P^{\textrm{Sym}}_\Lambda$ is $\GS^{\Spin}(\Ir_+,N)$, the existence
of such a vector $\psi''$ is impossible.
Thus we have a contradicition.
So, $N_\alpha$ does not equal any finite number infinitely often,
and this implies $N_\alpha \to \infty$.
\hspace{50pt}
\end{proof}

\begin{corollary}
With the hypotheses of the last lemma, 
$|\Lambda_\alpha| - (2\Spin)^{-1} N_\alpha \to \infty$.
\end{corollary}

\begin{proof}
Follows from the last lemma and simultaneous spin-flip/reflection
symmetry of $\tilde{H}^{\Spin}_{\Lambda}$.
\end{proof}

By choosing a subsequence, if necessary, we can assume 
$N_\alpha \nearrow \infty$
and $|\Lambda_\alpha| - (2S)^{-1} N_\alpha \nearrow \infty$.
Also, by taking an appropriate subsequence, we can assume that all
$N_\alpha$ are equivalent modulo $2\Spin$,
i.e. that $N_\alpha = 2 \Spin a_\alpha + N_0$ for $a_\alpha \in \Nl$
and a number $N_0 \in [0,2\Spin-1]$ independent of $\alpha$.
By taking a unitary transformation, we can assume 
$\Lambda_\alpha = [-a_\alpha+1,b_\alpha]$.
Since $N_\alpha \nearrow \infty$, it follows $a_\alpha \nearrow \infty$;
since $|\Lambda_\alpha| - (2S)^{-1} N_\alpha \nearrow \infty$,
it follows $b_\alpha \nearrow \infty$.
Let $\vec{e} \in \Nl^{2 \Spin}$ be the vector $(1,1,\dots,1)$.
Then every $\psi_\alpha$ can be written uniquely as
$$
\psi_\alpha = \sum_{\substack{\nvec \in \Ir^{2 \Spin}\\|\nvec| = N_0}}
  C_\alpha(\nvec) \Psi^{\Spin}_0(\Lambda_\alpha,\nvec + a_\alpha \vec{e})
  / \|\Psi^{\Spin}_0(\Lambda_\alpha,\nvec + a_\alpha \vec{e})\|\, ,
$$
with $\sum_{\nvec} |C_\alpha(\nvec)|^2 = 1$. 
(Recall that $|\nvec| := n_1 + \dots + n_{2\Spin}$ is the sum of the
parts of $\nvec$, not the $l^1$-norm.)

We observe that, for any fixed $\nvec \in \Ir^{2\Spin}$ with
$|\nvec| = N_0$, the sequence $\{C_\alpha(\nvec)\}$ is bounded-in-norm
by 1.
Thus we may choose a convergent subsequence
$C_{\alpha_\beta}(\nvec) \to C(\nvec)$.
By the Cantor diagonal trick we can, in fact, choose a subsequence
such that $C_{\alpha_\beta}(\nvec)$ converges for every $\nvec$
(since the set of $\nvec$ is countable).
We assume this is done from the outset, so that 
$C_\alpha(\nvec) \to C(\nvec)$ for all $\nvec$.
By Fatou's lemma $\sum_{\nvec} |C(\nvec)|^2 \leq 1$.
But there is no guarantee at the outset that the opposite inequality
holds, i.e.\ that $\sum_{\nvec} |C(\nvec)|^2 = 1$.
Next we will show that the coefficients $C_\alpha(\nvec)$ are small
whenever any part of $\nvec$ is too large.
This will allow the opposite inequality, and more.

\begin{lemma}
$$
\lim_{R \to \infty} \liminf_{\alpha \to \infty}
  \sum_{\substack{\nvec \in [-R+1,R]^{2\Spin} \\ |\nvec| = N_0}} 
  |C_\alpha(\nvec)|^2 = 1\, .
$$
\end{lemma}
\begin{proof}
The proof of this fact is the most technical part of the paper.
By hypothesis, 
$$
\lim_{\alpha \to \infty}
  \ip{\psi_\alpha}{P^{\textrm{Sym}}_{\Lambda_\alpha} \psi_\alpha} = 1\, .
$$
For any finite $\Lambda \subset \Ir$, and large enough $\alpha$,
$\Lambda \subset \Lambda_\alpha$.
In that case 
$$
\ip{\psi_\alpha}{P^{\textrm{Sym}}_{\Lambda_\alpha} \psi_\alpha} 
  = \ip{P^{\textrm{Sym}}_\Lambda \psi_\alpha}
  {P^{\textrm{Sym}}_{\Lambda_\alpha} P^{\textrm{Sym}}_\Lambda \psi_\alpha} 
  \leq \|P^{\textrm{Sym}}_\Lambda \psi_\alpha\|\, .
$$
So for every finite $\Lambda \subset \Ir$,
$\lim_{\alpha \to \infty} \|P^{\textrm{Sym}}_\Lambda \psi_\alpha\|
  = 1$.
Now, for any $R \in \Ir_+$, define
$\psi_\alpha^{R<}$ to be the sum of all those terms
$C_\alpha(\nvec) \Psi^{\Spin}_0(\Lambda_\alpha,\nvec + a_\alpha \vec{e})
  / \|\Psi^{\Spin}_0(\Lambda_\alpha,\nvec + a_\alpha \vec{e})\|$
for which $\nvec \in [-R+1,R]^{2\Spin}$,
and let $\psi_\alpha^{R>} = \psi_\alpha - \psi_\alpha^{R<}$.
We observe that
\begin{equation}
\label{Projection estimate}
\|P^{\textrm{Sym}}_\Lambda \psi_\alpha\| 
  \leq \|P^{\textrm{Sym}}_\Lambda \psi^{R<}_\alpha\|
  + \|P^{\textrm{Sym}}_\Lambda \psi^{R>}_\alpha\| 
  \leq \|\psi^{R<}_\alpha\| + \|P^{\textrm{Sym}}_\Lambda \psi^{R>}_\alpha\|,
\end{equation}
and
$$
\|P^{\textrm{Sym}}_\Lambda \psi^{R>}_\alpha\|^2
  = \sum_{\substack{\nvec,\nvec' \in \Ir^{2\Spin} \\
  \quad \setminus [-R+1,R]^{2\Spin}
  \\ |\nvec| = |\nvec'| = N_0}}
  C_\alpha(\nvec) \overline{C_\alpha(\nvec')}
  \ip{\frac{P^{\textrm{Sym}}_\Lambda \Psi^{\Spin}_0(\Lambda_\alpha,\nvec')} 
  {\|\Psi^{\Spin}_0(\Lambda_\alpha,\nvec')\|} }
  {\frac{P^{\textrm{Sym}}_\Lambda \Psi^{\Spin}_0(\Lambda_\alpha,\nvec)} 
  {\|\Psi^{\Spin}_0(\Lambda_\alpha,\nvec)\|} }
$$
But also
$$
\ip{P^{\textrm{Sym}}_\Lambda \Psi^{\Spin}_0(\Lambda_\alpha,\nvec')}
  {P^{\textrm{Sym}}_\Lambda \Psi^{\Spin}_0(\Lambda_\alpha,\nvec)}
  = 0
$$
unless $\|\nvec - \nvec'\|_1 \leq 2 \Spin |\Lambda|$,
and if $\|\nvec - \nvec'\|_1 \leq 2 \Spin |\Lambda|$, then
\begin{align*}
& \left| C_\alpha(\nvec) \overline{C_\alpha(\nvec')}
  \ip{\frac{P^{\textrm{Sym}}_\Lambda \Psi^{\Spin}_0(\Lambda_\alpha,\nvec')} 
  {\|\Psi^{\Spin}_0(\Lambda_\alpha,\nvec')\|} }
  {\frac{P^{\textrm{Sym}}_\Lambda \Psi^{\Spin}_0(\Lambda_\alpha,\nvec)} 
  {\|\Psi^{\Spin}_0(\Lambda_\alpha,\nvec)\|} } \right|\\
& \qquad  \leq \frac{1}{2} \cdot
  \frac{|C_\alpha(\nvec)|^2 \|P^{\textrm{Sym}}_\Lambda 
  \Psi^{\Spin}_0(\Lambda_\alpha,\nvec)\|^2}
  {\|\Psi^{\Spin}_0(\Lambda_\alpha,\nvec)\|^2} 
  + \frac{1}{2} \cdot
  \frac{|C_\alpha(\nvec')|^2 \|P^{\textrm{Sym}}_\Lambda 
  \Psi^{\Spin}_0(\Lambda_\alpha,\nvec')\|^2}
  {\|\Psi^{\Spin}_0(\Lambda_\alpha,\nvec')\|^2}\, ,
\end{align*}
by Cauchy-Schwarz.
Therefore,
\begin{align*}
\|P^{\textrm{Sym}}_\Lambda \psi^{R>}_\alpha\|^2
  &\leq 2 \Spin |\Lambda| 
  \sum_{\substack{\nvec \in \Ir^{2\Spin} \setminus [-R+1,R]^{2\Spin}
  \\ |\nvec| = N_0}} 
  |C_\alpha(\nvec)|^2
  \frac{\|P^{\textrm{Sym}}_\Lambda 
  \Psi^{\Spin}_0(\Lambda_\alpha,\nvec)\|^2}
  {\|\Psi^{\Spin}_0(\Lambda_\alpha,\nvec)\|^2} \\
  &\leq 2 \Spin |\Lambda| \|\psi^{R>}_\alpha\|^2 M(\Lambda,\Lambda_\alpha,R)\, ,
\end{align*}
where
$$
M(\Lambda,\Lambda_\alpha,R)^2 
  = \sup_{\substack{\nvec \in \Ir^{2\Spin} \setminus [-R+1,R]^{2\Spin}
  \\ |\nvec| = N_0}} 
  \frac{\|P^{\textrm{Sym}}_\Lambda 
  \Psi^{\Spin}_0(\Lambda_\alpha,\nvec)\|^2}
  {\|\Psi^{\Spin}_0(\Lambda_\alpha,\nvec)\|^2}\, .
$$
{\em It is understood that we only take the supremum over those
$\nvec$ for which $\Psi^{\Spin}_0(\Lambda_\alpha,\nvec)$ is defined.}
Thus $M(\Lambda,\Lambda_\alpha,R)^2$ actually has an implicit dependence on 
$\Lambda_\alpha$.
But we claim that $M(\Lambda,\Lambda_\alpha,R)^2$ has a bound, independent of 
$\Lambda_\alpha$.
\begin{claim}
$$
M(\Lambda,\Lambda_\alpha,R)^2 \leq 2^{\displaystyle -|\Lambda|} 
  + \mathcal{C} 
  q^{(\displaystyle \frac{4\Spin}{2\Spin-1}R - 2\sup\{|x|:x \in \Lambda\})}\, ,
$$
where $\mathcal{C}$ is a universal constant depending only on $q$
(not on $\Spin$, $N_\alpha$, $\Lambda_\alpha$ or $R$).
\end{claim}
\begin{proof}\hspace{-10pt} \textbf{(of Claim)}
Suppose that $\nvec$ is a vector in $\Ir^{2\Spin} \setminus [-R+1,R]^{2\Spin}$
such that $\Psi^{\Spin}_0(\Lambda_\alpha,\nvec)$ is well-defined.
Then for some $m_1 \in [1,2\Spin]$,
$$
\frac{1}{2} \pm (n_{m_1} - \frac{1}{2}) \geq R\, ,
$$
because $\nvec \not\in [-R+1,R]^{2\Spin}$,
and for some other $m_2 \in [1,2\Spin]$,
$$
\frac{1}{2} \mp (n_{m_2} - \frac{1}{2}) \geq (2\Spin-1)^{-1} (R - N_0)\, ,
$$
because $|\nvec| = N_0$.
We can estimate 
$\ip{\Psi^{\Spin}_0(\Lambda_\alpha,\nvec)}{P^{\textrm{Sym}}_\Lambda
\Psi^{\Spin}_0(\Lambda_\alpha,\nvec)}$
upwards by
$$
\ip{\Psi^{\Spin}_0(\Lambda_\alpha,\nvec)}{P^{\textrm{Sym},(m_1,m_2)}_{\Lambda}
\Psi^{\Spin}_0(\Lambda_\alpha,\nvec)}
$$
where $P^{\textrm{Sym},(m_1,m_2)}_\Lambda$ only symmetrizes in the two legs 
$m_1$, $m_2$.
Now, obviously, 
\begin{align*}
& \|P^{\textrm{Sym},(m_1,m_2)}_{\Lambda} 
  \Proj(\ket{\pm 1/2}_{\Lambda\times\{m_1\}})
  \Proj(\ket{\mp 1/2}_{\Lambda\times\{m_2\}})
\Psi^{\Spin}_0(\Lambda_\alpha,\nvec)\| \\
& \qquad \qquad  \leq 2^{-|\Lambda|} \|\Psi^{\Spin}_0(\Lambda_\alpha,\nvec)\|\, .
\end{align*}
So
\begin{align*}
& \frac{\|P^{\textrm{Sym},(m_1,m_2)}_{\Lambda} 
  \Psi^{\Spin}_0(\Lambda_\alpha,\nvec)\|}
  {\|\Psi^{\Spin}_0(\Lambda_\alpha,\nvec)\|}
  \leq 2^{-|\Lambda|} \\
&\quad  + \frac{\|[\unity - \Proj(\ket{\pm 1/2}_{\Lambda\times\{m_1\}})
  \Proj(\ket{\mp 1/2}_{\Lambda\times\{m_2\}})]
  \Psi^{\Spin}_0(\Lambda_\alpha,\nvec)\|}
  {\|\Psi^{\Spin}_0(\Lambda_\alpha,\nvec)\|}\, .
\end{align*}
And
\begin{align*}
&\frac{\|[\unity - \Proj(\ket{\pm 1/2}_{\Lambda\times\{m_1\}})
  \Proj(\ket{\mp 1/2}_{\Lambda\times\{m_2\}})]
  \Psi^{\Spin}_0(\Lambda_\alpha,\nvec)\|}
  {\|\Psi^{\Spin}_0(\Lambda_\alpha,\nvec)\|} \\
&\quad \leq 
  \frac{\|[\unity - \Proj(\ket{\pm 1/2}_{\Lambda\times\{m_1\}})]
  \Psi^{1/2}_0(\Lambda_\alpha\times\{m_1\},n_{m_1})\|}
  {\|\Psi^{1/2}_0(\Lambda_\alpha\times\{m_1\},n_{m_1})\|}  \\
&\qquad + 
  \frac{\|[\unity - \Proj(\ket{\mp 1/2}_{\Lambda\times\{m_2\}})]
  \Psi^{1/2}_0(\Lambda_\alpha\times\{m_2\},n_{m_2})\|}
  {\|\Psi^{1/2}_0(\Lambda_\alpha\times\{m_2\},n_{m_2})\|} \, .
\end{align*}
The estimate of the right-hand-side of the last display
is the type of calculation which may be carried out directly
from the definition \eq{ASW formula}.
For details of these types of calculations see, for example,
\cite{BCN}.
If $\Lambda = [-a+1,b]$, then the first of the two factors above is bounded by
a universal constant (depending only on $q$) times 
$q^{2R - \max(a,b)}$, and the second factor is bounded by the same
universal constant times $q^{2(2\Spin-1)^{-1}(R-N_0)- \max(a,b)}$.
Absorbing $q^{-N_0/(\Spin-1/2)}$, which is at most $q^{-1}$, 
into the universal constant-squared, we have the result.
\end{proof}
Now, if we let 
$$
\epsilon(R,\Lambda)^2 = 2 \Spin |\Lambda| \times 
 \text{ bound for $M(\Lambda,\Lambda_\alpha,R)$ }\, ,
$$
and if we let $x = \|\psi_\alpha^{R<}\|$ and 
$\delta = 1 - \|P^{\textrm{Sym}}_\Lambda \psi_\alpha\|$,
then by equation \eq{Projection estimate}, we have
$$
1 - \delta \leq x + \epsilon \sqrt{1 - x^2}\, .
$$
Solving for $x$, we see that $x$ lies between $x_{\pm}$, where
$$
x_\pm = \frac{1-\delta \pm \sqrt{(1-\delta)^2 -4(1+\epsilon^2)
  [(1-\delta)^2 - \epsilon^2]}}{1 + \epsilon^2}\, .
$$
If we let $x$ stand for $\liminf \|\psi_\alpha^{R<}\|$, instead,
then we can take $\delta \to 0$ (because 
$\ip{\psi_\alpha}{P^{\textrm{Sym}}_\Lambda \psi_\alpha} \to 1$). 
In this case, $x_- \leq x \leq x_+$, where
$$
x_{\pm} = \frac{1\pm \epsilon^2}{1+\epsilon^2}\, .
$$
Since $\epsilon(R,\Lambda)^2 \to 2 \Spin |\Lambda| 2^{-|\Lambda|}$
as $R \to \infty$, we have
$$
\lim_{R \to \infty}\ 
\liminf_{\alpha \to \infty} \|\psi_\alpha^{R<}\|
  \geq \frac{1 - 2 \Spin |\Lambda| 2^{-|\Lambda|}}
  {1 + 2 \Spin |\Lambda| 2^{-|\Lambda|}}\, .
$$
But since $\Lambda$ was arbitrary, we can take $|\Lambda| \to \infty$,
to obtain
$$
\lim_{R \to \infty}\ 
\liminf_{\alpha \to \infty} \|\psi_\alpha^{R<}\| \geq 1\, .
$$
The reverse inequality is trivial, so the lemma is proved.
\end{proof}
Here is an important application of the previous lemma:
\begin{corollary}
\label{Example corollary}
$\{C_\alpha(\nvec) : \nvec\}
  \stackrel{\|.\|_2}\longrightarrow \{C(\nvec) : \nvec\}$.
\end{corollary}
\begin{proof}
For any $\epsilon > 0$, we can find an $R$ such that 
$$
\liminf_{\alpha \to \infty} 
  \sum_{\substack{\nvec \in [-R+1,R]^{2\Spin} \\ |\nvec|=N_0}} 
  |C_\alpha(\nvec)|^2 \geq 1-\epsilon\, .
$$
Since the set of all $\nvec$ in the sum is finite, we see that 
$$
\sum_{\substack{\nvec \in [-R+1,R]^{2\Spin} \\ |\nvec|=N_0}} 
  |C_\alpha(\nvec)|^2 \to
  \sum_{\substack{\nvec \in [-R+1,R]^{2\Spin} \\ |\nvec|=N_0}} 
  |C(\nvec)|^2 
$$
as $\alpha \to \infty$.
Thus
$$
\sum_{\substack{\nvec \in [-R+1,R]^{2\Spin} \\ |\nvec|=N_0}} 
  |C(\nvec)|^2 \geq 1 - \epsilon
$$
which implies
\begin{align*}
\limsup_{\alpha \to \infty}
  \sum_{\substack{\nvec \in \Ir^{2\Spin} \\ |\nvec|=N_0}} 
  |C(\nvec) - C_\alpha(\nvec)|^2 
  &\leq 2 \epsilon + \limsup_{\alpha \to \infty}
  \sum_{\substack{\nvec \in [-R+1,R]^{2\Spin} \\ |\nvec|=N_0}} 
  |C(\nvec) - C_\alpha(\nvec)|^2  \\
  &= 2 \epsilon\, .
\end{align*}
Since $\epsilon$ was arbitrary, the corollary follows.
\end{proof}

We now define a new sequence of vectors
$$
\psi'_\alpha = \ket{\downarrow}_{(-\infty,-a_\alpha] \times [1,2\Spin]}
  \otimes \psi_\alpha \otimes \ket{\uparrow}_{[b_\alpha+1,\infty) \times 
  [1,2\Spin]}\, , 
$$
as well as the vector
$$
\psi'' = \sum_{\substack{\nvec \in \Ir^{2\Spin} \\
  |\nvec| = N_0}} C(\nvec) \Psi^{2\Spin}_0(\Ir,\nvec)
  /\|\Psi^{\Spin}_0(\Ir,\nvec)\|\, .
$$
Here, by $\Psi^{\Spin}_0(\Ir,\nvec)$ we mean the tensor product of all the
$\Psi^{1/2}_0(\Ir\times\{m\},n_m)$, $m=1,\dots,2\Spin$, where
$\Psi^{1/2}_0(\Ir,N)$ is given by \eq{Bi-infinite spin-1/2 ground state}.
It is trivial to check that for any fixed $\nvec$,
\begin{align*}
&  \ket{\downarrow}_{(-\infty,-a_\alpha] \times [1,2\Spin]} \otimes 
  \frac{\Psi^{\Spin}_0([-a_\alpha+1,b_\alpha],a_\alpha \vec{e} + \nvec)}
  {\|\Psi^{\Spin}_0([-a_\alpha+1,b_\alpha],a_\alpha \vec{e} + \nvec)\|}
  \otimes \ket{\uparrow}_{[b_\alpha+1,\infty) \times 
  [1,2\Spin]} \\
& \hspace{250pt}
  \stackrel{\|.\|_2}\longrightarrow
  \frac{\Psi^{\Spin}_0(\Ir,\nvec)}{\|\Psi^{\Spin}_0(\Ir,\nvec)\|}
\end{align*}
as $\alpha \to \infty$.
(It is a similar computation to that done in Lemma \ref{Infinite magnetization
lemma}.)
Thus, for any finite $R$, we have
$$
\limsup_{\alpha \to \infty}
  \|(\psi_\alpha')^{R<} - (\psi'')^{R<}\| = 0\, ,
$$
where putting the superscript $R<$ means the same thing as before,
namely truncating the terms to those involving only $\nvec$ with
$\nvec \in [-R+1,R]^{2\Spin}$.
By the lemma 
$$
\lim_{R \to \infty} \liminf_{\alpha \to \infty} 
  \|(\psi_\alpha')^{R<}\| = 1\, .
$$
Then following the argument in Corollary \ref{Example corollary}, 
$$
\|.\|_2-\lim_{\alpha \to \infty} \psi_\alpha' = \psi''\, .
$$

By its definition, $\psi''$ is a ground state for $\tilde{H}^{\Spin}_\Lambda$
for every finite $\Lambda \subset  \Ir$, because each 
$\Psi^{\Spin}_0(\Ir,\nvec)$ is.
Also, by hypothesis, for any $\Lambda \subset \Lambda_\alpha$,
$$
\lim_{\alpha \to \infty} \ip{\psi_\alpha}{P^{\textrm{Sym}}_\Lambda \psi_\alpha}
  \to 1\, .
$$
Since $\Lambda_\alpha \nearrow \Ir$, it is true that 
$P^{\textrm{Sym}}_\Lambda \psi'' = \psi''$ for any finite 
$\Lambda \subset \Ir$.
Then by equation \eq{Equivalent Hamiltonians}, $\psi''$ is a ground state
of $H^{\Spin}_\Lambda$ for every finite interval $\Lambda \subset \Ir$.
By Theorem \ref{Complete ground states}, $\psi''$ is a ground state.
Since it is pure it is $\omega_\uparrow$, $\omega_\downarrow$,
some kink,$\Psi^{\Spin}_0(\Ir,N)$, or some antikink.
The densely defined operator 
$$
S^3_{\textrm{Ren}}
  = \lim_{\Lambda \nearrow \Ir} \sum_{x \in \Lambda \cap [1,\infty)} 
  (\frac{1}{2} - S_x^3) 
  - \sum_{x \in \Lambda \cap (-\infty,0]} (\frac{1}{2} + S_x^3)
$$
distinguishes the different cases, and in particular all the infinite-volume
kink states are eigenvectors for $S^3_{\textrm{Ren}}$, with eigenvalue 
equal to $2 \Spin N$.
Similarly, each $\Psi^{\Spin}_0(\Ir,\nvec)$ is an eigenvector with eigenvalue
equal to $|\nvec|$.
From this we see that $\psi'' = \Psi^{\Spin}_0(\Ir,N_0)$.

But by hypothesis,
$$
\psi'_\alpha \perp \ket{\downarrow}_{(-\infty,-a_\alpha] \times [1,2\Spin]}
  \otimes \Psi^{\Spin}_0(\Lambda_\alpha,2 \Spin a_\alpha + N_0)
  \otimes \ket{\uparrow}_{[b_\alpha+1,\infty) \times [1,2\Spin]}\, .
$$
Another easy calculation is the fact that
\begin{align*}
&\|.\|-\lim_{\Lambda_\alpha \nearrow \Ir}
  \ket{\downarrow}_{(-\infty,-a_\alpha] \times [1,2\Spin]}
  \otimes \frac{\Psi^{\Spin}_0(\Lambda_\alpha,2 \Spin a_\alpha + N_0)}
  {\|\Psi^{\Spin}_0(\Lambda_\alpha,2 \Spin a_\alpha + N_0)\|}
  \otimes \ket{\uparrow}_{[b_\alpha+1,\infty) \times [1,2\Spin]} \\
& \hspace{225pt}  = \Psi^{\Spin}_0(\Ir,N_0)/\|\Psi^{\Spin}_0(\Ir,N_0)\|\, .
\end{align*}
So, this implies $\psi'' \perp \Psi^{\Spin}_0(\Ir,N_0)$, which is clearly a
contradiction.
Therefore the Proposition is proved.
\end{proof}

\begin{proof}\hspace{-10pt} \textbf{(of Theorem \ref{Our theorem})}
We will first prove that there is a nonzero gap above the infinite-volume
kink ground states.
The same will then hold for the infinite-volume antikink states by symmetry.
The gap above the translation invariant states will be calculated exactly,
using a different technique.

Let $\gamma$ be the largest number such that
$$
  \ip{\psi}{H^{\Spin}_\Lambda \psi} \geq \gamma \|\psi\|^2
$$
for all finite $\Lambda \subset \Ir$, and all $\psi \perp \GS^{\Spin}_\Lambda$.
By Corollary \ref{Useful corollary} and Proposition \ref{Finite proposition},
$\gamma > 0$.
Then we claim $\gamma$ is a lower bound for the spectral gap above any of the
kink states $\Psi^{\Spin}_0(\Ir,N)$.
To prove this, it suffices to show that for any kink state
$\Psi^{\Spin}_0(\Ir,N)$, and any local observable $X \in \Obs_\Lambda$,
\begin{equation}
\label{Condition for kink gap}
\ip{\Psi^{\Spin}_0(\Ir,N)}{X^* (H^{\Spin}_\Ir)^3 X \Psi^{\Spin}_0(\Ir,N)}
  \geq \gamma \ip{\Psi^{\Spin}_0(\Ir,N)}{X^* (H^{\Spin}_\Ir)^2 
  X \Psi^{\Spin}_0(\Ir,N)}\, .
\end{equation}
(This means that $H^{\Spin}_\Ir$ is greater than $\gamma \unity$ on its range
in the GNS Hilbert space of all excitations of $\Psi^{\Spin}_0(\Ir,N)$, since
the vectors $X \Psi^{\Spin}_0(\Ir,N)$, $X \in \Obs_{\textrm{loc}}$, 
are a core for $H^{\Spin}_\Ir$ and all its powers.)

We observe that since $H^{\Spin}_\Ir$ is a limit of a sum of nearest-neighbor
interactions, defining $\delta = [H^{\Spin}_\Ir,.]$,
$\delta^s(X) \in \Obs_{\Lambda\pm s}$.
Thus,
$$
\ip{\Psi^{\Spin}_0(\Ir,N)}{X^* (H^{\Spin}_\Ir)^3 X \Psi^{\Spin}_0(\Ir,N)}
  = \ip{\Psi^{\Spin}_0(\Ir,N)}{X^* (H^{\Spin}_{\Lambda\pm3})^3 X \Psi^{\Spin}_0(\Ir,N)}\, ,
$$
and 
\begin{align*}
\ip{\Psi^{\Spin}_0(\Ir,N)}{X^* (H^{\Spin}_\Ir)^2 
  X \Psi^{\Spin}_0(\Ir,N)}
  &= \ip{\Psi^{\Spin}_0(\Ir,N)}{X^* (H^{\Spin}_{\Lambda\pm2})^2 
  X \Psi^{\Spin}_0(\Ir,N)} \\
  &= \ip{\Psi^{\Spin}_0(\Ir,N)}{X^* (H^{\Spin}_{\Lambda\pm3})^2 
  X \Psi^{\Spin}_0(\Ir,N)}\, .
\end{align*}
Now define $\psi = H^{\Spin}_{\Lambda\pm3} X \Psi^{\Spin}_0(\Ir,N)$,
and $\omega = \ip{\psi}{\dots \psi}/\|\psi\|^2$.
Then $\omega$ restricted to $\Obs_{\Lambda\pm3}$ is a density matrix
$\omega = \sum_k \ip{\psi_k}{\dots \psi_k}$, where each 
$\psi_k \in \mathcal{D}^{\Spin}_{\Lambda\pm3}$ and
$\sum_k \|\psi_k\|^2 = 1$.
By the definition of $\gamma$,
$\ip{\psi_k}{H^{\Spin}_{\Lambda\pm3} \psi_k} \geq \gamma \|\psi_k\|^2$.
So 
$$
\omega(H^{\Spin}_{\Lambda\pm3}) \geq \gamma \omega(\unity)\, .
$$
This is equivalent to \eq{Condition for kink gap}.

This proves the existence of a positive spectral gap above the kink and 
antikink states, although the value of $\gamma$ has not been calculated.
For the translation-invariant ground states, the spectral gap can actually
be calculated by standard techniques.
First we obtain a lower bound.
For this, suppose that $\psi = X \ket{+S}_{\Ir}$, where
$\ket{\uparrow}$ is the all-up state in the Guichardet Hilbert space
$\otimes_{x \in \Ir} (\Cx^{2 \Spin+1},\ket{+S}_x)$, and 
$X \in \Obs_{\textrm{loc}}$.
Then, (the boundary-field term is irrelevant since $\psi$ is asymptotically
$\ket{+S}$ at $\pm \infty$), 
\begin{align*}
\ip{\psi}{H^{\Spin}_{\Ir} \psi}
  &= \sum_{x \in \Ir} 
  \ip{\psi}{(\Delta^{-1}[\Spin^2 - \vec{S}_x \cdot \vec{S}_{x+1}]
  + (1 - \Delta^{-1}) [\Spin^2 - S^3_x S^3_{x+1}]) \psi} \\
  &\geq (1 - \Delta^{-1})
  \sum_{x \in \Ir} \ip{\psi}{[\Spin^2 - S^3_x S^3_{x+1}] \psi}\ ,.
\end{align*}
Now $(1 - \Delta^{-1}) \sum_x [\Spin^2 - S^3_x S^3_{x+1}]$ is diagonal in
the basis 
$$
\phi_{\{n_x\}} =  \prod_{x \in \Ir} \frac{1}{n_x!} \binom{2\Spin}{n_x}^{-1/2} 
  (S_x^-)^{n_x}\ \ket{+\Spin}_\Ir
$$
(where $\{n_x\} \in [0,2\Spin]^{\Ir}$ with finite support),
and the lowest eigenvalue,
for any state other than $\phi_{\{0\}} = \ket{\uparrow}_\Ir$,
is $2 \Spin (1 - \Delta^{-1})$, which occurs whenever
$n_x = \delta_{xy}$ for some $y \in \Ir$.
This shows that $\gamma \geq 2 \Spin (1 - \Delta^{-1})$.

To obtain the reverse inequality, let $\ket{y}$ be the state
$\phi_{\{n_x\}}$ when $n_x = \delta_{xy}$.
We observe that 
$$
h^{\Spin}_{\{y,y\pm1\}} \ket{y} 
  = 2 \Spin \left[\frac{1}{2} (1 \pm \sqrt{1 - \Delta^{-2}}) \ket{y} 
  - \frac{1}{2 \Delta} \ket{y\pm1}\right]\, .
$$
Hence
$$
H^{\Spin}_\Ir \ket{y} 
  = 2 \Spin [\ket{y} - (2 \Delta)^{-1} \ket{y+1} - (2 \Delta)^{-1} \ket{y-1}]\, .
$$
Let $\chi_L = L^{-1/2} \sum_{y=1}^L \ket{y}$.
Then
$$
H^{\Spin}_\Ir \chi_L
  = 2 \Spin (1 - \Delta^{-1}) \chi_L
  + \frac{1}{2 \Delta \sqrt{L}} [-\ket{0} - \ket{L+1} + \ket{1} + \ket{L}]\, .
$$
So,
$$
\lim_{L \to \infty} \ip{\chi_L}{H^{\Spin}_\Ir \chi_L} 
  = 2 \Spin (1 - \Delta^{-1})\, ,
$$
which shows that $\gamma \leq 2 \Spin (1 - \Delta^{-1})$, as well.
\end{proof}

We observe that $2 \Spin (1 - \Delta^{-1})$ is not the exact value of $\gamma$
above the infinite-volume kink/antikink states.
To see this, consider the Ising limit $\Delta \to \infty$.
Then 
$$
\Psi_0 = \ket{\dots,+S,+S,+S,-S,-S,-S,\dots}
$$ 
is a ground state,
and 
$$
\Psi_1 = \ket{\dots,+S,+S,+S-1,-S+1,-S,-S,\dots}
$$ 
is an excitation, which is orthogonal to every ground state.
But the energy of the excitation is not $2 \Spin$, it is only 1.

\section{Numerical Approximation}
\label{Numerics}
The decomposition of the spin-$\Spin$ spin chain into a spin-$1/2$ spin ladder
was done just to prove the existence of a nonvanishing spectral gap.
However, in view of Lemma \ref{Key lemma}, we can obtain a lower bound for the
spectral gap of $H^{\Spin}_L$, in terms of the spectral gap of a 
much-reduced system.
This is useful from the point-of-view of a numerical method because, 
while $H^{\Spin}_L$ 
is sparse, even as a sparse matrix its dimension grows so quickly as a
function of $L$ and $\Spin$ that it poses serious memory problems even for a 
moderately large spin chain such as $\Spin = 3/2$, $L = 12$.
On the other hand, by the lemma we can obtain a lower bound for $\gamma_L$ 
in terms of $\delta_L$, where $\delta_L$ is the largest eigenvalue, less than
1, of the operator (notation, $\Lambda_L = [1,L]$)
$$\tilde{P}^{\textrm{Sym}}_L 
  := \Proj(\Hil^{\Spin}_0(\Lambda_L)) P^{\textrm{Sym}}_{\Lambda_L}
  \Proj(\Hil^{\Spin}_0(\Lambda_L))\, .
$$
The calculation of $\delta_L$ is possible because of the fact that the subspace
$\Hil^{\Spin}_0(\Lambda_L)$ has dimension $(L+1)^{2\Spin}$, which is much less than
the original dimension of $\Hil^{\Spin}(\Lambda_L)$ which is $2^{L(2\Spin+1)}$.

One may ask whether the determination of a lower bound for the spectral gap 
for finite systems really tells us something important about the infinite system.
We believe it does, since our main theorem shows that the spectral gap for the 
infinite-volume Hamiltonian is the {\em finite, non-zero} limit of the spectral
gap for the finite systems. 

The purpose of the next few paragraphs is to express the operator
$\tilde{P}^{\textrm{Sym}}_{\Lambda_L}$ in the notation of symmetric functions.
This is done partially just to obtain formulas for the matrix entries, which
can then be evaluated numerically.
But also, by this method we use the symmetries present to 
identify a large kernel for $\tilde{P}^{\text{Sym}}_L$ within 
$\Hil^{\Spin}_0(\Lambda_L)$.
We will see that the combinatorial formulas that we bring in are very relevant to
the problem, and arise naturally from direct analysis.
We begin by defining an orthonormal system for $\Hil^{\Spin}_{\Lambda_L}$.
A classical configuration would be specified by stating for exactly which pairs
$(x,m)$ there is a down spin.
From this point-of-view it is useful to define $\M(2\Spin,L;N)$ to be the set 
of all $2\Spin\times L$ matrices $A = (a(m,x))_{(m,x)}$ such that
$a(m,x) \in \{0,1\}$ for all $x$ and $m$, and such that the total number of $1$'s
is $N$.
Then we define the state
$$
\phi_A = \prod_{x=1}^L \prod_{m=1}^{2\Spin} (S_{(x,m)}^-)^{a(m,x)}\
  \ket{\uparrow}_{\Lambda_L \times [1,2\Spin]}\, .
$$
These states form an orthonormal basis for $\Hil^{\Spin}_0(\Lambda_L)$.
E.g.,
$$
A = \begin{pmatrix}0 & 1 & 1 \\ 0 & 0 & 1\end{pmatrix}\, ,\qquad
\phi_A = \left|{\begin{matrix} \uparrow & \downarrow & \downarrow \\
\uparrow & \uparrow & \downarrow \end{matrix}}
\right\rangle\, .
$$
For each matrix $A$ we define a length-$2\Spin$ vector of the row sums,
and a length-$L$ vector of the column sums:
$$
a(m,\Sigma) = \sum_{x=1}^L a(m,x)\, ,\qquad
a(\Sigma,x) = \sum_{m=1}^{2\Spin} a(m,x)\, .
$$
Then equations \eq{ASW formula} and \eq{Ladder state definition},
can be written as
\begin{equation}
\label{New ladder state formula}
\Psi_0(\Lambda_L,\nvec)
  = \sum_{\substack{A \in \M(2\Spin,L;N) \\ a(\cdot,\Sigma) = \nvec}}
  \phi_A \prod_{x=1}^L q^{x a(\Sigma,x)}\, .
\end{equation}
From this, we see that $\ip{\Psi_0(\Lambda_L,\nvec)}{\Psi_0(\Lambda_L,\mvec)}$
is zero unless $\mvec = \nvec$, and
$$
\|\Psi_0(\Lambda_L,\nvec)\|^2 = 
  \sum_{\substack{A \in \M(2\Spin,L;N) \\ a(\cdot,\Sigma) = \nvec}}
  \|\phi_A\|^2 \prod_{x=1}^L q^{2 x a(\Sigma,x)}
  = \sum_{\substack{A \in \M(2\Spin,L;N) \\ a(\cdot,\Sigma) = \nvec}}
  \prod_{x=1}^L q^{2 x a(\Sigma,x)}\, .
$$
We define
$$
M_{\nvec,\avec} = \sum_{\substack{B \in \M(2\Spin,L;N) \\ 
  b(\cdot,\Sigma) = \nvec,\
  b(\Sigma,\cdot) = \avec }} 1  \, ,
$$
with the result that
\begin{equation}
\label{First inner-product formula}
\begin{split}
\|\Psi_0(\Lambda_L,\nvec)\|^2 
  &= \sum_{\avec \in [0,2\Spin]^L} \prod_{x=1}^L q^{2 x a_x}\
  M_{\nvec,\avec} \\
  &= \sum_{\avec \in [0,2\Spin]^L} q^{2 \boldsymbol{x} \cdot \avec}\
  M_{\nvec,\avec}\, ,
\end{split}
\end{equation}
where $\boldsymbol{x} = (1,2,\dots,L) \in \Ir^L$.
We will come back to this equation later, to see how it can be made even
simpler.

We can also write
$$
\phi_A = \phi_{\{1\}\times a(\cdot,1)} \otimes \phi_{\{2\}\times a(\cdot,2)}
  \otimes \cdots \otimes \phi_{\{L\}\times a(\cdot,L)}\, ,
$$ 
which helps us to deduce
\begin{align*}
P^{\textrm{sym}}_{\Lambda_L} \phi_A
  &= \bigotimes_{x=1}^L P^{\textrm{sym}}_x \phi_{\{x\}\times a(\cdot,x)} \\
  &= \bigotimes_{x=1}^L \binom{2\Spin}{a(\Sigma,x)}^{-1} 
  \sum_{\substack{\boldsymbol{u} \in \{0,1\}^{2\Spin} \\
  |\boldsymbol{u}| = a(\Sigma,x)}} \phi_{\{x\}\times \boldsymbol{u}} \\
  &= \prod_{x=1}^L \binom{2\Spin}{a(\Sigma,x)}^{-1}\
  \sum_{\substack{B \in \M(2\Spin,L;N) \\ 
  b(\Sigma,\cdot) = a(\Sigma,\cdot)}} \phi_B\, .
\end{align*}
Therefore, combining this with \eq{New ladder state formula},
\begin{align*}
P^{\textrm{sym}}_{\Lambda_L} \Psi_0(\Lambda_L,\nvec)
  &= \sum_{\substack{A \in \M(2\Spin,L;N) \\ a(\cdot,\Sigma) = \nvec}}
  \prod_{x=1}^L\left[\binom{2\Spin}{a(\Sigma,x)}^{-1} q^{x a(\Sigma,x)}\right]
  \sum_{\substack{B \in \M(2\Spin,L;N) \\ 
  b(\Sigma,\cdot) = a(\Sigma,\cdot)}} \phi_B  \\
  &= \sum_{B \in \M(2\Spin,L;N)} \phi_B  
  \prod_{x=1}^L\left[\binom{2\Spin}{b(\Sigma,x)}^{-1} q^{x b(\Sigma,x)}\right]
  M_{\nvec,b(\Sigma,\cdot)} \\
  &= \sum_{\substack{\avec \in [0,2\Spin]^L \\ |\avec| = |\nvec|}}
  \prod_{x=1}^L\left[\binom{2\Spin}{a_x}^{-1} q^{x a_x}\right]
  M_{\nvec,\avec} 
  \sum_{\substack{B \in \M(2\Spin,L;N) \\ 
  b(\Sigma,\cdot) = \avec}} \phi_B  \, .
\end{align*}
Since
$$
\sum_{\substack{B \in \M(2\Spin,L;N) \\ b(\Sigma,\cdot) = \avec}}
  \|\phi_B\|^2
  = \sum_{\substack{B \in \M(2\Spin,L;N) \\ b(\Sigma,\cdot) = \avec}} 1
  = \prod_{x=1}^L \binom{2\Spin}{a_x}\, ,
$$
we then see that
\begin{equation}
\label{First sym inner-product}
\ip{P^{\textrm{sym}}_{\Lambda_L} \Psi_0(\Lambda_L,\nvec)}
{P^{\textrm{sym}}_{\Lambda_L} \Psi_0(\Lambda_L,\mvec)}
  = \sum_{\substack{\avec \in [0,2\Spin]^L \\ |\avec| = |\nvec|}}
  \prod_{x=1}^L\left[\binom{2\Spin}{a_x}^{-1} q^{2 x a_x}\right]
  M_{\nvec,\avec} M_{\mvec,\avec}\, .
\end{equation}

To further simplify equations \eq{First inner-product formula}
and \eq{First sym inner-product}, we observe that 
for any $\pi \in \mathfrak{S}_{2\Spin}$ and $\varrho \in \mathfrak{S}_L$,
$M_{\nvec,\avec} = M_{\pi(\nvec),\varrho(\avec)}$, where
$$
\pi(\nvec)_m = (\nvec)_{\pi(m)}\, ,\qquad
\varrho(\avec)_x = (\avec)_{\varrho(x)}\, .
$$
Moreover, for each orbit of $[0,2\Spin]^L$ under the action of 
$\mathfrak{S}_L$, there is a unique vector 
$\lambda = (\lambda_1,\dots,\lambda_L)$
such that $\lambda_1 \geq \lambda_2 \geq \dots \geq \lambda_L$.
For such a $\lambda$, one defines the {\em monomial symmetric function}
(in our case we consider it just as a function of $L$ variables
$\tvec \in \Cx^L$)
$$
m_\lambda(\tvec) = \sum_{\avec \sim \lambda} \prod_{x=1}^L t_x^{a_x}\, ,
$$
where $\sim$ means equivalent modulo the action of $\mathfrak{S}_L$.
We define $\Par(L,2\Spin;N)$ to be the subset of $[0,2\Spin]^L$ consisting of
those $\lambda$ such that $\lambda_1\geq\dots\geq\lambda_L$ and 
$\sum_{x=1}^L \lambda_x = N$.
They are called the partitions.
If $\mu,\nu \in \Par(2\Spin,L;N)$ are the partitions such that
$\mvec \sim \mu$, $\nvec \sim \nu$ modulo the action of $\mathfrak{S}_{2\Spin}$
on $[0,L]^{2\Spin}$, 
then equation \eq{First inner-product formula} can be rewritten as
\begin{equation}
\label{Second inner-product formula}
\|\Psi_0(\Lambda_L,\nvec)\|^2 
  = \sum_{\lambda \in \Par(2\Spin,L;N)} M_{\nu \lambda}
  m_\lambda(q^2,q^4,\dots,q^{2L})\, ,
\end{equation}
and equation \eq{First sym inner-product} can be written
\begin{equation}
\label{Second sym inner-product}
\ip{\Psi_0(\Lambda_L,\nvec)}
{\tilde{P}^{\textrm{sym}}_{\Lambda_L} \Psi_0(\Lambda_L,\mvec)}
  = \sum_{\lambda \in \Par(2\Spin,L;N)} \frac{M_{\nu \lambda} M_{\mu \lambda}
  m_\lambda(q^2,q^4,\dots,q^{2L})}
  {\prod_{x=1}^L \binom{2\Spin}{\lambda_x}}\, .
\end{equation}

It so happens that $m_\lambda(q^2,q^4,\dots,q^{2L})$ is difficult,
or at least messy, to calculate, particularly when $\lambda$ has many nonzero
parts.
In order to fix this difficulty, we bring in the {\em elementary symmetric 
functions}.
For $0\leq n\leq L$, define
$$
e_n(\tvec) = \sum_{1\leq x_1 < \dots < x_n \leq L} 
  t_{x_1} t_{x_2} \cdots t_{x_n}\, ,
$$
where $\tvec = (t_1,\dots,t_L)$,
and for $\mu \in \Par(2\Spin,L;N)$ define
$$
e_\mu(\tvec) = \prod_{m=1}^{2\Spin} e_{\mu_m}(\tvec)\, .
$$
These are the elementary symmetric functions (restricted to $L$ variables).
Two important properties of the elementary symmetric functions are
\begin{gather}
\label{Elementary symmetric property 1}
e_\mu(\tvec) = \sum_{\lambda \in \Par(L,2\Spin;N)} M_{\mu\lambda} 
  m_\lambda(\tvec)\, ; \\
\label{Elementary symmetric property 2}
e_\mu(q^2,q^4,\dots,q^{2L})
  = \prod_{m=1}^{2\Spin} q^{\mu_m(\mu_m+1)} \qbinom{L}{\mu_m}{q^2}\, .
\end{gather}
The second formula involves the well-known Gaussian polynomials
$$
\qbinom{n}{k}{q} = \prod_{j=1}^k \frac{1-q^{n-k+j}}{1-q^j}\, .
$$
Both of these formulas are proved in Richard Stanley's book \cite{Sta}.
With these formulas, equation \eq{Second inner-product formula} can be rewritten
as
\begin{equation}
\label{Norm of ladder state}
\|\Psi_0(\Lambda_L,\nvec)\|^2 
  = \prod_{m=1}^{2\Spin} q^{n_m(n_m+1)} \qbinom{L}{n_m}{q^2}\, .
\end{equation}
This simple formula can also be determined by other means.

For the matrix elements,
$\ip{\Psi_0(\Lambda_L,\nvec)}
{\tilde{P}^{\textrm{sym}}_{\Lambda_L} \Psi_0(\Lambda_L,\mvec)}\, ,$ 
we still seem to be stuck with the
calculation of $m_\lambda(q^2,\dots,q^{2L})$.
We can get around this by defining 
$$
(M^{\lambda\mu})_{\substack{\lambda \in \Par(L,2\Spin;N)\\
\mu \in \Par(2\Spin,L;N)}} 
  = \left((M_{\mu\lambda})_{\substack{\mu \in \Par(2\Spin,L;N) \\
\lambda \in \Par(L,2\Spin;N)}}\right)^{-1}\, .
$$
It is easy to convince oneself that the matrix elements $M^{\lambda\mu}$
are also integers, because $M_{\mu\lambda}$ has the triangularity property
that $M_{\mu\lambda} = 0$ unless $\lambda \leq \mu'$, where $\leq$
refers to dominance order.
This fact is also proved in \cite{Sta}.
Since the inverse of an upper-triangular matrix with integer components is 
also upper-triangular with integer components, we see that
$M^{\lambda\mu}$ is always an integer, and the integer is 0 unless $\mu' \leq \lambda$.
Then we see that
$$
m_\lambda(\tvec) = \sum_{\kappa \in \Par(2\Spin,L;N)} M^{\lambda\kappa} 
  e_\kappa(\tvec)\, .
$$
Along with the triangularity properties of 
$$
(M_{\mu\lambda})_{\substack{\mu \in \Par(2\Spin,L;N) \\
\lambda \in \Par(L,2\Spin;N)}}\quad \textrm{and}\quad
(M^{\lambda\mu})_{\substack{\lambda \in \Par(L,2\Spin;N)\\
\mu \in \Par(2\Spin,L;N)}}\, ,
$$
this last formula can be used to 
rewrite equation \eq{Second sym inner-product} as
\begin{equation}
\label{Third sym inner-product}
\ip{\Psi_0(\Lambda_L,\nvec)}
{\tilde{P}^{\textrm{sym}}_{\Lambda_L} \Psi_0(\Lambda_L,\mvec)}
  = \sum_{\substack{\kappa \in \Par(L,2\Spin;N)\\
  \kappa \geq \mu \vee \nu}} e_{\kappa}(q^2,q^4,\dots,q^{2L})
  \sum_{\substack{\lambda \in \Par(2\Spin,L;N)\\
  \mu \vee \nu \leq \lambda' \leq \kappa}} 
  \frac{M_{\nu \lambda} M_{\mu \lambda} M^{\lambda\kappa}}
  {\prod_{x=1}^L \binom{2\Spin}{\lambda_x}}
\end{equation}
This is compared with
\begin{equation}
\label{Third inner-product formula}
\|\Psi_0(\Lambda_L,\nvec)\|^2 
  = e_{\nu}(q^2,q^4,\dots,q^{2L})\, ,
\end{equation}
which we already derived.

We observe the fact that
$\ip{\Psi_0(\Lambda_L,\nvec)}
{\tilde{P}^{\textrm{sym}}_{\Lambda_L} \Psi_0(\Lambda_L,\mvec)}\, ,$ 
only depends on the orbit of $\nvec$ and $\mvec$ w.r.t.\ the action of
$\mathfrak{S}_{2\Spin}$.
We define the subspace 
$$
V_{\nu} = \Span\{\Psi_0(\Lambda_L,\nvec) : \nvec \sim \nu\}
$$
which has dimension
$$
m_\nu(1,1,\dots,1) = \# \textrm{ orbit of } \nu 
 = \binom{2\Spin}{\# 1'\textrm{s}(\nu), \# 2'\textrm{s}(\nu), \dots, 
\# L'\textrm{s}(\nu)}\, .
$$
Then cokernel of $\tilde{P}^{\textrm{sym}}_{\Lambda_L}$ intersects $V_{\nu}$
in the single-dimensional subspace spanned by the element
$$
\psi_\nu = \frac{1}{m_{\nu}(1,1,\dots,1)} \sum_{\nvec \sim \nu} \Psi_0(L,\nvec)\, .
$$
Then, from \eq{Third inner-product formula},
$$
\|\psi_\nu\|^2 = \frac{1}{m_{\nu}(1,1,\dots,1)} e_\nu(q^2,q^4,\dots,q^{2L})\, ,
$$
while, from \eq{Third sym inner-product}
$$
\ip{\psi_\mu}{\tilde{P}^{\textrm{sym}}_{\Lambda_L} \psi_\nu}
  = \sum_{\substack{\kappa \in \Par(L,2\Spin;N)\\
  \kappa \geq \mu \vee \nu}} e_{\kappa}(q^2,q^4,\dots,q^{2L})
  \sum_{\substack{\lambda \in \Par(2\Spin,L;N)\\
  \mu \vee \nu \leq \lambda' \leq \kappa}} 
  \frac{M_{\nu \lambda} M_{\mu \lambda} M^{\lambda\kappa}}
  {\prod_{x=1}^L \binom{2\Spin}{\lambda_x}}\, .
$$
Thus, computing the normalized matrix entries of 
$\tilde{P}^{\textrm{sym}}_{\Lambda_L}$, we have
\begin{equation}
\label{normalized matrix entry}
\begin{split}
\frac{\ip{\psi_\mu}{\tilde{P}^{\textrm{sym}}_{\Lambda_L} \psi_\nu}}
{\|\psi_\mu\|\, \|\psi_\nu\|}
  &= \sum_{\substack{\kappa \in \Par(L,2\Spin;N)\\
  \kappa \geq \mu \vee \nu}} 
  \frac{\ps_{L}(e_\kappa;q^2)}
  {\sqrt{\ps_{L}(e_\mu;q^2) \ps_{L}(e_\nu;q^2)}}\\
  &\quad \times \sum_{\substack{\lambda \in \Par(2\Spin,L;N)\\
  \mu \vee \nu \leq \lambda' \leq \kappa}} 
\frac{\sqrt{\ps_{2\Spin}(m_\mu;1) \ps_{2\Spin}(m_\nu;1)}}
{\ps_{2\Spin}(e_\lambda;1)}
M_{\nu \lambda} M_{\mu \lambda} M^{\lambda\kappa}\, ,
\end{split}
\end{equation}
where
$$
\ps_n(f,x) := f(1,x,x^2,\dots,x^{n-1})\, .
$$

One aspect of this formula is that a certain part of it is totally independent
of the value of $q$.
We may define 
$$
\Gamma_{\mu \nu}^\kappa = 
  \sum_{\substack{\lambda \in \Par(2\Spin,L;N)\\
  \mu \vee \nu \leq \lambda' \leq \kappa}} 
\frac{\sqrt{\ps_{2\Spin}(m_\mu;1) \ps_{2\Spin}(m_\nu;1)}}
{\ps_{2\Spin}(e_\lambda;1)}
M_{\nu \lambda} M_{\mu \lambda} M^{\lambda\kappa}\, ,
$$
so that
\begin{equation}
\label{Second normalized matrix entry}
\mathcal{M}_L(\mu,\nu)
= \frac{\ip{\psi_\mu}{\tilde{P}^{\textrm{sym}}_{\Lambda_L} \psi_\nu}}
{\|\psi_\mu\|\, \|\psi_\nu\|}
  = \sum_{\substack{\kappa \in \Par(L,2\Spin;N)\\
  \kappa \geq \mu \vee \nu}} 
  \frac{\ps_{L}(e_\kappa;q^2)}
  {\sqrt{\ps_{L}(e_\mu;q^2) \ps_{L}(e_\nu;q^2)}} \Gamma_{\mu \nu}^\kappa\, .
\end{equation}
One advantage of $\Gamma_{\mu\nu}^{\kappa}$ is that it stabilizes 
under $L$.
I.e.\ if $L$ and $L'$ are each large enough that $\mu$, $\nu$ and $\kappa$ are
in $\Par(L,2\Spin;N)$ as well as $\Par(L,2\Spin;N)$, then 
the value of $\Gamma_{\mu\nu}^\kappa$ agrees for both choices of $L$ or $L'$.
Also, the numbers $\Gamma_{\mu\nu}^\kappa$ have a certain translation invariance.
Specifically, adding one to each part of $\mu$, $\nu$ and $\kappa$ 
(i.e.\ $\mu_m \leftarrow \mu_m+1$, for $m=1,2,\dots,2\Spin$, etc.) does not change
the value of $\Gamma_{\mu\nu}^\kappa$, for $L$ large enough.
Also, the leading-order behavior in $q$ is easy to see, since
$$
\frac{\ps_{L}(e_\kappa;q^2)}
{\sqrt{\ps_{L}(e_\mu;q^2) \ps_{L}(e_\nu;q^2)}}
  = \prod_{m=1}^{2\Spin} \qbinom{L}{\kappa_m}{q^2} \qbinom{L}{\mu_m}{q^2}^{-1/2} 
  \qbinom{L}{\kappa_m}{q^2}^{-1/2}
  q^{\kappa_m^2 - \frac{1}{2} \mu_m^2 - \frac{1}{2} \nu_m^2}\, .
$$
Define $\Par(\Nl,2\Spin;N)$ to be the set of all partitions of $N$.
If one fixes $N$, and $\mu,\nu,\kappa \in \Par(\Nl,2\Spin;N)$, then taking 
$L \to \infty$ yields
$$
\lim_{L \to \infty} \frac{\ps_{L}(e_\kappa;q^2)}
{\sqrt{\ps_{L}(e_\mu;q^2) \ps_{L}(e_\nu;q^2)}}
  = \prod_{m=1}^{2\Spin} \frac{\sqrt{(q^2;q^2)_{\mu_m} (q^2;q^2)_{\nu_m}}}
  {(q^2;q^2)_{\kappa_m}} q^{\kappa_m^2 - \frac{1}{2} \mu_m^2 - \frac{1}{2} \nu_m^2}\, .
$$
This is only true for $q$ strictly less than 1.
However, for such $q$, 
this gives a lower bound for the finitely magnetized spin chain on the half-infinite
lattice $\Lambda = \Nl$:
\begin{equation}
\label{half-infinite model}
\begin{split}
\mathcal{M}_{\Nl}(\mu,\nu)
= \lim_{L \to \infty}
  \frac{\ip{\psi_\mu}{\tilde{P}^{\textrm{sym}}_{\Lambda_L} \psi_\nu}}
  {\|\psi_\mu\|\, \|\psi_\nu\|}
  &= \sum_{\substack{\kappa \in \Par(\Nl,2\Spin;N)\\
  \kappa \geq \mu \vee \nu}} 
\left(\prod_{m=1}^{2\Spin} \frac{\sqrt{(q^2;q^2)_{\mu_m} (q^2;q^2)_{\nu_m}}}
  {(q^2;q^2)_{\kappa_m}}\right) \\
  &\quad \times q^{\|\kappa\|^2 - \frac{1}{2} \|\mu\|^2 - \frac{1}{2} \|\nu\|^2}
  \Gamma_{\mu \nu}^\kappa\, ,
\end{split}
\end{equation}
which is defined for all $\mu,\nu \in \Par(\Nl,2\Spin;N)$.

This is almost of the form of a perturbation series, were it not for the $q$-shifted
factorials appearing in the formula.
In fact, there is a way to eliminate these last remnants of $q$-combinatorics, but 
at the expense of exchanging the finite matrices listed above for infinite matrices.
The way to do this is to take a sequence of spin chains with $N_i \to \infty$ and 
$L_i \to \infty$, but such that
$N_i \equiv N_j (\mod 2\Spin)$ for all $i$ and $j$, and such that
$L_i - N_i \to \infty$.
This is natural, anyway because it is by this limit that one obtains a state
on the bi-infinite spin chain $\Lambda = \Ir$.
Define $\Par(\Ir,2\Spin;N)$ to be the set of ``signed partitions'' or in other words
just all sequences $\mu$ of $2\Spin$ integers such that
$-\infty < \mu_1 \leq \mu_2 \leq \dots \leq \mu_{2\Spin} < +\infty$ such that
$\sum_{m=1}^{2\Spin} \mu_m = N$.
If one takes $\mu,\nu,\kappa \in \Par(\Ir,n;N)$ for $0 \leq n \leq 2\Spin-1$, then
as $N_i = 2\Spin k_i + n$, one can take $\mu_i = \mu + (k_i,k_i,\dots,k_i)$, etc.
Then one obtains 
\begin{equation}
\label{bi-infinite model}
\mathcal{M}_{\Ir}(\mu,\nu)
= \lim_{i \to \infty}
  \frac{\ip{\psi_{\mu_i}}{\tilde{P}^{\textrm{sym}}_{\Lambda_{L_i}} \psi_{\nu_i}}}
  {\|\psi_{\mu_i}\|\, \|\psi_{\nu_i}\|}
  = \sum_{\substack{\kappa \in \Par(\Ir,2\Spin;N)\\
  \kappa \geq \mu \vee \nu}} 
  q^{\|\kappa\|^2 - \frac{1}{2} \|\mu\|^2 - \frac{1}{2} \|\nu\|^2}
  \Gamma_{\mu \nu}^\kappa\, ,
\end{equation}
Note that $\Gamma_{\mu\nu}^\kappa$ is well-defined due to the translation-invariance
of $\Gamma$ with respect to its arguments.
The $q$-shifted factorials all approach $(q^2;q^2)_\infty$ in the limit $i\to \infty$,
so their ratio cancels out.
Again, this is only the case for $q$ strictly less than 1.

In the last two paragraphs we have taken limits, without commenting on whether it is 
valid to do so.
Nor will we provide any attempt at rigorous justification to these procedures.
The reason for our laziness is this: the entire point of this analysis is to obtain
a simple model from which we can extract lower-bounds, which requires numerical 
computation, for the original model, which is the XXZ spin chain.
We were, and are, interested in obtaining data from these lower-bounds, which hopefully
leads us to make conjectures about the original model.
It may in fact be easier to prove the conjectures for the XXZ model directly, rather than
for the lower-bounds.
Some of the conjectures will be presented in section \ref{section conjectures}.
We have already begun work on the proofs, and our results will be reported
in a later paper.
We don't want to undercut the importance of the lower bound derived in this section.
It has two important roles.
One is that it is the best, i.e.\ most efficient, numerical tool we have for looking
at the low spectrum of the XXZ spin chain for larger values of $\Spin$, so far.
The second is that it has an interesting definition in terms of representations of
quantum groups.
We present this now.

\subsection{Interpretation in terms of representations of quantum groups}
We observe the following interpretation for the spectrum of $\tilde{P}_{\Lambda_L}$.
We started by considering a spin ladder, which is really nothing more than an
$L \times 2\Spin$ array of 2-dimensional irreps of $\textrm{SU}(2)$,
$(\Cx_{(x,m)}^2)_{(x,m) \in [1,L]\times[1,2\Spin]}$.
Of course, the 2-dimensional representations of $\textrm{SU}_q(2)$ are identical 
to the 2-dimensional representation of $\textrm{SU}(2)$.
So we can also think of our spin ladder as an array of 2 dimensional representations
of $\textrm{SU}_q(2)$.
We define two sets of projectors.
For each $m \in [1,2\Spin]$, we define 
$P_q^{\textrm{top},(m)} \in \Obs(\bigotimes_{x=1}^L \Cx^2_{(x,m)})$ to be the projection
onto the top-dimensional irrep in the $\textrm{SU}_q(2)$ representation
$\bigotimes_{x=1}^L \Cx^2_{(x,m)}$.
For each $x \in [1,L]$, we define 
$Q_1^{\textrm{top},(x)} \in \Obs(\bigotimes_{m=1}^{2\Spin} \Cx^2_{(x,m)})$ 
to be the projection onto the top-dimensional 
irrep in the $\textrm{SU}(2)$ representation
$\bigotimes_{m=1}^{2\Spin} \Cx^2_{(x,m)}$.
Then we ask for the spectrum of the operator 
$$
\mathcal{Q}(L,2\Spin) = 
\bigotimes_{x=1}^L Q_1^{\textrm{top},(x)} \bigotimes_{m=1}^{2\Spin} 
P_q^{\textrm{top},(m)}\, .
$$
There are $L\Spin+1$ eigenvectors with eigenvalue 1, which correspond to the 
ground states of the spin-chain, which can also be interpreted as the totally
symmetric states with respect to some representation of the symmetric group
$\mathfrak{S}([1,L]\times [1,2\Spin])$.
(See Section 2.3.1 for more details.)
The operator can be block diagonalized according to the eigenvalues of 
$S^3_{\textrm{tot}} = \sum_{(x,m)} S^3_{(x,m)}$, which is well-defined since
the image of $S^3$ is the same whether we consider a representation of 
$\textrm{SU}(2)$ or $\textrm{SU}_q(2)$.
Then in each sector there is a unique next-highest eigenvalue, and we write this
as $1 - \delta_i(M)$.
Here $M$ is the eigenvalue of $S^3_{\textrm{tot}}$ which lies between 
$-L\Spin$ and $+L\Spin$.
The matrix defined in equation \eq{bi-infinite model} corresponds
to the limit of $\mathcal{Q}(L,2\Spin;n)$ as $L\to\infty$ with $n$ fixed.
For the special case that $q=1$, the question is entirely in terms of representations
of $\textrm{SU}(2)$, but we note that we expect that as $L \to \infty$ with $2\Spin$
fixed $\delta_i \to 0$.
For other values of $q$, this is not the anticipated behavior.
For example, for $q=0$ and $n=0$, we know that the answer is $\delta_i = 1/2\Spin$,
regardless of $L$.
The gap, $\gamma_i$, of the original spin chain is related to $\delta_i$ by 
$\gamma_i \geq 2 \Spin \frac{(1-q)^2}{1+q^2} \delta_i$.
An interesting limit is obtained by first taking $L \to \infty$, and 
then taking $2\Spin \to \infty$.
This is the classical limit of the spectral gap in the infinite-volume model
(thermodynamic limit).

\section{Numerical data}
\begin{figure}
\begin{center}
\resizebox{15truecm}{15truecm}{\includegraphics{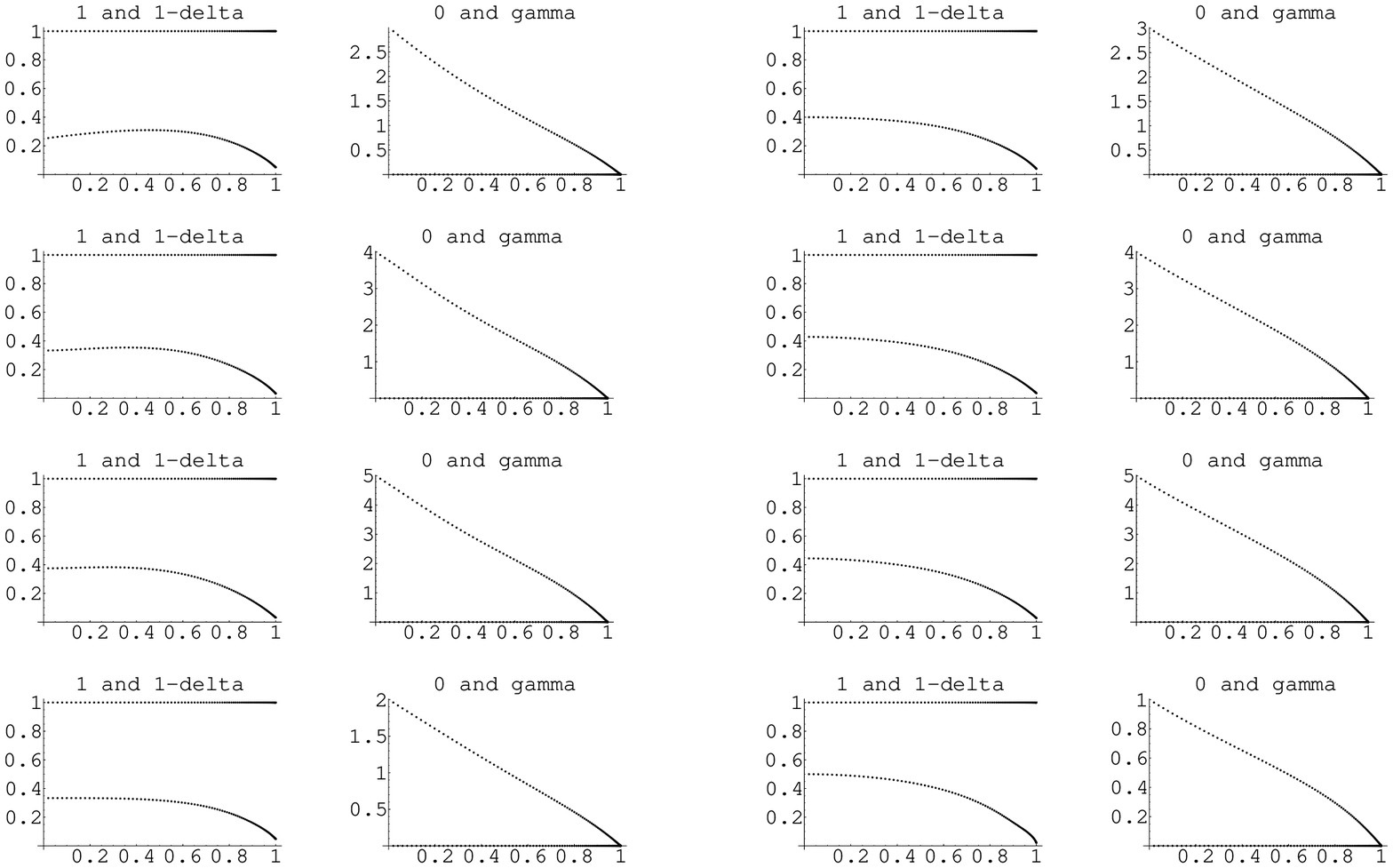}}\qquad
\end{center}
\caption{Plot of $1$, $1-\delta$ (left) and $0$, $\gamma$ (right) versus
$\Delta^{-1}$, for $(\Spin,n,L)$ equal to: First column  $(2, 2, 5)$, 
$(3, 3, 5)$, $(4, 4, 4)$, $(3/2, 1, 7)$;
Second column $(5/2, 2, 5)$, $(7/2, 3, 4)$, $(9/2, 4, 4)$, $(1, 0, 24)$}
\label{lowerbound1}
\end{figure}

\begin{figure}
\begin{center}
\resizebox{15truecm}{15truecm}{\includegraphics{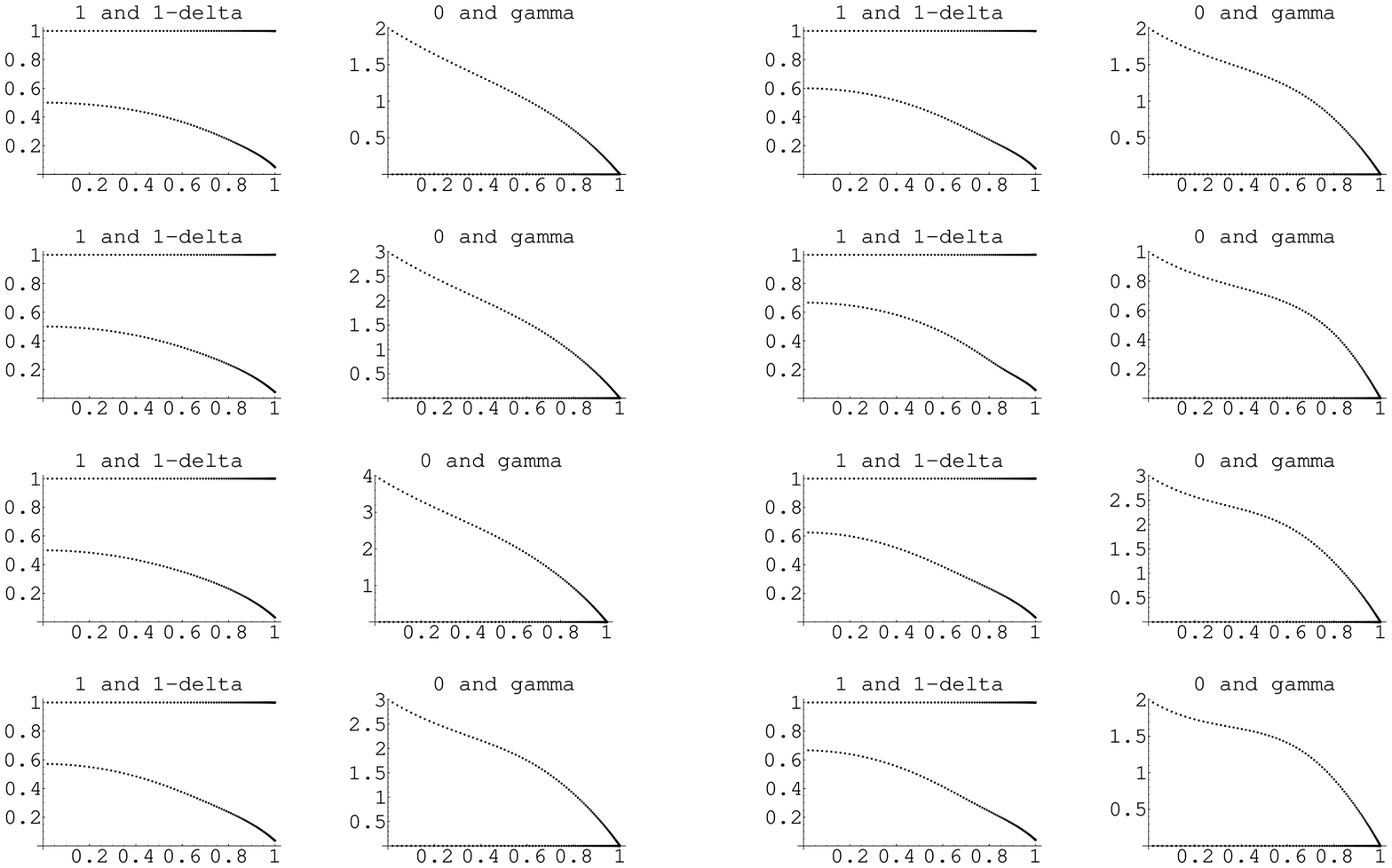}}\qquad
\end{center}
\caption{Plot of $1$, $1-\delta$ (left) and $0$, $\gamma$ (right) versus
$\Delta^{-1}$, for $(\Spin,n,L)$ equal to: First column  
$(2, 1, 5)$, $(3, 2, 4)$, $(4, 3, 4)$, $(7/2, 2, 4)$;
Second column $(5/2, 1, 5)$, $(3/2, 0, 6)$, $(4, 2, 4)$, $(3, 1, 4)$}
\label{lowerbound2}
\end{figure}

\begin{figure}
\begin{center}
\resizebox{15truecm}{15truecm}{\includegraphics{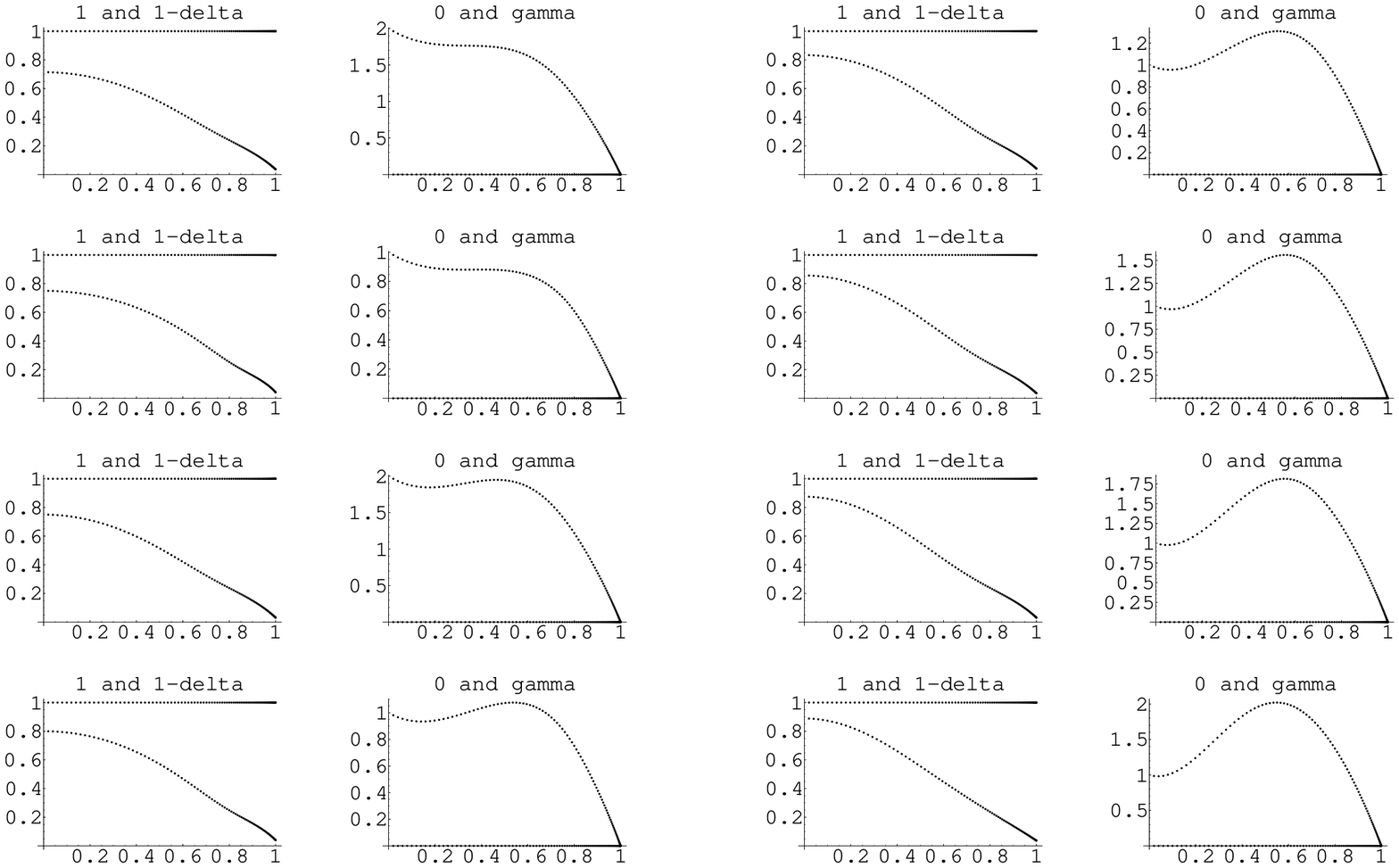}}\qquad
\end{center}
\caption{Plot of $1$, $1-\delta$ (left) and $0$, $\gamma$ (right) versus
$\Delta^{-1}$, for $(\Spin,n,L)$ equal to: First column  
$(7/2, 1, 4)$, $(2, 0, 6)$, $(4, 1, 4)$, $(5/2, 0, 5)$;
Second column $(3, 0, 4)$, $(7/2, 0, 4)$, $(4, 0, 4)$, $(9/2, 0, 3)$}
\label{lowerbound3}
\end{figure}

We present the results of our numerical experiments, now.
We begin by looking at the results of the lower bounds estimates.
For these estimates we considered the rigorous lower bounds obtained by equation
\eq{Second normalized matrix entry}.
More specifically, fixing a sector with, $N=2\Spin\floor{L/2}+n$ down spins, 
we looked at the highest two eigenvalues $1,1-\delta(L,n)$ of the matrix determined
by equation \eq{Second normalized matrix entry} in that sector.
This can be related to the spectral gap of the XXZ spin chain for the finite volume
$\Lambda_L$ by corollary \ref{Useful corollary}.
Since we expect the behavior to be nearly translation invariant for large enough
$L$ (i.e.\ in the limit that $L \to \infty$ we expect the system to be entirely 
translation invariant, taking the sector defined by $n \rightarrow n\pm 2\Spin$
should not change the gap), we only considered $n=0,1,2\dots,2\Spin$.
Moreover, since we expect the system to be invariant under simultaneous spin-flip
and reflection of the spin chain, we can map $n$ into $2\Spin-n$.
Therefore, we only conisdered $n=0,1,\dots,\floor{\Spin}$.
Also, for obvious reason we did not begin with $\Spin=1/2$, but with $\Spin=1$.
See Figures \ref{lowerbound1}, \ref{lowerbound2}, \ref{lowerbound3}.
It appears from the figures that in the sectors determined by $n=0$, as one
takes $\Spin \to \infty$, then $\delta$ approaches the functional form
$\delta = 1-\Delta^{-1}$.
In this case, one would have the lower bound for $\gamma$ equal to
$2\Spin \Delta^{-1} (1 - \Delta{^-1})$.
This means that the lower bound for the gap scales with $\Spin$ as $\Spin$,
and that $\Spin^{-1} \tilde{\gamma} \sim 2 \Delta^{-1} (1-\Delta^{-1})$.
This curve is significant because it has its maximum at a point other than
the Ising limit, specifically at $\Delta^{-1} = 1/2$, i.e.\ $\Delta = 2$.

\begin{figure}
\begin{center}
\resizebox{7truecm}{7truecm}{\includegraphics{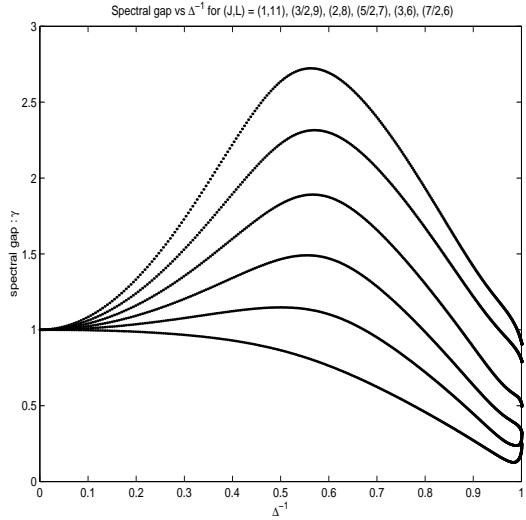}}\qquad
\end{center}
\caption{Some plots of the spectral gap using Lanczos iteration. For all three curves,
$n\equiv 0 \mod 2J$.}
\label{lanczospics}
\end{figure}

To verify some of the results we calculated the real spectral gap for the XXZ 
spin chain for some small
values of $\Spin$ and $L$ using Lancz\"os iteration.
See Figure \ref{lanczospics}.
This should give a very accurate answer, i.e. the numerical error is very low.
Unfortunately, the finite-size effects are more significant for the full XXZ
spin chain than for the lower bound, which appears to converge very rapidly in $L$.
For example, the Lancz\"os results at $q=1$ are certainly not the same as for
$L\to \infty$, because then we expect $\gamma = 0$ at $q=1$.
It is well-known that there are gapless excitations, spin-waves, for the
isotropic model in infinite-volume.

\subsection{Perturbation Series about Ising Limit}

\begin{figure}
\begin{center}
\resizebox{9truecm}{9truecm}{\includegraphics{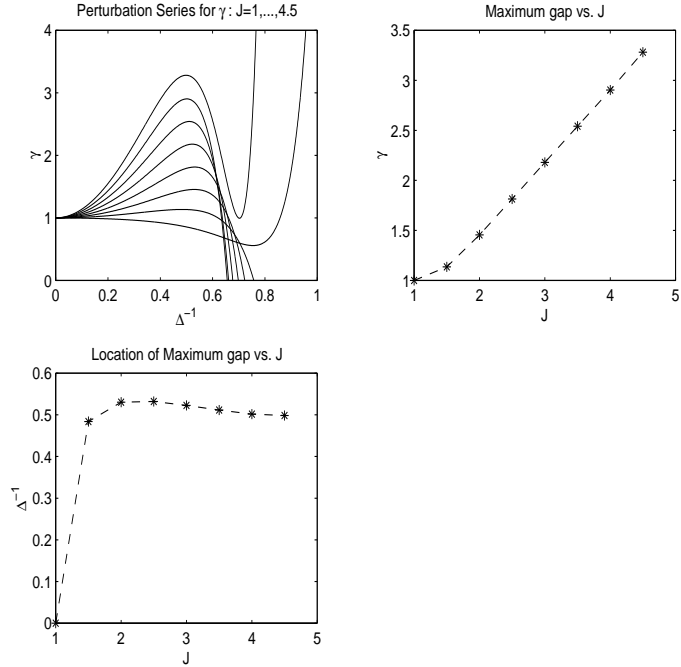}}\qquad
\end{center}
\caption{Some plots using $14$th-order perturbation series about $\Delta^{-1}=0$.}
\label{pertpics}
\end{figure}

We wish to perform a perturbation analysis for the first excited energy
in the small parameter $\Delta^{-1}$ about the Ising limit $\Delta^{-1}=0$.
We write
\begin{gather*}
H(\Delta^{-1}) = H^{(0)} + \Delta^{-1} H^{(1)} \\
H^{(0)} = \sum_{x=1}^{L-1} (\Spin + S_x^3)(\Spin - S_{x+1}^3) \\
H^{(1)} = \sum_{x=1}^{L-1} \left(-\frac{1}{2} S_x^+ S_{x+1}^- 
  - \frac{1}{2} S_x^- S_{x+1}^+\right))\, .
\end{gather*}
Before proceeding, we note that with our choice of normalizations, $H(\Delta^{-1})$
is symmetric about $\Delta^{-1}=0$.
The reason for this is that $H(\Delta^{-1})$ can be mapped unitarily to 
$H(-\Delta^{-1})$ via conjugation by the unitary operator
$$
U = \exp\left(2\pi i \sum_{j=1}^{\ceil{L/2}} S_{2j-1}^3\right)\, .
$$
Thus we expect that the spectral gap has always either a local maximum or a local 
minimum at the point $\Delta^{-1} = 0$.
In order to find which one, we perform second-order perturbation theory.
We ignore the questions of degeneracy of the excited state for the moment,
i.e.\ we begin by assuming that the excited state is nondegenerate.
In a sector with $M=2\Spin k +n$, and $n<\Spin$,
the ground state of the Ising limit is 
$\ket{\dots,-\Spin,-\Spin,-\Spin+n,+\Spin,+\Spin,+\Spin,\dots}$.
The first excited state is
$$
\psi^{(0)} = \ket{\dots,-\Spin,-\Spin+1,-\Spin+n-1,+\Spin,+\Spin,+\Spin,\dots}\, ,
$$
which has energy $E^{(0)} = n+1$.
First order perturbation theory yields
$$
H^{(0)} \psi^{(1)} + H^{(1)} \psi^{(0)} 
  = E^{(1)} \psi^{(0)} + E^{(0)} \psi^{(1)}\, ,
$$
which implies
$$
E^{(1)} 
  = \frac{\ip{\psi^{(0)}}{H^{(1)} \psi^{(0)}}}{\ip{\psi^{(0)}}{\psi^{(0)}}}\, .
$$
But $H^{(1)}\psi^{(0)} = 0$ because $\psi^{(0)}$ is an eigenstate of every
$S^3_x$.
So $E^{(1)} = 0$.
This then implies that
$$
(H^{(0)}-E^{(0)}) \psi^{(1)} = - H^{(1)} \psi^{(0)}\, .
$$
It is straightforward to calculate
\begin{align*}
H^{(1)} \psi^{(0)}
  &= \frac{1}{2} \sqrt{2\Spin(n+1)(2\Spin-n)} 
   \ket{\dots,-\Spin,-\Spin+1,-\Spin+n,+\Spin,+\Spin,+\Spin,\dots} \\
  &\hspace{-20pt} + \frac{1}{2} \sqrt{2(\Spin-2)(n+2)(2\Spin-n-1)}
   \ket{\dots,-\Spin,-\Spin,-\Spin+n+2,+\Spin-2,+\Spin,+\Spin,\dots} \\
  &\quad + \frac{1}{2} \sqrt{2\Spin(n+1)(2\Spin-n)}
   \ket{\dots,-\Spin,-\Spin,-\Spin+n,+\Spin,+\Spin,+\Spin,\dots} \\
  &\quad + \Spin 
  \ket{\dots,-\Spin,-\Spin+1,-\Spin+n+1,+\Spin,+\Spin-1,+\Spin,\dots}\, .
\end{align*}
Similarly, it is easy to calculate
\begin{align*}
&(E^{(0)} - H^{(0)})^{-1} 
  \ket{\dots,-\Spin,-\Spin+1,-\Spin+n,+\Spin,+\Spin,+\Spin,\dots} \\
&\qquad  = - \frac{1}{2\Spin-(n+1)}
  \ket{\dots,-\Spin,-\Spin+1,-\Spin+n,+\Spin,+\Spin,+\Spin,\dots} \\
&(E^{(0)}-H^{(0)})^{-1} 
  \ket{\dots,-\Spin,-\Spin,-\Spin+n+2,+\Spin-2,+\Spin,+\Spin,\dots} \\
&\qquad = - \frac{1}{n+3}
     \ket{\dots,-\Spin,-\Spin,-\Spin+n+2,+\Spin-2,+\Spin,+\Spin,\dots} \\
&(E^{(0)} - H^{(0)})^{-1} 
  \ket{\dots,-\Spin,-\Spin,-\Spin+n,+\Spin,+\Spin,+\Spin,\dots} \\
&\qquad  = \frac{1}{n+1}
  \ket{\dots,-\Spin,-\Spin,-\Spin+n,+\Spin,+\Spin,+\Spin,\dots} \\
&(E^{(0)} - H^{(0)})^{-1}   
  \ket{\dots,-\Spin,-\Spin+1,-\Spin+n+1,+\Spin,+\Spin-1,+\Spin,\dots} \\
&\qquad  = -\frac{1}{2(2\Spin-n-1)} 
    \ket{\dots,-\Spin,-\Spin+1,-\Spin+n+1,+\Spin,+\Spin-1,+\Spin,\dots}\, .
\end{align*}
By second-order perturbation theory, we obtain
\begin{gather*}
H^{(0)} \psi^{(2)} + H^{(1)} \psi^{(1)} 
  = E^{(2)} \psi^{(0)} + E^{(0)} \psi^{(2)}\\
\Rightarrow 
  E^{(2)} = \frac{\ip{\psi^{(0)}}{H^{(1)} \psi^{(1)}}}
  {\ip{\psi^{(0)}}{\psi^{(0)}}}
  = \frac{\ip{H^{(1)} \psi^{(0)}}{(E_0 - H_0)^{-1} H^{(1)} \psi^{(1)}}}
  {\ip{\psi^{(0)}}{\psi^{(0)}}}\, .
\end{gather*}
From our calculations, this gives
\begin{align*}
E_2 &= -\frac{1}{4} \frac{2\Spin(n+1)(2\Spin-n)}{2\Spin-n+1}
  - \frac{1}{4} \frac{2(\Spin-1)(n+2)(2\Spin-n-1)}{n+3}\\
  &\quad + \frac{1}{4} \frac{2\Spin(n+1)(2\Spin-n)}{n+1}
  - \frac{\Spin^2}{2(2\Spin-n-1)}\, .
\end{align*}
For $n > \Spin$, one can obtain the value of $E_2$ by the previous formula,
substituting $2\Spin-n$ for $n$.
For $\Spin \in \Ir$ and $n=\Spin$, there are two degenerate eigenvectors with
first excited energy equal to $\Spin+1$.
One can easily carry out the degenerate perturbation theory to obtain the 
second order correction to the spectral gap.
$$
E_2(n=\Spin \in \Nl) = 
  \frac{-4 \Spin^2}{\Spin-1} - \frac{(\Spin^2-4)(\Spin-1)}{2(\Spin+3)}
  - \frac{\Spin^2}{2(\Spin-1)}\, .
$$
A short table of $E_2$ values for some small values of $\Spin$ is given:
$$
\begin{array}{ccccccccccc}
&n=0&n=1&n=2&n=3&n=4&n=5&n=6&n=7&n=8&n=9\\
\Spin = 1 & -\frac{1}{6} & ** & -\frac{1}{6} \\
\Spin = \frac{3}{2} & \frac{43}{48} & -\frac{39}{16} 
  & -\frac{39}{16} & \frac{43}{48}\\
\Spin = 2 & 2 & -1 & -14 & -1 & 2 \\
\Spin = \frac{5}{2} & \frac{311}{96} & \frac{1}{16} & -\frac{307}{80} 
  & -\frac{307}{80} & \frac{1}{16} & \frac{311}{96} \\
\Spin = 3 & \frac{139}{30} & \frac{9}{8} & -\frac{27}{10} & -\frac{199}{12} 
  & -\frac{27}{10} & \frac{9}{8} & \frac{139}{30} \\
\Spin = \frac{7}{2} & \frac{99}{16} & \frac{181}{80} & -\frac{279}{160} & 
  -\frac{25}{4} &  -\frac{25}{4} & -\frac{279}{160} & \frac{181}{80}
  & \frac{99}{16} \\
\Spin = 4 &  \frac{166}{21} & \frac{7}{2} & -\frac{4}{5} & -\frac{16}{3} & 
   -\frac{446}{21} &  -\frac{16}{3} & -\frac{4}{5} & \frac{7}{2} & 
   \frac{166}{21}\\
\Spin = \frac{9}{2} & \frac{1879}{192} & \frac{543}{112} & \frac{3}{16} 
  & -\frac{68}{15} & -\frac{2157}{224} & -\frac{2157}{224} & -\frac{68}{15}
  & \frac{3}{16} & \frac{543}{112} & \frac{1879}{192}
\end{array}
$$
The blank entry in $\Spin=1$, $n=1$ is due to the fact that the excited state
is infinitely degenerate:
It is 
$$\ket{\dots,-1,-1,+1,-1,-1,\dots,-1,+1,+1,\dots}$$
 or
$$\ket{\dots,-1,-1,+1,\dots,+1,+1,-1,+1,+1,\dots}\, .$$
This is the only value of $(\Spin,n)$ other than $\Spin=1/2$ which is infinitely 
degenerate.
In the Ising limit, the minimum gap occurs for $n=0$.
One can see from the table above that for $n=0$ and $\Spin>1$, the gap is concave
up at $\Delta^{-1}=0$.
Thus, for finite volumes, the minimum gap attains its maximum value for some
$\Delta^{-1}$ other than 0, i.e.\ in some neighborhood of the Ising 
limit, as one increases the anisotropy, $\Delta^{-1}$, the spectral gap actually
raises.
The perturbation theory does not prove that in the thermodynamic limit the gap
is still maximized at some value other than $\Delta^{-1}=0$, since it is possible
this maximum value becomes smaller and smaller as $L \to \infty$.
However, one can carry the perturbation series to higher than second order.
We have calculated the perturbation series to 14th order for $n=0$.
The results are plotted in Figure \ref{pertpics}.
Unfortunately the perturbation series does not have radius of 
convergence equal to 1.
In fact the radius of convergence is apparently approximately $\Delta^{-1} = 1/2$.
One can see from these pictures that the maximum gap appears to occur for 
$\Delta^{-1}\approx 1/2$, as well, 
and the maximum gap seems to grow as a linear function of $\Spin$.

\section{Conjectures}
\label{section conjectures}
Based on the numerical evidence, we make the following conjectures.
\begin{conjecture}
Denoting the lower bound for the spectral gap by 
$\tilde{\gamma}(\Spin,n,\Delta^{-1})$,
$\tilde{\gamma}(\Spin,0,\Delta^{-1})$ scales linearly
with $\Spin$, and the functional dependence on $\Delta^{-1}$ is
$$
\lim_{\Spin \to \infty} \Spin^{-1} \tilde{\gamma}(\Spin,n=0,\Delta^{-1})
  = 2 \Delta^{-1} (1 - \Delta^{-1})\, .
$$
We also have strong evidence that the true spectral gap 
$\gamma(\Spin,n,\Delta)$ scales with
$\Spin$ and that there is some curve $u(\Delta^{-1})$, with maximum strictly
between $0$ and $1$ such that
$$
\lim_{\Spin \to \infty} \Spin^{-1} \gamma(\Spin,0,\Delta^{-1})
  = u(\Delta^{-1})\, .
$$
\end{conjecture}

\section{Matlab and Mathematica Notebooks}
Source code for these two scripts is available on the arXiv.
\cite{St}

{\center \Large Matlab Notebook for Lanczos Iteration}

{\small
\begin{verbatim}
This program defines the XXZ Hamiltonian with special boundary fields
%Also called the "Kink" Hamiltonian
%It restricts to the sector specified by the number "downspins" of down spins
%Using the differential equation for the first excited state as a function of Delta^(-1):
%$  h(x,y) = - S^3_x S^3_y - \Delta^{-1} (S^1_x S^1_y + S^2_x S^2_y) 
%  + A(Delta) (S_x^3 - S_y^3) $
%$ \frac{\partial h}{\partial \Delta^{-1}} = - (S^1_x S^1_y + S^2_x S^2_y)  \\
%  + \frac{\partial A(\Delta)}{\partial \Delta^{-1}} (S_x^3 - S_y^3) $
%  we solve the first excited energy level as a function of Delta^(-1).
%Performs a linear-solve method "bicg" : biconjugae gradient.

dq = 0.005                                 %Step-size for anisotropy parameter

L = 4;                                   %Spin chain length
J = 3/2;                                 %Spin-dimension 
downspins = floor(J*L)-1;                  %Number of downspins : sector 
N = 2*J+1                                %Dimension of single-site Hilbert space

S3 = sparse(1:N,1:N,J-(0:2*J),N,N);      %Spin-J, spin matrix about e3 axis
Splus = sparse(N,N);                     %Initialization of spin-raising operator
for j=0:(2*J-1)
  Splus = Splus + sparse(2*J-j,2*J-j+1,sqrt((2*J-j)*(j+1)),N,N);
end
Sminus = transpose(Splus);               %Spin-lowering operator
S1 = (Splus + Sminus)/2;                 %Spin matrix about e1 axis
S2 = (Splus - Sminus)/(2*i);             %Spin matrix about e2 axis


%IsingNN = Ising Nearest neighbor interaction S3(x)*S3(x)
IsingNN = kron(S3,S3);
%HopNN = Spin-hopping interaction = Splus(x)*Sminu(x+1) + Sminus(x)*Splux(x+1)
HopNN = - kron(Splus,Sminus) - kron(Sminus,Splus);  

%IsingH = Ising Hamiltonian
IsingH = sparse(N^L,N^L);
for x=1:(L-1)
 IsingH = IsingH + kron(eye(N^(x-1)),kron(IsingNN,eye(N^(L-1-x))));
end
%HopH = Hopping Hamiltonian
HopH = sparse(N^L,N^L);
for x=1:(L-1)
 HopH = HopH + kron(eye(N^(x-1)),kron(HopNN,eye(N^(L-1-x))));
end
%BdryH = Boundary-field terms
BdryH = kron(S3,eye(N^(L-1))) - kron(eye(N^(L-1)),S3);

%S3tot = total third-component of spin operator
S3tot = sparse(N^L,N^L);
for x=1:L, 
 S3tot = S3tot + kron(speye(N^(x-1)),kron(S3,speye(N^(L-x))));
end
%We now define the projection to the sector specified by
% S3tot = (J*L-downspins)
Proj = speye(N^L);
for n=0:(2*J*L)
 if ne(n,downspins),
   Proj = Proj*(S3tot - (J*L-n)*speye(N^L))/(n-downspins);
 end;
end;
%Proj is the orthogonal projection onto sector with specified number
% of downspins. Next we want to define a projection from this subspace
% to a vector space of the same dimension.
%The command "find" finds the nonzero elements of Proj, 
% which is just what we need.
[I,K] = find(Proj);
dim = length(I);
NewProj = sparse(I,1:dim,ones(dim,1),N^L,dim);

%We define new Hamiltonians which are conjugated by our projection
NewIsingH = transpose(NewProj)*IsingH*NewProj;
NewHopH = transpose(NewProj)*HopH*NewProj;
NewBdryH = transpose(NewProj)*BdryH*NewProj;

%We now find the first excited eigenstate of the Ising model
%IsingKinkH = Kink Hamiltonian in Ising limit
HKink = J^2*(L-1)*speye(dim) - NewIsingH - J*NewBdryH;
%We use Lanczos iteration via the command "eigs" to find the two smallest eigenvalues
%The matrix V has the eigenvectors, and D has the eigenvalues on its diagonal
[V,D] = eigs(HKink,2,'sm');
%The first eigenvalue should be 0. The second is the energy gap, which we call E
E = D(2,2);
gs = sparse(chop(V(:,1),6));
psi = sparse(chop(V(:,2),6));

%We now solve the differential-eigenvalue problem to continue the
%  first excited state to higher values of q
gaplist = [E];
for q=0:dq:(1-dq)
 dHKink = -((q+dq)/(1+(q+dq)^2)-q/(1+q^2))*NewHopH ...
  -J*((1-(q+dq)^2)/(1+(q+dq)^2) - (1-q^2)/(1+q^2))*NewBdryH;
 dE = full(transpose(psi)*dHKink*psi);
 dpsi = bicg((HKink-E*speye(dim)),(dE*speye(dim)-dHKink)*psi);
 psi = sparse(chop((psi+dpsi),6));
 psi = psi/norm(psi);
 E = E + dE
 HKink = HKink + dHKink;
 gaplist = [gaplist,E];
end

\end{verbatim}
}

\pagebreak

{\center \Large Mathematica Notebook for Lower Bound}
{\small

Norm returns the sum of the parts of the list. If the parts are all positive, as is usual for a partition, then this is equal to the L\(\RawWedge\)1-norm.
However in our application, we consider "signed parititions", i.e. lists which may be positive or negative integers but with the property of being
weakly decreasing.

Norm[List\_{}] :\(=\) Fold[Plus,0,List]

MinPar returns the minimum partition, with respect to reverse lexicographic order (rlo), subject to the conditions of having prescribed length and
presecribed total sum. (I.e. prescribed "norm".) The minimum partition satisfying these conditions is as flat as possible.

Clear[MinPar];

MinPar[Length\_{},Number\_{}]:\(=\) Table[Quotient[Number,Length],\{Length\}]\(+\)Sum[Table[KroneckerDelta[j-k],\{j,1,Length\}],\{k,1,Mod[Number,Length]\}]

OrdParList creates a list of partitions, ordered by lro. The length, i.e. number of parts, for the partitions is specified by Length. Number specifies
the number the partitions are partitioning, i.e. the sum of the parts, and Card says how many partitions should be in the list. The algorithm is
simple. One starts with the minimum partition using MinPar. At the next step it creates the next larger partition by 1) adding one to the last part
which is strictly less than all the following parts; 2) fixing all the parts before that one; and 3) updating the remaining parts by creating the
minimum partition with that many parts and with Number equal to their sum minus one.

OrdParList[Length\_{},Number\_{},Card\_{}] :\(=\) Module[\{ParOrder ,LastPar,NewPar\},

    LastPar \(=\) MinPar[Length,Number];

    ParOrder\(=\)\{LastPar\};

    For[n\(=\)1,n\(\leq \)Card-1,n\(+\)\(+\),

      j \(=\) Length-1;

      While[LastPar[[j]]\(=\)\(=\)LastPar[[j-1]],j\(=\)j-1];

      NewPar\(=\)LastPar;

      NewPar[[j]]\(=\)LastPar[[j]]\(+\)1;

      NewPar[[Range[j\(+\)1,Length]]]\(=\)MinPar[Length-j,Norm[LastPar[[Range[j\(+\)1,Length]]]]-1];

      ParOrder \(=\) Append[ParOrder,NewPar];

      LastPar\(=\)NewPar];

    ParOrder]

OrdParList2 is the same as OrdParList, but with the variables Length, Number and Card replaced by Depth, FFactor and L. Actually it is different
because instead of finding an ordered partition list with a certain cardinality, it finds the ordered list of all partitions such that none of the
parts is \(<\)\(=\)-L/2 and none of the parts is \(>\)L/2.

OrdParList2[Depth\_{},FFactor\_{},L\_{}] :\(=\) Module[\{ParOrder ,LastPar,NewPar\},

    LastPar \(=\) MinPar[Depth,FFactor];

    ParOrder\(=\)\{\};

    While[LastPar[[1]]\(\leq \)Ceiling[L/2],

      ParOrder\(=\)Append[ParOrder,LastPar];

      j\(=\)Depth-1;

      While[LastPar[[j]]\(=\)\(=\)LastPar[[j-1]],j\(=\)j-1];

      NewPar\(=\)LastPar;

      NewPar[[j]]\(=\)LastPar[[j]]\(+\)1;

      NewPar[[Range[j\(+\)1,Depth]]]\(=\)MinPar[Depth-j,Norm[LastPar[[Range[j\(+\)1,Depth]]]]-1];

      LastPar\(=\)NewPar];

    Select[ParOrder,\#{}[[-1]]\(\geq \)-Floor[L/2]\&{}]]

LEQ checks whether Par1 is less than or equal to Par2 with respect to dominance order.

LEQ[Par1\_{},Par2\_{}] :\(=\) Module[\{l,NewPar1,NewPar2,PSum1,PSum2\},

    l \(=\) Max[Length[Par1],Length[Par2]];

    NewPar1\(=\)Join[Par1,Table[0,\{l-Length[Par1]\}]];

    NewPar2\(=\)Join[Par2,Table[0,\{l-Length[Par2]\}]];

    PSum1\(=\)FoldList[Plus,0,NewPar1];

    PSum2\(=\)FoldList[Plus,0,NewPar2];

    Fold[And,1\(==\)1,Table[PSum1[[j]]\(<\)\(=\)PSum2[[j]],\{j,1,l\(+\)1\}]]]

ParGraph is a Module which produces a graphic of the Young diagram of a partition.

ParGraph[Par\_{}] :\(=\)Module[\{L\(=\)Length[Par],n\},

    Grph\(=\)\{Line[\{\{0,0\},\{0,-L\}\}]\};

    For[n\(=\)1,n\(<\)\(=\)L,n\(+\)\(+\),

      If[Par[[n]]\(=\)\(=\)0,NewGrph\(=\)\{Disk[\{0,-n\(+\)0.5\},0.5]\},

        If[Par[[n]]\(>\)0,NewGrph\(=\)

             Table[Disk[\{j,-n\(+\)0.5\},0.5],\{j,1,Par[[n]]\}],

          NewGrph\(=\)Table[Disk[\{j-1,-n\(+\)0.5\},0.5],\{j,Par[[n]]\(+\)1,0\}]]];

      Grph\(=\)Join[Grph,NewGrph]];

    Grph]

ParSimplex produces a 3d simplex of partitions (Par1,Par2,Par3) satisfying Par1\(<\)\(=\)Par2\(<\)\(=\)Par3 w.r.t. dominance order.

ParSimplex[Length\_{},Number\_{},Card\_{}] :\(=\) Module[\{ParOrder\},

    ParOrder \(=\) OrdParList[Length,Number,Card];Flatten[Table[Table[Table[\{Part[ParOrder,i],\\
    Part[ParOrder,j],Part[ParOrder,k]\},\{k,j,Card\}],\{j,i,Card\}],\{i,1,Card\}],2]]

Simplex produces the usual simplex in \(\CapitalNu \)\(\RawWedge\)3 consisting of all triples (i,j,k) such that 0\(<\)i\(<\)\(=\)j\(<\)\(=\)k\(<\)\(=\)Card.

Simplex[Card\_{}] :\(=\) Flatten[Table[Table[Table[\{i,j,k\},\{k,j,Card\}],\{j,i,Card\}],\{i,1,Card\}],2]

SimplexPartitionPartition partitions up the ParSimplex according to the "norm"  2 \(\Vert \)Par3\(\Vert \)\(\RawWedge\)2 - \(\Vert \)Par1\(\Vert
\)\(\RawWedge\)2 - \(\Vert \)Par2\(\Vert \)\(\RawWedge\)2 .  

SimplexPartitionPartition[Depth\_{},FFactor\_{},L\_{}] :\(=\) Module[\{Card,Simp,Simp2,Simp3\},

    OPL \(=\) OrdParList2[Depth,FFactor,L];

    Card \(=\) Length[OPL];

    Simp \(=\) Simplex[Card];

    Simp2 \(=\) Map[\{\#{},2*OPL[[\#{}[[3]]]].OPL[[\#{}[[3]]]]-OPL[[\#{}[[2]]]].OPL[[\#{}[[2]]]]-\\
  OPL[[\#{}[[1]]]].OPL[[\#{}[[1]]]]\}\&{},Simp];

    For[n\(=\)0,n\(<\)\(=\)Max[Map[\#{}[[2]]\&{},Simp2]],n\(+\)\(+\),

      Simp3[n] \(=\) Map[\#{}[[1]]\&{},Select[Simp2,\#{}[[2]]\(==\)n\&{}]]];

    Simp3[-1] \(=\) n-1;

    Simp3]

Mon produces the monomial symmetric function of Par.

Clear[Mon];

Mon[Par\_{}] :\(=\) Fold[Plus,0,Map[Fold[Times,1,Table[x[n],\{n,Length[Par]\}]\(\RawWedge\)\#{}]\&{},Permutations[Par]]]

MonOne produces the monomial symmetric function of Par specialized to 1. This is the same as the number of elelements in the orbit of Par under the
action of permuting the parts.

MonOne[Par\_{}] :\(=\) Fold[Times,1,Table[x[n],\{n,Length[Par]\}]\(\RawWedge\)Par]

Trans takes the transpose of a partition.

Trans[Par\_{}] :\(=\) Module[\{m,M,NewPar,Div\},

    m \(=\) Par[[-1]];

    M \(=\) Par[[1]];

    NewPar \(=\) \{\};

    For[n\(=\)Min[1,m],n\(<\)\(=\)Max[1,M],n\(+\)\(+\),NewPar\(=\)Append[NewPar,Length[Select[Par,\#{}\(\geq \)n\&{}]]]];

    \{NewPar,1-Min[1,m]\}]

El gives the elementary symmetric function associated with a partition. Note that because these are signed partitions, this is generally not a polynomial.
It is a rational function. Thus the output form is \{numerator,denominator\}.

El[Par\_{}] :\(=\) Module[\{L,TPar,DenExp,Num\},

    L\(=\)Length[Par];

    TPar \(=\) Trans[Par][[1]];

    DenExp \(=\) Trans[Par][[2]];

    Num\(=\)Fold[Times,1,Map[Mon[Join[Table[1,\{\#{}\}],Table[0,\{L-\#{}\}]]]\&{},TPar]];

    \{Num,Mon[Table[1,\{L\}]]\(\RawWedge\)DenExp\}]

General::spell1: Possible spelling error: new symbol name "TPar" is similar to existing symbol "Par".

MatBase produces a basic matrix of size d\(\times \)d with zeros at all entries except the (m,n) entry. 

MatBase[m\_{},n\_{},d\_{}] :\(=\) Table[KroneckerDelta[k,n]*Table[KroneckerDelta[j,m],\{j,1,d\}],\{k,1,d\}]

PSMon produces the same thing as MonOne, but more efficiently.

PSMon[Par\_{}] :\(=\) Length[Permutations[Par]]

PSEl produces the specialization of El to all ones. \\
This equals the numtinomial coefficient 
B[Length[Par];Number of 1's, Number of 2's,...].

PSEl[Par\_{}] :\(=\) Module[\{TPar,Len\},

    Len \(=\) Length[Par];

    TPar \(=\) Trans[Par][[1]];

    Fold[Times,1,Map[Binomial[Len,\#{}]\&{},TPar]]]

Here we begin the actual calculations. We time the procedure. The initial time is TimeStart. AbsoluteTime[] is the total number of seconds since
January 1, 1900.

TimeStart\(=\)AbsoluteTime[]

Here we initialize the number of parts for the partitions, the filling factor (which is called n in the paper) and the length of the partition (which
is the length of the spin chain in the paper). The number of parts is 2J, where \${}J\${} is the spin from the paper.

parts\(=\)4;

ff\(=\)3;

length\(=\)5;

We define OPL to be the ordered list of signed partitions with specified number of parts, and all parts between -Floor[L/2] and  \(+\)Ceiling[L/2],

which partition ff.           

OPL \(=\) OrdParList2[parts,ff,length]

Length[\%{}]

We go through some procedures to display Young diagrams of the included partitions. The vertical line is zero, and all dots between 0 and the actual
value of the part are drawn. Blank spots in the GraphicsArray are filled with black rectangles.

GraphPrim \(=\) Join[Map[ParGraph,OPL],Table[Rectangle[\{-Floor[length/2],0\},\\
\{Ceiling[length/2],-parts\}],\{Mod[5-Length[OPL],5]\}]];

Show[GraphicsArray[Table[Table[Graphics[GraphPrim[[m\(+\)5*n]], AspectRatio\(\rightarrow \)Automatic,\\
PlotRange\(\rightarrow \)\{\{-Floor[length/2]-0.5,Ceiling[length/2]\(+\)0.5\},\{0.5,-parts-0.5\}\}],\{m,1,5\}],\\
\{n,0,Length[GraphPrim]/5-1\}]]]

We define SPP to be the partitioned simplex partition.

SPP \(=\) SimplexPartitionPartition[parts,ff,length]

SPP[-1] equals the length of SPP.

SPP[-1]

Elem is the list of all elementary symmetric functions for the partitions in OPL

Elem \(=\) Map[El,OPL];

Monom is the list of all monomial symmetric functions for the partitions in OPL

Monom \(=\) Map[MonOne,OPL];

Msub is the transition matrix between the monomial and elementary symmetric functions corresponding to the restricted partitions of OPL. This typically
is one of the most time-consuming parts of the calculations.

Msub \(=\)Table[Join[Table[Coefficient[Elem[[i]][[1]] ,Elem[[i]][[2]]*Monom[[j]]],

        \{j,1,i\}],Table[0,\{Length[OPL]-i\}]],\{i,1,Length[OPL]\}] 

Msup is the inverse matrix to Msub. It is denoted by \inlineTFinmath{{M^{\Mvariable{\lambda \mu }}}} in the paper, while Msub is \inlineTFinmath{{M_{\Mvariable{\mu
\lambda }}}}. It is very quick to calculate one Msub has been done.

Msup \(=\) Inverse[Msub]

PSMonList gives the list of principle stable specializations of the monomial symmetric functions in OPL.

PSMonList \(=\) Map[PSMon,OPL]

PSMonList gives the list of principle stable specializations of the elementary symmetric functions in OPL.

PSElList \(=\) Map[PSEl,OPL]

F[i,j,k] is the "tensor" \inlineTFinmath{{{{{\Gamma }_{\Mvariable{\mu \nu }}}}^{\kappa }}} from the paper, where \(\mu \) \(=\) OPL[[i]], \(\nu \)\(=\)OPL[[j]]
and \(\kappa \)\(=\)OPL[[k]]. Note (i,j,k).

F[i\_{},j\_{},k\_{}] :\(=\) Sqrt[PSMonList[[i]]*PSMonList[[j]]]Sum[Msub[[l,i]]*Msub[[l,j]]\\
*Msup[[k,l]]/PSElList[[l]],\{l,j,k\}]

This line is really redundant. It defines the basic matrices with d specified to be Length[OPL]. 

For[m\(=\)1,m\(\leq \)Length[OPL],m\(+\)\(+\),

  For[n\(=\)1,n\(\leq \)Length[OPL],n\(+\)\(+\),

    ElemMat[m,n] \(=\) MatBase[m,n,Length[OPL]]]]

F[1,2,3]

SPP[-1]

This defines the partition Epar\(=\)Floor[L/2]*(1,1,dots,1). Adding Epar essentially translates the partitions so that  all the parts are nonnegative.

Epar \(=\) Floor[length/2]*Table[1,\{parts\}]

OPL2 is the set of all the partitions translated to be nonnegative.

OPL2 \(=\) Map[Epar \(+\) \#{}\&{},OPL]

Clear[q]

qtab is the q\(\RawWedge\)2-specialization of the elementary symmetric polynomial specified by n.  It is q\(\RawWedge\)(n(n-1)) time the q\(\RawWedge\)2-binomial
coefficient of L choose n.

qtab \(=\) Table[q\(\RawWedge\)(n*(n-1))*Product[1-q\(\RawWedge\)(2k),\{k,length\}]/(Product[1-q\(\RawWedge\)(2k),\{k,n\}]*\\
Product[1-q\(\RawWedge\)(2k),\{k,length-n\}]),\{n,0,length\}];

PSELq is the table of q\(\RawWedge\)2-specializations of all the elementary symmetric ploynomials labelled by partitions from OPL2.

PSELq \(=\) Table[Fold[Times,1,Map[qtab[[\#{}\(+\)1]]\&{},OPL2[[n]]]],\{n,1,Length[OPL2]\}];

This part of the program calculates the matrix components of \(\backslash\)tilde\{P\}from the paper. It is usually the second longest part of the
computations. We break up the computation by the leading-order power of \${}q\${}. This is the purpose of SPP. We list which part has been done as
the calculation proceeds.

For[n\(=\)0,n\(\leq \)SPP[-1],n\(=\)n\(+\)2,M[n] \(=\)Fold[Plus,0,Map[(1-(1/2)*KroneckerDelta[\#{}[[1]],\#{}[[2]]])*\\
(ElemMat[\#{}[[1]],\#{}[[2]]]\(+\)ElemMat[\#{}[[2]],\#{}[[1]]])*F[\#{}[[1]],\#{}[[2]],\#{}[[3]]]*PSELq[[\#{}[[3]]]]\\
/Sqrt[PSELq[[\#{}[[1]]]]*PSELq[[\#{}[[2]]]]]\&{},SPP[n]]];

  Print[n]]

MasterM[q] is the sum of all the leading-order-homogeneous matrices corresponding to the powers of q calculated above.

MasterM[q\_{}] \(=\) Sum[M[n],\{n,0,SPP[-1],2\}];

We save the data of each calculation. (What data we save exactly is listed below.) This line opens the file where the data is saved, os that results
of past calculations can be accessed.

\(<\)\(<\)Gapdata ;

We calculate the top two eigenvalues of MasterM[q] for 101equally psaced values of q between 0 and 1. Note the top eigenvalue shopuld be identically
1. Calculating it anyway gives a check that our code is doing what it should be doing.

dq \(=\) 0.01;

Eiglist \(=\) Module[\{Eigs,LocalList\},

      LocalList\(=\)\{\};

      For[q\(=\)dq,q\(<\)1,q\(=\)q\(+\)dq,

        Eigs \(=\) Sort[Re[Eigenvalues[N[MasterM[q]]]],Greater];

        LocalList \(=\) Append[LocalList,\{q,Eigs[[1]],Eigs[[2]]\}]];

      LocalList];

In the following two lines we append the data from our current calculation, which is the list of the top two eigenvalues at all 100 q-points, to
the save file. 

EL[parts/2,length,ff] \(=\) Eiglist;

Save["Gapdata",EL]

In this line we define the function that maps between the parameters q and \inlineTFinmath{{{\Delta }^{-1}}}.

Clear[q];

DeltaInv[q\_{}] \(=\) 2*q/(1\(+\)q\(\RawWedge\)2)

In this line we plot the top two eigenvalues as a function of q.

ListPlot[Join[Map[\{\#{}[[1]],\#{}[[2]]\}\&{},Eiglist],Map[\{\#{}[[1]],\#{}[[3]]\}\&{},Eiglist]]]

\(\SkeletonIndicator \)Graphics\(\SkeletonIndicator \)

We plot the lower bound for the bottom two eigenvalues of the XXZ spin chain as a function of q.

ListPlot[Join[Map[\{\#{}[[1]],parts*(1-\#{}[[1]])\(\RawWedge\)2/(1\(+\)\#{}[[1]]\(\RawWedge\)2)*(1-\#{}[[2]])\}\&{},Eiglist],\\
Map[\{\#{}[[1]],parts*(1-\#{}[[1]])\(\RawWedge\)2/(1\(+\)\#{}[[1]]\(\RawWedge\)2)*(1-\#{}[[3]])\}\&{},Eiglist]]];

We concatenate the two images for export to a picture file.

Show[GraphicsArray[\{\{\%{}\%{},\%{}\}\}]]

In this line we plot the top two eigenvalues of MaterM as a function of \inlineTFinmath{{{\Delta }^{-1}}}.

ListPlot[Join[Map[\{DeltaInv[\#{}[[1]]],\#{}[[2]]\}\&{},Eiglist],Map[\{DeltaInv[\#{}[[1]]],\#{}[[3]]\}\&{},Eiglist]]]

In this line we plot the bottom two eigenvalues of \inlineTFinmath{{H^{\Mvariable{XXZ}}}} as a function of \inlineTFinmath{{{\Delta }^{-1}}}.

ListPlot[Join[Map[\{DeltaInv[\#{}[[1]]],parts*(1-\#{}[[1]])\(\RawWedge\)2/(1\(+\)\#{}[[1]]\(\RawWedge\)2)*(1-\#{}[[2]])\}\&{},Eiglist],\\
Map[\{DeltaInv[\#{}[[1]]],parts*(1-\#{}[[1]])\(\RawWedge\)2/(1\(+\)\#{}[[1]]\(\RawWedge\)2)*(1-\#{}[[3]])\}\&{},Eiglist]]];

We concatentate the two images for export to an image file.

Show[GraphicsArray[\{\{\%{}\%{},\%{}\}\}]]

\(\SkeletonIndicator \)GraphicsArray\(\SkeletonIndicator \)

We list J, n and L again for cataloging purposes.

parts/2

ff

length

We stop the timer and calcuate the total time for the computation.

TimeFinish \(=\) AbsoluteTime[];

TimeFinish - TimeStart
}


\newpage
\pagestyle{myheadings} 
\markright{  \rm \normalsize CHAPTER 4. \hspace{0.5cm} 
Droplet States for the 1d, Spin-$\frac{1}{2}$ model}
\chapter{Droplet States for the 1d, Spin-$\frac{1}{2}$ model}
\thispagestyle{myheadings}

\Section{Summary}
In this chapter we present two models for droplet states in the XXZ quantum spin chain.
Droplets have been an important part of classical spin systems since the derivation
of the phenomenalogical Wullf construction for crystal growth by \cite{DKS}.
In the Ising spin system, putting boundary fields which force up spins at the edge of a 
one-dimensional or two-dimensional domain leads to a specific geometry for the equilibrium
states at low temperatures.
In the one-dimensional Ising model, the ground states with such a boundary field
consist of a connected subinterval of down spins, strictly in the interior of the spin
chain, and all up spins on the complement. 
In the two-dimensional Ising model, the low-temperature equilibrium states
are mixtures of pure states where a large domain of nearly all down-spins lies in the
center of the box, with an up-spin sea surrounding it, and the shape of the separating 
contour is given by the Wulff-construction.

For the quantum system the situation is complicated, even in one-dimension.
We leave the problem of a quantum spin droplet in two dimensions entirely alone,
although it is obviously a subject we hope someday to come back to.
Returning for a moment to the one-dimensional kink Hamiltonian, the ground state has
asymptotically all down-spins at $-\infty$, and all up-spins at $+\infty$, with a 
quantum interface separating the two regions.
The droplet Hamiltonian is defined similarly to the kink Hamiltonian,
except that instead of placing boundary fields of opposite signs on the
opposite edges of the spin chain, we place boundary fields of the same sign
to force up-spins at the edges.
The structure of the quantum interface becomes an important issue for the one-dimensional
droplet, as is the nature of the broken translation invariance of the ground state.
For the quantum kink ground state, the location of the interface is determined by the 
conserved quantity $S^3_{\textrm{tot}}$, $[H,S^3_{\textrm{tot}}]=0$.
For one-dimensional Ising droplet states, there is a greater degeneracy than this,
because there are $L+1-n$ locations for a subinterval of length 
$n$ inside an interval of length $L$.
This extra degeneracy persists in the quantum picture.
One can remove the degeneracy if one places an external magnetic field at some site of the
spin chain.
We call this a pinning field since it forces the droplet of down spins to be centered 
about the site where the new field occurs.

The first model we solve is the one with the pinning field.
This is the subject of Section 4.2.
It is particularly simple if we choose the correct magnitudes for the pinning field.
Namely, we choose the pinning field to have twice the magnitude
of the boundary fields, and opposite orientation.
Then
$$
H^{\textrm{AK}}_{[-L,L]} 
  = H^{\textrm{XXZ}}_{[-L,L]} + A(\Delta) S^3_1 
  - A(\Delta)\left(S^3_{-L} + S^3_{L} - 2 S^3_0\right)
  = H^{+-}_{[-L,0]} + H^{-+}_{[0,L]}\, .
$$
We call this the antikink-kink Hamiltonian, because it is the sum of an antikink Hamiltonian
on the left and a kink Hamiltonian on the right.
Alternatively, one can think of it as a ``kink'' Hamiltonian in the general sense of 
Section 2.3.1, where the graph is the usual graph on $[-L,L]$, the one with an edge for every 
nearest-neighbor pair, and the height function is $l(x) = |x|$ instead of $l(x)=x$.
Then by Theorem \ref{ASWTheorem}, there is a unique ground state in each sector of fixed
total magnetization.
Morevoer, there is a closed formula for this state.
The main question then becomes one of deciphering the formula to obtain the behavior
for the droplet interfaces in the limit that the number of droplet down-spins and background 
up-spins both approach infinity.
By exploiting some formulas from Section \ref{ExactCalculationsSection}, we are able to
give an exact formula for the asymptotic form
in Proposition \ref{KAInterface}.
The answer is that right droplet interface can be realized as a convex
combination of kink states,
and the left interface can be realized as a mixture of antikink interfaces.

In Sections 4.3--4.10, we present a more realistic model of a droplet Hamiltonian,
$$
H^{++}_{[-L,L]} = H^{\textrm{XXZ}}_{[-L,L]} + A(\Delta) S^3_1 
  - A(\Delta)(S^3_{-L} + S^3_{L})\, ,
$$
which has no pinning field.
The extra degeneracy of the ground states complicates the analysis, but the 
same basic ideas from the kink-antikink model are stil true.
If the number of down-spins, $n$ is large enough then there is a large interval of nearly
all down spins surrounded by two large intervals of nearly all up spins.
Moreover, the right and left interfaces
are well approximated by mixtures of the kink and antikink ground states.
For this reason, we hypothesize that the ground states of this model are nothing more 
than convex combinations of states formed by tensoring an antikink state
to the left with a kink state to the right.
This very simple picture turns out to be exactly true in the limit that $n \to \infty$
with exponentially small corections for finite volumes,
which is the subject of Theorem \ref{main:theorem}.

Our proof relies heavily on the existence and calculation of the spectral gap for the
one-dimensional kink and antikink states, which is the subject of \cite{KN1}.
In order to make use of that result we must relate $H^{++}_{[-L,L]}$
to a combination of kink and antikink Hamiltonians.
We do this by employing a fundamental lemma,
Propostion \ref{IHS:Prop}, which states that any low-energy state, $\omega$,
of a Hamiltonian which is a finite perturbation of the boundary-field-free
XXZ Hamiltonian on a long enough spin chain must have a large interval $I$ such
that $\omega\vert_I$ is close to a convex combination of the fully-polarized state
$\omega_\uparrow$ and $\omega_\downarrow$ on $I$.
This lemma is powerful, and uses the same basic idea which is key in proving the 
completeness of ground states for the infinite-volume XXZ Hamiltonian in \cite{KN3}.
The rest of the argument is a somewhat technical induction argument.
An easy induction argument is available if one assumes that the density of down-spins,
$\rho=n/L$ is bounded from below by a positive number as $L \to \infty$.
However, one then finds that the results are all independent of the choice
for $\rho_{\textrm{min}}$.
So it is natural to seek a proof which allows $n$ to approach infinity much more slowly than
$L$, which means $\rho$ may approach $0$ in the limit.
Instead of an induction argument based on density, one must use induction on the number
$n$ of down-spins, directly. 
This involves some complicated estimates,
most of which are relegated to the Appendices (Section 4.9 and 4.10).
We would be happier to find a simpler argument, although the basic idea of the proof,
which is to cut the spin chain at appropriate sites, is completely
straightforward.

A consequence of our analysis of ground states for $H^{++}_{[-L,L]}$ is that we determine
the ground states of the periodic XXZ spin chain, as well as the lowest energy vectors
of the infinite-volume XXZ spin chain in the GNS representation w.r.t.\ the translation
invariant all up-spin ground state, when we restrict attention to sector with large,
but finite, numbers of down spins.

\Section{The Antikink-Kink Hamiltonian}
In this section we consider a toy model for ``droplet states''
on a spin-$\frac{1}{2}$ chain of length $2L+1$,  which we obtain as the ground states
of a certain Hamiltonian
\begin{align*}
H^{\textrm{AK}}_{[-L,L]} 
  &= H^{\textrm{XXZ}}_{[-L,L]} + A(\Delta) S^3_1 
  - A(\Delta)\left(S^3_{-L} + S^3_{L} - 2 S^3_0\right)\\
  &= H^{+-}_{[-L,0]} + H^{-+}_{[0,L]}\, .
\end{align*}
We recall that $H^{+-}$ and $H^{-+}$ are called the antikink, and kink Hamiltonians,
respectively, hence we call this Hamiltonian the anitkink-kink Hamitlonian.
Also recall
$$
H^{\textrm{XXZ}}_{[-L,L]}
  = \sum_{x=-L}^{L-1} \left(\frac{1}{4} - S_x^3 S_{x+1}^3 - 
  \frac{1}{\Delta} S_x^1 S_{x+1}^1 - \frac{1}{\Delta} S_x^2 S_{x+1}^2\right)\, ,
$$
is the Hamiltonian with no external fields.
So the antikink-kink Hamiltonian represents the energy of a state due to the anisotropic
Heisenberg interactions of $H^{\textrm{XXZ}}$ plus a positive boundary field at both
ends of the sample, and with a pinning field of opposite sign, and double the magnitude
as the boundary fields, right at the center of the spin chain.
Our real interest is in the Hamiltonian with no pinning field, which we view as a good first
model of a quantum spin chain with ``droplet'' boundary conditions.

One would certainly expect the effect of the pinning field, $2 A(\Delta) S_0^3$, is to
force a down-spin at the site $x=0$ in the ground state, which is why we call it a 
``pinning field''.
Of course, even without the pinning field, one would expect the boundary fields 
attract up-spins towards, and repel down-spins from, the edge of the spin chain, 
and this should have the subsequent effect of forcing any down spins which live on the
chain to aggregate in the middle, which is why we call the Hamiltonian
$$
H^{++}_{[-L,L]} 
  = H^{\textrm{XXZ}}_{[-L,L]} + A(\Delta) S^3_1 
  - A(\Delta)\left(S^3_{-L} + S^3_{L}\right)
$$ 
the ``droplet Hamiltonian''.
If this hypothesis is true of the ground states for the droplet Hamiltonian, 
then the effect of the pinning field should simply be add a constant term to 
the Hamiltonian, because it is more-or-less guaranteed that the spin at the site
$x=0$ would be down due to the droplet-boundary conditions anyway.
Actually this is not entirely true, as we will see in the subsequent sections of 
this chapter, since the droplet Hamiltonian has a higher ground state degeneracy 
than the antikink-kink, Hamiltonian, and in particular, it possesses an approximate
symmetry (the Hamiltonian projected to low-energy states commutes with an exponentially
small error) by discrete translations.
The translation symmetry is natural to expect since the infinite-volume Hamiltonian
is certainly translation-invariant.
However what occurs for the antikink-kink Hamiltonian is that the pinning field serves
to split the ground state degeracy, very slightly, so that the unique ground state
has a ``droplet'' of down spins centered exactly at the origin.
(So the droplet is pinned.)
This is a definite flaw as far as extrapolating information from $H^{\textrm{HA}}$
to $H^{\textrm{AG}}$.
However, one could hope to study the droplet Hamiltonian by considering the linear span
of the ground state of antikink-kink Hamiltonian along with all its finite translates.
This can only be truly carried out in infinite-volume, but if the kink-antikink ground
state
is approximately ``all spins up'' at the two edges, then one could make sense of cutting
out one site on the left or right and adding another site at the opposite end.
Also, one could consider the kink-antikink Hamiltonian with pinning field at sites
other than the origin. This turns out to be a good idea as we show in the following
sections, although the arguments for the droplet Hamiltonian are logically independent
of the results we present in this section.

Although there are certain drawbacks to studying the kink-antikink Hamiltonian,
posed by the fact that the physical situation of the Hamiltonian is a little unnatural,
this Hamiltonian has the nice property that the ground states
are uniquley defined in each sector of a fixed number of down spins, and have a simple
closed-form expression.
This is a result of Theorem \ref{ASWTheorem},
since $H^{\textrm{ka}}_{[-L,L]}$ may be viewed as a kink Hamiltonian on the
graph of $[-L,L]$ but with height function $l(x) = |x|$ instead of $l(x)=x$.
In other words, the oriented bonds are all $(x,y)$ such that $|x-y|=1$ and 
$|y|=|x|+1$.
The ground state is then given by
\begin{equation}
\label{KA gs def}
\psi_L^{\textrm{ak}}(n) = \sum_{-L\leq x_1<x_2<\dots<x_n\leq L} q^{\sum_{j=1}^n |x_j|}
  \prod_{j=1}^n S_{x_j}^-\, \ket{\textrm{all up}}_{[-L,L]}\, .
\end{equation}
Altenatively, if one defines the kink and antikink ground states as
\begin{equation}
\label{K,A gs def1}
\begin{split}
\psi^{\textrm{k}}_{[a,b]}(n) 
  = \sum_{a\leq x_1<\dots<x_n\leq b} q^{\sum_{j=1}^n (x_j-a+1)}
	\prod_{j=1}^n S_{x_j}^-\, \ket{\textrm{all up}}_{[a,b]}\, ,\\
 \psi^{\textrm{a}}_{[a,b]}(n) 
  = \sum_{b\geq x_1>\dots>x_n\geq a} q^{\sum_{j=1}^n (b+1-x_j)}
	\prod_{j=1}^n S_{x_j}^-\, \ket{\textrm{all up}}_{[a,b]}\, ,
\end{split}
\end{equation}
then
$$
\psi_L^{\textrm{ak}}(n) 
  = \sum_{k=0}^n q^{-k} \psi^{\textrm{a}}_{[-L,0]}(k)\otimes
  \psi^{\textrm{k}}_{[1,L]}(n-k)\, .
$$

The next question is how these droplet states behave in the thermodynamic
limit? 
In particular, how the interface changes as one takes a larger and larger 
droplet?
The interface is the transition region between having nearly-all up spins
and nearly-all down spins.
For the ground state $\psi_0^{\textrm{ak}}(n)$, there are two interfaces 
at $\pm n/2$.
Is each interface exponentially localized, independently of $L$, as it is for the kink and 
antikink states, or is there a scaling behavior with $L$ and $n$?
It turns out that the interface is exponentially localized.
And that is the main result we will show in this section.

In order to describe the thermodynamic limit for the antikink-kink ground states,
we take a few moments to describe the thermodynamic limits for the kink and antikink
ground states.
Since we are considering one-dimensional spin systems, there are very few types
of infinite connected domains.
Specifically there are only three types: rays which point from left to right,
rays which point from right to left, and the whole line.
(All domains are interpreted as the being subdomains of the integer lattice, where 
``connected'' has the graph-theoretic meaning.)
For each of these domains, one can define the ground states of the kink and antikink
Hamiltonians working in the Guichardet Hilbert space (or Incomplete Tensor Product).
Let
\begin{equation}
\label{KA gs def2}
\begin{split}
\widetilde{\psi}^{\textrm{k}}_{[a,b]}(n) 
  = \sum_{b\geq x_1>\dots>x_n\geq a} q^{\sum_{j=1}^n (b+1-x_j)}
	\prod_{j=1}^n S_{x_j}^+\, \ket{\textrm{all down}}_{[a,b]}\, ,\\
\widetilde{\psi}^{\textrm{a}}_{[a,b]}(n) 
  = \sum_{a\leq x_1<\dots<x_n\leq b} q^{\sum_{j=1}^n (x_j-a+1)}
	\prod_{j=1}^n S_{x_j}^+\, \ket{\textrm{all down}}_{[a,b]}\, .
\end{split}
\end{equation}
be the kink and antikink ground states enumerated from the vacuum state 
$\ket{\textrm{all down}}$ instead of the vacuum state $\ket{\textrm{all up}}$.
For $[a,b]$ a finite interval, there are two unitaries on the Hilbert space
$\Hil_{[a,b]} = (\Cx^2)^{\otimes [a,b]}$, spin-flip and reflection:
\begin{align*}
\mathcal{F} &= \bigotimes_{x \in [a,b]} \left(S_x^+ + S_x^-\right)\, ,\\
\mathcal{R}\left(\frac{a+b}{2}\right) &: \bigotimes_{x \in [a,b]} \ket{\phi(x)}_x \mapsto
  \bigotimes_{x \in [a,b]} \ket{\phi(b+a-x)}_x\, .
\end{align*}
Both are involutions, and we have the diagram
$$
\begin{CD}
\psi^{\textrm{k}}_{[a,b]}(n) 
  @>\mathcal{R}\left(\frac{a+b}{2}\right)>> \psi^{\textrm{a}}_{[a,b]}(n)\\
@V{\mathcal{F}}VV @VV\mathcal{F}V\\
\widetilde{\psi}^{\textrm{a}}_{[a,b]}(n) @>\mathcal{R}\left(\frac{a+b}{2}\right)>> 
\widetilde{\psi}^{\textrm{k}}_{[a,b]}(n)
\end{CD}
$$
(Note that for the states $\psi^{\textrm{k,a}}_{[a,b]}(n)$, $n$ refers to the number 
of down-spins, while for the states $\widetilde{\psi}^{\textrm{k,a}}_{[a,b]}(n)$,
$n$ refers to the number of up-spins.)
For the half-infinite interval, $[a,\infty)$,
the GNS Hilbert space for the kink ground states is 
the Guichardet Hilbert space consisting
of those vectors obtained as quasilocal perturbations of the all up-spin state,
and the GNS space for the antikink states is the space of those vectors obtained by quasilocal 
perturbations of the all down-spin state.
The GNS vectors are
\begin{align*}
\psi^{\textrm{k}}_{[a,\infty)}(n) = \sum_{a\leq x_1<\dots<x_n} q^{\sum_{j=1}^n (x_j-a+1)}
	\prod_{j=1}^n S_{x_j}^-\, \ket{\textrm{all up}}_{[a,\infty)}\, ,\\
\widetilde{\psi}^{\textrm{a}}_{[a,\infty)}(n) 
  = \sum_{a\leq x_1<\dots<x_n} q^{\sum_{j=1}^n (x_j-a+1)}
	\prod_{j=1}^n S_{x_j}^+\, \ket{\textrm{all down}}_{[a,\infty)}\, .
\end{align*}
(Of course for the kink Hamiltonian there is one other ground state, which has
a different GNS space, the all-down spin state.
Similarly for the antikink Hamiltonian, the all-up spin state is orthogonal to the
entire GNS space of the antikink ground states.)
Similarly on $(-\infty,a]$ the GNS space for the kink is the Guichardet Hilbert space
based on the infinite tensor product of all down-spins, and the GNS space of the antikink
comes from the infinite tensor product of all up-spins, and the GNS vectors are
\begin{align*}
\widetilde{\psi}^{\textrm{k}}_{(-\infty,b]}(n) 
  = \sum_{b\geq x_1>\dots>x_n} q^{\sum_{j=1}^n (b+1-x_j)}
	\prod_{j=1}^n S_{x_j}^+\, \ket{\textrm{all down}}_{(-\infty,b]}\, ,\\
\psi^{\textrm{a}}_{(-\infty,b]}(n) 
  = \sum_{b\geq x_1>\dots>x_n} q^{\sum_{j=1}^n (b+1-x_j)}
	\prod_{j=1}^n S_{x_j}^-\, \ket{\textrm{all up}}_{(-\infty,b]}\, .
\end{align*}
The maps $\mathcal{F}$ and $\mathcal{R}$ still exist, as unitary transformations
between different Hilbert spaces
\begin{alignat}{2}
\mathcal{F} &: 
  \Hil^{\textrm{GNS,k}}_{[a,\infty)} \leftrightarrow \Hil^{\textrm{GNS,a}}_{[a,\infty)}\, ,
&\qquad \mathcal{F} &: 
  \Hil^{\textrm{GNS,k}}_{(-\infty,b]} \leftrightarrow \Hil^{\textrm{GNS,a}}_{(-\infty,b]}\, ,\\
\mathcal{R}\left(\frac{a+b}{2}\right) &: 
  \Hil^{\textrm{GNS,k}}_{[a,\infty)} \leftrightarrow \Hil^{\textrm{GNS,a}}_{(\infty,b]}\, ,
&\qquad \mathcal{R}\left(\frac{a+b}{2}\right) &: 
  \Hil^{\textrm{GNS,k}}_{(-\infty,b]} \leftrightarrow \Hil^{\textrm{GNS,a}}_{[a,\infty)}\, .
\end{alignat}
and one has the commutative diagram
$$
\begin{CD}
\psi^{\textrm{k}}_{[a,\infty)}(n) 
  @>\mathcal{R}\left(\frac{a+b}{2}\right)>> \psi^{\textrm{a}}_{(-\infty,b]}(n)\\
@V{\mathcal{F}}VV @VV\mathcal{F}V\\
\widetilde{\psi}^{\textrm{a}}_{[a,\infty)}(n) @>\mathcal{R}\left(\frac{a+b}{2}\right)>> 
\widetilde{\psi}^{\textrm{k}}_{(-\infty,b]}(n)
\end{CD}
$$
Finally, for the bi-infinite interval $(-\infty,\infty)$, the kink states are in the
Guichardet Hilbert space based on the infinite product
$$
\Omega^{\textrm{k}} = \bigotimes_{x=-\infty}^{\infty} \ket{\Omega^{\textrm{k}}(x)}_x\, 
\qquad \ket{\Omega^{\textrm{k}}(x)} = \begin{cases} \ket{\uparrow} & x>0\, ,\\
\ket{\downarrow} & x\leq 0\, ,\end{cases}
$$
with ground state vectors
$$
\widehat{\psi}^{\textrm{k}}_{(-\infty,\infty)}(n) 
  = q^{-n(n+1)/2} \sum_{k=0}^\infty q^{-k} \widetilde{\psi}^{\textrm{k}}_{(-\infty,0]}(k) 
  \otimes \psi^{\textrm{k}}_{[1,\infty)}(n+k)\, .
$$
In this equation $\widetilde{\psi}^{\textrm{k}}_{(-\infty,0]}(k)$ and
$\psi^{\textrm{k}}_{[1,\infty)}(k)$ are interpreted as zero whenever $k<0$.
(Now there are two extra ground states, the translation invariant states 
consisting of all up-spins and all down-spins on $(-\infty,\infty)$.)
Likewise, the antikink Hilbert space is based on
$$
\Omega^{\textrm{a}} = \bigotimes_{x=-\infty}^{\infty} \ket{\Omega^{\textrm{a}}(x)}_x\, 
\qquad \ket{\Omega^{\textrm{a}}(x)} = \begin{cases} \ket{\downarrow} & x>0\, ,\\
\ket{\uparrow} & x\leq 0\, ,\end{cases}
$$
and
$$
\widehat{\psi}^{\textrm{a}}_{(-\infty,\infty)}(n) 
  = q^{-n(n+1)/2} \sum_{k=0}^\infty q^{-k} \psi^{\textrm{a}}_{(-\infty,0]}(k) 
  \otimes \widetilde{\psi}^{\textrm{a}}_{[1,\infty)}(n+k)\, .
$$
The maps $\mathcal{F}$ and $\mathcal{R}$ are unitary maps between
$\Hil^{\textrm{GNS,k}}_{(-\infty,\infty)}$ and $\Hil^{\textrm{GNS,a}}_{(-\infty,\infty)}$,
and in particular there is a unitary
$\mathcal{N} = \mathcal{F} \mathcal{R}(\frac{1}{2})$
which acts on both Hilbert spaces, and
$$
\mathcal{N}(\widehat{\psi}^{\textrm{k,a}}_{(-\infty,\infty)}(n))
  = \widehat{\psi}^{\textrm{k,a}}_{(-\infty,\infty)}(-n)\, .
$$
More useful than this, for our present purposes, is the translation automorphism
$\tau = \mathcal{R}\left(\frac{1}{2}\right) \mathcal{R}(0)$.
Instead of thinking of $\tau$ as an isometry of the Hilbert spaces 
$\Hil^{\textrm{k,a}}_{(-\infty,\infty)}$, we prefer to think of it as a unitary
transformation on the algebra of quasilocal observables 
$\overline{\Obs}_{\textrm{loc}}$, and more specifically as a unitary transformation
from the subalgebra $\Obs_{\Lambda}$ to $\Obs_{\Lambda+1}$ for any finite
$\Lambda \subset \Ir$.
In this way we can make perfect sense of the quantity
$$
\ip{\psi^{\textrm{k}}_{[a,b]}(n)}{\tau^* X \tau \psi^{\textrm{k}}_{[a,b]}}
  = \ip{\psi^{\textrm{k}}_{[a,b]}(n)}{\textrm{Ad}\,\tau(X) \psi^{\textrm{k}}_{[a,b]}}\, ,
$$
as long as $X \in \Obs_{\Lambda}$ and both $\Lambda$ and $\Lambda+1$ are subsets
of $[a,b]$.
(We remind ourselves that $\Obs_{\Lambda} \subset \Obs_{[a,b]}$
by the canonical map $X \mapsto X_\Lambda \otimes \unity_{[a,b] \setminus \Lambda}$.)

We now state the main result of this section.
We wish to consider only the simplest case of a droplet possible, and that is that
both $L$ and $n$ approach infinity, but in such a way that $L$ approaches much faster
than $n$.
One way to do this is to observe that there is an infinite-volume, i.e.\ $L \to \infty$,
limit of the finite-volume antikink-kink states $\psi^{\textrm{ak}}_L(n)$.
Namely, considering the vector
$\psi^{\textrm{ak}}_L(n)$ as a state on $\Obs_{[-L,L]}$, one can take
the limit of the states acting on all the local observables, and the limit also 
exists in the algebra of quasilocal observables, and equals the pure state given by the
vector
\begin{equation}
\label{InfiniteKAstate}
\psi^{\textrm{ak}}_\infty(n) 
  = \sum_{k=0}^n q^{-k} \psi^{\textrm{a}}_{(-\infty,0]}(k)\otimes
  \psi^{\textrm{k}}_{[1,\infty)}(n-k)\, .
\end{equation}
which exists as a vector in the Guichardet Hilbert space based on the translation-invariant
all up-spin vector.
By taking the limit $n \to \infty$ of the states determined by the vectors
$\psi^{\textrm{ak}}_\infty(n)$, one has the most extreme case of $L$ converging to
infinity faster than $n$, namely, $L \to \infty$ first, then $n \to \infty$.
\begin{proposition}
\label{KAInterface}
For any local observable $X \in \Obs_{\Lambda}$,
one has
\begin{equation}
\label{lim1}
  \lim_{\substack{n \to \infty \\ \text{$n$ even}}}
  \frac{\langle{\psi^{\textrm{ak}}_\infty(n), \textrm{Ad}\,\tau^{n/2}(X)
  \psi^{\textrm{ak}}_\infty(n)}\rangle}
  {\langle{\psi^{\textrm{ak}}_\infty(n),\psi^{\textrm{ak}}_\infty(n)}\rangle} 
  = \frac{\sum_{k \in \Ir} q^{2k(k+1)} \ip{\widehat{\psi}^{\textrm{k}}_{(-\infty,\infty)}(k)}
  {X \widehat{\psi}^{\textrm{k}}_{(-\infty,\infty)}(k)}}
  {\sum_{k \in \Ir} q^{2k(k+1)} (q^2;q^2)_\infty}\, ,
\end{equation}
and
\begin{equation}
\label{lim2}
  \lim_{\substack{n \to \infty \\ \text{$n$ odd}}}
  \frac{\langle{\psi^{\textrm{ak}}_\infty(n), \textrm{Ad}\,\tau^{(n-1)/2}(X)
  \psi^{\textrm{ak}}_\infty(n)}\rangle}
  {\langle{\psi^{\textrm{ak}}_\infty(n),\psi^{\textrm{ak}}_\infty(n)}\rangle} 
  = \frac{\sum_{k \in \Ir} q^{2k^2} \ip{\widehat{\psi}^{\textrm{k}}_{(-\infty,\infty)}(k)}
  {X \widehat{\psi}^{\textrm{k}}_{(-\infty,\infty)}(k)}}
  {\sum_{k \in \Ir} q^{2k^2} (q^2;q^2)_\infty}\, .
\end{equation}
\end{proposition}
\textrm{Note:}
The size of the droplet, i.e.\ the number of downspins in the state 
$\psi^{\textrm{ak}}_\infty(n)$ is $n$.
So translating by $n/2$, for even $n$, $X \in \Obs_{\Lambda}$ 
becomes an observable $\textrm{Ad}\,\tau^{n/2} X \in \Obs_{\Lambda + \frac{n}{2}}$.
Thus, we are tracking the right interface of the droplet.
The results for the left interface are obtained by conjugating by $\mathcal{R}(0)$,
(since $\mathcal{R}(0) \psi^{\textrm{ak}}_\infty(n) = \psi^{\textrm{ak}}_\infty(n)$,)
to obtain
\begin{corollary}
For any local observable $X \in \Obs_{\Lambda}$,
one has
\begin{equation*}
  \lim_{\substack{n \to \infty \\ \text{$n$ even}}}
  \frac{\langle{\psi^{\textrm{ak}}_\infty(n), \textrm{Ad}\,\tau^{-n/2}(X)
  \psi^{\textrm{ak}}_\infty(n)}\rangle}
  {\langle{\psi^{\textrm{ak}}_\infty(n),\psi^{\textrm{ak}}_\infty(n)}\rangle} 
  = \frac{\sum_{k \in \Ir} q^{2k(k+1)} \ip{\widehat{\psi}^{\textrm{a}}_{(-\infty,\infty)}(k)}
  {X \widehat{\psi}^{\textrm{a}}_{(-\infty,\infty)}(k)}}
  {\sum_{k \in \Ir} q^{2k(k+1)} (q^2;q^2)_\infty}\, ,
\end{equation*}
and
\begin{equation*}
  \lim_{\substack{n \to \infty \\ \text{$n$ odd}}}
  \frac{\langle{\psi^{\textrm{ak}}_\infty(n), \textrm{Ad}\,\tau^{-(n-1)/2}(X)
  \psi^{\textrm{ak}}_\infty(n)}\rangle}
  {\langle{\psi^{\textrm{ak}}_\infty(n),\psi^{\textrm{ak}}_\infty(n)}\rangle} 
  = \frac{\sum_{k \in \Ir} q^{2k^2} \ip{\widehat{\psi}^{\textrm{a}}_{(-\infty,\infty)}(k)}
  {X \widehat{\psi}^{\textrm{a}}_{(-\infty,\infty)}(k)}}
  {\sum_{k \in \Ir} q^{2k^2} (q^2;q^2)_\infty}\, .
\end{equation*}
\QED
\end{corollary}
To prove these limits we need a fact about the kink system itself which is
\begin{lemma}
For any local observable $X$,
\begin{equation}
\lim_{k \to \infty} \frac{\ip{\psi^{\textrm{k}}_{[-k+1,\infty)}(n+k)}
  {X \psi^{\textrm{k}}_{[-k+1,\infty)}(n+k)}}
  {\|\psi^{\textrm{k}}_{[-k+1,\infty)}(n+k)\|^2}
  = \frac{\ip{\psi^{\textrm{k}}_{(-\infty,\infty)}(n)}
  {X \psi^{\textrm{k}}_{(-\infty,\infty)}(n)}}
  {\|\psi^{\textrm{k}}_{(-\infty,\infty)}(n)\|^2}\, .
\end{equation}
\end{lemma}
\begin{proof}\textbf{(of lemma)}
We prove the lemma for $n=0$, then the more general result holds by conjugating 
$X$ by $\tau^n$
(since $\tau \widehat{\psi}^{\textrm{k}}_{(-\infty,\infty)}(n) = 
\widehat{\psi}^{\textrm{k}}_{(-\infty,\infty)}(n-1)$). 
Suppose that $X \in \Obs_{[-N,N]}$, which is true for some $N$ because
$X$ is local. 
Define
$$
\widehat{\psi}^{\textrm{k}}_{[-N,N]}(n)
  = \sum_k q^{-k} \widetilde{\psi}^{\textrm{k}}_{[-N,0]}(k) 
  \otimes \psi^{\textrm{k}}_{[1,N]}(n+k)\, .
$$
Then  
\begin{equation*}
\widehat{\psi}^{\textrm{k}}_{(-\infty,\infty)}(0) 
  = \sum_{n=-N-1}^{N} \widehat{\psi}^{\textrm{k}}_{[-N,N]}(n) 
  \otimes \psi''_{\Ir \setminus [-N,N]}(-n)\, ,
\end{equation*}
where
\begin{equation*}
\psi''_{\Ir \setminus [-N,N]}(n) 
  = \sum_k q^{N(2n+k)} \widetilde{\psi}^{\textrm{k}}_{(-\infty,-N+1]}(k)
  \psi^{\textrm{k}}_{[N+1,\infty)}(n+k)\, .
\end{equation*}
What is most imporant is the value of 
$Z_{\Ir\setminus [-N,N]}(n) = \|\psi''_{\Ir \setminus [-N,N]}(n)\|^2$:
\begin{equation*}
Z''_{\Ir\setminus [-N,N]}(-n) = \begin{cases}
  \sum_{k=0}^\infty \sum_{\substack{x_{-k}<\dots<x_{-1}<-N \\
  N<x_1<\dots<x_{n+k}}} q^{2 \sum_{j=1}^{n+k} x_k - 2 \sum_{j=1}^k x_{-j}} &
  \text{for $n\geq 0$} \\
  \sum_{k=0}^\infty \sum_{\substack{x_{-k+n}<\dots<x_{-1}<-N \\
  N<x_1<\dots<x_k}} q^{2 \sum_{j=1}^k x_j - 2 \sum_{j=1}^{k-n} x_{-j}} &
  \text{for $n\leq 0$}
\end{cases}
\end{equation*}
Then
$$
\frac{\ip{\widehat{\psi}^{\textrm{k}}_{(-\infty,\infty)}(0)}
{X \widehat{\psi}^{\textrm{k}}_{(-\infty,\infty)}(0)}}
{\|\widehat{\psi}^{\textrm{k}}_{(-\infty,\infty)}(0)\|^2}
  = \frac{\sum_{n=-N-1}^{N} \ip{\widehat{\psi}^{\textrm{k}}_{[-N,N]}(n)}
  {X \widehat{\psi}^{\textrm{k}}_{[-N,N]}(n)} Z''_{\Ir\setminus [-N,N]}(-n)}
  {\sum_{n=-N-1}^{N} \|\widehat{\psi}^{\textrm{k}}_{[-N,N]}(n)\|^2
  Z''_{\Ir\setminus [-N,N]}(-n)}\, .
$$
Similarly, for $r$ large enough
\begin{equation*}
 \psi^{\textrm{k}}_{(-r,\infty)}(r+1) 
  = C(r) \sum_{n=-N-1}^{N} \widehat{\psi}^{\textrm{k}}_{[-N,N]}(n) 
  \otimes \psi''_{[-r,\infty) \setminus [-N,N]}(-n)\, ,
\end{equation*}
where $C(r)$ is a normalizing constant and 
\begin{equation*}
\psi''_{[r,\infty) \setminus [-N,N]}(n) 
  = \sum_k q^{N(2n+k)} \widetilde{\psi}^{\textrm{k}}_{[-r,-N+1]}(k)
  \psi^{\textrm{k}}_{[N+1,\infty)}(n+k)\, .
\end{equation*}
Then defining $Z''_{[-r,\infty)\setminus [-N,N]}(n) 
= \|\psi''_{[r,\infty) \setminus [-N,N]}(n)\|^2$,
\begin{equation*}
Z''_{[-r,\infty)\setminus [-N,N]}(n)
  = \begin{cases}
  \sum_{k=0}^\infty \sum_{\substack{-r\leq x_{-k}<\dots<x_{-1}<-N \\
  N<x_1<\dots<x_{n+k}}} q^{2 \sum_{j=1}^{n+k} x_k - 2 \sum_{j=1}^k x_{-j}} &
  \text{for $n\geq 0$} \\
  \sum_{k=0}^\infty \sum_{\substack{-r\leq x_{-k+n}<\dots<x_{-1}<-N \\
  N<x_1<\dots<x_k}} q^{2 \sum_{j=1}^k x_j - 2 \sum_{j=1}^{k-n} x_{-j}} &
  \text{for $n\leq 0$}
\end{cases}
\end{equation*}
and
\begin{align*}
&\frac{\ip{\psi^{\textrm{k}}_{[-r,\infty)}(r+1)}
  {X \psi^{\textrm{k}}_{[-r,\infty)}(r+1)}}
  {\|\widehat{\psi}^{\textrm{k}}_{[-r,\infty)}(r+1)\|^2}\\
  &\qquad\qquad  
  = \frac{\sum_{n=-N-1}^{N} \ip{\widehat{\psi}^{\textrm{k}}_{[-N,N]}(n)}
  {X \widehat{\psi}^{\textrm{k}}_{[-N,N]}(n)} Z''_{[-r,\infty)\setminus [-N,N]}(-n)}
  {\sum_{n=-N-1}^{N} \|\widehat{\psi}^{\textrm{k}}_{[-N,N]}(n)\|^2
  Z''_{[-r,\infty)\setminus [-N,N]}(-n)}\, .
\end{align*}
The lemma will be proved if we show that
$$
\lim_{r \to \infty} Z''_{[-r,\infty)\setminus[-N,N]}(n)
  = Z''_{\Ir\setminus[-N,N]}(n)\, ,
$$
for each $n$.
But $Z''_{[-r,\infty)\setminus[-N,N]}(n)$, is just the same 
as $Z''_{\Ir\setminus[-N,N]}(n)$, except that summands where
some $x_{-j}$ is less than $-r$ are excluded.
Then by MCT (or DCT since $Z''_{\Ir\setminus[-N,N]}(n)$ 
is finite), the sums converge, and this proves the lemma.
\end{proof}
\begin{proof}\textbf{(of Proposition)}
To prove the limits (\ref{lim1}) and (\ref{lim2}), start by noticing that
\begin{equation}
  \|\psi^{\textrm{k}}_{[1,\infty)}(n)\|^2 
  = \|\psi^{\textrm{a}}_{(-\infty,0]}(n)\|^2 
  = \frac{q^{n(n+1)}}{(q^2;q^2)_n} 
\end{equation}
(C.f.\ Section \ref{ExactCalculationsSection} for proof.)
Suppose $X \in \Obs_{[-N,N]}$.
If $n$ is even and $\geq N$, then by \eq{InfiniteKAstate}
\begin{align*}
&\ip{\psi^{\textrm{ak}}_\infty(n)}{\textrm{Ad}\, \tau^{n/2}(X) \psi^{\textrm{ak}}_\infty(n)}\\
&\hspace{50pt}
  = \sum_{k=0}^n q^{-2k} \|\psi^{\textrm{a}}_{(-\infty,0]}(k)\|^2
    \ip{\psi^{\textrm{k}}_{[1,\infty)}(n-k)}
  {\textrm{Ad}\, \tau^{n/2}(X) \psi^{\textrm{k}}_{[1,\infty)}(n-k)}\\
&\hspace{50pt}
  = \sum_{k=0}^n \frac{q^{-2k+k(k+1)+(n-k)(n-k+1)}}{(q^2;q^2)_k (q^2;q^2)_{n-k}}
  \frac{\ip{\psi^{\textrm{k}}_{[1,\infty)}(n+k)}
  {\textrm{Ad}\, \tau^{n/2}(X) \psi^{\textrm{k}}_{[1,\infty)}(n)}}
  {\|\psi^{\textrm{k}}_{[1,\infty)}(n-k)\|^2} \\
&\hspace{50pt}
  = \frac{q^{n^2/2}}{(q^2;q^2)_n} 
  \sum_{k=-n/2}^{n/2} \qbinom{n}{\frac{n}{2}+k}{q^2} q^{2k(k+1)}
  \frac{\ip{\psi^{\textrm{k}}_{[1-\frac{n}{2},\infty)}(\frac{n}{2}+k)}
  {X \psi^{\textrm{k}}_{[1-\frac{n}{2},\infty)}(\frac{n}{2}+k)}}
  {\|\psi^{\textrm{k}}_{[1-\frac{n}{2},\infty)}(\frac{n}{2}+k)\|^2} 
\end{align*}
By the lemma we know the summand converges to
\begin{equation}
  \frac{q^{2k(k+1)}}{(q^2;q^2)_\infty}  
  \frac{\ip{\psi^{\textrm{k}}_{(-\infty,\infty)}(k)}
  {X \psi^{\textrm{k}}_{(-\infty,\infty)}(k)}}
  {\|\psi^{\textrm{k}}_{(-\infty,\infty)}(k)\|^2}\, ,
\end{equation}
for each $k$, as $n  \to \infty$.
But we may also bound the summand by a summable sequence.
Namely, for any $\psi$, 
$\frac{\langle\psi,X\psi\rangle}{\langle\psi,\psi\rangle} \leq \|X\|_{op}$
and $\begin{bmatrix}n\\k\end{bmatrix} \leq (q^2;q^2)_\infty^{-1}$ for any
$n$ and $k$.
So the summand is bounded (uniformly in $n$) by 
$(q^2;q^2)_\infty^{-1} q^{2k(k+1)} \|X\|_{op}$, which is summable.
Thus by DCT the limit exists, i.e.
\begin{align*}
&\lim_{n \to \infty}
\left(\frac{q^{n^2/2}}{(q^2;q^2)_n} \right)^{-1}
\ip{\psi^{\textrm{ak}}_\infty(n)}{\textrm{Ad}\, \tau^{n/2}(X) \psi^{\textrm{ak}}_\infty(n)} \\
&\hspace{75pt}
  = \frac{1}{(q^2;q^2)_\infty}
  \sum_{k \in \Ir} q^{2k(k+1)} \frac{\ip{\psi^{\textrm{k}}_{(-\infty,\infty)}(k)}
  {X \psi^{\textrm{k}}_{(-\infty,\infty)}(k)}}
  {\|\psi^{\textrm{k}}_{(-\infty,\infty)}(k)\|^2}\, .
\end{align*}
Hence
$$
\lim_{n \to \infty}
\frac{\ip{\psi^{\textrm{ak}}_\infty(n)}
{\textrm{Ad}\, \tau^{n/2}(X) \psi^{\textrm{ak}}_\infty(n)}}
{\|\psi^{\textrm{ak}}_\infty(n)\|^2}
= \frac{\sum_{k \in \Ir} q^{2k(k+1)} \frac{\ip{\psi^{\textrm{k}}_{(-\infty,\infty)}(k)}
  {X \psi^{\textrm{k}}_{(-\infty,\infty)}(k)}}
  {\|\psi^{\textrm{k}}_{(-\infty,\infty)}(k)\|^2}}
{\sum_{k \in \Ir} q^{2k(k+1)}}\, .
$$
Since
$$
\|\psi^{\textrm{k}}_{(-\infty,\infty)}(k)\|^2 = \frac{1}{(q^2;q^2)_\infty}\, ,
$$
(c.f.\ Section \ref{ExactCalculationsSection} keeping in mind that here we rescaled
the bi-infinite kink states so that $\tau \psi^{\textrm{k}}_{(-\infty,\infty)}(k)
= \psi^{\textrm{k}}_{(-\infty,\infty)}(k-1)$),
we have the desired result when $n$ is even.
The other limit is proved just the same, except using odd integers and the
translation $\tau^{(n-1)/2}$.
\end{proof}  
Note: We have proved that the interface for the kink-antikink state is exponentially
localized, and we have calculated the exact form in the limit that the two interfaces 
become infinitely far from one another, as well as from the edges of the spin chain.
Although we took $L \to \infty$ before $n$, this was just a technicality to simplify
the proof.
One could take any sequence $L,n \to \infty$ such that $L-n \to \infty$ as well.
The most significant consequence of the localization is that we know that
for a droplet state, the bulk of the sites of the spin chain will find their spins polarized
to almost entirely all up or almost entirely all down.
In other words, the intuition one would take from the Ising model, that in the scaling
limit (that the distance between sites decreases like $1/L$, with $q$ fixed and 
$n \propto L$) the ground state looks just like a well-defined interval of down-spins
surrounded by up spins, is correct to first approximation.
In the next sections we present a more realistic model of a droplet state, one without
pinning, and our main tool there is the existence of large intervals of nearly 
completely polarized spins, which is true for any Hamiltonian which is a finite perturbation
of the free-boundary XXZ Hamiltonian.
The intuition for this theorem came to us as a result of the calculations of the
present section.

\pagebreak

{\baselineskip=10pt \thispagestyle{empty} {{\small Originally published
Comm. Math. Phys. {\bf 218}, 569--607. (2001),\qquad
arXiv:math-ph/0009002}
\hspace{\fill}}

\vspace{20pt}

\addcontentsline{toc}{chapter}{\textit{ Droplet States in the XXZ Heisenberg Chain}}
\begin{center}
{\LARGE \bf Droplet States in the XXZ Heisenberg Chain\\[27pt]}
{\large \bf Bruno Nachtergaele 
and Shannon Starr\\[10pt]}
{\large  Department of Mathematics\\
University of California, Davis\\
Davis, CA 95616-8633, USA\\[15pt]}
{\normalsize bxn@math.ucdavis.edu, sstarr@math.ucdavis.edu}\\[30pt]
\end{center}

{\bf Abstract:}
We consider the ground states of the ferromagnetic XXZ chain with 
spin up boundary conditions in sectors with a fixed number of down
spins. This forces the existence of a droplet of down spins in the
system. We find the exact energy and the states that describe these
droplets in the limit of an infinite number of down spins. We prove 
that there is a gap in the spectrum above the droplet states. 
As the XXZ Hamiltonian has a gap above the fully magnetized ground 
states as well, this means that the droplet states (for sufficiently
large droplets) form an isolated band. The width of this band tends
to zero in the limit of infinitely large droplets.
We also prove the analogous results for finite chains with periodic 
boundary conditions and for the infinite chain.

\vspace{8pt}
{\small \bf Keywords:} Anisotropic Heisenberg ferromagnet, XXZ chain,
droplet states, excitations, spectral gap.
\vskip .2 cm
\noindent
{\small \bf PACS 1999 numbers:} 05.70.Np, 75.10.Jm, 75.30.Kz, 75.70.Kw 
\newline
{\small \bf MCS 2000 numbers:} 82B10, 82B24, 82D40 
\vfill
\hrule width2truein \smallskip {\baselineskip=10pt \noindent Copyright
\copyright\ 2000 by the authors. Reproduction of this article in its entirety, 
by any means, is permitted for non-commercial purposes.\par }}

\newpage
\section{Introduction}

Droplet states have been studied in considerable detail for the Ising model
\cite{DKS,Pfi,BIV}, where they play an important role in understanding 
dynamical phenomena \cite{SS}. In this paper we  consider the
spin-$\frac{1}{2}$ ferromagnetic XXZ Heisenberg chain and prove that the bottom
of its spectrum consists of an isolated nearly flat band of droplet states in 
a sense made precise below.

The Hamiltonian for a chain of $L$ spins acts on the Hilbert space
$$
  \Hil_L = \Cx_1^2 \otimes \dots \otimes \Cx_L^2
$$
as the sum of nearest-neighbor interactions
$$
  \HXXZ_{[1,L]} = \sum_{x=1}^{L-1} \HXXZ_{x,x+1}\, 
$$
of the form
\begin{equation}
\label{XXZ:Ham}
\HXXZ_{x,x+1} 
  = - \Delta^{-1} (\vec{S}_x \cdot \vec{S}_{x+1} - \frac{1}{4})
  - (1 - \Delta^{-1}) (S_x^3 S_{x+1}^3 - \frac{1}{4})\, .
\end{equation}
Here $S_x^i$ ($i=1,2,3$) are the spin matrices,
acting on $\Cx^2_x$, extended by unity to $\Hil_L$,
and normalized so that they have eigenvalues $\pm 1/2$.
The anisotropy parameter, $\Delta$, is always assumed to be
$> 1$. To formulate the results and also for the proofs,
we need to consider the following combinations of boundary fields for
systems defined on an arbitrary interval:
for $\alpha,\beta=\pm 1,0$, and $[a,b]\subset\Ir$, define
\begin{equation}
H^{\alpha\beta}_{[a,b]}= \sum_{x=a}^{b-1}\HXXZ_{x,x+1} - A(\Delta)(\alpha
S^{3}_a+\beta S^{3}_b)\quad, 
\label{Hab}\end{equation}
where $A(\Delta) = \frac{1}{2} \sqrt{1 - \Delta^{-2}}$. Note that
$H^{00}_{[1,L]}=\HXXZ_{[1,L]}$.

As all the Hamiltonians $H^{\alpha\beta}_{[a,b]}$ commute with the 
total third component of the spin, it makes sense to study their
ground states restricted to a subspace of fixed number of down spins. 
The subspace for a chain of $L$ spins consisting of the states with
$n$ down spins will be denoted by $\Hil_{L,n}$, for $0\leq n\leq L$.
In all cases the ground state is then unique. 
The Hamiltonians with $+-$ and $-+$ boundary fields have been studied 
extensively and have kink and antikink ground states respectively 
\cite{ASW,GW,KN1,Mat,KN3,BCN,BM}. The unique ground states for a chain
on $[a,b]\subset\Ir$, in the sector with $n$ down spins, will be
denoted by $\psi^{\alpha\beta}_{[a,b]}(n), 0\leq n\leq b-a+1$.
For $\alpha\beta=+-, -+$, they are given by 
\begin{eqnarray}
\label{Intro:+-}
\psi^{+-}_{[a,b]}(n) = 
  \sum_{a \leq x_1 < \dots < x_n \leq b} q^{\sum_{k=1}^n (b+1-x_k)}
  \left( \prod_{k=1}^n S_{x_k}^- \right) \ket{\uparrow \dots 
  \uparrow}_{[a,b]} \\
\label{Intro:-+}
\psi^{-+}_{[a,b]}(n) = 
  \sum_{a \leq x_1 < \dots < x_n \leq b} q^{\sum_{k=1}^n (x_k+1-a)}
  \left( \prod_{k=1}^n S_{x_k}^- \right) \ket{\uparrow \dots \uparrow}_{[a,b]} 
\end{eqnarray}
where $\Delta = (q+q^{-1})/2$.
Note that the norm of these vectors depends on the length (but not on the
position) of the interval $[a,b]$ (see \eq{App:kink-norm}). There is a uniform 
lower bound for the spectral gap above these ground states \cite{KN1}, a 
property that will be essential in the proofs. 

Here, we are interested in the ground states of the Hamiltonian
with $++$ boundary fields, which we refer to as the {\it droplet
Hamiltonian}, in the regime where there are a sufficently large number
of down spins. This includes, but is not limited 
to, the case where there is a fixed density $\rho$, $0<\rho\leq 1$,
of down spins in a system with $++$ boundary conditions. We 
prove that under these conditions the ground states contain one droplet of
down spins in a background of up spins.

From the mathematical point-of-view there is an important
distinction between the kink Hamiltonian and the droplet Hamiltonian,
which is that the droplet Hamiltonian does not possess $SU_q(2)$ symmetry. 
In contrast to the kink Hamiltonian where explicit formulae
are known for the ground states in finite volumes, no such explicit
analytic formulae are known for the droplet Hamiltonian for general $L$.
Therefore, we rely primarily on energy estimates, and our main
results are formulated as estimates that become exact only in 
the limit $n, L\to\infty$. This is natural as, again unlike for the kink 
ground states, there is no immediate infinite-volume description of the 
droplet states. We find the exact energy of an infinite droplet and an 
approximation of the droplet ground states that becomes exact in the 
thermodynamic limit. We also prove that all states with the energy of the
droplet are necessarily droplet states, again, in the thermodynamic limit. For
the droplet Hamiltonians this means that the droplet states are all the ground
states, and that there is a gap above them. One can also interpret this as
saying that all excitations of the fully magnetized ground states of the XXZ
chain, with sufficiently many overturned spins and not too high an energy, are
droplet states.

\subsection{Main Result}
The main result of this paper is the approximate
calculation of the ground state
energy, the ground state space, and a lower bound for the
spectral gap of the operator
$H^{++}_{[1,L]}$ restricted to the sector $\Hil_{L,n}$.
If the results were exact, we would have an eigenvalue $E_0$, a subspace
$\Hil_{L,n}^0 \subset \Hil_{L,n}$, and a positive number $\gamma$, such that
$$
H^{++}_{[1,L]} \textrm{Proj}(\Hil^0_{L,n}) = E_0 \textrm{Proj}(\Hil^0_{L,n})
$$
and
$$
H^{++}_{[1,L]} \textrm{Proj}(\Hil_{L,n})
  \geq E_0 \textrm{Proj}(\Hil_{L,n}) + \gamma (\textrm{Proj}(\Hil_{L,n})
  - \textrm{Proj}(\Hil^0_{L,n}))\, .
$$
We will always use the notation ${\rm Proj}(V)$ to mean orthogonal projection
onto a subspace $V$.

\begin{figure}
\begin{center}
\resizebox{12truecm}{2truecm}{\includegraphics{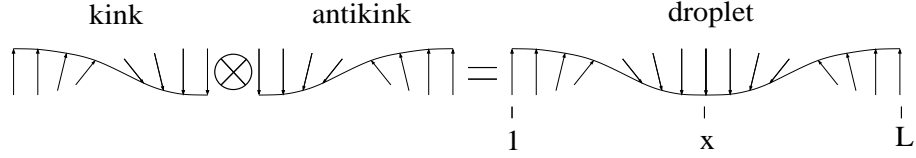}}
\parbox{11truecm}{\caption{\baselineskip=5 pt\small
\label{Fig:typical-drop}
Diagram of a typical droplet as the tensor product of a kink
and antikink.}
}
\end{center}
\end{figure}
Our results are approximations, with increasing accuracy as
$n$ tends to infinity, independent of $L$.
First, we identify the proposed ground state space.
For $n\geq 0$ and $\floor{n/2} \leq x \leq L - \ceil{n/2}$
define
\begin{equation}
\xi_{L,n}(x) \
  = \psi^{+-}_{[1,x]}(\floor{n/2})
  \otimes \psi^{-+}_{[x+1,L]}(\ceil{n/2})\, .
\label{def_drop}
\end{equation}
For any real number $x$, $\floor{x}$ is the greatest integer $\leq x$,
and $\ceil{x}$ is the least integer $\geq x$.
The typical magnetization profile of $\xi_{L,n}(x)$ is shown in 
Figure \ref{Fig:typical-drop}.
We define the space of approximate ground states as follows: 
$$
\calK_{L,n} = \Span \{ \xi_{L,n}(x) : \, 
\floor{n/2}\leq x \leq L-\ceil{n/2} \}\, .
$$
$\calK_{L,n}$ is the space of ``approximate'' droplet states 
with $n$ down spins for a finite chain of length $L$.
An interval of length $n$ can occur in $L-n+1$ positions inside 
a chain of length $L$.
This explains why $\dim \mathcal{K}_{L,n} = L-n+1$.

Alternatively, we could use the following definitions of
approximate droplet states:
$$
\xi'_{L,n}(x)=[S^{\rm{antikink},+}_{[1,L]}]^{x-\floor{n/2}}
[S^{\rm{kink},+}_{[1,L]}]^{L-\ceil{n/2}-x}\alldown
$$
where $S^{\rm{kink},+}_{[1,L]}$ is the $SU_q(2)$ raising operator
(see, e.g., (2.5b) of \cite{KN1}), and $S^{\rm{antikink},+}_{[1,L]}$ 
is the left-right reflection of $S^{\rm{kink},+}_{[1,L]}$. Yet another
option for the droplet states is to take the exact ground states of
the Hamiltonians $H_{[1,L]}=H_{[1,x]}^{+-}+ H_{[x,L]}^{-+}$, which
have a pinning field at position $x$, and for which exact expressions
for the ground states can be obtained. One can show that suitable linear
combinations of these states differ in norm from the $\xi_{L,n}(x)$ by 
no more
than $O(q^n)$. We will only use the states $\xi_{L,n}(x)$ defined in 
\eq{def_drop}, as they have a more intuitive interpretation as a tensor
product of a kink and an antikink state.

\begin{theorem} $\vspace{1mm}$
\label{main:theorem}

a) There exists a constant $C < \infty$ such that
$$
\| (H^{++}_{[1,L]} - A(\Delta)) \Proj(\calK_{L,n}) \|
  \leq C q^n\, .
$$
The constant $C$ depends only on $q$, not on $L$ or $n$.

b) There exists a sequence $\epsilon_n$,
with $\lim_{n \to \infty} \epsilon_n = 0$,
such that
\begin{eqnarray*}
&&
H^{++}_{[1,L]} \Proj(\Hil_{L,n}) \geq (A(\Delta) - 2 C q^n) \Proj(\Hil_{L,n}) \\
&& \hspace{4cm}
  + (\gamma - \epsilon_n)(\Proj(\Hil_{L,n}) - \Proj(\calK_{L,n}))\, ,
\end{eqnarray*}
where $\gamma = 1 - \Delta^{-1}$.
The sequence $\epsilon_n$ can be chosen to decay at least as fast as $n^{-1/4}$,
independent of $L$.
\end{theorem}

For $H^{XXZ}_{[1,L]}$, which is the one without boundary terms, the
large-droplet states are not separated in the spectrum from other excitations
such as the spin waves, i.e., the band of continuous spectrum due to spin wave
excitations overlaps with the states of droplet type.
Although similar results should hold for boundary fields of larger magnitude
the value, $A(\Delta)$, of the boundary fields in the droplet
Hamiltonian, is particularly convenient for at least two reasons: 1)
it allows us to write the Hamiltonian as a sum of kink and anti-kink
Hamiltonians, which is the basis for many of our arguments, 2) the
energy of a droplet in the center of the chain is the same as for a
droplet attached to the boundary. This allows us to construct explicitly
the subspace of all droplet states asymptotically in the thermodynamic
limit.

\begin{figure}
\begin{center}
\resizebox{6truecm}{6truecm}{\includegraphics{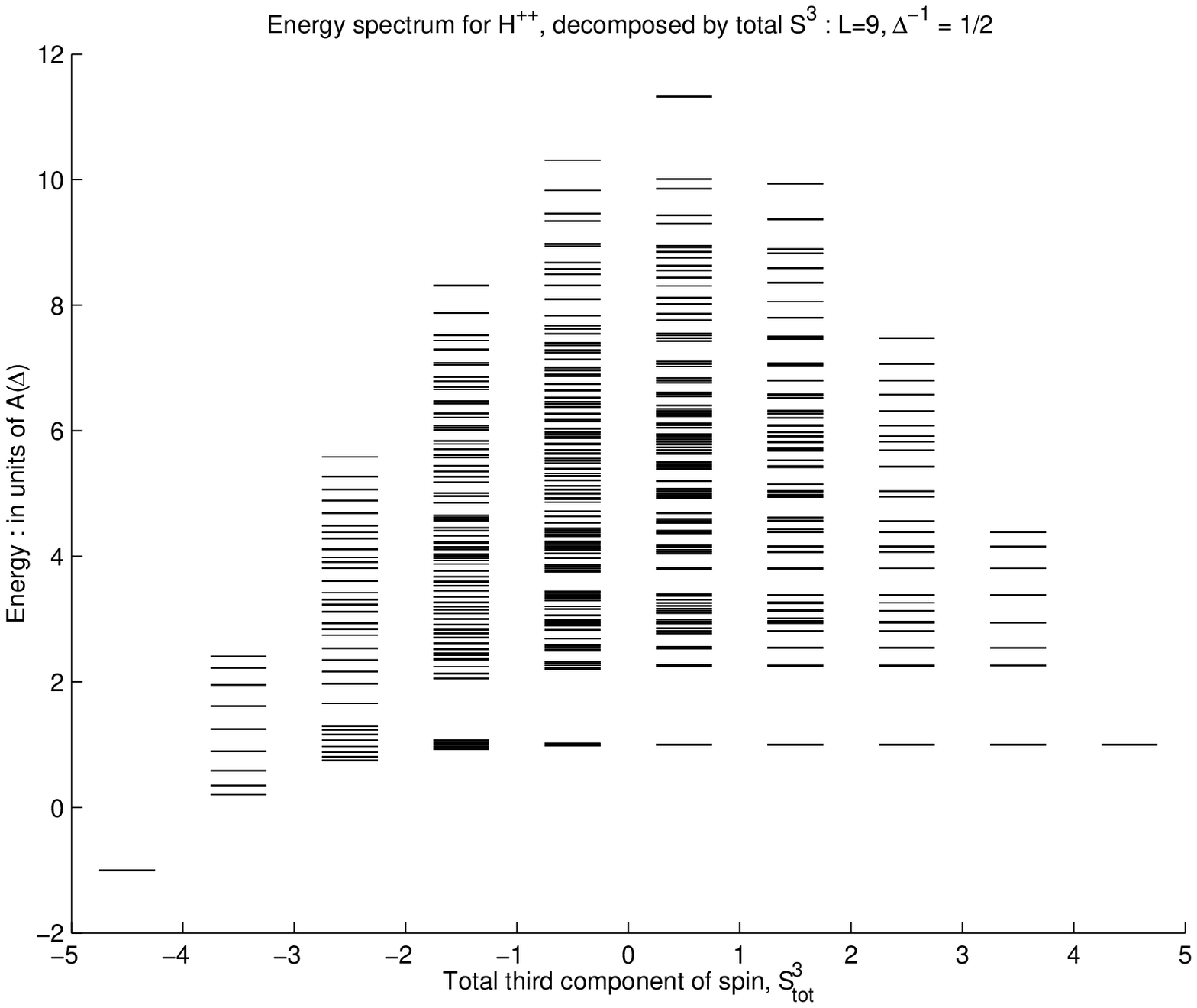}} \hspace{25pt}
\resizebox{6truecm}{6truecm}{\includegraphics{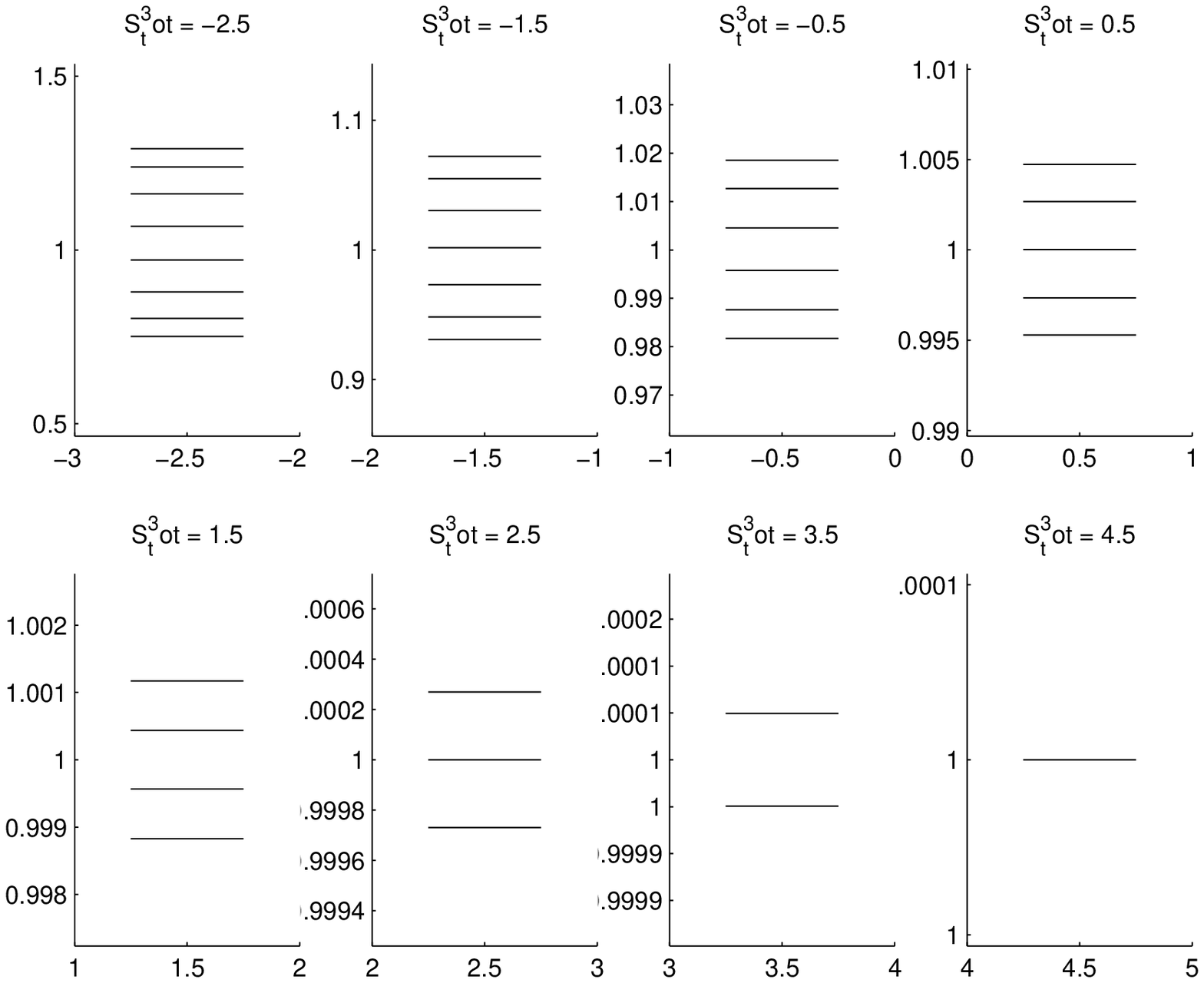}}
\parbox{11truecm}{\caption{\baselineskip=5 pt\small
\label{fig:dropspec}
(a) Spectrum for $H^{++}_{[1,9]}$ when $\Delta = 2$ ($q=2-\sqrt{3}$),
(b) Enlargement of spectrum about $A(\Delta)$ for fixed sectors $S^3_{\textrm{tot}} = -5/2,\dots,+9/2$}
}
\end{center}
\end{figure}

Although our main results are about infinite droplets, i.e., they are 
asymptotic properties of finite droplets in the limit of their size
tending to infinity, we can extract from our proofs estimates of the 
corrections for finite size droplets. This allows the following reformulation
of the main result in terms of the eigenvalues near the bottom
of the spectrum and the corresponding eigenprojection.
Let 
$\lambda_{L,n}(1) \leq \lambda_{L,n}(2) \leq \dots $
be the eigenvalues of $H^{++}_{[1,L]}$ restricted to the sector $\Hil_{L,n}$.
Let $\psi^{++}_{L,n}(1), \psi^{++}_{L,n}(2), \dots $ 
be the corresponding eigenstates, and define
$$
  \Hil^{k}_{L,n} = \Span \{ \psi^{++}_{L,n}(j) : 1 \leq j \leq k\}\, .
$$
\begin{theorem} 
\label{main:theorem2} $\vspace{1mm}$

a) We have the following information about the spectrum of $H^{++}_{[1,L]}$
restricted to $\Hil_{L,n}$:
$$
\lambda_{L,n}(1),\dots \lambda_{L,n}(L-n+1) 
  \in  [A(\Delta) - O(q^n),A(\Delta) + O(q^n)]\, ,
$$
and

b) $\lambda_{L,n}(L-n+2) \geq A(\Delta) + \gamma - O(n^{-1/4})$.

c) We have the following information about the eigenspace for the low-energy
states, $\lambda_{L,n}(1),\dots,\lambda_{L,n}(L-n+1)$:
$$
\|\Proj(\calK_{L,n}) - \Proj(\Hil^{L-n+1}_{L,n})\|
= O(q^{n/2})\, . 
$$
Equivalently
\begin{eqnarray*}
\sup_{0 \neq \psi \in \calK_{L,n}} 
  \left( \inf_{\psi' \in \Hil^{L-n+1}_{L,n}} 
  \frac{\|\psi - \psi'\|^2}{\|\psi\|^2} \right) = O(q^n)\, , \\
\sup_{0 \neq \psi' \in \Hil^{L-n+1}_{L,n}}
  \left( \inf_{\psi \in \calK_{L,n}}
  \frac{\|\psi - \psi'\|^2}{\|\psi'\|^2} \right) = O(q^n)\, .
\end{eqnarray*}
\end{theorem}
Figure \ref{fig:dropspec} illustrates the spectrum for a specific choice
of $L$ and $q$.
Note that Theorem \ref{main:theorem2} also implies that, for any sequence of
states with energies converging to $A(\Delta)$, we must have that the distances
of these states to the subspaces $\calK_{L,n}$ converges to zero.
The remainder of the paper is organized as follows.

Section \ref{sec:Ham} reviews some preliminary properties of the 
Hamiltonians that appear in the paper: a simple estimate for the gap
above the ground state of the XXZ Hamiltonian on an open chain without 
boundary terms, the spectral gap for the Hamiltonian with kink and
antikink boundary terms, and a preliminary lower bound for the energy
of a droplet state.

The proof of the main theorems is given in Sections 
\ref{Sect:Eval}, \ref{sec:polarized}, and \ref{Sec:ROP}. 
First, in Section \ref{Sect:Eval}, we calculate the energy of the 
proposed droplet states $\xi_{L,n}(x)$, defined in 
\eq{def_drop}. We also prove that these states are approximate
eigenstates. 

In Section \ref{sec:polarized}, we prove a basic estimate on the 
probability that an interval $J\subset [1,L]$ is fully polarized 
(i.e., {\em all up} or {\em all down}). It turns out that this 
probability can be bounded by 
$$
\mbox{Prob[{\small the spins in $J$ are all up or all down}]}
\geq 1-\mbox{Constant}\times \vert J\vert\times\frac{E}{L},
$$
where $E$ is a bound on the energy of the state. The meaning of this
bound is clear. For fixed energy $E$, as $L$ increases it becomes 
more and more likely that any given interval $J$ is in the {\em all up}
or {\em all down} state. The spectral gap of the model enters through
the constant. This fact should be expected for any ferromagnetic model 
with a gap, as the interaction encourages like spins to aggregate.

Section \ref{Sec:ROP} contains the most intricate part of the proof. We
implement the idea that the presence of an interval of all up or all down spins
in a state, allows one to decouple the action of  the Hamiltonians on the
subsystems to the left and the right of this  interval. If the spins in the
interval are {\em down}, the Hamiltonian decouples into a sum of a kink and an
antikink Hamiltonian, for which it is  known that there is spectral gap. If the
spins in the interval are {\em up}, we do not immediately obtain an estimate for
the gap, but  we can repeat the argument for the two decoupled subsystems and so
on.  If there are a sufficiently large number of down spins in the original 
system, this procedure must eventually lead to an interval of {\em down} 
spins.  

We will also prove, in Section \ref{Sec:Periodic-Infinite}, 
the analogous statements for rings and for the infinite chain with a large 
but finite number of down spins. Some calculations that are used in the 
proofs are collected in two appendices.


\section{Properties of the XXZ Hamiltonians}
\label{sec:Ham}

In this section, we collect all the Hamiltonians that appear
in the paper, and describe some of their properties.
The first Hamiltonian we consider is
\begin{equation}
\label{xxzhamdef}
  \HXXZ_{[1,L]} = \sum_{x=1}^{L-1} \HXXZ_{x,x+1}\, 
\end{equation}
where
\begin{equation}
\label{XXZ:Ham2}
\HXXZ_{x,x+1} 
  = - \Delta^{-1} (\vec{S}_x \cdot \vec{S}_{x+1} - \frac{1}{4})
  - (1 - \Delta^{-1}) (S_x^{(3)} S_{x+1}^{(3)} - \frac{1}{4})\, .
\end{equation}
$\Delta > 1$ is the anisotropy parameter. 
Note that for $\Delta=1$ it is the isotropic
Heisenberg model, and for $\Delta = \infty$ it is the Ising model.

The diagonalization of $\HXXZ_{x,x+1}$, considered as an operator on
the four dimensional space $\Cx_x^2 \otimes \Cx_{x+1}^2$ is
\begin{equation}
\label{XXZ-nn:Diag}
\HXXZ_{x,x+1} : 
\qquad
\begin{array}{|c|c|}
  \rm{eigenvalue} & \rm{eigenvector} \\
  \hline 
  0 & \ket{\uparrow \uparrow},\, 
  \ket{\downarrow \downarrow} \\
  \frac{1}{2}(1 - \Delta^{-1})  & \frac{1}{\sqrt{2}} 
  (\ket{\uparrow \downarrow} + \ket{\downarrow \uparrow})\\
  \frac{1}{2}(1 + \Delta^{-1})  & \frac{1}{\sqrt{2}} 
  (\ket{\uparrow \downarrow} - \ket{\downarrow \uparrow})
\end{array}
\end{equation}
Let us define  
\begin{equation}
\label{pdef}
P^\sigma_{x,x+1} = \unity_1 \otimes \dots \otimes \unity_{x-1}
\otimes \ket{\sigma \sigma}\bra{\sigma \sigma} \otimes \unity_{x+2} \otimes
\dots \otimes \unity_L
\end{equation}
for $\sigma = \uparrow,\downarrow$,
and $P_{x,x+1} = P^\uparrow_{x,x+1} + P^\downarrow_{x,x+1}$.
Then, clearly,  
\begin{equation}
\label{localxxzgap} 
\HXXZ_{x,x+1} \geq \frac{1}{2}(1 - \Delta^{-1}) (\unity - P_{x,x+1})\, .
\end{equation}
\begin{lemma}
\label{XXZ-gap:Lemma}
The ground state energy for $\HXXZ_{[1,L]}$ is $0$,
and the ground state space is 
$\Span \{ \allup, \alldown \}$.
The following bounds hold
\begin{equation}
\label{XXZ-gap:bound}
  \HXXZ_{[1,L]} \geq \frac{1}{2}(1 - \Delta^{-1}) 
\Big(\unity - \Proj(\Span \{\allup,\alldown\})\Big)\, .
\end{equation}
\end{lemma}

\begin{proof}
The fact that $\allup$ and $\alldown$ are annihilated by $\HXXZ_{[1,L]}$
follows trivially from the fact that $\allup$ and $\alldown$ are annihilated
by each pairwise interaction $\HXXZ_{x,x+1}$.
So, in fact these states are frustration-free ground states.
Next,
$$
\HXXZ_{[1,L]}
  \geq \frac{1}{2}(1 - \Delta^{-1}) \sum_{x=1}^{L-1} (\unity - P_{x,x+1})\, ,
$$
by \eq{xxzhamdef} and \eq{localxxzgap}.
We observe that each $P_{x,x+1}$ is an orthogonal projection.
Moreover $P_{x,x+1}$ commutes with $P_{y,y+1}$ for every $x$ and $y$.
So
$$
\unity - \prod_{x=1}^{L-1} P_{x,x+1} 
  = \sum_{x=1}^{L-1} \left(\prod_{y=1}^{x-1} P_{y,y+1} \right) 
  (\unity - P_{x,x+1}) 
  \leq \sum_{x=1}^{L-1} (\unity - P_{x,x+1})\, .
$$
But $\prod_{x=1}^{L-1} P_{x,x+1} = \Proj(\Span\{\allup,\alldown\})$,
which proves \eq{XXZ-gap:bound}.
\end{proof}

All the other Hamiltonians we consider, namely $H^{\alpha \beta}_{[1,L]}$ 
for $\alpha,\beta=\pm 1,0$, defined in \eq{Hab}, are perturbations of 
$\HXXZ_{[1,L]}$ by boundary fields.
The Hamiltonian $H^{+-}_{[1,L]}$ is known as the kink Hamiltonian, 
and $H^{-+}_{[1,L]}$ is the antikink Hamiltonian.
These two models are distinguished 
because they each possess a quantum group symmetry, for the quantum group
$SU_q(2)$.
It should be mentioned that the representation of $SU_q(2)$ on $\Hil_L$
which commutes with $H^{+-}_{[1,L]}$ is different than the representation
which commutes with $H^{-+}_{[1,L]}$.
These Hamiltonians are also distinguished because, like $\HXXZ_{[1,L]}$,
they can be written as sums of nearest-neighbor interactions
and all their ground states are frustration-free.
We will give a formula, sufficient for our purposes,
for the ground states of $H^{+-}_{[1,L]}$ and $H^{-+}_{[1,L]}$, respectively.
First define the sectors of fixed total down-spins so that
$\Hil_{L,0} = \Span\{\allup\}$, and for $n=1,\dots,L$
$$
  \Hil_{L,n} = \Span \{ \left(\prod_{i=1}^n S_{x_i}^{-}\right) \allup : 
  1 \leq x_1 < x_2 < \dots < x_n \leq L \}\, .
$$
Thus, $S^3_{tot} {\rm Proj}(\Hil_{L,n}) 
= (\frac{L}{2} - n) {\rm Proj}(\Hil_{L,n})$.
Then $H^{+-}_{[1,L]}$ and $H^{-+}_{[1,L]}$ each have $L+1$ ground states,
one for each sector.
Let $\psi^{+-}_{[1,L]}(n)$ and $\psi^{-+}_{[1,L]}(n)$ be these ground states,
normalized as given in \eq{Intro:+-} and \eq{Intro:-+}. 
The spectral gap is known to exist
for each sector $\Hil_{L,n}$, $n=1,\dots,L-1$, and to be independent
of $n$.
Specifically, in \cite{KN1} the following was proved
\begin{proposition}
\label{kink-gap:Prop}
For the $SU_q(2)$ invariant Hamiltonian $H^{+-}_{[1,L]}$, $L \geq 2$, 
and $\Delta \geq 1$ one has
\begin{eqnarray*}
  \gamma_L &:=& \inf \left\{ \frac{\ip{\psi}{H^{+-}_{[1,L]} \psi}}
  {\ip{\psi}{\psi}}\, :\, 
  \psi \in \Hil_{L,n}\, , \psi \neq 0\, ,
  \ip{\psi}{\psi^{+-}_{[1,L]}}=0 \right\} \\
  &=& 1 - \Delta^{-1} \cos(\pi/L)\, .
\end{eqnarray*}
In particular
$$
  \gamma_L \geq 1 - \Delta^{-1},
$$
for all $L \geq 2$, and in addition the spectral gap above any of the
ground state representations of the GNS
Hamiltonian for the infinite chain is exactly $1 - \Delta^{-1}$.
\QED \quad
\end{proposition}

We will define $\gamma = 1 - \Delta^{-1}$ 
which is the greatest lower bound of all $\gamma_L$, and the 
spectral gap for the infinite chain.
A result identical with this one holds for the $H^{-+}_{[1,L]}$ spin chain,
which may be obtained using spin-flip or reflection symmetry.

There are important differences between the droplet Hamiltonian, 
$H^{++}_{[1,L]}$, and the kink Hamiltonian, which we briefly explain.
Since $H^{++}_{[1,L]}$ commutes with $S^3_{tot}$, it makes sense to
block diagonalize it with respect to the sectors $\Hil_{L,n}$,
$n=0,\dots,L$.
If we consider the spectrum of $H^{++}_{[1,L]}$ on the sector $\Hil_{L,n}$
for $L$ and $n$ both large, we will see that there are $L+1-n$ eigevalues
in a very small interval about $A(\Delta)$.
Then there is a gap above $A(\Delta)$ of width approximately $\gamma$,
with error at most $O(n^{-1/4})$,
which is free of any eigenvalues.
This is different from the case of the kink and antikink Hamiltonians where
the ground state in each sector is nondegenerate, with a uniform spectral
gap above. In our case, the ground state is non-degenerate only because
the translation invariance is broken in the finite systems. As $L\to\infty$,
the translation invariance is restored and the lowest eigenvalue in each sector
becomes infinitely degenerate. Therefore, as is done in Theorem 
\ref{main:theorem2}, it is natural to consider the spectral projection 
corresponding to the $L+1-n$ lowest eigenvalues as opposed to just the 
ground state space. 

Before beginning to prove the main theorem, we will observe some simple facts
about the droplet Hamiltonian.
First, the two site Hamiltonian $H^{++}_{x,x+1}$ restricted to 
$\Cx_x^2 \otimes \Cx_{x+1}^2$ is diagonalized as follows
\begin{equation}
\label{Hpp-2site:Diag}
H^{++}_{x,x+1} : 
\qquad
\begin{array}{|c|c|}
\rm{eigenvalue} & \rm{eigenvector} \\
  \hline 
-A(\Delta) & \ket{\uparrow \uparrow} \\
 \frac{1}{2}(1 - \Delta^{-1}) & \frac{1}{\sqrt{2}}
  (\ket{\uparrow \downarrow} + \ket{\downarrow \uparrow}) \\
  A(\Delta) & \ket{\downarrow \downarrow} \\
  \frac{1}{2}(1 + \Delta^{-1}) & \frac{1}{\sqrt{2}} 
  (\ket{\uparrow \downarrow} - \ket{\downarrow \uparrow})
\end{array}
\end{equation}
Note that it is \textit{not} true that $\Hpp_L$ is the sum of
$\Hpp_{x,x+1}$ for all nearest neighbor pairs $x,x+1 \in [1,L]$
as was the case for $\HXXZ_L$ and $H^{+-}_L$.
Instead the following identities are true:
\begin{eqnarray}
\label{useful1}
\Hpp_L &=& H^{+-}_{[1,x]} + H^{++}_{x,x+1} + H^{-+}_{[x+1,L]}\, ,\\
\label{useful2}
  &=& H^{+-}_{[1,x]} + H^{++}_{[x,L]}\, ,\\
\label{useful3}
  &=& H^{++}_{[1,x]} + H^{-+}_{[x,L]}\, ,
\end{eqnarray}
for $1\leq x\leq L-1$. These identities should be kept in mind
since they allow us to \textit{cut} the droplet spin chain at the
sites $x,x+1$.
This vague notion will be explained in detail in Section \ref{Sec:ROP}.
The diagonalization of $H^{--}_{x,x+1}$ is the same as the 
diagonalization of $H^{++}_{x,x+1}$ above, except that $\uparrow$
and $\downarrow$ are interchanged for each of the eigenvectors.

Now we state an obvious (but poor) preliminary lower bound for 
$\lambda_{L,n}(1)$.
\begin{proposition}
\label{APBound:Prop}
The ground state energy of $\Hpp_L$ on $\Hil_L$ is $-A(\Delta)$,
and the ground state space is $\Span\{\allup\}$.
Moreover, 
\begin{equation}
\label{hard:APBound}
\Rayleigh{\psi}{H^{++}_{[1,L]}} 
  \geq - A(\Delta) + \frac{1}{2}(1 - \Delta^{-1})
  \quad \textrm{for all nonzero}
  \  \psi \perp \allup\, .
\end{equation}
\end{proposition}

\begin{proof}
First, $H^{++}_{[1,L]} \geq -A(\Delta) \unity$ 
because $\HXXZ_{[1,L]} \geq 0$ and
$-A(\Delta) (S_1^{(3)} + S_L^{(3)}) \geq -A(\Delta) \unity$.
It is also clear that $H^{++}_{[1,L]} \allup = -A(\Delta) \allup$, and
$\Hpp_L \alldown = A(\Delta) \alldown$,
in agreement with \eq{hard:APBound}. 
Because $\allup$ and $\alldown$ are eigenvectors of the self-adjoint
operator $H^{++}_{[1,L]}$, all that remains is to check that
\eq{hard:APBound} holds on $\Span\{\allup,\alldown\}^\perp$.
But this is true by Lemma \ref{XXZ-gap:Lemma},
since $\Hpp_L \geq -A(\Delta) + \HXXZ_L$ and 
$\HXXZ_L \geq \frac{1}{2}(1 - \Delta^{-1})$
on $\Span\{\allup,\alldown\}^\perp$.
\end{proof}
We now begin the actual proof of the Theorems \ref{main:theorem}
and \ref{main:theorem2}.

\section{Evaluation of $H^{++}_{[1,L]}$ on droplet states.}
\label{Sect:Eval}

We begin by proving part (a) of Theorem \ref{main:theorem}.
This is straightforward because we have closed expressions for
each $\xi_{L,n}(x)$ and for $H^{++}_{[1,L]}$.
The heart of the proof is a number of computations which show that
$\xi_{L,n}(x)$ and $\xi_{L,n}(y)$ are approximately
orthogonal with respect to the inner product $\ip{*}{*}$ as well
as $\ip{*}{H^{++}_{[1,L]} *}$ and $\ip{*}{(H^{++}_{[1,L]})^2 *}$,
when $x \neq y$ and $n$ is large enough.
Specifically,
\begin{eqnarray}
\label{Refer:app:res1}
\frac{|\ip{\xi_{L,n}(x)}{\xi_{L,n}(y)}|}
  {\|\xi_{L,n}(x)\| \cdot \|\xi_{L,n}(y)\|}
  &\leq& \frac{q^{n|y-x|}}{\pn \infty} 
  \quad \textrm{for all} \quad x,y\, ;\\
\label{Refer:app:res2}
\frac{|\ip{\xi_{L,n}(x)}{H^{++}_{[1,L]} \xi_{L,n}(y)}|}
  {\|\xi_{L,n}(x)\| \cdot \|\xi_{L,n}(y)\|}
  &\leq& \frac{q^{n|y-x|}}{\pn \infty} 
  \quad \textrm{if} \quad x\neq y\, ;\\
\label{Refer:app:res3}
\frac{|\ip{\xi_{L,n}(x)}{(H^{++}_{[1,L]})^2 \xi_{L,n}(y)}|}
  {\|\xi_{L,n}(x)\| \cdot \|\xi_{L,n}(y)\|}
  &\leq& \frac{q^{n|y-x|}}{\pn \infty} 
  \quad \textrm{if} \quad |x-y| \geq 2\, .
\end{eqnarray}
Here $\pn \infty$ is a number arising in partition theory \cite{And},
$$
\pn \infty = \prod_{n=1}^\infty (1 - q^{2n})\, .
$$
(It is usually written as $(q^2;q^2)_\infty$.)
The important fact is that $\pn \infty \in (0,1]$ for
$q \in [0,1)$.

We need one more piece of information, which is that
\begin{equation}
\label{nec:detail}
\frac{\|(H^{++}_{[1,L]} - A(\Delta)) \xi_{L,n}(x)\|^2}
  {\|\xi_{L,n}(x)\|^2}
  \leq \frac{2 q^{2 \floor{n/2}}}{1 - q^{2 \floor{n/2}}}\, .
\end{equation}
To prove this, we refer to equation (6.7) of \cite{BCN}.
In that paper, it is proved that
$$
\frac{\|P^\downarrow_{L} \psi^{-+}_{[1,L]}(n)\|^2}{\|\psi^{-+}_{[1,L]}(n)\|^2} 
  < q^{2(L-n)} \frac{1-q^{2n}}{1- q^{2L}} 
  \leq \frac{q^{2(L-n)}}{1-q^{2(L-n)}}\, ,
$$
where 
$$
P^\sigma_{x} = \unity_1 \otimes \dots \otimes \unity_{x-1}
  \otimes \ket{\sigma}\bra{\sigma} \otimes \unity_{x+1} \otimes
  \dots \otimes \unity_L
$$
for $\sigma = \uparrow,\downarrow$.
Using spin-flip and reflection symmetry, we obtain
$$
\frac{\|P^\uparrow_{L} \psi^{+-}_{[1,L]}(n)\|^2}
  {\|\psi^{+-}_{[1,L]}(n)\|^2} 
  < \frac{q^{2n}}{1-q^{2n}}\, , \qquad
\frac{\|P^\uparrow_{1} \psi^{-+}_{[1,L]}(n)\|^2}
  {\|\psi^{-+}_{[1,L]}(n)\|^2} 
  < \frac{q^{2n}}{1-q^{2n}}\, .
$$
Since $\xi_{L,n}(x) = \psi^{+-}_{[1,x]}(\floor{n/2}) \otimes
\psi^{-+}_{[x+1,L]}(\ceil{n/2})$, we then have the bounds
\begin{equation}
\label{bcnsbounds}  
 \frac{\|P^\uparrow_{x} \xi_{L,n}(x)\|^2}{\|\xi_{L,n}(x)\|^2} 
  \leq \frac{q^{2\floor{n/2}}}{1 - q^{2\floor{n/2}}}\, ,\quad 
 \frac{\|P^\uparrow_{x+1} \xi_{L,n}(x)\|^2}{\|\xi_{L,n}(x)\|^2} 
  \leq \frac{q^{2\ceil{n/2}}}{1 - q^{2\ceil{n/2}}}\, .
\end{equation}
Now $H^{++}_{[1,L]} \xi_{L,n}(x) = H^{++}_{x,x+1} \xi_{L,n}(x)$, because
of the identity \eq{useful1}, and the fact that
$$
H^{+-}_{[1,x]} \xi_{L,n}(x) = H^{-+}_{[x+1,L]} \xi_{L,n}(x) = 0\, .
$$
By \eq{Hpp-2site:Diag}, we estimate
$$
0 \leq (H^{++}_{x,x+1} - A(\Delta))^2
  \leq  P^\uparrow_x + P^\uparrow_{x+1}\, ,
$$
which, together with \eq{bcnsbounds}, proves \eq{nec:detail}.

We are now poised to prove Theorem \ref{main:theorem} (a).
We state the argument, which is very simple, as a lemma.
It is useful to do it this way, because we will repeat the argument
twice more in the proofs of Theorems \ref{Thm:Periodic} and \ref{Thm:Infinite}.
\begin{lemma}
\label{Lem:OrthStates}
Let $\{f_n : n \in \Ir\}$ be a family of states, normalized so that
$\|f_n\| = 1$ for all $n$, but not necessarily orthogonal.
Suppose, however, that there are constants $C < \infty$ and $\epsilon < 1$
such that $|\ip{f_n}{f_m}| \leq C \epsilon^{|n-m|}$ for all $m,n$. 
If  $(1 + 2 C) \epsilon < 1$, then
\begin{equation}
\label{Lem:orth1}
\left\| \sum_{n\in\Ir} \Proj(f_n) 
  - \Proj(\Span(\{f_n : n\in \Ir\})) \right\|
  \leq \frac{2 C \epsilon}{1 - \epsilon}\, . 
\end{equation}
Suppose that $X$ is a self-adjoint operator such that for some $r<\infty$
we have $\|X f_n\| \leq r$ for all $n$, 
and for some $C'<\infty$, $N \in \Nl$ we have
$|\ip{X f_n}{X f_m}| \leq C' \epsilon^{|n-m|}$ whenever $|n-m| \geq N$.
Then
\begin{equation}
\label{Lem:orth2}
\left\| X \cdot \Proj(\Span(\{f_n : n \in \Ir\})) \right\| 
  \leq \left[ \frac{(2N-1) r^2 + \frac{2 C' \epsilon^N}{1-\epsilon}}
  {1 - \frac{2 C \epsilon}{1 - \epsilon}} \right]^{1/2} \, .
\end{equation}
The same results hold if $\{f_n\}$ is a finite family, 
in which case the bounds are even smaller.
\end{lemma}

\begin{proof}
Define $F = \sum_{n=-\infty}^\infty \ket{f_n}\bra{f_n}$.
Define $E$ an infinite matrix such that $E_{mn} = \ip{f_m}{f_n}$.
Let $\{e_n : n \in \Ir\}$ be an orthonormal family in any Hilbert
space, and let $A = \sum_n \ket{f_n} \bra{e_n}$.
Then $E = A^* A$ and $F = A A^*$. 
For simplicity let $\mathcal{F}=\textrm{cl}(\Span(\{f_n : n \in \Ir\}))$,
and let $\mathcal{E}=\textrm{cl}(\Span(\{e_n : n \in \Ir\}))$.
We consider $A : \mathcal{E} \to \mathcal{F}$.
Then we calculate
$$
\|A^* A - \unity_{\mathcal{E}}\|
  \leq \sup_m \sum_{\substack{n \\ n\neq m}} |E_{mn}|
  \leq \frac{2 C \epsilon}{1 - \epsilon}\, .
$$
Since $2 C \epsilon < 1 - \epsilon$,
this shows that $A$ is bounded and $A^* A$ is invertible.
Under the invertibility condition, it is true that $A A^*$ is also
invertible on $\mathcal{F}$, and considering this as its domain,
$\sigma(A A^*) = \sigma(A^* A)$.
If we let $E$ and $F$ operate on proper superspaces of
$\mathcal{E}$ and $\mathcal{F}$,
then they will be identically zero on the orthogonal complements.
But it is still true that
$$
\sigma(E) \setminus \{0\} 
  = \sigma(A A^*) 
  = \sigma(A^* A) 
  = \sigma(F) \setminus \{0\}\, .
$$
In particular, if we let $P_{\mathcal{F}}$ be the orthogonal projection
onto $\mathcal{F}$, 
then
$$
\|F - P_{\mathcal{F}}\| = \|A^*A - \unity_{\mathcal{E}}\|
\leq \frac{2 C \epsilon}{1 - \epsilon}\, .
$$
This proves \eq{Lem:orth1}.

To prove the second part, let $\psi = \sum_n \alpha_n f_n$
be a state in $\mathcal{F}$.
Let $\phi = \sum_n \alpha_n e_n$.
Then 
\begin{equation}
\label{Proof:orth1}
\|\psi\|^2 = \ip{\phi}{A^*A\phi} 
  \geq (1 - \frac{2C\epsilon}{1-\epsilon}) \sum_n |\alpha_n|^2\, .
\end{equation}
We calculate 
\begin{eqnarray*}
\|X \psi\| 
  = \sum_{m,n} \overline{\alpha}_m \alpha_n \ip{X f_m}{X f_n} 
  \leq \sum_n |\alpha_n|^2 \cdot \sup_m 
  \sum_{n} |\ip{X f_m}{X f_n}|\, .
\end{eqnarray*}
Breaking the sum into two pieces yields,
for any $m \in \Ir$,
\begin{eqnarray*}
\sum_n |\ip{X f_m}{X f_n}|
  &\leq& \sum_{\substack{n \\ |m-n| < N}} |\ip{X f_m}{X f_n}|
  + \sum_{\substack{n \\ |m-n| \geq N}} |\ip{X f_m}{X f_n}| \\
  &\leq& (2N-1) r^2 + \frac{2 C' \epsilon^N}{1-\epsilon}\, .
\end{eqnarray*}
So, using \eq{Proof:orth1}, we have
$$
\frac{\|X \psi\|^2}{\|\psi\|^2}
  \leq \frac{(2N-1) r^2 + \frac{2 C' \epsilon^N}{1 - \epsilon}}
  {1 - \frac{2 C \epsilon}{1 - \epsilon}} 
$$
for any nonzero $\psi \in \mathcal{F}$.
This proves \eq{Lem:orth2}.
\end{proof}

Now to prove Theorem \ref{main:theorem}(a), we note that
the hypotheses of the lemma are met.
Namely, take $f_x = \xi_{L,n}(x)$.
By \eq{Refer:app:res1}, we have $|\ip{f_x}{f_y}| \leq C \epsilon^{|x-y|}$,
where $C = f_q(\infty)^{-1}$ and $\epsilon = q^n$.
We set $X = H^{++}_{[1,L]} - A(\Delta)$.
Then by \eq{Refer:app:res1}, \eq{Refer:app:res2} and \eq{Refer:app:res3},
we have $\ip{X f_x}{X f_y} \leq C' \epsilon^{|x-y|}$, for $|x-y| \geq 2$,
where $C' = 4/\pn{\infty}$. (Since $A(\Delta) \leq 1$,
$1 + 2 A(\Delta) + A(\Delta)^2 \leq 4$.)
By \eq{nec:detail}, we have $\|X \xi_x\| \leq r$ for all $x$, where
$r^2 = 2 q^{2 \floor{n/2}}/(1 - q^{2 \floor{n/2}})$.
Therefore, by Lemma \ref{Lem:OrthStates}, and some trivial estimations
\begin{equation}
\label{Eval:result1}
\|(H^{++}_{[1,L]} - A(\Delta))\cdot \Proj(\calK_{L,n})\| 
  \leq  \frac{2 \sqrt{2} q^{\floor{n/2}}}
  {\sqrt{(1 - 3 q^{2 \floor{n/2}}) f_q(\infty)}}\, .
\end{equation}
The lemma also gives us the following result
\begin{equation}
\label{Eval:result2}
\|\Proj(\calK_{L,n}) 
  - \sum_{x=\floor{n/2}}^{L-\ceil{n/2}} \Proj(\xi_{L,n}(x))\|
  \leq \frac{2 q^n}{(1-q^n) f_q(\infty)}\, .
\end{equation}
This will prove useful in Section \ref{Sec:ROP}, 
because it is a precise statement of  
just how orthogonal our proposed states $\xi_{L,n}(x)$ are
to each other.


\section{Existence of fully polarized intervals}
\label{sec:polarized}

We know that the ground states of the kink Hamiltonian exhibit a 
localized interface such that to the left of the interface 
nearly all spins are observed in the $\downarrow$ state,
and to the right nearly all spins are observed in the
$\uparrow$ state. 
The interface has a thickness due to quantum fluctuations.
A similar phenomenon occurs with the antikink Hamiltonian but with left
and right reversed or alternatively with $\uparrow$ and $\downarrow$ 
reversed.
We might hope that the ground state of the droplet Hamiltonian will
also contain an interval (or several intervals) 
with nearly all $\uparrow$- or all $\downarrow$-spins.
This is the case, and we prove it next.

\begin{definition}
For any finite interval $J \subset  \Ir$
define the orthogonal projections
\begin{eqnarray*}
P^\uparrow_J  &=& \ket{\uparrow \dots \uparrow} 
  \bra{\uparrow \dots \uparrow}_J \otimes \unity_{I \setminus J}\, ,\\
P^\downarrow_J  &=& \ket{\downarrow \dots \downarrow} 
  \bra{\downarrow \dots \downarrow}_J \otimes \unity_{I \setminus J}\, ,\\
P_J &=& P^\uparrow_J + P^\downarrow_J\, .  
\end{eqnarray*}

We also define for any operator $X$ and any nonzero state $\psi$,
the Rayleigh quotient
$$
\rho(\psi,X) = \Rayleigh{\psi}{X}\, .
$$
\end{definition}

\begin{proposition}
\label{IHS:Prop}
Suppose $\psi \in \Hil_L$ is a nonzero state, and let
$$
E = \rho(\psi,\HXXZ_L)\, .
$$
Given $l < L$, there is a subinterval $J = [a,a+l-1] \subset [1,L]$
satisfying the bound
\begin{equation}
\label{proj:prox}
  \frac{\|P_J \psi\|^2}{\|\psi\|^2}
  \geq 1 - \frac{2 E}{\gamma \lfloor{L/l}\rfloor}\, .
\end{equation}
Moreover denoting 
$$
\epsilon := \frac{2E}{\gamma \lfloor{L/l}\rfloor}\, ,
$$ 
then as long as $\epsilon < 1$, we have the following bound
\begin{equation}
\label{ihs:en:comp}
\rho(P_J \psi,\HXXZ_{[1,L]})
  \leq \frac{E}{1 - \epsilon} + 
  2 \Delta^{-1} \sqrt{\frac{\epsilon}{1-\epsilon}}\, .
\end{equation}
\end{proposition}

\begin{proof}
Partition $[1,L]$ into $r=\lfloor{L/l}\rfloor$ intervals $J_1,\dots,J_r$
each of length $\geq l$.
If $J_i = [a_i,a_{i+1}-1]$ then 
$$
\HXXZ_L = \sum_{i=1}^r \HXXZ_{J_i} + \sum_{i=2}^r \HXXZ_{a_i-1,a_i}\,
  \geq \sum_{i=1}^r \HXXZ_{J_i}\, .
$$ 
By Lemma \ref{XXZ-gap:Lemma}, 
$$
\rho(\psi,\HXXZ_{J_i}) 
  \geq \frac{\gamma}{2} (1 - \rho(\psi,P_{J_i}))\, .
$$
So
\begin{eqnarray*}
E \geq \frac{\gamma}{2} \sum_{i=1}^r (1 - \rho(\psi,P_{J_i}))
  \geq r \frac{\gamma}{2} \min_i (1 - \rho(\psi,P_{J_i}))\, .
\end{eqnarray*}
In other words,
$$
\rho(\psi,P_{J_i}) \geq 1 - \frac{2 E}{\gamma r}\, ,
$$
for some $i$.
Since $[a_i,a_i+l] \subset J_i$, $P_{J_i} \leq P_{[a_i,a_i+l+1]}$.
Let $J = [a_i,a_i+l-1]$, then \eq{proj:prox} holds.

Note that for any orthogonal projection $P$ and any operator $H$ we have 
the decomposition
$$
H = P H P + (1-P) H (1-P) + [P,[P,H]]\, .
$$
If $H$ is nonnegative, then $(1-P)H(1-P)$ is as well.
Hence
$$
P H P  \leq H  - [P,[P,H]]\, .
$$
On the other hand, it is obvious that
$$
P [P,[P,H]] P = (1-P) [P,[P,H]] (1-P) = 0\, ,
$$
which implies
$$
\rho(\psi,P H P) \leq \rho(\psi,H) 
  + 2 \|[P,[P,H]]\| 
  \frac{\|P \psi\|\, \|(1-P) \psi\|}{\|\psi\|^2}
$$
for any nonzero $\psi$.

Moreover,
\begin{equation}
\label{ray:ineq}
\rho(P \psi,H) = \frac{\rho(\psi,P H P)}{\rho(\psi,P)}
  \leq \frac{\rho(\psi,H)}{\rho(\psi,P)} 
  + 2 \|[P,[P,H]]\| \sqrt{\frac{\rho(\psi,1-P)}{\rho(\psi,P)}}\, .
\end{equation}
In our particular case, where
$H = \HXXZ_L$ and $P = P_J$, 
\eq{ray:ineq} and \eq{proj:prox} imply
\begin{equation}
\label{ihs:en:ineq}
\rho(P_J \psi,\HXXZ_L) 
  \leq \frac{E}{1-\epsilon} 
  + 2 \|[P_J,[P_J,\HXXZ_L]]\| \sqrt{\frac{\epsilon}{1-\epsilon}}\, .
\end{equation}
All that remains is to calculate $\|[P_J,[P_J,\HXXZ_{[1,L]}]]\|$.

Notice that 
$$
[P_J,[P_J,\HXXZ_{[1,L]}]] 
  = \sum_{\substack{x \in [1,L-1] \\ \alpha,\beta \in \{\uparrow,\downarrow\}}}
  [P^\alpha_J,[P^\beta_J,\HXXZ_{x,x+1}]]\, ,
$$
and that $\HXXZ_{x,x+1}$
commutes with $P^\beta_J$ for all $x,x+1$ except $a-1,a$
and $b,b+1$. (We define $b=a+l-1$.)
Straightforward computations yield
$$
[P^{\beta}_J,\HXXZ_{a-1,a}] 
  = - \frac{1}{2 \Delta} \unity_{[1,a-2]} 
  \otimes (\ket{\beta \beta'} \bra{\beta' \beta} 
  - \ket{\beta' \beta} \bra{\beta \beta'}) \otimes P^{\beta}_{[a+1,b]}
  \otimes \unity_{[b+1,L]} 
$$
and
$$
[P^{\beta}_J,\HXXZ_{b,b+1}]
  = - \frac{1}{2 \Delta} \unity_{[1,a-1]} 
  \otimes P^\beta_{[a,b-1]} 
  \otimes (\ket{\beta \beta'} \bra{\beta' \beta} 
  - \ket{\beta' \beta} \bra{\beta \beta'})
  \otimes \unity_{[b+2,L]}\, ,
$$
where $\uparrow' = \downarrow$ and $\downarrow' = \uparrow$.
It is easy to deduce that $[P^\alpha_J,[P^\beta_J,\HXXZ_L]]$
is zero unless $\alpha=\beta$.
($[P^\beta_J,\HXXZ_{a-1,a}]$ has a tensor factor $P^\beta_{[a+1,b]}$
and $P^\alpha_J$ has a tensor factor $P^\alpha_{[a+1,b]}$, which
implies $[P^\alpha_J,[P^\beta_J,\HXXZ_{a-1,a}]]$ is zero unless
$\alpha = \beta$.
The term $[P^\alpha_J,[P^\beta_J,\HXXZ_{b,b+1}]]$ is treated similarly.)
Another straightforward computation yields
$$
[P^\beta_J,[P^{\beta}_J,\HXXZ_{a-1,a}]]
  = - \frac{1}{2 \Delta} \unity_{[1,a-2]} 
  \otimes (\ket{\beta \beta'} \bra{\beta' \beta} 
  + \ket{\beta' \beta} \bra{\beta \beta'}) \otimes P^{\beta}_{[a+1,b]}
  \otimes \unity_{[b+1,L]} 
$$
and
$$
[P^\beta_J,[P^{\beta}_J,\HXXZ_{b,b+1}]]
  = - \frac{1}{2 \Delta} \unity_{[1,a-1]} \otimes P^\beta_{[a,b-1]} 
  \otimes (\ket{\beta \beta'} \bra{\beta' \beta} 
  + \ket{\beta' \beta} \bra{\beta \beta'})
  \otimes \unity_{[b+2,L]}\, .
$$
So
$$
\begin{array}{l}
\displaystyle
[P_J,[P_J,\HXXZ_L]]
  = -\frac{1}{2 \Delta} \Big(
  \unity_{[1,a-2]} 
  \otimes A_{a-1,a}
  \otimes P_{[a+1,b]}
  \otimes \unity_{[b+1,L]} \\
\hspace{125pt} \displaystyle
  + \unity_{[1,a-1]} \otimes P_{[a,b-1]} 
  \otimes A_{b,b+1}
  \otimes \unity_{[b+2,L]} \Big)\, ,
\end{array}
$$
where
$A =  \ket{\uparrow \downarrow} \bra{\downarrow \uparrow} 
  + \ket{\downarrow \uparrow} \bra{\uparrow \downarrow}$.
In particular $\|A\| = 1$, so that
$$
\|\unity_{[1,a-2]} 
  \otimes A_{a-1,a}
  \otimes P_{[a+1,b]}
  \otimes \unity_{[b+1,L]}\|
  = 1\, ,
$$
and 
$$
\|\unity_{[1,a-1]} \otimes P_{[a,b-1]} 
  \otimes A_{b,b+1}
  \otimes \unity_{[b+2,L]} \| = 1\, .
$$
Thus
$\|[P_J,[P_J,\HXXZ_L]]\| \leq \Delta^{-1}$,
which along with \eq{ihs:en:ineq} proves \eq{ihs:en:comp}.
\end{proof}


In the following corollary, we show that essentially the 
same results hold for any bounded perturbation of $\HXXZ_{[1,L]}$.

\begin{corollary}
\label{IHS:Cor}
\label{IHS:Prop2}
Suppose $H_L$ is a bounded operator on $\Hil_L$ with
$$
M = \|H_L - \HXXZ_{[1,L]}\|\, .
$$ 
Let $E<\infty$ and $\psi \in \Hil_L$ be a nonzero state with
$$
\rho(\psi,H_L) \leq E\, .
$$
Given any subinterval $K \subset [1,L]$ and $l < |K|$, there is a 
sub-subinterval $J \subset K$ of length $l$, satisfying the bound
\begin{equation}
\label{cor:proj:prox}
  \|\psi - P_J \psi\|^2
  \leq \epsilon \|\psi\|^2\, ,
\end{equation}
where
$$
\epsilon = \frac{2(E+M)}{\gamma \floor{|K|/|J|}}\, .
$$
This statement is nonvacuous when $\epsilon < 1$.
Also under the assumption that $\epsilon<1$, we have the bound
\begin{equation}
\label{cor:en:comp}
\ip{\psi}{H_L \psi}
  \geq \ip{P_J \psi}{H_L P_J \psi} -
 \left(M \epsilon
  + 2 (\Delta^{-1} + 2 M) \sqrt{\epsilon(1-\epsilon)}\right)\, .
\end{equation}
\end{corollary}


\begin{proof}
Since $\|H_L - \HXXZ_{[1,L]}\| = M$, it is clear that
$$
\rho(\psi,\HXXZ_K) \leq \rho(\psi,\HXXZ_{[1,L]}) \leq E + M\, .
$$
So Proposition \ref{IHS:Prop} implies \eq{cor:proj:prox}.
To prove \eq{cor:en:comp} notice that for any operator $H$,
any orthogonal projection $P$, and any nonnegative operator $\tilde H$,
\begin{eqnarray*}
H - PHP 
  &=& (1-P)H(1-P) + [P,[P,H]] \\
  &=& (1-P)\tilde{H}(1-P) + (1-P)(H - \tilde{H})(1-P) \\
  && \qquad + [P,[P,\tilde{H}]] + [P,[P,H-\tilde{H}]] \\
  &\geq& (1-P)(H - \tilde{H})(1-P) + [P,[P,\tilde{H}]] \\
  && \qquad + [P,[P,H-\tilde{H}]]\, .
\end{eqnarray*}
So, for any nonzero $\psi$,
\begin{eqnarray*}
&& \rho(\psi,H - PHP)
  \geq - \|H - \tilde{H}\| \rho(\psi,1-P)  \\
  && \hspace{50pt}
  - 2 (\|[P,[P,\tilde{H}]]\| + 2 \|H - \tilde{H}\|) 
  \rho(\psi,P)^{1/2} \rho(\psi,1-P)^{1/2}\, .
\end{eqnarray*}
Setting $H = H_L$, $\tilde{H} = \HXXZ_L$ and $P = P_J$ we have
$$
\rho(\psi,H_L) - \rho(\psi,P_J H_L P_J)
  \geq - M \epsilon - 2 (\Delta^{-1} + 2 M) \sqrt{\epsilon (1-\epsilon)}\, .
$$
Since
$$
\rho(P_J \psi,H_L) = \frac{\rho(\psi,P_J H_L P_J)}{\rho(\psi,P_J)}
	\leq \frac{\rho(\psi,P_J H_L P_J)}{1-\epsilon},
$$
the corollary is proved.
\end{proof}


\section{Remainder of the proof}
\label{Sec:ROP}

We will now prove Theorem \ref{main:theorem}(b).
Let us henceforth denote $\Proj(\Span\{\phi\})$ simply by
$\Proj(\phi)$ for any nonzero state $\phi$.
We observe by \eq{Eval:result2} that there are constants $C_0(q)$
and $N_0(q)$, such that
$$
\|\Proj(\calK_{L,n}) 
  - \sum_{x=\floor{n/2}}^{L-\ceil{n/2}} \Proj(\xi_{L}(x,n))\|
  \leq C_0(q) q^n\, .
$$
whenever $n \geq N_0(q)$.
By \eq{Eval:result2}, 
$N_0(q) = 1$ and $C_0(q) = (1-q)^{-1} \pn \infty ^{-1}$. 
Suppose we exhibit a sequence
$\epsilon_n$, with $\lim_{n \to \infty} \epsilon_n = 0$, such that
\begin{equation}
\begin{array}{l}
\displaystyle H^{++}_{[1,L]} \Proj(\Hil_{L,n}) \geq \vspace{1mm}\\
\quad \displaystyle
(A(\Delta) - \epsilon_n) \Proj(\Hil_{L,n})
  + \gamma [\Proj(\Hil_{L,n}) 
- \sum_{x=\floor{n/2}}^{L - \ceil{n/2}} \Proj(\xi_{L}(x,n))]\, .
\label{ROP:piece1}
\end{array}
\end{equation}
We know, by Theorem \ref{main:theorem}(a), that 
$(H^{++}_{[1,L]} - A(\Delta)) \Proj(\calK_{L,n})$ is bounded above
and below by $\pm C q^n \unity$.
Then we would know
\begin{eqnarray*}
&&H^{++}_{[1,L]} \Proj(\Hil_{L,n}) 
  \geq (A(\Delta) - 2 C q^n) \Proj(\Hil_{L,n}) + \\
&&\qquad \qquad  (\gamma - \epsilon_n) (\Proj(\Hil_{L,n}) - 
\Proj(\calK_{L,n}))\, .
\end{eqnarray*}
So to prove Theorem \ref{main:theorem}(b), it suffices to 
verify that there is a sequence $\epsilon_n$ satisfying 
\eq{ROP:piece1}.

We will prove this fact in this section.
We find it convenient to consider an arbitrary gap $\lambda$,
$0 \leq \lambda<\gamma$.
Define $\epsilon_\lambda(L,n)$ to be the smallest nonnegative
number such that 
$$
\ip{\psi}{H^{++}_{[1,L]} \psi}
  \geq (A(\Delta) - \epsilon_\lambda(L,n)) \|\psi\|^2
  + \lambda \ip{\psi}{[\unity - 
  \sum_{x=\floor{n/2}}^{L - \ceil{n/2}} \Proj(\xi_{L}(x,n))] \psi}
$$
holds for all $\psi \in \Hil_{L,n}$.
We also define 
\begin{eqnarray*}
\epsilon_\lambda'(L,n) &=& \max_{\substack{L' \\ n \leq L' \leq L}} 
\epsilon_\lambda(L',n) \\
\epsilon_\lambda'(\infty,n) &=& \lim_{L \to \infty} 
\epsilon_\lambda'(L,n) \\
\epsilon_\lambda''(n) &=& \sup_{\substack{n' \\ n' \geq n}} 
\epsilon_\lambda'(\infty,n')
\end{eqnarray*}
If we can prove that for every $\lambda<\gamma$,
$\lim_{n \to \infty} \epsilon_\lambda''(n) = 0$,
then we will have proved Theorem \ref{main:theorem}(b).

Given $0\leq q<1$, define
$$
N_1(q) = \left(\frac{5-4q+\sqrt{(6-5q)(4-3q)}}{1-q}\right)^2\, .
$$
Suppose $n > N_1(q)$ and $L \geq n$.
(The requirement that $n > N_1(q)$ allows us to apply Corollary \ref{IHS:Cor}
effectively, i.e.\ with $\epsilon < 1$.)
Define an interval $K = [\ceil{\frac{1}{4}L},\floor{\frac{3}{4}L}]$,
and suppose $\psi \in \Hil_{L,n}$ is a nonzero state with
$\rho(\psi,H^{++}_{[1,L]}) \leq A(\Delta) + \gamma$.
Then by Corollary \ref{IHS:Cor} and the requirement that $n>N_1(q)$, 
we can find an interval $J \subset K$ such that $|J| = \floor{L^{1/2}}$,
\begin{equation}
\label{ROP:corres1}
\|\psi - P_J \psi\|^2 \leq C_1(q) L^{-1/2} \|\psi\|^2\, ,
\end{equation}
and
\begin{equation}
\label{ROP:corres2}
\ip{\psi}{H^{++}_{[1,L]} \psi} 
  \geq \ip{P_J \psi}{H^{++}_{[1,L]} P_J \psi}
  - C_2(q) L^{-1/4} \|\psi\|^2\, ,
\end{equation}
where
$$
\begin{array}{l}
\displaystyle 
C_1(q) = \frac{8}{1-q} (1 - 2 n_1(q)^{-1/2} - n_1(q)^{-1})^{-1}\, ,
\vspace{2mm}\\
\displaystyle
C_2(q) = \frac{(1+3q)(3-q)}{2(1+q^2)} C_1(q)^{1/2}\, .
\end{array}
$$
Let $J = [a,b]$.

We need to extend our definition of $\Hil_{L,n}$ in the following way.
For integers $s \leq t$, let 
$$
\Hil_{[s,t]} = \Cx_s^2 \otimes \Cx_{s+1}^2 \otimes \cdots \otimes\Cx_t^2.
$$
For $0 \leq r \leq s-t+1$, let
$$
\Hil_{[s,t],r} = \Span\{ \left(\prod_{i=1}^r S_{x_i}^-\right)
  \ket{\uparrow \dots \uparrow}_{[s,t]} :
  s \leq x_1 < x_2 < \dots <x_r \leq t \}\, .
$$
So $\Hil_L = \Hil_{[1,L]}$ in the new notation,
and $\Hil_{L,n} = \Hil_{[1,L],n}$.
We are free to decompose
$$
\psi = \sum_{n_1,n_2,n_3} \psi(n_1,n_2,n_3) 
$$
where $\psi(n_1,n_2,n_3) \in \Hil_{[1,a-1],n_1} \otimes
\Hil_{[a,b],n_2} \otimes \Hil_{[b+1,L],n_3}$.
The condition that $\psi \in \Hil_{L,n}$ implies
$\psi(n_1,n_2,n_3) \neq 0$ only if
$(n_1,n_2,n_3) \in [0,a-1] \times [0,b-a+1] \times[0,L-b]$,
and $n_1+n_2+n_3 = n$.
Also, since the range of $P_J$ is precisely the direct sum of all those 
triples $\Hil_{[1,a-1],n_1} \otimes \Hil_{[a,b],n_2} \otimes \Hil_{[b+1,L],n_3}$
such that $n_2\in\{0,|J|\}$, we can restrict attention to 
those states $\psi(n_1,n_2,n_3)$ satisfying the same condition.
Therefore, let $\psi^\uparrow(j) = \psi(j,0,n-j)$,
and $\psi^\downarrow(j) = \psi(j,|J|,n-j-|J|)$.
Then $\psi^\uparrow(j)$ lies in the range of $P^\uparrow_J$
and $\psi^\downarrow(j)$ lies in the range of $P^\downarrow_J$,
and 
$$
P_J \psi = \sum_{j = 0}^{n} \psi^\uparrow(j)
    + \sum_{j=0}^{n-|J|} \psi^\downarrow(j)\, .
$$
Let $Q(n_1,n_2,n_3) = \Proj(\Hil_{[1,a-1]}^{n_1} \otimes
\Hil_{[a,b]}^{n_2} \otimes \Hil_{[b+1,L]}^{n_3})$.
Then it is easy to see that
$$
Q(n_1,n_2,n_3) \, H^{++}_{[1,L]} \, Q(m_1,m_2,m_3) = 0
$$ 
except when
$(n_1-m_1,n_2-m_2,n_3-m_3)$ equals $(\pm 1,\mp 1,0)$
or $(0,\pm 1,\mp 1)$.
But if $n_2,m_2 \in \{0,|J|\}$
(and $|J| > 1$), then the condition of the previous line can never
be met.
Therefore
\begin{equation}
\label{rop:1}
\ip{P_J \psi}{H^{++}_{[1,L]} P_j \psi}
  = \sum_{j=0}^n 
  \ip{\psi^\uparrow(j)}{H^{++}_{[1,L]} \psi^\uparrow(j)}
  + \sum_{j = 0}^{n-|J|}
  \ip{\psi^\downarrow(j)}{H^{++}_{[1,L]} \psi^\downarrow(j)}\, ,
\end{equation}
just as
\begin{equation}
\label{rop1a}
\|P_J \psi\|^2  
  = \sum_{j=0}^n \|\psi^\uparrow(j)\|^2
  + \sum_{j = 0}^{n-|J|} \|\psi^\downarrow(j)\|^2\, .
\end{equation}
We will next bound each of the terms on the right hand side
of \eq{rop:1}.

Let $x = a + \floor{|J|/2} = \floor{(a+b+1)/2}$.
Since $x,x+1 \in J$, consulting \eq{Hpp-2site:Diag}, we have
$$
H^{++}_{x,x+1} \psi^\downarrow(j) 
  = A(\Delta) \psi^\downarrow(j)\, .
$$
Then, by \eq{useful1}, it is clear
$$
\begin{array}{rcl}
\displaystyle
  \ip{\psi^\downarrow(j)}{H^{++}_{[1,L]} \psi^\downarrow(j)} 
  &\geq& A(\Delta) \|\psi^\downarrow(j)\|^2 \vspace{2mm}\\
\displaystyle  &&\quad + \ip{\psi^\downarrow(j)}
  {(H^{+-}_{[1,x]}+H^{-+}_{[x+1,L]}) \psi^\downarrow(j)} \, .
\end{array}
$$
By Proposition \ref{kink-gap:Prop}
$$
\begin{array}{l}
\displaystyle
\ip{\psi^\downarrow(j)}
  {(H^{+-}_{[1,x]}+H^{-+}_{[x+1,L]}) \psi^\downarrow(j)} 
  \geq \gamma \bra{\psi^\downarrow(j)} \vspace{2mm}\\
\hspace{1cm} \displaystyle
  \left(\unity - \Proj(\psi^{+-}_{[1,x]}(j') \otimes
  \psi^{-+}_{[x+1,L]}(n-j'))\right) \ket{\psi^\downarrow(n_1)}
\end{array}
$$
where $j' = j + \floor{|J|/2} + 1$.
Also, defining $\tilde{x}_j = a-1+\floor{n/2}-j$,
we know by \eq{App:result1}
$$
\|\Proj(\psi^{+-}_{[1,x]}(j') \otimes
  \psi^{-+}_{[x+1,L]}(n-j')) - \Proj(\xi_{L,n}(\tilde{x}_j)\|
  \leq C_3(q) q^{|J|/2} \, ,
$$
where
$C_3(q) = 4 (1 - q^2)^{-1/2}$.
Therefore,
\begin{equation}
\label{rop:2}
\begin{array}{l}
\displaystyle
\ip{\psi^\downarrow(n_1)}{H^{++}_{[1,L]} \psi^\downarrow(j)}
  \geq (A(\Delta) - C_3 q^{|J|/2}) \|\psi^\downarrow(j)\|^2 
\vspace{2mm}\\
\hspace{3cm} \displaystyle
+ \gamma \ip{\psi^\downarrow(j)}
  {\Big(\unity - \Proj(\xi_{L,n}(\tilde{x}_j))\Big)\psi^\downarrow(j)}\, .
\end{array}
\end{equation}

Next, we bound
$\ip{\psi^\uparrow(j)}{H^{++}_{[1,L]} \psi^\uparrow(j)}$
in the case that $1 \leq j \leq \floor{n/2}$.
The case $\floor{n/2} \leq j \leq n-1$, will be the same by symmetry.
Referring to \eq{useful2},
$$
\ip{\psi^\uparrow(j)}{H^{++}_{[1,L]} \psi^\uparrow(j)}
  = \ip{\psi^\uparrow(j)}{H^{+-}_{[1,x]} \psi^\uparrow(j)} 
  + \ip{\psi^\uparrow(j)}{H^{++}_{[x,L]} \psi^\uparrow(j)}\, .
$$
Now, since $\psi^\uparrow(j) \in \Hil_{[1,x-1],j} \otimes 
\Hil_{[x,L],n-j}$, we may bound
$$
\ip{\psi^\uparrow(j)}{H^{++}_{[x,L]} \psi^\uparrow(j)}
  \geq (A(\Delta) - \epsilon_\lambda(L-x+1,n-j)) \|\psi^\uparrow(j)\|^2\, .
$$
By the definition of $\epsilon_\lambda'(.)$ and $\epsilon_\lambda''(.)$, 
$$
\epsilon_\lambda(L-x+1,n-j)
 \leq \epsilon_\lambda'(n-j)
 \leq \epsilon_\lambda''(\ceil{n/2})\, ,
$$
since $n-j \geq \ceil{n/2}$.
So
\begin{equation}
\label{ROP:aneq}
\ip{\psi^\uparrow(j)}{H^{++}_{[x,L]} \psi^\uparrow(j)}
  \geq (A(\Delta) - \epsilon_\lambda''(\ceil{n/2})) \|\psi^\uparrow(j)\|^2\, .
\end{equation}
By Proposition \ref{kink-gap:Prop}, 
$$
\ip{\psi^\uparrow(j)}{H^{+-}_{[1,x]} \psi^\uparrow(j)} 
  \geq \gamma \ip{\psi^\uparrow(j)}
  {\Big((\unity - \Proj(\psi^{+-}_{[1,x]}(j))\otimes \unity_{[x+1,L]}\Big)
  \psi^\uparrow(j)}\, .
$$
We can prove
\begin{equation}
\label{rop:little}
\ip{\psi^\uparrow(j)}{\Proj(\psi^{+-}_{[1,x]}(j)) \otimes \unity_{[x+1,L]}
  \psi^\uparrow(j))} 
  \leq \frac{q^{|J|}}{\pn \infty} \|\psi^\uparrow(j)\|^2\, .
\end{equation}
Indeed, since $\psi^\uparrow(j) \in \Hil_{[1,a-1],j} \otimes \Hil_{[a,x],0} 
\otimes \Hil_{[x+1,L],n-j}$
we have
\begin{eqnarray*}
&&\ip{\psi^\uparrow(j)}{\Proj(\psi^{+-}_{[1,x]}(j)) \otimes \unity_{[x+1,L]}
  \psi^\uparrow(j)} 
  \leq \|\psi^\uparrow(j)\|^2 \\
&&\quad\times
  \|\Proj(\psi^{+-}_{[1,x]}(j)) 
  \Proj(\Hil_{[1,a-1],j} \otimes \Hil_{[a,x],0}) \|^2\, ;
\end{eqnarray*}
so it suffices to check
$$
\|\Proj(\psi^{+-}_{[1,x]}(j)) 
  \Proj(\Hil_{[1,a-1],j} \otimes \Hil_{[a,x],0})\|^2 
  \leq \frac{q^{|J|}}{\pn \infty}\, .
$$
But, by a computation,
\begin{eqnarray*}
&&\|\Proj(\psi^{+-}_{[1,x]}(j))
  \Proj(\Hil_{[1,a-1],j} \otimes \Hil_{[a,x],0}) \| \\
&&\hspace{2cm} \displaystyle
  = \frac{\|\Proj(\Hil_{[1,a-1]}^{j} \otimes \Hil_{[a,x]}^{0}) 
  \psi^{+-}_{[1,x]}(j) \|^2}{\|\psi^{+-}_{[1,x]}(j)\|^2} \\
&&\hspace{2cm} \displaystyle
  = \frac{\|q^{j(x-a+1)} \psi^{+-}_{[1,a-1]}(j) \otimes 
  \psi^{+-}_{[a,x]}(0)\|^2}{\|\psi^{+-}_{[1,x]}(j)\|^2} \\
&&\hspace{2cm}
  = \qbinom{a-1}{j}{q^2} q^{2 j(\floor{|J|/2}+1)} \Big/
  \qbinom{x}{j}{q^2} \\
&&\hspace{2cm} \displaystyle
  \leq \frac{q^{|J|}}{\pn \infty}\, .
\end{eqnarray*}
The last calculation is deduced from equations \eq{App:+-coprod}
and \eq{App:-+coprod}, and note that it is necessary that $j \geq 1$.
From this we conclude
\begin{equation}
\ip{\psi^\uparrow(j)}{H^{+-}_{[1,x]} \psi^\uparrow(j)} 
  \geq \gamma(1 - \frac{q^{|J|}}{\pn \infty}) 
  \|\psi^\uparrow(j)\|^2\, .
\end{equation}
Combining this with \eq{ROP:aneq}, we have
\begin{equation}
\label{rop:3}
\begin{array}{l}
\ip{\psi^\uparrow(j)}{H^{++}_{[1,L]} \psi^\uparrow(j)} \vspace{2mm}\\
\hspace{1cm} \displaystyle
  \geq \left(A(\Delta) - \epsilon_\lambda''(\ceil{n/2})
  + \gamma \left(1 - \frac{q^{|J|}}{\pn \infty}\right)\right)
  \|\psi^\uparrow(j)\|^2
\end{array}
\end{equation}
as long as $1 \leq j \leq \floor{n/2}$.
A symmetric argument yields the same bound for the case
that $\ceil{n/2} \leq j \leq n-1$.

For $j = 0$, note that 
$\psi^\uparrow(0) = \ket{\uparrow \dots \uparrow}_{[1,x]} \otimes 
\psi'_{[x+1,L]}$, for some $\psi'_{[x+1,L]} \in \Hil_{[x+1,L],n}$.
Also, by \eq{useful1},
$$
H^{++}_{[1,L]} = H^{+-}_{[1,x+1]} + H^{++}_{[x+1,L]}
  \geq H^{++}_{[x+1,L]}\, .
$$
So
\begin{eqnarray*}
&\ip{\psi^\uparrow(0)}{H^{++}_{[1,L]} \psi^\uparrow(0)}
  &\geq \ip{\psi'_{[x+1,L]}}{H^{++}_{[x+1,L]} \psi'_{[x+1,L]}} \\
  &&\geq (A(\Delta) - \epsilon_\lambda(L-x,n)) \|\psi'_{[x+1,L]}\|^2 \\
  &&  + \lambda \ip{\psi'_{[x+1,L]}}{\Big(\unity 
  - \sum_{\tilde{x}=x+\floor{n/2}}^{L-\ceil{n/2}} 
  \Proj(\xi_{[x+1,L],n}(\tilde{x})\Big) \psi'_{[x+1,L]}}\, .
\end{eqnarray*}
We can replace $\|\psi'_{[x+1,L]}\|^2$ by $\|\psi^\uparrow(0)\|^2$.
Also, since 
$$
\psi'_{[x+1,L]} \in \Hil_{[x+1,b],0} \otimes \Hil_{[b+1,L],n}\, ,
$$
it is true that
$$
  \Proj(\xi_{[x+1,L]}(\tilde{x},n)) \psi'_{[x+1,L]}=0
$$
unless $\tilde{x} \geq b+\floor{n/2}$.
Furthermore, 
$$
\Proj(\Hil_{[1,x],0} \otimes \Hil_{[x+1,L],n})
\xi_{[x+1,L],n}(\tilde{x}) = 
\ket{\uparrow \dots \uparrow}_{[1,x]} \otimes \xi_{[x+1,L],n}(\tilde{x})\, .
$$
Therefore
\begin{eqnarray*}
&&\ip{\psi'_{[x+1,L]}}{\sum_{\tilde{x}=b+\floor{n/2}}^{L-\ceil{n/2}}
  \Proj(\xi_{[x+1,L],n}(\tilde{x})) \psi'_{[x+1,L]}}\\
&&\hspace{1cm}
  = \ip{\psi^\uparrow(0)}{\sum_{\tilde{x}=b+\floor{n/2}}^{L-\ceil{n/2}}
  \frac{\|\xi_{[x+1,L],n}(\tilde{x})\|^2}
  {\|\xi_{[1,L],n}(\tilde{x})\|^2}\Proj(\xi_{[1,L],n}(\tilde{x})) 
  \psi^\uparrow(0)}\, ,
\end{eqnarray*}
But it is very easy to see that 
$\|\xi_{[x+1,L],n}(\tilde{x})\|^2 \leq \|\xi_{[1,L],n}(\tilde{x})\|^2$.
So
\begin{equation}
\label{rop:4}
\begin{array}{l}
\displaystyle
\ip{\psi^\uparrow(0)}{H^{++}_{[1,L]} \psi^\uparrow(0)}
  \geq (A(\Delta) - \epsilon_\lambda'(\frac{3}{4}L,n))
  \|\psi^\uparrow(0)\|^2 \vspace{2mm}\\
\displaystyle
\hspace{1cm}
  + \lambda
  \ip{\psi^\uparrow(0)}
  {\Big(\unity - \sum_{\tilde{x} = b+\floor{n/2}}^{L-\ceil{n/2}} 
  \Proj(\xi_L(\tilde{x}))\Big) \psi^\uparrow(0)}\, .
\end{array}
\end{equation}
By an analogous argument
\begin{equation}
\label{rop:5}
\begin{array}{l}
\displaystyle
\ip{\psi^\uparrow(n)}{H^{++}_{[1,L]} \psi^\uparrow(n)}
  \geq (A(\Delta) - \epsilon_\lambda'(\frac{3}{4}L,n))
  \|\psi^\uparrow(n)\|^2  \vspace{2mm}\\
\displaystyle
\hspace{1cm}
  + \lambda
  \ip{\psi^\uparrow(n)}
  {\Big(\sum_{\tilde{x} = \floor{n/2}}^{a-1-\ceil{n/2}} 
  \Proj(\xi_{L,n}(\tilde{x}))\Big) \psi^\uparrow(n)}\, .
\end{array}
\end{equation}

Let us summarize the proof so far.
We began with a state $\psi \in \Hil_{L,n}$.
By Corollary \ref{IHS:Cor}, we found an interval $J$
such that $P_J \psi$ is a good approximation to $\psi$.
We decomposed $P_J \psi$ according to whether $\psi$ is in the
range of $P_J^\uparrow$ or $P_J^\downarrow$, and by the number of
downspins to the left of $J$.
We split the states $\psi^\sigma(j)$ into five classes
($\sigma = \downarrow$; $\sigma = \uparrow$, $j=0$;
$\sigma = \uparrow$, $1 \leq j \leq \floor{n/2}$;
$\sigma = \uparrow$, $\floor{n/2} \leq j \leq n-1$;
$\sigma = \uparrow$, $j=n$) and gave some spectral
gap estimates for each.
The only piece of the proof left is an induction argument, 
and one other thing: a proof that all of the spectral gap
estimates for each of the states $\psi^\sigma(j)$ can
be combined to a single spectral gap estimate for $P_J \psi$.
Specifically, while the $\psi^{\sigma}(j)$ are orthogonal
with respect to $\ip{*}{*}$ and $\ip{*}{H^{++}_{[1,L]}*}$,
it is not true that they are orthogonal with respect to
$\ip{*}{\Proj(\xi_{L,n}(\tilde{x}))*}$ for every $\tilde{x}$.
The trick is that they are nearly orthogonal with respect
to the projection for specific choices of $\tilde{x}$:
namely, if $\tilde{x} \in I_1 \cup I_2 \cup I_3$,
where $I_1 = [\floor{n/2},a-1-\ceil{n/2}]$, 
$I_2 = [a-1-\floor{n/2}+|J|,b+\floor{n/2}-|J|]$
and $I_3 = [b+\floor{n/2},L-\ceil{n/2}]$.
We will prove in Appendix B that,
in fact
\begin{eqnarray*}
&& \ip{P_J \psi}{\sum_{\tilde{x} \in I_1 \cup I_2 \cup I_3} 
  \Proj(\xi_{L,n}(x)) P_J \psi} \\
&& \qquad \geq - C_4(q) q^{|J|} \|P_J \psi\|^2
  + \sum_{j=0}^{n-|J|} \ip{\psi^\downarrow(j)}
  {\Proj(\xi_{L,n}(\tilde{x}_j)) \psi^\downarrow(j)} \\
&& \qquad + \sum_{\tilde{x}=b+\floor{n/2}}^{L-\ceil{n/2}}
  \ip{\psi^\uparrow(0)}{\Proj(\xi_{L,n}(\tilde{x})) \psi^\uparrow(0)} \\
&& \qquad + \sum_{\tilde{x}=\floor{n/2}}^{a-1-\ceil{n/2}}
  \ip{\psi^\uparrow(n)}{\Proj(\xi_{L,n}(\tilde{x})) \psi^\uparrow(n)}\, ,
\end{eqnarray*}
for some $C_4(q) < \infty$, as long as $n \geq N_4(q)$.

Equations \eq{rop:1}--\eq{rop:5} together with the result of
Appendix B imply
$$
\begin{array}{l}
\displaystyle
\ip{P_j \psi}{H^{++}_{[1,L]} P_J \psi}
  \geq (A(\Delta) - \eta) \|P_J \psi\|^2
\vspace{2mm}\\
\hspace{1cm} \displaystyle
  + \lambda \ip{P_J \psi}{\Big(\unity - \sum_{\tilde{x} \in I_1 + I_2 + I_3}
  \Proj(\xi_L(\tilde{x},n))\Big) P_J \psi}\, ,
\end{array}
$$
where
$$
\eta \leq (C_3(q) + C_4(q)) q^{|J|/2} 
  + \max\{0,\epsilon_\lambda''(\ceil{n/2}) - (\gamma-\lambda),
  \epsilon_\lambda'(\frac{3}{4}L,n)\}\, .
$$
Since each term $-\lambda \Proj(\xi_{L,n}(\tilde{x}))$, for
$\tilde{x} \in (I_1 \cup I_2 \cup I_3)'$
gives a negative contribution to the expectation, we can 
add those terms to the inequality:
\begin{equation}
\label{rop:7}
\begin{array}{rcl}
\displaystyle
\ip{P_j \psi}{H^{++}_{[1,L]} P_J \psi}
  &\geq&\displaystyle (A(\Delta) - \eta) \|P_J \psi\|^2
\vspace{1mm}\\
&&\displaystyle
  + \lambda \ip{P_J \psi}{\Big(\unity - \sum_{\tilde{x} = \floor{n/2}}
  ^{L - \ceil{n/2}}
  \Proj(\xi_L(\tilde{x},n))\Big) P_J \psi}\, ,
\end{array}
\end{equation}
Using \eq{Eval:result2}, and the fact that
$\|\unity - P\| \leq 1$, for any projection $P$, 
we have
$$
\Big\|\unity - \sum_{\tilde{x} = \floor{n/2}}^{L-\ceil{n/2}}
  \Proj(\xi_{L,n}(\tilde{x}))\Big\| 
  \leq 1 + \frac{2 q^n}{(1-q) \pn{\infty}}\, .
$$
This and \eq{ROP:corres1}, \eq{ROP:corres2}
and \eq{rop:7} imply 
\begin{eqnarray*}
\ip{\psi}{H^{++}_{[1,L]} \psi}
  &\geq& (A(\Delta) - \epsilon_\lambda(L,n)) \|\psi\|^2 \\
  && + \lambda \ip{\psi}{\Big(\unity - \sum_{\tilde{x} = \floor{n/2}}
  ^{L- \ceil{n/2}} \Proj(\xi_{L,n}(\tilde{x})) \Big) \psi}
\end{eqnarray*}
where, for some $C_5(q)$ and $C_6(q)$,
\begin{eqnarray*}
\epsilon_\lambda(L,n) 
  &\leq& \eta + A(\Delta) C_1(q) L^{-1/2} + C_2(q) L^{-1/4} \\
  && + 2 \lambda(1 + \frac{2q^n}{(1-q^n)f_q(\infty)}) 
  C_1(q)^{1/2} L^{-1/4} \\
  &\leq& C_5(q) q^{\frac{1}{2} \sqrt{L}} + C_6(q) L^{-1/4} \\ 
  && + \max\{0,\epsilon_\lambda''(\ceil{n/2}) - (\gamma-\lambda),
  \epsilon_\lambda'(\frac{3}{4}L,n)\}\, .
\end{eqnarray*}
We have not stated the exact dependence of $C_5(q)$ and $C_6(q)$ on
$q$, though it can be deduced from our previous calculations.
The important fact is that there exists $N_5(q)$, such that
if $n \geq N_5(q)$, then the above holds with $C_5(q)$ and $C_6(q)$
both finite, positive numbers.
From this, it follows
$$
\epsilon_\lambda'(\frac{4}{3}L,n)
  \leq C_5(q) q^{\frac{1}{2} \sqrt{L}}
  + C_6(q) L^{-1/4} 
  + \max\{0,\epsilon_\lambda''(\ceil{n/2}) + \lambda - \gamma,
  \epsilon_\lambda'(L,n)\}\, ,
$$
and
\begin{eqnarray*}
\epsilon_\lambda'((\frac{4}{3})^k n,n)
  &\leq& C_5(q) q^{\frac{1}{2} \sqrt{n}} 
  \sum_{r=1}^{k-1} q^{[(4/3)^{r/2}-1]\sqrt{n}}
  + C_6(q) n^{-1/4} \sum_{r=1}^{k-1} (\frac{3}{4})^{r/4}\\
  &&\quad+ \max\{0,\epsilon_\lambda''(\ceil{n/2}) + \lambda - \gamma,
  \epsilon_\lambda'(n,n)\}\, .
\end{eqnarray*}
Note $\epsilon_\lambda'(n,n)=0$, because $\Hil_{n,n}$ is one-dimensional,
and the single vector 
$$\xi_{n,n}(\floor{n/2}) = \ket{\downarrow \dots \downarrow}$$ 
satisfies
$H^{++}_{[1,L]} \xi_n(\floor{n/2},n) 
= A(\Delta) \xi_n(\floor{n/2},n)$.
Therefore,
\begin{eqnarray*}
\epsilon_\lambda'(\infty,n)
  &\leq& C_5 q^{\frac{1}{2} \sqrt{n}} \sum_{k=1}^{\infty} 
  q^{[(4/3)^{k/2}-1]\sqrt{n}}
  + C_6 n^{-1/4} \sum_{k=1}^{\infty} (\frac{3}{4})^{k/4}\\
  &&\quad + \max\{0,\epsilon_\lambda''(\ceil{n/2}) + \lambda - \gamma\}\, .
\end{eqnarray*}
Taking the $\limsup$ as $n \to \infty$, we find
$$
\epsilon_\lambda''(\infty)
  \leq \max\{0,\epsilon_\lambda''(\infty) + \lambda - \gamma\}\, .
$$
For $\lambda < \gamma$ this implies $\epsilon_\lambda''(\infty)$ either
equals zero or $+\infty$.
But, by Proposition \ref{APBound:Prop},
$\epsilon''_\lambda(\infty)<A(\Delta)$.
So $\epsilon''_\lambda(\infty) = 0$, as desired, for every $\lambda<\gamma$.

By the Cantor diagonal argument, there is a sequence
$\epsilon_n$ satisfying \eq{ROP:piece1},
constructed from the $\epsilon_\lambda(n)$, with $\lambda \to \gamma$
and $n \to \infty$.
So Theorem \ref{main:theorem}(a) is proved.
Theorem \ref{main:theorem2} is a reformulation of the same result, 
so it needs no proof.

\section{Results for the Ring and the Infinite Chain}
\label{Sec:Periodic-Infinite}

\subsection{The Spin Ring}

The spin ring (periodic spin chain) has state space
$\Hil_{L}$ and is defined by the Hamiltonian
$
\HXXZ_{\Ir/L} = \sum_{x=1}^{L-1} \HXXZ_{x,x+1} + \HXXZ_{L,1}\, .
$
We define a periodic droplet with $n$ down spins
$$
\xi_{\Ir/L,n}(0) = \xi_{L,n}(\floor{L/2})
  = \psi^{+-}_{[1,\floor{L/2}]}(\floor{n/2})
  \otimes \psi^{-+}_{[\floor{L/2}+1,L]}(\ceil{n/2})\, .
$$
There are $L-1$ additional droplet states
$$
\xi_{\Ir/L,n}(x) = T^x \xi_{\Ir/L,n}(0)
\qquad (x = 1,\dots,L-1)
$$
where $T$ is the unitary operator on $\Hil_L$
such that
$$ 
T(v_1 \otimes v_2 \otimes \dots \otimes v_L)
  = v_L \otimes v_1 \otimes \dots \otimes v_{L-1}\, .
$$
Let $\mathcal{K}_{\Ir/L,n}$ be the span of
$\xi_{\Ir/L,n}(x),\dots,\xi_{\Ir/L}(L-1,n)$.
Let 
$$
 \lambda_{\Ir/L,n}(1) \leq \lambda_{\Ir/L,n}(2) \leq 
 \dots \lambda_{\Ir/L,n}(\sbinom{L}{n})
$$
be the ordered eigenvalues of $\HXXZ_{\Ir/L}$ acting on the invariant
subspace $\Hil_{L,n}$, and let $\Hil_{\Ir/L,n}^k$ be the span of the
first $k$ eigenvectors.

\begin{theorem}
\label{Thm:Periodic}
For $1 \leq n \leq L-1$
$$
  \lambda_{\Ir/L,n}(1),\dots,\lambda_{\Ir/L,n}(L)
  \in [2 A(\Delta) - O(q^{n}+q^{L-n}),2 A(\Delta) + O(q^n + q^{L-n})]\, .
$$
Also,
$$
\liminf_{\substack{n,L \\ \min(n,L-n) \to \infty}}
  \lambda(L,n,L+1) \geq 2 A(\Delta) + \gamma\, .
$$
Finally,
$$
\|\Proj(\mathcal{K}_{\Ir/L,n}) - \Proj(\Hil_{\Ir/L,n}^L)\|
  = O(q^n + q^{L-n})\, .
$$
\end{theorem}

\begin{proof}
We first prove that 
\begin{equation}
\label{pc:1}
\|(\HXXZ_{\Ir/L} - 2 A(\Delta)) \Proj(\calK_{\Ir/L,n})\| 
= O(q^{n} + q^{L-n})\, .
\end{equation}
It is easy to see that, just as for the droplets on an interval,
\begin{eqnarray}
\label{Per:1}
\frac{|\ip{\xi_{\Ir/L,n}(x)}{\xi_{\Ir/L,n}(y)}|}
  {\|\xi_{\Ir/L,n}(x)\| \cdot \|\xi_{\Ir/L,n}(y)\|}
  &\leq& \frac{q^{n\cdot d(x,y)}}{\pn \infty} 
  \quad \textrm{for all} \quad x,y\, ;\\
\label{Per:2}
\frac{|\ip{\xi_{\Ir/L,n}(x)}{\HXXZ_{\Ir/L} \xi_{\Ir/L,n}(y)}|}
  {\|\xi_{\Ir/L,n}(x)\| \cdot \|\xi_{\Ir/L,n}(y)\|}
  &\leq& \frac{q^{n\cdot d(x,y)}}{\pn \infty} 
  \quad \textrm{if} \quad x\neq y\, ;\\
\label{Per:3}
\frac{|\ip{\xi_{\Ir/L,n}(x)}{(\HXXZ_{\Ir/L})^2 \xi_{\Ir/L,n}(y)}|}
  {\|\xi_{\Ir/L,n}(x)\| \cdot \|\xi_{\Ir/L,n}(y)\|}
  &\leq& \frac{q^{n\cdot d(x,y)}}{\pn \infty} 
  \quad \textrm{if} \quad d(x,y) \geq 2\, ;
\end{eqnarray}
where $d(x,y) = \min(|x-y|,|x+y-L|)$.
In fact, using the same tools as in Appendix A, 
we can calculate exactly, for $0\leq x\leq \floor{L/2}$,  
$$
\rho(\xi_{\Ir/L,n}(0),T^x)
  = q^{n x} \sum_k \frac{\sqbinom{\floor{L/2}-x}{\floor{n/2}-k}{q^2}
  \sqbinom{\ceil{L/2}-x}{\ceil{n/2}-k}{q^2}}
  {\sqbinom{\floor{L/2}}{\floor{n/2}}{q^2}
  \sqbinom{\ceil{L/2}}{\ceil{n/2}}{q^2}}
  \binom{x}{k}^2 q^{k(L+2k)}\, .
$$
It is verifiable that this satisfies the bounds above.
The other expectations 
$$\rho(\xi_{\Ir/L,n}(0),\HXXZ_{\Ir/L,n} T^x)\quad
\textrm{and}\quad \rho(\xi_{\Ir/L,n}(0),(\HXXZ_{\Ir/L,n})^2 T^x)$$
are similar.
Applying Lemma \ref{Lem:OrthStates}, proves \eq{pc:1}.

Now we prove that, considering $\HXXZ_{\Ir/L}$ acting on the invariant subspace
$\Hil_{\Ir/L,n}$,
\begin{equation}
\label{pc:2}
\HXXZ_{\Ir/L} \geq (2 A(\Delta) - \epsilon_n - \epsilon_{L-n}) \unity
  + \gamma (\unity - \Proj(\calK_{L,n}))\, ,
\end{equation}
where $\lim_{n \to \infty} \epsilon_n = 0$.
To do this, we use Corollary \ref{IHS:Cor}.
There exists an $L_0(q)$ and $C_0(q)$
such that, if  $L > L_0(q)$ 
then for any 
$\psi \in \Hil_{L,n}$ with $\rho(\psi,\HXXZ_{\Ir/L}) \leq 2 A(\Delta) + \gamma$,
Corollary \ref{IHS:Cor} guarantess the existence of a ``subinterval'' 
$J \subset \Ir/L$ satisfying
$|J| = 2 \floor{L^{1/2}}$, $\|P_J \psi - \psi\| \leq C_0(q) L^{-1/2}$,
and
$$
\ip{\psi}{\HXXZ_{\Ir/L} \psi}
  \geq \ip{P_J \psi}{\HXXZ_{\Ir/L} P_J \psi}
  - C_0(q) L^{-1/4} \|\psi\|^2\, .
$$
We can take
$L_0(q) = (7-6q+3q^2)^2/(1-q)^4$ 
and
$C_0(q) =(5 + 18 q + 5 q^2) L_0(q)^{1/4}/(2+2q^2)$.
By ``subinterval'', we mean that there exists an interval $J' \subset \Ir$,
such that $J \equiv J' ({\rm mod} L)$.
Without loss of generality,
we assume $J = [1,\dots,\floor{L^{1/2}}] \cup [L+1-\floor{L^{1/2}},L]$.
Next,
$$
\ip{P_J \psi}{\HXXZ_{\Ir/L} P_J \psi}
  = \ip{P_J^\uparrow \psi}{\HXXZ_{\Ir/L} P_J^\uparrow \psi}
  + \ip{P_J^\downarrow \psi}{\HXXZ_{\Ir/L} P_J^\downarrow \psi}\, 
$$
and $\|P_J \psi\|^2 = \|P_J^\uparrow \psi\|^2 + \|P_J^\downarrow \psi\|^2$.

We estimate $\ip{P_J^\uparrow \psi}{\HXXZ_{\Ir/L} P_J^\uparrow \psi}$, first.
Of course, $\HXXZ_{\Ir/L} = H^{--}_{L,1} + H^{++}_{[1,L]}$,
and since $H^{--} \ket{\uparrow \uparrow} = A(\Delta) \ket{\uparrow \uparrow}$,
we see that
$
\HXXZ_{\Ir/L} P_J^\uparrow \psi 
  = (A(\Delta) + H^{++}_{[1,L]}) P_J^\uparrow \psi
$.
Then using Theorem \ref{main:theorem}(b),
\begin{eqnarray*}
&&\ip{P_J^\uparrow \psi}{\HXXZ_{\Ir/L} P_J^\uparrow \psi}
  \geq (A(\Delta) - \epsilon(n)) \|P_J^\uparrow \psi\|^2 \\
&&\qquad  \qquad + \gamma
  \ip{P_J^\uparrow \psi}
  {(\unity - \Proj(\calK_{L,n}))P_J^\uparrow \psi}\, ,
\end{eqnarray*}
where $\lim_{n \to \infty} \epsilon(n) = 0$.
But 
\begin{eqnarray*}
P_J^\uparrow \Proj(\calK_{L,n}) P_J^\uparrow
  &=& P_J^\uparrow \sum_{x=\floor{n/2}}^{L-\ceil{n/2}} \Proj(\xi_{L,n}(x))
  P_J^\uparrow + O(q^n) \\
  &=& P_J^\uparrow \sum_{x=\floor{L^{1/2}}+\floor{n/2}}
  ^{L+1-\floor{L^{1/2}}-\ceil{n/2}} \Proj(\xi_{L,n}(x)) P_J^\uparrow
  + O(q^n) \\
  &\leq& P_J^\uparrow \sum_{x=0}^{L-1} \Proj(\xi_{\Ir/L,n}(x)) P_J^\uparrow
  - O(q^n) \\
  &=& P_J^\uparrow \Proj(\calK_{\Ir/L,n}) P_J^\uparrow - O(q^n)\, ,
\end{eqnarray*}
where by $A = B + O(q^n)$, we mean $\|A - B\| = O(q^n)$,
and by $A \geq B - O(q^n)$, we mean $B-A \leq O(q^n) \unity$.
We omit the calculations here.
So
\begin{eqnarray*}
&& \ip{P_J^\uparrow \psi}{\HXXZ_{\Ir/l} P_J^\uparrow \psi}
  \geq (2A(\Delta) - \epsilon(n) - O(q^n)) \|P_J^\uparrow \psi\|^2 \\
&&\qquad   + \gamma \ip{P_J^\uparrow \psi}{(\unity - \Proj(\calK_{\Ir/L,n})) 
  P_J^\uparrow \psi}\, .
\end{eqnarray*}
Symmetrically,
\begin{eqnarray*}
&& \ip{P_J^\downarrow \psi}{\HXXZ_{\Ir/l} P_J^\downarrow \psi}
  \geq (2A(\Delta) - \epsilon(L-n) - O(q^{L-n})) \|P_J^\downarrow \psi\|^2 \\
&&\qquad   + \gamma \ip{P_J^\downarrow \psi}{(\unity - 
  \Proj(F \calK_{\Ir/L,L-n})) P_J^\downarrow \psi}\, ,
\end{eqnarray*}
where $F : \Hil_{L,L-n} \to \Hil_{L,n}$ denotes the spin-flip.
But $\calK_{\Ir/L,n} = F \calK_{\Ir/L,L-n}$.
Also,
$\|P_J^\downarrow \Proj(\calK_{\Ir/L,n}) P_J^\uparrow\| = O(q^n + q^{L-n})$.
So, for any $\psi \in \Hil_{L,n}$,
\begin{equation}
\label{pc:3}
\begin{array}{l}
\displaystyle
\ip{\psi}{\HXXZ_{\Ir/l} \psi} \vspace{2mm}\\
\qquad \displaystyle
  \geq (2A(\Delta) - [\epsilon_n + \epsilon_{L-n} 
  + O(q^n+q^{L-n}) + O(L^{-1/4})]) 
  \|\psi\|^2 \vspace{2mm}\\
\qquad \qquad \displaystyle  
+ \gamma \ip{\psi}{(\unity - \Proj(\calK_{\Ir/L,n})) \psi}\, .
\end{array}
\end{equation}
Equations \eq{pc:1} and \eq{pc:3} together imply the corollary.
\end{proof}

\subsection{The Infinite Spin Chain}

Let $\ket{\Omega} = \ket{\dots \uparrow \uparrow \uparrow \dots}_{\Ir}$ be a 
vacuum state, and define 
$$
\Hil_{\Ir,n} = {\rm cl}( 
\Span\{S_{x_1}^- S_{x_2}^- \dots S_{x_n}^- \ket{\Omega}
  : x_1 < x_2 < \dots < x_n\})\, ,
$$
where ${\rm cl}(.)$ is the $l^2$-closure.
This is a separable Hilbert space, and 
$$
\HXXZ_\Ir = \sum_{x=-\infty}^\infty \HXXZ_{x,x+1}
$$
is a densely defined, self-adjoint operator.
This Hamiltonian defines the infinite spin chain.
We check that the series does converge.
In fact
$$
0 \leq \HXXZ_{x,x+1} \leq \frac{1}{2}(1+\Delta^{-1})(\hat{N}_x + \hat{N}_{x+1})
$$
where $\hat{N}_x = (\frac{1}{2} - S_x^3)$ counts the number of down spins at 
$x$. But $\sum_{x=-\infty}^\infty \hat{N}_x \equiv n$ on $\Hil_{\Ir,n}$.
So the series does converge, and
$\HXXZ_\Ir \leq n (1+\Delta^{-1})$.
We define the droplet states
$$
\xi_{\Ir,n}(x) = \psi^{+-}_{(-\infty,x]}(\floor{n/2})
  \otimes \psi^{-+}_{[x,\infty)}(\ceil{n/2});
$$
and let $\calK_{\Ir,n}$ be the $l^2$ closure of 
$\Span\{\xi_{\Ir,n}(x) : x \in \Ir\})$.

\begin{theorem}
\label{Thm:Infinite}
The following bounds exist for the infinite spin chain
$$
\|(\HXXZ_\Ir - 2 A(\Delta)) \Proj(\calK_{\Ir,n})\|
  = O(q^n)\, ,
$$
and, considering $\HXXZ_\Ir$ as an operator on $\Hil_{\Ir,n}$,
$$
\HXXZ_\Ir \geq (2 A(\Delta) - \epsilon_n) \unity
  + \gamma (\unity - \Proj(\calK_{\Ir,n}))\, ,
$$
where $\epsilon_n$ is a sequence with $\lim_{n \to \infty} \epsilon_n = 0$.
\end{theorem}

\begin{proof}
The proof that 
\begin{equation}
\label{ic:1}
\|(\HXXZ_{\Ir} - 2 A(\Delta)) \Proj(\calK_{\Ir,n})\| = O(q^n)\,
\end{equation}
is essentially the same as in Section \ref{Sect:Eval}.
One fact we should check is that for each $\xi_{\Ir,n}(x)$,
$\|(\HXXZ_\Ir - 2 A(\Delta)) \xi_{\Ir,n}(x)\|^2 = O(q^n)$.
We observe that
\begin{eqnarray*}
\HXXZ_{[-L,L]} \xi_{\Ir,n}(0)
  &=& (H^{+-}_{[-L,0]} + H^{-+}_{[1,L]} + H^{++}_{0,1} 
  + A(\Delta) (S_{-L}^3 + S_L^3))
  \xi_{\Ir,n}(0) \\
  &=& (H^{++}_{0,1} + A(\Delta) (S_{-L}^3 + S_L^3)) \xi_{\Ir,n}(0)\, .
\end{eqnarray*}
But as before, 
$$
\|(H^{++}_{0,1} - A(\Delta)) \xi_{\Ir,n}(0)\|^2 
  \leq O(q^n) \|\xi_{\Ir,n}(0)\|^2\, .
$$
An obvious fact is
$$
\|(S_{-L}^3 + S_L^3 - 1) \xi(0,n)\|^2 \leq O(q^{L-n}) \|\xi(0,n)\|^2\, .
$$
Taking $L \to \infty$, yields the desired result.
We have the usual orthogonality estimates 
\begin{eqnarray*}
\frac{|\ip{\xi_{\Ir,n}(x)}{\xi_{\Ir,n}(y)}|}
{\|\xi_{\Ir,n}(x)\| \cdot \|\xi_{\Ir,n}(y)\|}
  &\leq& \frac{q^{n|x-y|}}{\pn{\infty}} ,\\
\frac{|\ip{\xi_{\Ir,n}(x)}{\HXXZ_\Ir \xi_{\Ir,n}(y)}\|}
{\|\xi_{\Ir,n}(x)\| \cdot \|\xi_{\Ir,n}(y)\|}
  &\leq& \frac{q^{n|x-y|}}{\pn{\infty}}\quad
\textrm{for } x\neq y\, ,\\
\frac{|\ip{\xi_{\Ir,n}(x)}{(\HXXZ_\Ir)^2 \xi_{\Ir,n}(y)}|}
{\|\xi_{\Ir,n}(x)\| \cdot \|\xi_{\Ir,n}(y)\|}
  &\leq& \frac{q^{n|x-y|}}{\pn{\infty}}\quad
\textrm{for } |x-y| \geq 2\, .\\
\end{eqnarray*}
In fact, the estimate of $\ip{\xi_{\Ir,n}(x)}{\xi_{Ir,n}(y)}$
follows by \eq{App:result1}, taking the limit that $L \to \infty$,
and the other estimates are consequences.
Applying Lemma \ref{Lem:OrthStates} proves \eq{ic:1}.

For the second part, suppose $\psi \in \Hil_{\Ir,n}$.
Then 
$$
\rho(\psi,\HXXZ_\Ir) = \lim_{L \to \infty} \rho(\psi,\HXXZ_{[-L,L]})\, .
$$
Furthermore 
$\HXXZ_{[-L,L]} = H^{++}_{[-L,L]} + A(\Delta) (S_{-L}^3 + S_L^3)$,
and 
$$
\lim_{L \to \infty} \ip{\psi}{(S_{-L}^3 + S_L^3) \psi} = \|\psi\|^2
$$ 
by virtue of the fact that $n$, the total number of down spins 
in the state $\psi$,
is finite.
Essentially the same fact is restated as
$ \lim_{L \to \infty} \psi_L = \psi$, where
$$
\psi_L = 
  \Proj(\Hil_{(-\infty,-L-1],0} \otimes \Hil_{[-L,L],n} 
  \otimes \Hil_{[L+1,\infty),0}) \psi\, .
$$
Let us define 
$$
\Xi_{L,n} = 
  \Proj(\Hil_{(-\infty,-L-1],0} \otimes \mathcal{K}_{[-L,L],n} 
  \otimes \Hil_{[L+1,\infty),0}) \psi\, ,
$$
where $\mathcal{K}_{[-L,L],n}$ is the droplet state subspace
for the finite chain.
By Theorem \ref{main:theorem}(b),
\begin{eqnarray*}
&& \ip{\psi_L}{H^{++}_{[-L,L]} \psi_L} 
  \geq (2 A(\Delta) - \epsilon(n)) \|\psi_L\|^2 \\
&&\qquad \qquad  + \gamma \ip{\psi_L}
  {(\unity - \Xi_{L,n})\psi_L}\, .
\end{eqnarray*}
Since $\psi_L \to \psi$ in the norm-topology, as $L \to \infty$, 
all we need to check is that  
$\Xi_{L,n}$ converges weakly to $\Proj(\calK_{\Ir,n})$.

It helps to break up $\Xi_{L,n}$ into two pieces, 
\begin{eqnarray*}
&&
\Xi'_{L,n} = 
  \Proj(\Span\{ \ket{\dots \uparrow}_{(-\infty,-L-1]} 
  \otimes \xi_{[-L,L],n}(x)
  \otimes \ket{\uparrow \dots}_{[L+1,\infty)} :\\
&& \hspace{4cm}
  - \floor{L/2} + \floor{n/2} \leq x \leq \ceil{L/2} - \ceil{n/2} \})\, ,
\end{eqnarray*}
and $\Xi''_{L,n} = \Xi_{L,n} - \Xi'_{L,n}$.
Define
$$
\phi_{L,n}(x) = \ket{\dots \uparrow}_{(-\infty,-L-1]} 
  \otimes \xi_{[-L,L],n}(x)
  \otimes \ket{\uparrow \dots}_{[L+1,\infty)}\, .
$$
Note that for any sequence $x_L$ such that
$x_L \in [-\floor{L/2}+\floor{n/2},\ceil{L/2}-\ceil{n/2}]$,
we have
$$
\lim_{L \to \infty} \rho(\phi_{L,n}(x_L),\calK_{\Ir,n}) = 1\, .
$$
The reason is that $\|\phi_{L,n}(x) - \xi_\Ir(x,n)\| = O(q^{L/2})$ 
because the the left and right interfaces of the droplet in
$\phi_{L,n}(x)$ are a distance at least $L/2$ from the left and right endpoints
of the interval $[-L,L]$, and 
the probability of finding an overturned spin decays
$q$-exponentially with the distance from the inteface.
For the same reason, for any fixed $x \in \Ir$,
$\lim_{L,\to \infty} \rho(\xi_\Ir(x,n),\Xi'_{L,n}) = 1$.
These two facts imply that $\Xi'_{L,n}$ converges weakly to 
$\Proj(\calK_{\Ir,n})$.
Now $\Xi''_{L,n}$ converges weakly to zero, 
because every state in $\Xi''_{L,n}$ has over
half its downspins concentrated in the annulus
$[-L,L] \setminus [-\floor{L/2}+\floor{n/2},\ceil{L/2}-\ceil{n/2}]$,
and the inner radius tend to infinity.
This means that $\textrm{w}-\lim_{L \to \infty} \Xi_{L,n} = 
\Proj(\calK_{\Ir,n})$,
as claimed.

Thus, taking the appropriate limits,
\begin{eqnarray*}
&&\ip{\psi}{\HXXZ_\Ir \psi}
  \geq (2 A(\Delta) - \epsilon(n)) \|\psi\|^2 \\
&& \qquad  
  + \gamma \ip{\psi}{(\unity - \Proj(\calK_{\Ir,n})) \psi}\, ,
\end{eqnarray*}
which finishes the proof of the theorem.
\end{proof}


\section{Appendix A}
\label{App}
In this section we carry out several calculations,
whose results are needed in the main body of the paper,
but whose proofs are not very enlightening for understanding
the main arguments.
The definitions of the kink states, $\psi^{+-}_{[a,b]}(n)$, and the
antikink states, $\psi^{-+}_{[a,b]}(n)$, are given in 
\eq{Intro:+-} and \eq{Intro:-+}.
One nice feature of these states is that they are governed by a quantum
Clebsh-Gordan formula, due to the $SU_q(2)$ symmetry of
$H^{\alpha \beta}_{[a,b]}$, $\alpha \beta = +-,-+$. 
By this we mean the following: Suppose $a\leq x\leq b$.
Then, 
\begin{eqnarray}
\label{App:+-coprod}
\psi^{+-}_{[a,b]}(n) &=& \sum_k \psi^{+-}_{[a,x]}(k) \otimes 
  \psi^{+-}_{[x+1,b]}(n-k) q^{(b-x)k}\, ,\\
\label{App:-+coprod}
\psi^{-+}_{[a,b]}(n) &=& \sum_k \psi^{-+}_{[a,x]}(k) \otimes 
  \psi^{-+}_{[x+1,b]}(n-k) q^{(x+1-a)(n-k)} \, .
\end{eqnarray}
We let the sum in $k$ run over all integers $k$, with the understanding that
$\psi^{+-}_{[a,b]}(n) = \psi^{-+}_{[a,b]}(n)$ if $n<0$ or $n>b-a+1$.
One need not refer to the quantum group to understand this decomposition,
it is enough just to check the definitions.
We can also see from the definitions that 
\begin{eqnarray}
\label{App:kink-norm}
\ip{\psi^{\alpha \beta}_{[a,b]}(m)}{\psi^{\alpha \beta}_{[a,b]}(n)}
  &=& \delta_{m,n} \qbinom{b-a+1}{n}{q^2} q^{n(n+1)} \\
\label{App:mixed-ip}
  \ip{\psi^{\alpha \beta}_{[a,b]}(m)}{\psi^{\beta \alpha}_{[a,b]}(n)}
  &=& \delta_{m,n} \binom{b-a+1}{n} q^{b-a+2}\, ,
\end{eqnarray}
for $\alpha \beta = +-,-+$.

The combinatorial prefactor in \eq{App:kink-norm}
is a $q$-binomial coefficient (in this case a $q^2$-binomial coefficient), 
also known as a Gauss polynomial.
The most important feature, for us, is the $q$-binomial formula
$$
\prod_{k=1}^L (1 + q^{2k} x) = \sum_{n=0}^L \qbinom{L}{n}{q^2} q^{n(n+1)}x^n \, .
$$
At this point let us introduce another useful combinatorial quantity,
$\pn n$, defined for $n=0,1,2,\dots,\infty$:
$$
\pn n = \prod_{k=1}^n (1 - q^{2k})\, .
$$
For a fixed $q \in [0,1)$, the sequence $\pn n$ is clearly montone decreasing,
and $\pn \infty > 0$. 
We note that 
$$
\qbinom{n}{k}{q^2} = \frac{\pn{n}}{\pn{k} \pn{n-k}}
$$
which means that for $0 \leq k \leq n$,
$$
1 \leq \qbinom{n}{k}{q^2} \leq \frac{1}{\pn \infty}\, .
$$
The first result we wish to prove is that
\begin{equation}
\label{App:prelim}
\begin{array}{l}
\displaystyle
\ip{\psi^{+-}_{[1,x]}(m) \otimes \psi^{-+}_{[x+1,x+y+r]}(n+k)}
{\psi^{+-}_{[1,x+r]}(m+k) \otimes \psi^{-+}_{[x+r+1,x+y+r]}(n)}
\vspace{2mm}  \\
\hspace{3cm} 
= \binom{r}{k} \qbinom{x}{m}{q^2} \qbinom{y}{n}{q^2}
  q^{m(m+k+1) + n(n+k+1) + k(r+1)}\, .
\end{array}
\end{equation}
This is very simple.
From \eq{App:+-coprod} and \eq{App:-+coprod},
\begin{eqnarray*}
&& \ip{\psi^{+-}_{[1,x]}(m) \otimes \psi^{-+}_{[x+1,x+y+r]}(n+k)}
  {\psi^{+-}_{[1,x+r]}(m+k) \otimes \psi^{-+}_{[x+r+1,x+y+r]}(n)} \\
&& \quad =
  \sum_{j,l} \bra{q^{r(n+k-j)} \psi^{+-}_{[1,x]}(m) \otimes 
  \psi^{-+}_{[x+1,x+r]}(j)
  \otimes \psi^{+-}_{[x+r+1,x+y+r]}(n+k-j)} \\
&& \qquad \qquad
  \ket{q^{r(m+k-l)} \psi^{+-}_{[1,x]}(l) \otimes \psi^{+-}_{[x+1,x+r]}(m+k-l)
  \otimes \psi^{-+}_{[x+r+1,x+y+r]}(n)} \\
&& \quad = \sum_{j,l} q^{r(m+n+2k-l-j)} 
  \ip{\psi^{+-}_{[1,x]}(m)}{\psi^{+-}_{[1,x]}(l)} \\
&& \qquad \qquad
  \times \ip{\psi^{-+}_{[x+1,x+r]}(j)}{\psi^{+-}_{[x+1,x+r]}(m+k-l)} \\
&& \qquad \qquad
  \times \ip{\psi^{-+}_{[x+r+1,x+y+r]}(n+k-j)}{\psi^{-+}_{[x+r+1,x+y+r]}(n)}\, .
\end{eqnarray*}
Consulting \eq{App:kink-norm} and \eq{App:mixed-ip}, we see that the only 
choice
of $j$ and $l$ for which none of the inner-products vanishes is $j=l=k$.
Plugging in these values for $j$ and $l$ and using the formulae for the 
inner-products yields \eq{App:prelim}.
We can use \eq{App:kink-norm} to normalize the inner-product in the following 
way,
\begin{equation}
\label{App:prelim2}
\begin{array}{l}
\displaystyle
\frac{ \ip{\psi^{+-}_{[1,x]}(m) \otimes \psi^{-+}_{[x+1,x+y+r]}(n+k)}
{\psi^{+-}_{[1,x+r]}(m+k) \otimes \psi^{-+}_{[x+r+1,x+y+r]}(n)} }
{\|\psi^{+-}_{[1,x]}(m) \otimes \psi^{-+}_{[x+1,x+y+r]}(n+k)\| \cdot
\|\psi^{+-}_{[1,x+r]}(m+k) \otimes \psi^{-+}_{[x+r+1,x+y+r]}(n)\|}
\vspace{2mm}  \\
\hspace{1cm} 
= \binom{r}{k} \sqrt{\qbinom{x}{m}{q^2} \qbinom{y}{n}{q^2}
 \Big/ \qbinom{x+r}{m+k}{q^2} \qbinom{y+r}{n+k}{q^2}}\,
  q^{(m+n+k)(r-k)}\, .
\end{array}
\end{equation}

We wish to specialize this formula in two ways.
First, by setting $k=r$ we have
\begin{equation}
\label{App:special1}
\begin{array}{l}
\displaystyle
\frac{ \ip{\psi^{+-}_{[1,x]}(m) \otimes \psi^{-+}_{[x+1,x+y+r]}(n+r)}
{\psi^{+-}_{[1,x+r]}(m+r) \otimes \psi^{-+}_{[x+r+1,x+y+r]}(n)} }
{\|\psi^{+-}_{[1,x]}(m) \otimes \psi^{-+}_{[x+1,x+y+r]}(n+r)\| \cdot
\|\psi^{+-}_{[1,x+r]}(m+r) \otimes \psi^{-+}_{[x+r+1,x+y+r]}(n)\|}
\vspace{2mm}  \\
\hspace{1cm} 
= \sqrt{\qbinom{x}{m}{q^2} \qbinom{y}{n}{q^2}
 \Big/ \qbinom{x+r}{m+r}{q^2} \qbinom{y+r}{n+r}{q^2}}\, .
\end{array}
\end{equation}
Second, by setting $k=0$, we have
\begin{equation}
\label{App:special2}
\begin{array}{l}
\displaystyle
\frac{ \ip{\psi^{+-}_{[1,x]}(m) \otimes \psi^{-+}_{[x+1,x+y+r]}(n)}
{\psi^{+-}_{[1,x+r]}(m) \otimes \psi^{-+}_{[x+r+1,x+y+r]}(n)} }
{\|\psi^{+-}_{[1,x]}(m) \otimes \psi^{-+}_{[x+1,x+y+r]}(n)\| \cdot
\|\psi^{+-}_{[1,x+r]}(m) \otimes \psi^{-+}_{[x+r+1,x+y+r]}(n)\|}
\vspace{2mm}  \\
\hspace{1cm} 
= \sqrt{\qbinom{x}{m}{q^2} \qbinom{y}{n}{q^2}
 \Big/ \qbinom{x+r}{m}{q^2} \qbinom{y+r}{n}{q^2}}\, 
  q^{(m+n)r}\, .
\end{array}
\end{equation}
To estimate \eq{App:special1}, we notice that
$$
\begin{array}{l}
\qbinom{x}{m}{q^2} \qbinom{y}{m}{q^2} \Big/
  \qbinom{x+r}{m+r}{q^2} \qbinom{y+r}{n+r}{q^2}
\vspace{2mm} \\
\qquad \displaystyle
  = \frac{\pn{x}}{\pn{x+r}} \cdot \frac{\pn{m+r}}{\pn{m}}
  \cdot \frac{\pn{y}}{\pn{y+r}} \cdot \frac{\pn{n+r}}{\pn{n}}
\end{array}
$$
This quantity is at most 1 (when $r=0$).
To get a lower bound we observe that the first and third ratios on the right
hand side are greater than 1, while the product of the second and third
is easily bounded
\begin{eqnarray*}
\frac{\pn{m+r}}{\pn{m}} \cdot \frac{\pn{n+r}}{\pn{n}} 
  &\geq& \prod_{k=1}^r (1 - q^{2(m+k)})^{-1} (1-q^{2(n+k)})^{-1} \\
  &\geq& \left(1 - \frac{q^{2(m+1)}}{1-q^2}\right)^{-1}
  \left(1 - \frac{q^{2(n+1)}}{1-q^2}\right)^{-1}\, .
\end{eqnarray*}
Inserting the inequality to \eq{App:special1}
$$
\begin{array}{l}
\displaystyle
\frac{ \ip{\psi^{+-}_{[1,x]}(m) \otimes \psi^{-+}_{[x+1,x+y+r]}(n+r)}
{\psi^{+-}_{[1,x+r]}(m+r) \otimes \psi^{-+}_{[x+r+1,x+y+r]}(n)} }
{\|\psi^{+-}_{[1,x]}(m) \otimes \psi^{-+}_{[x+1,x+y+r]}(n+r)\| \cdot
\|\psi^{+-}_{[1,x+r]}(m+r) \otimes \psi^{-+}_{[x+r+1,x+y+r]}(n)\|}
\vspace{2mm}  \\
\hspace{1cm} \displaystyle
\geq \left(1 - \frac{q^{2(m+1)}}{1-q^2}\right)^{-1/2}
  \left(1 - \frac{q^{2(n+1)}}{1-q^2}\right)^{-1/2}
\end{array}
$$
This leads to a useful formula. 
If $\psi$ and $\phi$ are normalized states then 
$\|\Proj(\psi) - \Proj(\phi)\| = \sqrt{1 - |\ip{\psi}{\phi}|^2}$.
Thus,
$$
\begin{array}{l}
\|\Proj(\psi^{+-}_{[1,x]}(m) \otimes \psi^{-+}_{[x+1,x+y+r]}(n+r))
  \vspace{2mm} \\
\displaystyle \hspace{25pt}
  - \Proj(\psi^{+-}_{[1,x+r]}(m+r) \otimes \psi^{-+}_{[x+r+1,x+y+r]}(n))\| 
  \leq \sqrt{\frac{8 q^2}{1 - q^2} (q^{2m} + q^{2n})}\, .
\end{array}
$$
In particular, changing notation to match the body of the paper,
\begin{equation}
\label{App:result1}
\|\Proj(\psi^{+-}_{[1,x]}(n_1)\otimes \psi^{-+}_{[x+1,L]}(n_2)) - 
  \Proj(\xi_{L,n_1+n_2}(\tilde{x}))\|
  \leq \frac{4 q^{\min(n_1,n_2)+1}}{\sqrt{1-q^2}}
\end{equation}
where $\tilde{x} = x + \floor{(n_2-n_1)/2}$.

To estimate \eq{App:special2}, we begin again by observing
\begin{eqnarray*}
&&\qbinom{x}{a}{q^2} \qbinom{y}{b}{q^2}
  \Big/ \qbinom{x+r}{m}{q^2} \qbinom{y+r}{n}{q^2}  \\
&& \hspace{75 pt}
  = \frac{\pn {x}}{\pn {x+r}} \cdot
  \frac{\pn {x-m+r}}{\pn {x-m}} \cdot
  \frac{\pn {y}}{\pn {y+r}} \cdot
  \frac{\pn{y-n+r}}{\pn{y-n}}\, .
\end{eqnarray*}
By the monotonicity of $\pn{x}$ in $x$, we have
$$
\pn \infty^2 \leq 
\qbinom{x}{m}{q^2} \qbinom{y}{n}{q^2}
  \Big/ \qbinom{x+r}{a}{q^2} \qbinom{y+r}{b}{q^2}  
\leq \frac{1}{f_q(\infty)^2}\, .
$$
From this it follows
\begin{equation}
\label{cf:n=0:bounds}
\begin{array}{l}
\displaystyle
\frac{\ip{\psi^{+-}_{[1,x]}(m) \otimes \psi^{-+}_{[x+1,x+y+r]}(n)}
{\psi^{+-}_{[1,x+r]}(m) \otimes \psi^{-+}_{[x+r+1,x+y+r]}(n)}}
{ \|\psi^{+-}_{[1,x]}(m) \otimes \psi^{-+}_{[x+1,x+y+r]}(n)\| 
\cdot
\|\psi^{+-}_{[1,x+r]}(m) \otimes \psi^{-+}_{[x+r+1,x+y+r]}(n)\| }
\vspace{2mm}  \\
\hspace{1cm} 
= C(x,y,m,n,r) q^{(m+n)r}\, ,
\end{array}
\end{equation}
where 
$$
f_q(\infty) \leq C(x,y,m,n,r) \leq \frac{1}{f_q(\infty)}\, .
$$
In particular, we have the useful bound
\begin{equation}
\label{App:result2}
\frac{|\ip{\xi_{L,n}(x)}{\xi_{L,n}(y)}|}
  {\|\xi_{L,n}(x)\| \cdot \|\xi_{L,n}(y)\|}
  \leq \frac{q^{n|y-x|}}{f_q(\infty)}\, .
\end{equation}
This is the first in a series of three inequalities needed for 
Section \ref{Sect:Eval}.

Next, we need a bound for 
$$
\frac{|\ip{\xi_{L,n}(x)}{H^{++}_{[1,L]} \xi_{L,n}(y)}|}
  {\|\xi_{L,n}(x)\| \cdot \|\xi_{L,n}(y)\|}.
$$
It turns out that the is well approximated by the normalized
inner-product above. 
The reason is that, while $H^{++}_{[1,L]}$ is not a 
small operator in general, when acting on the droplet states
it reduces to just one nearest-neighbor interaction:
$H^{++}_{[1,L]} \xi_{L,n}(x) = H^{++}_{x,x+1} \xi_{L,n}(x)$.
To exploit this we return to the notation above, and observe 
that as long as $r\geq 1$
\begin{equation}
\label{App:prelim3}
\begin{array}{l}
\displaystyle
\ip{\psi^{+-}_{[1,x]}(m) \otimes \psi^{-+}_{[x+1,x+y+r]}(n+k)}
{\psi^{+-}_{[1,x+r]}(m+k) \otimes \psi^{-+}_{[x+r+1,x+y+r]}(n)}
\vspace{2mm}  \\
\hspace{1cm} \displaystyle
= \sum_{j,l} q^{(r+2) m + n + k - 3 j + (r - 2) l}
  \ip{\psi^{+-}_{[1,x-1]}(m-j)}{\psi^{+-}_{[1,x-1]}(m-j)} \vspace{2mm}\\
\hspace{1cm} \displaystyle
  \times \ip{\psi^{+-}_{\{x\}}(j) \otimes \psi^{-+}_{\{x+1\}}(l)}
  {\psi^{+-}_{[x,x+1]}(j+l)} \vspace{2mm}\\
\hspace{1cm} \displaystyle
  \times \ip{\psi^{-+}_{[x+12,x+y+r]}(n+k-l)}
  {\psi^{+-}_{[x+2,x+r]}(k-l) \otimes 
  \psi^{-+}_{[x+r+1,x+y+r]}(n)}\, .
\end{array}
\end{equation}
This is derived just as before, using equations \eq{App:+-coprod} --
\eq{App:mixed-ip}. 
Note 
$$
\psi^{+-}_{\{x\}}(j) = \psi^{-+}_{\{x\}}(j) = q^j (S_x^{-})^j 
\ket{\uparrow}_x\, .
$$
The usefulness of this formula is in the fact that 
\begin{eqnarray*}
&& |\ip{\psi^{+-}_{\{x\}}(j) \otimes \psi^{-+}_{\{x+1\}}(l)}
  {H^{++}_{x,x+1} \psi^{+-}_{[x,x+1]}(j+l)}| \\
&& \qquad \leq \ip{\psi^{+-}_{\{x\}}(j) \otimes \psi^{-+}_{\{x+1\}}(l)}
  {\psi^{+-}_{[x,x+1]}(j+l)}\, .
\end{eqnarray*}
Indeed, the formula for the right-hand-side
is
$$
\ip{\psi^{+-}_{\{x\}}(j) \otimes \psi^{-+}_{\{x+1\}}(l)}
  {\psi^{+-}_{[x,x+1]}(j+l)}
  = q^{2j+3l}\, ,
$$
while the left-hand-side is
$$
\begin{array}{c|c|l}
\displaystyle
j & l & \ip{\psi^{+-}_{\{x\}}(j) \otimes \psi^{-+}_{\{x+1\}}(l)}
  {H^{++}_{x,x+1} \psi^{+-}_{[x,x+1]}(j+l)}  \\
\hline
\displaystyle
\rule{0mm}{5mm} 
0 & 0 & - A(\Delta) \\
\rule{0mm}{6mm} 
\displaystyle
0 & 1 & \displaystyle  \frac{q^2 (1-q)^2}{2(1+q^2)} \\
\rule{0mm}{6mm} 
\displaystyle
1 & 0  & \displaystyle - \frac{q^4(1-q^2)}{2(1+q^2)} \\
\rule{0mm}{5mm} 
\displaystyle
1 & 1 & A(\Delta) q^5
\end{array}
$$
Thus,
$$
\begin{array}{l}
\displaystyle
|\ip{\psi^{+-}_{[1,x]}(m) \otimes \psi^{-+}_{[x+1,x+y+r]}(n+k)}
{H^{++}_{x,x+1} \psi^{+-}_{[1,x+r]}(m+k) \otimes \psi^{-+}_{[x+r+1,x+y+r]}(n)}|
\vspace{2mm}  \\
\hspace{0.5cm} \displaystyle
\leq \sum_{j,l} q^{(r+2) m + n + k - 3 j + (r - 2) l}
  \ip{\psi^{+-}_{[1,x-1]}(m-j)}{\psi^{+-}_{[1,x-1]}(m-j)} \vspace{2mm}\\
\hspace{1.0cm} \displaystyle
  \times |\ip{\psi^{+-}_{\{x\}}(j) \otimes \psi^{-+}_{\{x+1\}}(l)}
  {H^{++}_{x,x+1} \psi^{+-}_{[x,x+1]}(j+l)}| \vspace{2mm}\\
\hspace{1.0cm} \displaystyle
  \times \ip{\psi^{-+}_{[x+12,x+y+r]}(n+k-l)}
  {\psi^{+-}_{[x+2,x+r]}(k-l) \otimes 
  \psi^{-+}_{[x+r+1,x+y+r]}(n)} \vspace{2mm}\\
\hspace{0.5cm} \displaystyle
\leq \sum_{j,l} q^{(r+2) m + n + k - 3 j + (r - 2) l}
  \ip{\psi^{+-}_{[1,x-1]}(m-j)}{\psi^{+-}_{[1,x-1]}(m-j)} \vspace{2mm}\\
\hspace{1.0cm} \displaystyle
  \times {\psi^{+-}_{\{x\}}(j) \otimes \psi^{-+}_{\{x+1\}}(l)}
  {\psi^{+-}_{[x,x+1]}(j+l)} \vspace{2mm}\\
\hspace{1.0cm} \displaystyle
  \times \ip{\psi^{-+}_{[x+12,x+y+r]}(n+k-l)}
  {\psi^{+-}_{[x+2,x+r]}(k-l) \otimes 
  \psi^{-+}_{[x+r+1,x+y+r]}(n)} \vspace{2mm}\\
\hspace{0.5cm} \displaystyle 
  = \ip{\psi^{+-}_{[1,x]}(m) \otimes \psi^{-+}_{[x+1,x+y+r]}(n+k)}
  {\psi^{+-}_{[1,x+r]}(m+k) \otimes \psi^{-+}_{[x+r+1,x+y+r]}(n)}\, .
\end{array}
$$
This result, in conjunction with \eq{App:result2}, gives
\begin{equation}
\label{App:result3}
\frac{|\ip{\xi_{L,n}(x)}{H^{++}_{[1,L]} \xi_{L,n}(y)}|}
  {\|\xi_L(x,n)\| \cdot \|\xi_L(y,n)\|}
  \leq \frac{q^{n|y-x|}}{f_q(\infty)}\, ,
\end{equation}
whenever $|x-y|\geq 1$.
The requirement that $|x-y|\geq 1$ comes from the fact that $r$ must be 
at least one for \eq{App:prelim3} to hold true.

Similarly, we note
$$
\ip{\xi_{L,n}(x)}{(H^{++}_{[1,L]})^2 \xi_{L,n}(y)}
  = \ip{\xi_{L,n}(x)}{(H^{++}_{x,x+1} H^{++}_{y,y+1} \xi_{L,n}(y)}
$$
as long as $|x-y| \geq 2$.
Then the same argument as above can show that
$$
|\ip{\xi_{L,n}(x)}{(H^{++}_{x,x+1} H^{++}_{y,y+1} \xi_{L,n}(y)}|
  \leq \ip{\xi_{L,n}(x)}{\xi_{L,n}(y)}\, .
$$
Thus we have
\begin{equation}
\label{App:result4}
\frac{|\ip{\xi_{L,n}(x)}{(H^{++}_{[1,L]})^2 \xi_{L,y}(n)}|}
  {\|\xi_{L,n}(x)\| \cdot \|\xi_{L,y}(n)\|}
  \leq \frac{q^{n|y-x|}}{f_q(\infty)}\, ,
\end{equation}
whenever $|x-y|\geq 2$.


\section{Appendix B}
\label{Sec:AppB}

In this section we derive a single result.
We need the following definitions, some of which appeared previously
in the paper.
Given an arbitrary finite subset $\Lambda \subset \Ir$,
let $\Hil_\Lambda$ be the $|\Lambda|$-fold
tensor product $\bigotimes_{x\in \Lambda} \Cx_x^2$,
the space of all spin states on $\Lambda$.
The subspace of all vectors $\psi \in \Hil_\Lambda$ with exactly $n$
down spins is denoted $\Hil_{\Lambda,n}$.
For any subset $\Lambda_1 \subset \Lambda$,
we can define $Q_{\Lambda_1,n}$ to be the projection onto the subspace of 
$\Hil_\Lambda$ consisting of those vectors with exactly $n$ down spins in 
$\Lambda_1$.
So, $Q_{\Lambda_1,n} = \Proj(\Hil_{\Lambda_1,n} \otimes 
\Hil_{\Lambda\setminus \Lambda_1})$.
We also define $P_{\Lambda_1} = Q_{\Lambda_1,0} + Q_{\Lambda_1,|\Lambda_1|}$. 
It is the projection onto the span of vectors such that on $\Lambda_1$ they 
have all up spins or all down spins, but nothing else.

Now, let $0 \leq n < L$. 
Suppose $J = [a,b]$ is a subinterval of $[1,L]$.
We define the projections:
\begin{eqnarray*}
  G^\uparrow_j &=& Q_{[1,a-1],j}\, Q_{J,0}\, Q_{[b+1,L],n-j}\, ,\\
  G^\downarrow_j &=& Q_{[1,a-1],j}\, Q_{J,|J|}\, Q_{[b+1,L],n-j-|J|}\, .
\end{eqnarray*}
Then, for any $\psi \in \Hil_{[1,L],n}$,
$$
P_J \psi = \sum_{j=0}^n G^\uparrow_j \psi 
  + \sum_{j=0}^{n-|J|} G^\downarrow_j \psi\, .
$$
We recall the definition of droplet states:
For $\floor{n/2} \leq x \leq L-\ceil{n/2}$, 
$$
\xi_{L,n}(x) 
  = \psi^{+-}_{[1,x]}(\floor{n/2}) \otimes \psi^{-+}_{[x+1,L]}(\ceil{n/2})\, ,
$$
where $\psi^{+-}_{[1,x]}(\floor{n/2})$ and $\psi^{-+}_{[x+1,L]}(\ceil{n/2})$
are the kink and antikink states defined in \eq{Intro:+-}
and \eq{Intro:-+}.
Let $\Xi_{x} = \Proj(\xi_{L,n}(x))$.
Define the intervals 
\begin{eqnarray*}
I_1 &=& [\floor{n/2},a-\ceil{n/2}-1]\, ,\\
I_2 &=& [b-\ceil{n/2},a-1+\floor{n/2}]\, ,\\  
I_3 &=& [b+\floor{n/2},L-\ceil{n/2}]\, .
\end{eqnarray*}
Some of these intervals may be empty.
We have the following result.
There exists an $N(q) \in \Nl$ and a $C(q) < \infty$,
such that as long as $n \geq N(q)$
\begin{eqnarray*}
\sum_{x \in I_1 \cup I_2 \cup I_3} P_J \Xi_x P_J
  &\geq&  \sum_{x \in I_1} G^\uparrow_n \Xi_x G^\uparrow_n
  + \sum_{x \in I_2} G^\downarrow_{a-1+\floor{n/2}-x} \Xi_x 
  G^\downarrow_{a-1+\floor{n/2}-x} \\
  &&+ \sum_{x \in I_3} G^\uparrow_0 \Xi_x G^\uparrow_0
  -C(q) q^{|J|} P_J \Proj(\Hil_{[1,L],n}) \, .
\end{eqnarray*}

To prove this we group certain projections, $G_j^\sigma$,
and certain projections, $\Xi_x$, together.
Let
$$
\begin{array}{rclrcl}
\displaystyle
\mathcal{G}_1 &=
&\displaystyle \sum_{j=0}^{n-|J|} G^\downarrow_j\, ,
&\displaystyle \mathcal{X}_1 &=
&\displaystyle \sum_{j=0}^{n-|J|} \Xi_{a-1+\floor{n/2}-j}\, ;
\vspace{1mm}\\
\displaystyle \mathcal{G}_2 &= 
&\displaystyle G^\uparrow_0\, ,
&\displaystyle \mathcal{X}_2 &= 
&\displaystyle \sum_{x=b+\floor{n/2}}^{L-\ceil{n/2}} \Xi_x\, ;
\vspace{1mm}\\
\mathcal{G}_3 &=
&\displaystyle  \sum_{j=1}^{\floor{n/2}-1} G^\uparrow_j\, ;
\vspace{1mm}\\
\mathcal{G}_4 &=
&\displaystyle  G^\uparrow_{\floor{n/2}}\, ;
\vspace{1mm}\\
\mathcal{G}_5 &= 
&\displaystyle \sum_{j=\floor{n/2}+1}^{n-1} G^\uparrow_j\, ;
\vspace{1mm}\\
\mathcal{G}_6 &= 
&\displaystyle G^\uparrow_n\, ,
&\displaystyle \mathcal{X}_6 &= 
&\displaystyle \sum_{x=\floor{n/2}}^{a-1-\ceil{n/2}} \Xi_x\, .
\end{array}
$$
To prove the claim it suffices to prove 
$\|\mathcal{X}_i \mathcal{G}_j\| \leq O(q^{|J|})$ for 
$i \neq j$, and 
\begin{equation}
\label{AppB:1}
\| \mathcal{G}_1 \mathcal{X}_1 \mathcal{G}_1 
  - \sum_{j=0}^{n-|J|} G^\downarrow_j \cdot \Xi_{a-1+\floor{n/2}-j} 
  \cdot G^\downarrow_j \|
  \leq O(q^{|J|})\, .
\end{equation}
We will explain how this may be done now.

By our definition, each $\mathcal{G}_i$ may be written
$\sum_{k \in E_i} G^{\sigma_i}_k$, and each $\mathcal{X}_j$ 
may be written $\sum_{x \in F_j} \Xi_x$, for intervals
$E_i, F_j$, possibly empty, and $\sigma_i \in \{\uparrow,\downarrow\}$.
Thus, letting $\sigma = \sigma_i$,
\begin{eqnarray*}
&& (\mathcal{X}_j \mathcal{G}_i)^* (\mathcal{X}_j \mathcal{G}_i)
  \, =\,  \sum_{k,l \in E_i} \sum_{x,y \in F_j}
  G^\sigma_k \Xi_x \Xi_y G^\sigma_l \\
&& \quad =  \sum_{k,l \in E_i} \sum_{x,y \in F_j}
  G^\sigma_k \cdot
  \frac{\ket{\xi_{L,n}(x)}\bra{\xi_{L,n}(x)}}
  {\ip{\xi_{L,n}(x)}{\xi_{L,n}(x)}} \cdot
  \frac{\ket{\xi_{L,n}(y)}\bra{\xi_{L,n}(y)}}
  {\ip{\xi_{L,n}(y)}{\xi_{L,n}(y)}} \cdot
  G^\sigma_l \\
&& \quad =  \sum_{k,l \in E_i} \sum_{x,y \in F_j}
  G^\sigma_k \cdot
  \frac{\ket{G^\sigma_k \xi_{L,n}(x)}}{\|\xi_{L,n}(x)\|} \cdot
  \frac{\ip{\xi_{L,n}(x)}{\xi_{L,n}(y)}}
  {\|\xi_{L,n}(x)\| \cdot \|\xi_{L,n}(y)\|} \cdot
  \frac{\bra{G^\sigma_l \xi_{L,n}(y)}}{\|\xi_{L,n}(y)\|}
  \cdot G^\sigma_l
\end{eqnarray*}
Applying Cauchy-Schwarz we deduce that 
$$
\| \mathcal{X}_j \mathcal{G}_i \psi \|^2
  \leq \sum_{k,l \in E_i} \|G_k \psi\| \, \|G_l \psi\|  
  M^{j \sigma_i}_{kl}\, ,
$$
where 
$$
M^{j \sigma}_{kl} = \sum_{xy \in F_j}
  \frac{\|G^\sigma_k \xi_{L,n}(x)\|}{\|\xi_{L,n}(x)\|} \cdot
  \frac{|\ip{\xi_{L,n}(x)}{\xi_{L,n}(y)}|}
  {\|\xi_{L,n}(x)\| \cdot \|\xi_{L,n}(y)\|} \cdot
  \frac{\|G^\sigma_l \xi_{L,n}(y)\|}{\|\xi_{L,n}(y)\|}\, .
$$
Since the projections $G^\sigma_k$ are mutually orthogonal to
one another,
$$ 
\|\mathcal{G}_i \psi\|^2 = \sum_{k \in E_i} \|G_k^\sigma \psi\|^2\, .
$$
Thus,
$$
\| \mathcal{X}_j \mathcal{G}_i \psi \|^2
  \leq \|\mathcal{G}_i \psi\|^2 \cdot 
  \|(M^{j \sigma_i}_{kl \in E_i})_{kl}\|\, .
$$
Of course, $\|\mathcal{G}_i \psi\|^2 \leq \|\psi\|^2$,
because $\mathcal{G}_i$ is a projection.
So 
$$
\| \mathcal{X}_j \mathcal{G}_i \| \leq
\|(M^{j \sigma_i}_{kl})_{kl \in E_i}\|^{1/2}\, .
$$

We now discuss how to bound $\|(M^{j \sigma_i}_{kl})_{kl \in E_i}\|$.
We can bound the inner-product 
$\ip{\xi_{l,n}(x)}{\xi_{l,n}(y)}$ by \eq{App:result1}.
So
$$
M^{j \sigma}_{kl} 
  \leq \sum_{xy \in F_j} \frac{q^{n|x-y|}}{\pn{\infty}}
  \cdot \frac{\|G^\sigma_k \xi_{L,n}(x)\|}{\|\xi_{L,n}(x)\|} \cdot
  \frac{\|G^\sigma_l \xi_{L,n}(y)|}{\|\xi_{L,n}(y)\|}\, .
$$
Then, using the operator norm with respect $l^\infty$,
\begin{eqnarray*}
\|(M^{j \sigma}_{kl})_{kl \in E_i}\|
  &\leq& \|(M^{j \sigma}_{kl})_{kl \in E_i}\|_\infty \\
  &\leq& 
  \sup_{k \in E_i}\, \sum_{l \in E_i} \sum_{x,y \in F_j}
  \frac{q^{n|x-y|}}{\pn{\infty}}
  \cdot \frac{\|G^\sigma_k \xi_{L,n}(x)\|}{\|\xi_{L,n}(x)\|} \cdot
  \frac{\|G^\sigma_l \xi_{L,n}(y)|}{\|\xi_{L,n}(y)\|}\, .
\end{eqnarray*}
To proceed, we need to estimate 
$\|G^\sigma_l \xi_{L,n}(x)\|/\|\xi_{L,n}(x)\|$
for each $\sigma$, $l$ and $x$.
In fact, no estimation is required, we can perform the computation 
exactly.
Let us explain how this is done.
The operator $G^\sigma_l$ falls in the following class of projections.
Suppose we have some partition $\mathcal{P}$ of $[1,L]$,
composed of intervals $[x_{j-1}+1,x_j]$ where $0=x_0<x_1<\dots<x_r=L$,
and suppose we have a vector $\vec{n} = (n_1,\dots,n_r)$,
where $0 \leq n_j \leq x_j-x_{j-1}$ and $\sum_{j=1}^r n_j = n$.
Then we can define the projection 
$$
Q_{\mathcal{P},\vec{n}} := \prod_{j=1}^r Q_{[x_{j-1}+1,x_j],n_j}\, .
$$  
The operators $G^\sigma_l$ are of this form, where the partition
has three intervals $[1,a-1]$, $[a,b]$ and $[b+1,L]$, and 
$\vec{n} = (j,0,n-j)$ or $\vec{n}=(j,|J|,n-j-|J|)$, depending on whether
$\sigma$ is $\uparrow$ or $\downarrow$.
We can reduce the problem of computing 
$Q_{\mathcal{P},\vec{n}} \xi_{L,n}(x)$ to one of computing
$Q_{\mathcal{P}_1,\vec{n}_1} \psi^{+-}_{[1,x]}(\floor{n/2})$,
and
$Q_{\mathcal{P}_2,\vec{n}_2} \psi^{+-}_{[x+1,L]}(\ceil{n/2})$
for some partitions and vectors $\mathcal{P}_1$,$\mathcal{P}_2$,
$\vec{n}_1$ and $\vec{n}_2$.
To accomplish this, let $k$ be the integer such that $x_{k-1}+1 \leq x < x_k$.
Define the partition $\mathcal{P}'$ where $x_j'=x_j$ for $j<k$,
$x_k = x$, and $x'_j = x_{j-1}$ for $j>k$, and
define the $r+1$-vector $\vec{n}'$ by
$n'_j = n_j$ for $j <k$, $n'_k = \floor{n/2} - \sum_{j=1}^{k-1} n_j$,
$n_{k+1} = n_k - n'_k$, and $n'_j = n_{j-1}$ for $j > k+1$.
Since $\xi_{L,n}(x)$ has a definite number of downspins,
$\floor{n/2}$, to the left of $x$ and a definite number of 
downspins, $\ceil{n/2}$, to the right of $x+1$, the vector 
$Q_{\mathcal{P},\vec{n}} \xi_{L,n}(x)$ is the same as
$Q_{\mathcal{P}',\vec{n}'} \xi_{L,n}(x)$.
In fact, since $\xi_{L,n}(x) = \psi^{+-}_{[1,x]}(\floor{n/2}) \otimes 
\psi^{-+}_{[x+1,L]}(\ceil{n/2})$, we know
$$
Q_{\mathcal{P},\vec{n}} \xi_{L,n}(x) 
  = (Q_{\mathcal{P}_1,\vec{n}_1} \psi^{+-}_{[1,x]}(\floor{n/2}))
  \otimes (Q_{\mathcal{P}_2,\vec{n}_2} \psi^{-+}_{[x+1,L]}(\ceil{n/2}))\, ,
$$
where $\mathcal{P}_1$ is the partition consisting of the first $k$
parts of $\mathcal{P}'$,
$\mathcal{P}_2$ is the remainder partition, 
$\vec{n}_1 = (n'_1,\dots,n'_k)$ and 
$\vec{n}_2 = (n'_{k+1},\dots,n'_r)$.
Therefore,
$$
\frac{\|Q_{\mathcal{P},\vec{n}} \xi_{L,n}(x)\|}
{\|\xi_{L,n}(x)\|}
  = \frac{\|Q_{\mathcal{P}_1,\vec{n}_1} \psi^{+-}_{[1,x]}(\floor{n/2})\|}
  {\|\psi^{+-}_{[1,x]}(\floor{n/2})\|} \cdot
  \frac{Q_{\mathcal{P}_2,\vec{n}_2} \psi^{-+}_{[x+1,L]}(\ceil{n/2})\|}
  {\|\psi^{-+}_{[x+1,L]}(\ceil{n/2})\|}\, .
$$
We now present the formula for the two quantities on the right-hand-side 
of the equation.

The key to the computation is the decomposition formulae of
\eq{App:+-coprod} and \eq{App:-+coprod}.
These have trivial generalizations.
Specifically, for $x_0 < x_1 < \dots <x_r$, 
\begin{eqnarray}
\label{CB+-:gen}
\psi^{+-}_{[x_0+1,x_r]}(n)
  &=& \sum_{\substack{(n_1,\dots,n_r) \\ n_1+\dots+n_r=n}}
  q^{n x_r - (n_1 x_1 + \dots n_r x_r)}
  \bigotimes_{j=1}^r \psi^{+-}_{[x_{j-1}+1,x_j]}(n_j) \\
\label{CB-+:gen}
\psi^{-+}_{[x_0+1,x_r]}(n)
  &=& \sum_{\substack{(n_1,\dots,n_r) \\ n_1+\dots+n_r=n}}
  q^{(n_1 x_0 + \dots n_r x_{r-1}) - n x_0}
  \bigotimes_{j=1}^r \psi^{-+}_{[x_{j-1}+1,x_j]}(n_j) 
\end{eqnarray}
From this one can easily calculate
\begin{eqnarray}
\label{projform+-}
\frac{\|Q_{\mathcal{P},\vec{n}} \psi^{+-}_{[x_0+1,x_r]}(n)\|^2}
{\|\psi^{+-}_{[x_0+1,x_r]}(n)\|^2}
  = \frac{\prod_{j=1}^r \qbinom{x_j - x_{j-1}}{n_j}{q^2}}
  {\qbinom{x_r-x_0}{n}{q^2}}\, 
  q^{\sum_{j=1}^r n_j(2(x_r-x_j)-(n-n_j))} \\
\label{projform-+}
\frac{\|Q_{\mathcal{P},\vec{n}} \psi^{-+}_{[x_0+1,x_r]}(n)\|^2}
{\|\psi^{-+}_{[x_0+1,x_r]}(n)\|^2}
  = \frac{\prod_{j=1}^r \qbinom{x_j - x_{j-1}}{n_j}{q^2}}
  {\qbinom{x_r-x_0}{n}{q^2}}\, 
  q^{\sum_{j=1}^r n_j(2(x_{j-1}-x_0)-(n-n_j))} \\
\end{eqnarray}
We notice the following interesting fact.
The exponent of $q$ in the formulas above has the following 
interpretation.
The most probable locations of the downspins for 
kink state $\psi^{+-}_{[1,L]}(n)$ are in the interval $[L+1-n,L]$.
Suppose we place marbles in these places and ask for
the minimum transport required to move these marbles
so that $n_j$ of the marbles lie in the 
bin $[x_{j-1}+1,x_j]$ for each $j$.
Then this is precisely the exponent of $q$ in \eq{projform+-}.
To state this in symbols
\begin{eqnarray*}
&&\sum_{j=1}^r n_j(2(x_r - x_1) - (n-n_j)) = 
\min\{\sum_{x=1}^L |f(x) - x| : f \in \textrm{Perm}([1,L]), \\ 
&&\qquad \#\big(f([L+1-n,L]) \cap [x_{j-1}+1,x_j]\big) = n_j, j=1,\dots,r\} 
\end{eqnarray*}
The exponent of $q$ in \eq{projform-+} has a similar interpretation,
except that the marbles initially occupy the sites of
$[1,n]$ instead of $[L+1-n,L]$.

Having said how one can perform the computations of 
$\|G^\sigma_j \xi_{L,n}(x)\|$, we now state our results.
The following notation is convenient:
$$
\EXP{*}{L,n,x} := \frac{\ip{\xi_{L,n}(x)}{* \xi_{L,n}(x)}}
  {\ip{\xi_{L,n}(x)}{\xi_{L,n}(x)}}\, .
$$
This is the expectation value of an observable with respect to
the droplet state $\xi_{L,n}(x)$. 
\begin{itemize}
\item
If $0 \leq x \leq a-1$ and $\sigma = \uparrow$ let 
$r = a-1-x-j+\floor{n/2}$. Then
$$
\EXP{G^\uparrow_j}{L,n,x}
  = \frac{\qbinom{a-1-x}{r}{q^2} \qbinom{L-b}{n-j}{q^2}}
  {\qbinom{L-x}{\ceil{n/2}}{q^2}} q^{2(n-j)(|J|+r)}\, .
$$
We make the convention that 
$$
\qbinom{n}{k}{q^2} = 0
\quad
\textrm{if}
\quad k<0 
\quad
\textrm{or}
\quad k>n\, .
$$
Thus the formula above is zero unless $0\leq r\leq a-1-x$.
\item
If $0 \leq x \leq a-1$ and $\sigma = \downarrow$ let 
$r = a-1-x-j+\floor{n/2}$. Then
$$
\EXP{G^\downarrow_j}{L,n,x}
  = \frac{\qbinom{a-1-x}{r}{q^2} \qbinom{L-b}{n-j-|J|}{q^2}}
  {\qbinom{L-x}{\ceil{n/2}}{q^2}} q^{2(n-j)r}\, .
$$
\item
If $a \leq x \leq b$ and $\sigma = \uparrow$, 
the answer is zero unless $j = \floor{n/2}$, and
$$
\EXP{G^\uparrow_{\floor{n/2}}}{L,n,x}
  = \frac{\qbinom{a-1}{\floor{n/2}}{q^2} \qbinom{L-b}{\ceil{n/2}}{q^2}}
  { \qbinom{x}{\floor{n/2}}{q^2} \qbinom{L-x}{\ceil{n/2}}{q^2} }
  q^{2[\floor{n/2}(x-a+1) + \ceil{n/2}(b-x)]}\, .
$$
\item
If $a \leq x \leq b$ and $\sigma = \downarrow$, the answer is zero
unless $j = \floor{n/2}-x+a-1$, and
$$
\EXP{G^\downarrow_{\floor{n/2}}}{L,n,x}
  = \frac{\qbinom{a-1}{x-\floor{n/2}}{q^2} \qbinom{L-b}{L-x-\ceil{n/2}}{q^2}}
  { \qbinom{x}{\floor{n/2}}{q^2} \qbinom{L-x}{\ceil{n/2}}{q^2} }\, .
$$
\item
If $b+1 \leq x \leq L$ and $\sigma = \uparrow$, let 
$r = x-b-\floor{n/2}+j$.
Then
$$
\EXP{G^\uparrow_j}{L,n,x}
  = \frac{\qbinom{a-1}{j}{q^2} \qbinom{x-b}{r}{q^2}}
  {\qbinom{x}{\floor{n/2}}{q^2}} q^{2j(|J|+r)}\, .
$$
\item
If $b+1\leq x\leq L$ and $\sigma = \downarrow$, let
$x-a+1-\floor{n/2}+j$.
Then
$$
\EXP{G^\uparrow_j}{L,n,x}
  = \frac{\qbinom{a-1}{j}{q^2} \qbinom{x-b}{r}{q^2}}
  {\qbinom{x}{\floor{n/2}}{q^2}} q^{2j(|J|+r)}\, .
$$
\end{itemize}

The rest of the computations proceed directly from these observations.
Note that each $q^2$-binomial coefficient can be bounded above
by $\pn \infty ^{-1}$, but one should remember to restrict the
indices $j$ and $x$ to those for which none of the 
$q^2$-binomial coefficients vanish.
Our results are the following:
\begin{itemize}
\item
As mentioned above, it is easy to check that 
$$
\mathcal{X}_1 \mathcal{G}_2 
  = \mathcal{X}_1 \mathcal{G}_6 
  = \mathcal{X}_2 \mathcal{G}_6 
  = \mathcal{X}_2 \mathcal{G}_5 
  = \mathcal{X}_6 \mathcal{G}_2 
  = \mathcal{X}_6 \mathcal{G}_3
  = 0\ .
$$
Simply put, if one consults the formulae in the paragraph,
each of the products above is composed of $\Xi_x G^\sigma_j$
for which the $q$-binomial coefficients vanish.
\item
A simultaneous bound for $\|\mathcal{X}_1 \mathcal{G}_3\|^2$
and $\|\mathcal{X}_1 \mathcal{G}_5\|^2$ is $C(q) q^{2|J|}$,
where
$$
C(q) = \frac{2+8q}{(1-q)^4 \pn{\infty}^3}\, .
$$
\item 
$$\|\mathcal{X}_1 \mathcal{G}_4\|^2
  \leq \frac{1}{\pn{\infty}^3}
  \left(|J| + \frac{1+q^{\floor{n/2}}}{1-q^{\floor{n/2}}}\right)^2
  q^{2 |J| \floor{n/2}}\, .
$$
\item
We bound $\|\mathcal{X}_2 \mathcal{G}_1\|^2$
and $\|\mathcal{X}_6 \mathcal{G}_1\|^2$, simultaneously,
by $C(q) q^{2(|J|-1)^2}$, where
$$
C(q) = \frac{1}{\pn{\infty}^3(1-q^{|J|})^2 (1-q^{2(|J|-1)})}\, .
$$
The reason the bound is so small is that it is actually equal to zero,
if $|J| > n$,
as can be understood by counting downspins to the left and right of
$x$.
\item
Both 
$\|\mathcal{X}_2 \mathcal{G}_3\|^2$ and
$\|\mathcal{X}_6 \mathcal{G}_5\|^2$ can each be bounded
by $C(q) q^{2|J|}$, where
$$
C(q) = \frac{q^2}{\pn{\infty}^3 (1 - q)^2 (1 - q^{|J|+2})}\, .
$$
\item
Both $\|\mathcal{X}_2 \mathcal{G}_4\|^2$ and
$\|\mathcal{X}_6 \mathcal{G}_4\|^2$ can each be bounded
by 
$$
\frac{1}{\pn{\infty}^3} \left( \frac{1+q^{2\ceil{n/2}}}
  {1 - q^{2\ceil{n/2}}} + \frac{1+q^{2\floor{n/2}}}{1-q^{2\floor{n/2}}}
  \right)
q^{4 \ceil{n/2} (|J| + \ceil{n/2})}\, .
$$
\end{itemize}
That accounts for all of the necessary computations except one,
which we now carry out.

We show in this paragraph that
\begin{equation}
\label{example}
\left\|\mathcal{G}_1\mathcal{X}_1 \mathcal{G}_1 - 
  \sum_{j=0}^{n-|J|} G^\downarrow_{j} 
  \Xi_{a-1+\floor{n/2}+j} G^\downarrow_{j}
  \right\| \leq \frac{4 q^{|J|}}{\pn{\infty}^3 (1 - q^{|J|})^2}\, .
\end{equation}
In this case we can define $x_j = a-1+\floor{n/2}+j$, 
for each $0 \leq j \leq n-|J|$, and we have
$$
\EXP{G_j^\downarrow}{L,n,x} 
  \leq \frac{1}{\pn{\infty}} q^{|J|\cdot|x-x_j|}\, .
$$
This is understood because $|x-x_j|$ downspins must be moved all the way 
across the droplet in order to change the basic interval for
$\xi(x)$ into a state compatible with $\mathcal{G}^\downarrow_j$.
Thus, proceeding in the same way as before, we obtian
$$
\left\|\mathcal{G}_1\mathcal{X}_1 \mathcal{G}_1 - 
  \sum_{j=0}^{n-|J|} G^\downarrow_{j} 
  \Xi_{a-1+\floor{n/2}+j} G^\downarrow_{j}
  \right\| \leq \|\mathcal{M}\|\ .
$$
where $\mathcal{M}_{jj}=0$ for each $j$, and
$$
\mathcal{M}_{jk} \leq \frac{1}{\pn{\infty}^2} \sum_{x\in I_2}
  q^{|J|\cdot |x-x_j| + |J|\cdot |x-x_k|}
$$
when $j \neq k$.
By extending the indices $x$ to cover all integers, and
by translating so that $x_j$ is the new origin of $x$,
we have
$$
\mathcal{M}_{jk} \leq \frac{1}{\pn{\infty}^3} \sum_{x}
  q^{|J|\cdot |x| + |J|\cdot |x+j-k|}\, .
$$
The series is easily calculated as
$$
\sum_{x}
  q^{|J|\cdot |x| + |J|\cdot |x+j-k|}
  = q^{|J|\cdot|j-k|} \left(|j-k| + \frac{1+q^{2|J}}{1-q^{2|J|}}\right)\, .
$$
So, for any fixed $j$, we have
$$
\sum_{\substack{k\in \Ir \\ k \neq j}} \mathcal{M}_{jk}
  \leq \frac{2}{\pn{\infty}^2}
  \sum_{l=1}^\infty 
  q^{|J|l} \left(l + \frac{1+q^{2|J}}{1-q^{2|J|}}\right)\, .
$$
This sum is then easily computed as
$$
\sum_{l=1}^\infty 
  q^{|J|l} \left(l + \frac{1+q^{2|J|}}{1-q^{2|J|}}\right)
  = \frac{4 q^{|J|}}{(1-q^{|J|})^2}\, .
$$
From this we obtain \eq{example}.  


\newpage
\pagestyle{myheadings} 
\markright{  \rm \normalsize CHAPTER 5. \hspace{0.5cm} 
Bounds on the spectral gap for the $d$-dimensional model : $d > 1$}
\chapter{Bounds on the spectral gap for the $d$-dimensional model : $d > 1$}
\thispagestyle{myheadings}

In this chapter we present work on interface states in dimensions greater than 1.
It is a general theorem that in the thermodynamic limit, the spectral gap
above the interface states must vanish for all dimensions greater than 1.
This was originally presented in \cite{KN2} to prove that there is no 
spectral gap in two-dimensions, and later in \cite{Mat} it was repeated to 
prove that there are gapless excitations also for all dimensions greater than 2.
The proof is based on the Goldstone theorem as presented in \cite{LPW} and exploits
the continuous $\textrm{U}(1)$ symmetry-breaking of the interface ground states
in the phenomena sometimes known as ``phase-locking''.
Although by these general techniques, it is known that the spectral gap must vanish
in all dimensions greater than one, the exact rate for the vanishing of the gap
was still open. 
Also, we were interested in determining which variational states, from a specified
submanifold which is orthogonal to the ground states, minimizes the energy.
In the two papers which will be presented next, we solve the problem in the sense 
that we prove an upper-bound for the spectral gap which vanishes like $O(1/R^2)$,
and we determined the states, from among some small-dimensional class of
variational states, which give the lowest excitations.
To set-up the problem a little, we note that  the spectral gap above an 
infinite-volume ground state,
$\omega$, is defined to be the largest constant $\gamma\geq 0$ such that
$$
\langle{X^* H^3 X}\rangle =\langle{X^* [H^3,X]}\rangle 
  \geq \gamma \langle{X^*[H^2,X]}\rangle 
  = \gamma \langle{X^* H^2 X}\rangle
$$
for all local observables $X$, where $H=H_{\textrm{GNS}}$ is the densely-defined 
self-adjoint operator
obtained from the GNS representation, and $\langle{.}\rangle$ is the expectation with
respect to the GNS vector $\Omega_{\textrm{GNS}}$.
One may ask what is $\gamma(R)$, the largest number such that the above inequality
holds for any 
$$
X \in \bigcup_{\substack{\Lambda \subset \Ir^d\\
\textrm{diameter}(\Lambda) \leq R}} \Obs_\Lambda\, .
$$
It is shown, in the two papers to follow, that $\gamma(R)$ is at largest $O(R^{-2})$.
Recently Pietro Caputo and Fabio Martinelli \cite{CapMart}
have obtained lower bounds for the spectral gap
in a cylinder of size $L$ which behave as $1/L^2$, which shows that our upper-bounds are
of the correct order, at least.
The results of Caputo and Martinelli are very interesting also because they map the
quantum spin system onto a Markov process and use results from that field \cite{CM}
to bound the spectral gap.
Also their results are true uniformly in $n$, the total number of down-spins,
although we point out that the dependence of the gap on $n$, in particular through the 
filling factor of down spins in the interface plane, is an interesting point of our
results.

There are two main ideas in the proofs of the papers to follow.
The first is that due to the broken continuous symmetry, we can introduce
a continuous perturbation to the ground state, analagous to a spin-wave.
Only, because there is broken translation symmetry in the 111 direction, the
spin wave must be localized to a small region about the interface plane.
In other words, the wave is effectively a surface wave on the plane of the interface.
The second idea is to prove an \textit{equivalence of ensembles} result relating
two different types of ground state: the ground states with a fixed density of down
spins, and ground states with a fluctuating number of down spins, which are
exponential generating functions for the former.
We call the first type of ground state the canonical ground states, and the second
type the grand-canonical ground states, in analogy with classical statistical mechanics
at nonzero temperature.
The first paper, which occupies Sections 5.2--5.5
does not involve equivalence of ensembles at all.
Because the grand-canonical ensemble is closer to a classical system (since
the states one considers are simple-tensor states, and the manifold of all such 
states is $(\Cx P^1)^{|\Lambda|} \cong (S^2)^{|\Lambda|}$)
the idea of the surface wave perturbation is more clearly 
explained in terms of a standard differential equation.
The disadvantage is that the perturbation produced is not shown to be orthogonal to
all ground states of the XXZ model, only to those in a particular tangent plane of the 
two-sphere of grand-canonical ground states.
Therefore, one does not obtain a rigorous bound on the spectral gap.
In Sections 5.6--5.11, a second paper is presented
for calculating the spectral gap in the canonical ensemble.
The second paper is self-contained, and has the benefit of containing an 
equivalence of ensembles result which should have further applications.
Also the bound on the spectral gap is rigorous, and one obtains information on how 
the spectral gap depends on the partial filling of the interface
plane for different numbers of down-spins.

There is one fault with the result of Sections 5.6--5.11.
For purely technical reasons, the Equivalence of Ensembles result presented there
is only proved for dimensions three and higher, even though it is true for
two-dimensions as well.
(It cannot be true for one-dimension, as is well-known.)
The same proof from the paper works in dimension two, if it is considered more carefully.
This is what we do in Sections 5.12 and 5.13.
In Section 5.12, we give an alternative to the Local-Central-Limit-Theorem proof
of activity bounds.
Our proof relies on the method of steepest descents, and more specifically Hayman's method
(c.f.\ \cite{Wilf} for a nice elementary description).
This same method is also used to prove the local central limit theorem, so one may ask
what is the point?
The reason for appealing to the more basic technique is to obtain nearly optimal error
bounds which are needed for section 5.13.
Section 5.13 is a terse derivation of an inequality analogous to 
Theorem \ref{thm:eqv}, which is the first EOE result, but better.
With this fix, one can proceed through the remaining arguments of Sections 5.9 and
5.10, applying it to two-dimensions instead of three.
We leave this as an exercise for the enthusiastic reader.

\pagebreak

{\baselineskip=10pt \thispagestyle{empty} {{\small 
Originally Pulished: Electronic Journal of Differential Equations, Conf.\ 04, 1--10. (2000)\\
\indent arXiv:cond-mat/9909018} 
\hspace{\fill}}

\vspace{20pt}

\begin{center}
\addcontentsline{toc}{chapter}{\textit{A continuum approximation for the excitations
of the $(1,1,\dots,1)$  interface in the quantum Heisenberg model}}
{\LARGE \bf A continuum approximation for the excitations
of the $(1,1,\dots,1)$
interface in the quantum Heisenberg model\\[27pt]}
{\large \bf Oscar Bolina, Pierluigi Contucci, Bruno Nachtergaele 
and Shannon Starr\\[10pt]}
{\large  Department of Mathematics\\
University of California, Davis\\
Davis, CA 95616-8633, USA\\[15pt]}
{\normalsize bolina@math.ucdavis.edu, contucci@math.ucdavis.edu, 
bxn@math.ucdavis.edu, sstarr@math.ucdavis.edu}\\[30pt]
\end{center}

\noindent
{\bf Abstract:}
It is shown that, with an appropriate scaling, the energy of low-lying 
excitations of the $(1,1,\dots,1)$ interface in the $d$-dimensional
quantum Heisenberg model are given by the spectrum of the $d-1$-dimensional 
Laplacian on an suitable domain.

\vspace{8pt}
{\small \bf Keywords:} Anisotropic Heisenberg ferromagnet, XXZ model,
interface excitations, 111 interface.
\vskip .2 cm
\noindent
{\small \bf MCS2000 numbers:} 82B10, 82B24, 82D40 
\vfill
\hrule width2truein \smallskip {\baselineskip=10pt \noindent Copyright
\copyright\ 1999 Bolina, Contucci, Nachtergaele, and Starr. Reproduction of 
this article in its entirety, by any means, is permitted for non-commercial 
purposes.\par }}

\newpage

\Section{Introduction and main results}

We consider the spin 1/2 XXZ Heisenberg model on the $d$-dimensional
lattice $\Ir^d$. For any finite volume $\Lambda\subset\Ir^d$, the 
Hamiltonian is given by 
\be
H_\Lambda = - \sum_{\substack{x,y\in\Lambda\\ \vert x-y\vert=1}}
\Delta^{-1} (S_x^{(1)} S_y^{(1)} + S_x^{(2)} S_y^{(2)}) 
+ S_x^{(3)} S_y^{(3)},
\ee
where $\Delta>1$ is the anisotropy. We refer to the next section for more
precise definitions. By adding an appropriate boundary term one can 
insure that the ground states of this model describe an interface
in the $(1,1,\dots,1)$ direction between two domains with opposite magnetization.
For a particular choice of boundary term, the model has exactly one
ground state $\psi_n$ for each fixed number of down spins, $n$. We call these
the canonical ground states. In analogy with statistical mechanics of
particle systems one can introduce the grand canonical ground states
of the form
$$
\Psi =\sum_n z^n \psi_n
$$
It turns out that these states are inhomogeneous product states
\cite{GW}. In this paper, we consider a class of perturbations of these
product states, of which we calculate the energy. By the variational 
principle this leads to bounds for the energy of the first excited state
of the model. As the excitation spectrum above the interface states
is gapless \cite{KN2, Mat2}, this bound should vanish as the volume 
tends to infinity. This is indeed the case (see \eq{energy:norm}).

The perturbations we consider are in correspondence with functions 
$f : \Lambda \to \Cx$.
Furthermore, we consider functions which are slowly-varying
in all directions perpendicular to $(1,1,\dots,1)$ though they may have 
discrete jumps parallel to this direction.
In other words $\|\nabla f \cdot v\|_\infty \ll \|f\|_\infty$ for all
$v \perp (1,1,\dots,1)$.
We consider general perturbations of this type and
conclude that the optimal perturbations, in the sense of minimizing energy,
are localized near the interface.
With this restriction, the Hamiltonian, projected to and restricted to the 
appropriate subspace, is just the Laplacian

This result may be compared to the recent bound of
\cite{BCNS}.
The main dif\-ference is that there we considered a canonical ensemble,
for which there were a fixed number of down-spins (hence a fixed number
of up-spins).
We developed a version of equivalence of ensembles whereby we
estimated the canonical expectation of a gauge invariant observable
by a grand canonical expectation, provided that the interfaces of the 
canonical and grand canonical states occupied the same position.

In the present paper, we begin with the grand canonical ensemble, 
so that we make no reference to equivalence of ensembles.
Specifically, we consider a cylindrical region of total height $L+1$
and whose cross-section is a region $\Omega_R$ with linear size $R$.
Then a class of excitations is parametrized by smooth functions $\Phi$
on a fixed domain $\Omega = R^{-1} \Omega_R$.

{\bf Main Result:}
{\em Excitations on $\Lambda$ have a normalized energy
\be
\label{energy:norm}
\frac{\op{\psi^f}{H}{\psi^f}}{\ip{\psi^f}{\psi^f}}
  \approx\frac{1}{2 \Delta R^2} \cdot 
  \frac{\|\nabla\Phi\|^2_{L^2(\Omega)}}{\|\Phi\|^2_{L^2(\Omega)}}
  \cdot g(\Delta, \mu)
\ee
where 
$$
g(\Delta,\mu)=\frac{\sum_{l=-L/2}^{L/2-1} \sech(\alpha[l-\mu]) 
  \sech(\alpha[l+1-\mu])}{\sum_{l=-L/2}^{L/2} \sech(\alpha[l-\mu]) 
  \sech(\alpha[l-\mu])}.
$$
Here, $\mu$ is a real parameter of the grand canonical ground
state describing the location of the interface between the regions 
of homogeneous up and down spins.
As $\mu \to -\infty$, the ground state has all spins up, 
and for $\mu \to \infty$, all spins are down.
For all $\mu\in \Rl$, and sufficiently large $L$, $g$ satisfies the bounds
$$
\frac{1}{2\Delta}\leq \Delta-\sqrt{\Delta^2-1}\leq g(\Delta,\mu)\leq 1
$$}

\begin{rem}{}
The normalized energy of \eq{energy:norm}
is exactly the same as that for the Laplacian.
Equating the first variation to zero, we see that
the local extrema of the normalized energy are precisely the solutions of
$\nabla^2 \Phi = - \lambda \Phi$ (here $\nabla^2$ is the Laplacian),
and $\lambda = \|\nabla\Phi\|^2_{L^2(\Omega)}/\|\Phi\|^2_{L^2(\Omega)}$.
The space of excitations we consider does not form an invariant subspace
of $H$,
so that the eigenvectors of the Laplacian are not truly eigenvectors of $H$.
But, using the variational inequality, we see that the spectral gap of
$H$ is bounded thus:
$$
\gamma_1 \leq 
  \frac{\lambda_1}{2 \Delta R^2} \cdot g(\Delta, \mu) (1 + O(\frac{1}{R^2})),
$$
where $\lambda_1$ is the first positive eigenvalue of $-\nabla^2$ with
Dirichlet boundary conditions on the domain $\Omega$.
\end{rem}

\Section{The Spin-$\frac{1}{2}$ Heisenberg XXZ Ferromagnet}

A quantum spin model, such as the Heisenberg XXZ ferromagnet,
is defined in terms of a family of local Hamiltonians $H_\Lambda$, acting
as self-adjoint linear operators on a Hilbert space $\HH_\Lambda$. This
family is parametrized by finite subsets $\Lambda\subset \Ir^d$. 

We choose $\Lambda$ to be ``cylindrical'' in the following sense:
Let $\{e_j\}_{j=1}^d$ be the set of coordinate unit vectors
and define the vector $e_* = \sum_{j=1}^d e_j = (1,1,\dots,1)$,
which is the axial direction for the cylinder.
Define the functional  $l(x) = x \cdot e_* = \sum_{j=1}^d x^j$,
where $x = \sum_{j=1}^d x^j e_j$.
Observe that the kernel of $l$ in $\Ir^3$ is a $(d-1)$-dimensional 
sublattice perpendicular to the axial direction.
Take for the base of $\Lambda$ a finite subset of this $(d-1)$-dimensional 
sublattice, and call it $\Gamma$. 
A discrete approximation to the line of all scalar mutliples of $e_*$
is the \textit{one-dimensional stick} $\Sigma$.
$\Sigma$ is a bi-infinite sequence of points $\{x_n\}_{n=-\infty}^\infty$
such that $x_0=0$ and all other points $x_n$ are specified by the relation
$x_n - x_{n-1} = e_{n \mod d}$.
So 
\begin{eqnarray*}
\Sigma &=& \{\ldots,-(e_d+e_{d-1}+\cdots+e_1+e_d), -(e_d+e_{d-1}+\cdots+e_1), 
  \ldots, -e_d, \\
  &&  0, e_1, (e_1+e_2), \ldots,
  (e_1+e_2+\cdots+e_d), (e_1+e_2+\cdots+e_d+e_1), \ldots\}.
\end{eqnarray*}
A finite stick of length $L+1$, where $L$ is even, is 
$\Sigma_{L} = \{x \in \Sigma : -L/2 \leq l(x) \leq L/2\}$.
Now define $\Lambda$ to be the translates of $\Gamma$ along $\Sigma_{L}$,
i.e.\ 
\be
\Lambda = \Gamma + \Sigma_L = \{x + y : x\in \Gamma, y \in \Sigma_L\}.
\ee

Let us now define \textit{nearest neighbors} to be points $x,y \in \Ir^d$
such that $|l(x) -  l(y)|=1$ and $\|x - y\|_{l^1} = 1$.
Also, we define oriented bonds between nearest neighbors as 
ordered pairs $(x,y)$ satisfying $l(y) = l(x) + 1$
and $\|x - y\|_{l^1} = 1$.
Hence $\{(x,x + e_j)\}_{j=1}^d$ is the set of all
oriented bonds with lower point $x$. 
The collection of all oriented bonds with both points in $\Lambda$,
will be called $B(\Lambda)$.

The local Hilbert spaces are $\HH_\Lambda = (\Cx^2)^{\otimes |\Lambda|}$.
Each copy of $\Cx^2$ comes with an ordered basis $(\up,\down)$
and a spin-$\frac{1}{2}$ representation
of $SU(2)$ defined by the Pauli matrices:
\be
S^{(1)} = \left(\begin{array}{cc} 0 & 1/2 \\
1/2 & 0 \end{array}\right),\quad
S^{(2)} = \left(\begin{array}{cc} 0 & -i/2 \\
i/2 & 0 \end{array}\right),\quad
S^{(3)} = \left(\begin{array}{cc} 1/2 & 0 \\
0 & -1/2 \end{array}\right).		
\ee
(So, for example, $S^{(3)} \up = \frac{1}{2} \up$ and
$S^{(3)} \down = -\frac{1}{2} \down$.)
We consider a family of Hamiltonians parametrized by a real
number $\Delta \geq 1$.
In order to define the total Hamiltonian, we first define 
local Hamiltonians $h_{x y}$ for each oriented bond $(x,y)$:
\be
h_{x,y} = - \Delta^{-1} (S_{x}^{(1)} S_{y}^{(1)}
+ S_{x}^{(2)} S_{y}^{(2)}) - S_{x}^{(3)} S_{y}^{(3)} 
+ \frac{1}{4} + \frac{1}{4} A(\Delta) (S_{y}^{(3)} - S_{x}^{(3)}),
\ee
where $A(\Delta) = \frac{1}{2} \sqrt{1 - 1/\Delta^2}$.
The total Hamiltonian is
\be\label{hamCCC}
H_\Lambda = \sum_{(x,y)\in B(\Lambda)} h^{q}_{x,y}.
\ee
$\Delta$ parametrizes ``anisotropic coupling''.
The case $\Delta = 1$ is the isotropic model, also known as the 
Heisenberg XXX ferromagnet, which exhibits $SU(2)$ symmetry 
(because $H_\Lambda$ commutes with $S^1$, $S^2$ and $S^3$).

We find it convenient to introduce a positive constant
$\alpha$, which solves $\Delta = \cosh(\alpha)$.
We note that the nearest neighbor interaction $h_{x y}$ is an 
orthogonal projection 
\be
h_{x y} = \ket{\xi_{x y}} \bra{\xi_{x y}} 
  \otimes \unity_{\Lambda \setminus (x,y)},
\ee
where
\be
\xi_{x y} = \frac{e^{-\alpha/2} \du - e^{\alpha/2} \ud}
  {\sqrt{2 \cosh(\alpha)}}. \label{def:xiCCC}
\ee
This also shows that each $h_{x y}$ is a  nonnegative
self-adjoint operator, hence $H_\Lambda$ is, as well.
To simplify the notation we will often drop the subscript 
$\Lambda$ when the volume is obvious from the context.

\Section{Ground States and a Perturbation}

The ground states of the XXZ ferromagnet can be calculated exactly \cite{ASW}.
We will choose a particular ground state and construct  an orthogonal subspace
(but not the entire orthogonal complement)  which is parametrized by
$H^1$-functions on a compact domain  $\Omega_0 \subset \Rl^{d-1}$. 
The inner product becomes approximately the $L^2$ inner-product and the  
orthogonal projection of the Hamiltonian is approximately the Laplacian.

The lowest eigenvalue for $H$, which is zero, has a 
$(|\Lambda|+1)$-fold degeneracy in the eigenspace.
This space of ground states is spanned by the simple tensor ground states,
which we will call grand canonical states.
Specifically, let $z$ be any complex number, and $\mu = \real(z)$.
Define the vector
\be
v_x(z) = \frac{e^{\alpha (l_x-z)/2} \up 
  + e^{-\alpha (l_x-z)/2} \down}
  {\sqrt{2 \cosh(\alpha[l_x - \mu])}} ,
\ee
for each site $x \in \Lambda$.
We define the product of these vectors
\be
\psi_0(z) = \bigotimes_{x \in \Lambda} v_x(z) ,
\ee
and we may quickly establish that it is a ground state.
Indeed, the oriented bonds are defined between points $x$ and $y$ with 
$l(y) = l(x) + 1$,
from which we see 
\be
\ip{\uparrow \downarrow}{v_{x}(z) \otimes v_{y}(z)} 
  = e^{\alpha} \ip{\downarrow \uparrow}{v_{x}(z) \otimes v_{y}(z)}.
\ee
This implies $v_{x}(z) \otimes v_{y}(z)$ is orthogonal to $\xi_{x y}$,
for each $(x,y) \in B(\Lambda)$, which proves that 
$\psi_0(z)$ is a ground state.
As we have said, the states $\psi_0(z)$ span the entire ground state
space, as $z$ ranges over all the complex numbers \cite{GW}. 
(More than this can be said.
The simple tensor ground states are parametrized by elements of 
$\Cx P^1$, so that the submanifold of all such states in $\HH$
is topologically a sphere.
But to obtain the north and south poles of the sphere, it is necessary to
take the limits $z \to \infty$ and $z \to -\infty$.) 

Let us now fix $z$, and for simplicity we will just write 
$\psi_0$ and $v_x$ without explicit reference to $z$.
For each site $x$ we define a vector orthogonal to $v_x$, 
\be
w_x = \frac{e^{-\alpha (l_x-\bar{z})/2} \up 
  - e^{\alpha (l_x-\bar{z}/2} \down}
  {\sqrt{2 \cosh(\alpha[l_x - \mu])}} .
\ee
We will make use of $w_x$ to define an orthonormal system
of states
\be
\psi^x = w_x \otimes \bigotimes_{y \in \Lambda \setminus x}
  v_y ,
\ee
where $x$ ranges over $\Lambda$.
Each of these states is also orthogonal to $\psi_0$,
let us call their span $V$. An arbitrary state in $V$ is
characterized by a function $f:\Lambda \to \Cx$.
Explicitly, $\psi^f = \sum_{x \in \Lambda} f(x) \psi^x$.
It is then clear that
$\ip{\psi^f}{\psi^g} = \sum_{x \in \Lambda} \overline{f(x)} g(x)$.

Our interest is the case that $\Lambda \nearrow \Ir^d$,
i.e.\ the thermodynamic limit.
In terms of $v_x$ and $w_x$, we see that
the local interaction $h_{x y}$ 
describes a nearest-neighbor interaction.
It may be interpreted as a bilinear form, which is a first order 
finite-difference operator in each variable.
To be clear, a straightforward calculation gives
\begin{eqnarray}
\label{local:en:exact}
\op{\psi^f}{h_{x y}}{\psi^g}
  &=& \frac{1}{2} \sech(\alpha) \sech(\alpha[l_{x}-\mu]) 
  \sech(\alpha[l_{y}-\mu]) \nonumber \\
  && \times \big(\cosh(\alpha[l_{y}-\mu]) \overline{f(y)}
  - \cosh(\alpha[l_{x}-\mu]) \overline{f(x)}\big) \nonumber \\
  && \times \big(\cosh(\alpha[l_{y}-\mu]) g(y)
  - \cosh(\alpha[l_{x}-\mu]) g(x)\big) .
\end{eqnarray}
Recall that $\mu = \real(z)$) and the energy is
\be
\label{total:en}
\op{\psi^f}{H}{\psi^g}
  = \sum_{l = -L/2}^{L/2-1}\ \sum_{x \in \Gamma_l}\ \sum_{j=1}^d
  \op{\psi^f}{h_{x,x+e_j}}{\psi^g},
\ee
where $\Gamma_l$ refers to the set of points $x \in \Lambda$
with $l(x) = l$.
In the thermodynamic limit, 
we may scale the plane $e_*^\perp = \{v \in \Rl^d : v \cdot e_* = 0\}$
so that $H$ becomes, to first order, a differential 
operator with respect to each direction of the plane.
However, the inhomogeneity in the $e_*$ direction admits no such scaling
for that coordinate, so that $H$ is genuinely a finite-difference operator
even in the thermodynamic limit.

This intuitive description of the last paragraph
is made precise, now.
Let $\Omega$ be a bounded, open subset of $e_*^\perp$
with a $C^1$ boundary.
Let $\Omega_R$ be the dilation $R \cdot \Omega = \{R x : x \in \Omega\}$,
and let $\Gamma = \Omega_R \cap \Ir^d$ be the discrete approximation
to $\Omega_R$.
As before, $\Gamma$ is the base of $\Lambda$.
Now we choose a smooth, complex-valued function $\Phi$ 
on $\Omega$, and extend it to the infinite cylinder 
$\Omega \times \Rl e_*$ so that $\nabla \Phi \cdot e_* = 0$.
(In other words, $\Phi$ is constant along the direction $e_*$.)
Let $\phi(x) = \Phi(x/R)$, which is defined on 
$\Omega_R \times \Rl e_*$ with the 
property that $\nabla \phi \cdot e_* = 0$.
Finally, let $f(x) = F(l_x) \phi(x)$, where
$F$ is a sequence $F(-L/2), \dots, F(L/2)$.
Note that $f$ is not the most general form possible
for a function on $\Lambda$, most notably because it is the product 
of functions which vary on perpendicular subspaces.
However, the span of such functions does correspond to all of 
$V$ for a fixed value of $L$ and $R$.

Next we consider the norm and energy for such a state.
We will introduce estimates for these quantities, but we will
postpone the actual error terms until the next section.
First we replace the sum over $\Gamma$ with the integral
over $\Omega$, and thus obtain an expression for the norm:
\begin{eqnarray}
\label{norm:approx}
\ip{\psi^f}{\psi^f} 
  &=& \sum_{l=-L/2}^{L/2} \ \sum_{x \in \Gamma_l} |f(x)|^2 \nonumber\\
  &\approx& |\Gamma| \ \sum_{l=-L/2}^{L/2} |F(l)|^2 \cdot
  \frac{1}{m(\Omega_R)} \int_{\Omega_R} |\phi(x)|^2\, dx \nonumber\\
  &=& |\Gamma| \ \sum_{l=-L/2}^{L/2} |F(l)|^2 \cdot
  \frac{1}{m(\Omega)} \int_{\Omega} |\Phi(x)|^2\, dx .
\end{eqnarray}
To obtain an approximation for $\op{\psi^f}{H}{\psi^f}$, we decompose
a step of $f$ along a coordinate direction into a step parallel 
to $e_*$ and a step perpendicular to $e_*$,
\begin{eqnarray*}
f(x+e_j) 
  &=& F(l_x+1) \phi(x+e_j) \\
  &\approx& F(l_x+1) \phi(x) + F(l_x+1) \nabla \phi(x) \cdot e_j.
\end{eqnarray*}
Then using the fact that
$$\sum_{j=1}^d \nabla \phi(x) \cdot e_j = \nabla \phi(x) \cdot e_* = 0,$$
and referring to \eq{local:en:exact} and \eq{total:en}
we have the apparently cumbersome expression
\begin{eqnarray*}
\op{\psi^f}{H}{\psi^f}
  &\approx& \frac{3 |\Gamma|}{2 \cosh(\alpha)} \cdot 
  \frac{1}{m(\Omega)} \int_\Omega |\Phi(x)|^2 dx \\
  && \quad \times \sum_{l=-L/2}^{L/2} 
  \Big[\sech(\alpha[l-\mu]) \sech(\alpha[l+1-\mu]) \\
  && \qquad |\cosh(\alpha[l+1-\mu]) F(l+1) 
  - \cosh(\alpha[l-\mu]) F(l)|^2\Big] \\
  && + \frac{|\Gamma|}{2 R^2 \cosh(\alpha)} 
  \cdot \frac{1}{m(\Omega)} \int_\Omega |\nabla \Phi(x)|^2 dx \\
  && \quad \times \sum_{l=-L/2}^{L/2} 
  \sech(\alpha[l-\mu]) \cosh(\alpha[l+1-\mu]) |F(l+1)|^2 .
\end{eqnarray*}

We notice that the first summand is order 1, while the second summand 
is order $1/R^2$.
We wish to minimize the energy in the limit $R \to \infty$,
so it seems sensible to eliminate the order 1 summand.
This is accomplished by letting $F(l) = \frac{1}{2} \sech(\alpha[l-\mu])$,
or any constant multiple thereof.
One point of interest is that the perturbation takes place 
primarily in a neighborhood of the interface.
The expression for the energy is
\begin{eqnarray}
\label{energy:approx}
\op{\psi^f}{H}{\psi^f}
  &\approx& \frac{|\Gamma|}{8 R^2 \cosh(\alpha)}
  \cdot \frac{1}{m(\Omega)} \int_\Omega |\nabla \Phi(x)|^2 dx \nonumber \\
  && \quad \times \sum_{l=-L/2}^{L/2-1} 
  \sech(\alpha[l-\mu]) \sech(\alpha[l+1-\mu]) .
\end{eqnarray}
Similarly, \eq{norm:approx} may be rewritten as
\begin{eqnarray}
\label{norm:approx2}
\ip{\psi^f}{\psi^f}
  &\approx& \frac{|\Gamma|}{4}
  \cdot \frac{1}{m(\Omega)} \int_\Omega |\Phi(x)|^2 dx 
  \cdot \sum_{l=-L/2}^{L/2} \sech^2(\alpha[l-\mu]) .
\end{eqnarray}
Taking the ratio, we arrive at a normalized energy
\begin{eqnarray}
\frac{\op{\psi^f}{H}{\psi^f}}{\ip{\psi^f}{\psi^f}}
  &\approx& \frac{\sech(\alpha)}{2 R^2} \cdot 
  \frac{\|\nabla\Phi\|^2_{L^2(\Omega)}}{\|\Phi\|^2_{L^2(\Omega)}} \nonumber\\
  && \times
  \frac{\sum_{l=-L/2}^{L/2-1} \sech(\alpha[l-\mu]) 
  \sech(\alpha[l+1-\mu])}{\sum_{l=-L/2}^{L/2} \sech(\alpha[l-\mu]) 
  \sech(\alpha[l-\mu])}.
\end{eqnarray}

Let $P$ be the orthogonal projection to the subspace of perturbations 
considered so far, i.e.\ the span of $\psi^f$, where
$f(x) = \frac{1}{2} \sech(\alpha[l_x - \mu]) \phi(x)$.
Then the projection of $H$ to this subspace is $PHP$.
We have determined that 
$PHP \psi^f = \psi^g$ where $g$ has in place of $\Phi$
\be
\Psi = - \frac{\sech(\alpha)}{2 R^2} \cdot 
  \frac{\sum_{l=-L/2}^{L/2-1} \sech(\alpha[l-\mu]) 
  \sech(\alpha[l+1-\mu])}{\sum_{l=-L/2}^{L/2} \sech(\alpha[l-\mu]) 
  \sech(\alpha[l-\mu])} \nabla^2 \Phi.
\ee
(We write $\nabla^2$ for the Laplacian.
The symbol $\Delta$ is reserved for the anisotropy.)
We should note that it really is necessary to consider $PHP$ instead 
of $H$.
The reason for this is that
\be
\label{newdef:xi}
\xi_{xy} = \frac{-2 \cosh(\alpha[l_x-\mu]) w_x \otimes v_y 
+ 2 \cosh(\alpha[l_y-\mu]) v_x \otimes w_y
+ 2 \sinh(\alpha) w_x \otimes w_y}
{\sqrt{2 \cosh(\alpha[l_x-\mu]) \cdot
2 \cosh(\alpha[l_y-\mu]) \cdot
2 \cosh(\alpha)}} ,
\ee
which means that $H$ does not preserve the total number of $v_x$'s or
$w_x$'s.
Thus the perturbations we have considered (those with a single
$w_x$) do not form an invariant subspace of $H$.

\section{Error Terms}
We now come to the task of tying-up some loose ends,
in order that non-rigorous approximations can be replaced by 
rigorous bounds.
We start with a simple lemma.

\begin{lemma}
Let $\Gamma$ be a finite subset of a lattice $L$.
Let $\Omega$ be the Voronoi domain of $\Gamma$ with respect to
$L$, and let $\Omega_0$ be the Voronoi domain for the
single site $0 \in L$.
Then, for a smooth function $\phi : \Omega \to \C$,
\be
\Big| \frac{1}{|\Gamma|} \sum_{x \in \Gamma} u(x)
  - \frac{1}{m(\Omega)} \int_{\Omega} \phi(x)\, dx \Big| 
  < \|\partial^2 \phi\|_{op,\infty} \cdot \frac{1}{m(\Omega_0)} 
  \int_{\Omega_0} \frac{|x|^2}{2} dx ,
\ee
where $\partial^2 \phi$ is the second-derivative matrix and
\be
\|\partial^2 \phi\|_{op,\infty}
  = \sup_{x \in \Omega} \, \sup_{v \in \Rl^d \setminus 0} 
  \frac{v \cdot \partial^2 u(x) v}{v \cdot v} .
\ee
Note that the second moment $m(\Omega_0)^{-1} \int |x|^2\, dx$
is bounded by the radius of the Voronoi domain, which is in turn
bounded by the distance of nearest neighbors of $L$.
\end{lemma}

\begin{proof}
For the Voronoi domain $\Omega_0$ of $0$, we observe that
\begin{eqnarray*}
\frac{1}{m(\Omega_0)} \int_{\Omega_0} \phi(x)\, dx - \phi(0)
  &=& \frac{1}{m(\Omega_0)} \int_{\Omega_0} [\phi(x) - \phi(0)]\, dx \\
  &=& \frac{1}{m(\Omega_0)} \int_{\Omega_0} \int_0^1
    \nabla \phi(t x) \cdot x\, dt\, dx \\
  &=& \frac{1}{m(\Omega_0)} \int_{\Omega_0} \int_0^1 \int_0^t
  x \cdot \nabla^2 \phi(s x) x\, ds\, dt\, dx \\
  && + \frac{1}{m(\Omega_0)} \int_{\Omega_0} \nabla \phi(0) \cdot x\, dx\\
  &=& \frac{1}{m(\Omega_0)} \int_{\Omega_0} \int_0^1 (1-s)
  x \cdot \partial^2 \phi(s x) x\, ds\, dx \\
  && + \nabla \phi(0) \cdot \frac{1}{m(\Omega_0)} \int_{\Omega_0} x\, dx .
\end{eqnarray*}
But the centroid of $\Omega_0$ is 0.
Thus
\be
\label{dim}
\Big| \frac{1}{m(\Omega_0)} \int_{\Omega_0} \phi(x)\, dx - \phi(0) \Big|
  \leq \frac{1}{m(\Omega_0)} \int_{\Omega_0} \frac{|x|^2}{2} dx
  \times \|\partial^2 \phi\|_{op,\infty} .
\ee
The lemma follows by decomposing $\Omega$ into the $|\Gamma|$ affine copies
of $\Omega_0$, one for each site, and adding the inequalities obtained
from \eq{dim}.
\end{proof}
Using the result of this lemma, we make rigorous the approximation of
\eq{norm:approx}.
Thus
\be
\ip{\psi^f}{\psi^f}
  = |\Gamma| \ \sum_{l=-L/2}^{L/2} |F(l)|^2 \cdot
  \left(\frac{1}{m(\Omega)} \int_{\Omega} |\Phi(x)|^2\, dx + \epsilon_1
  \right),
\ee
where
\be
|\epsilon_1| \leq \frac{1}{R^2} 
  \|\partial^2 |\Phi|^2 \|_{op,\infty}.
\ee
(We have used the fact that the distance between nearest-neighbors
for $\Gamma$ is $\sqrt{2}$.)
In order to fix the approximation of \eq{energy:approx}, we begin
with the elementary bound
$|\phi(x+e_j) - \phi(x) - \nabla \phi(x) \cdot e_j| 
  < \frac{1}{2} \|\partial^2 \phi\|_{op,\infty}$
and its natural successor
\be
\Big| \sum_{j=1}^d |\phi(x+e_j) - \phi(x)|^2 
  - \|\nabla \phi(x)\|^2 \Big|
  < d \Big( \|\nabla \phi\|_{\infty}
  + \frac{1}{4} \|\partial^2 \phi\|_{op,\infty} \Big)
  \|\partial^2 \phi\|_{op,\infty} .
\ee
Using this estimate, as well as the lemma, we may replace 
\eq{energy:approx} with
\begin{eqnarray}
\label{energy:boundCCC}
\op{\psi^f}{H}{\psi^f}
  &\approx& \frac{|\Gamma|}{8 R^2 \cosh(\alpha)} \sum_{l=-L/2}^{L/2} 
  \sech(\alpha[l-\mu]) \sec(\alpha[l+1-\mu]) \nonumber \\
  && \Bigg( \frac{1}{m(\Omega)} \int_\Omega |\nabla \Phi(x)|^2 dx 
  + \epsilon_2 + \epsilon_3 \Bigg),
\end{eqnarray}
where
\be 
|\epsilon_2| \leq \frac{d}{R} \Big(\|\nabla \Phi\|_{\infty}
  + \frac{1}{4 R} \|\partial^2 \Phi\|_{op,\infty} \Big)
  \|\partial^2 \Phi\|_{op,\infty} ,
\ee
and
\be
|\epsilon_3| \leq \frac{1}{R^2} 
  \|\partial^2 |\nabla \Phi|^2 \|_{op,\infty}.
\ee

\section*{Acknowledgements}

O.B. was supported by Fapesp under grant 97/14430-2. B.N. was partially
supported by the National Science Foundation under grant \# DMS-9706599.


\pagebreak

{\baselineskip=10pt \thispagestyle{empty} {{\small Originally published
Comm. Math. Phys. {\bf 212}, 63--91. (2000),\qquad
arXiv:math-ph/9908018} 
\hspace{\fill}}

\vspace{20pt}

\addcontentsline{toc}{chapter}{\textit{Finite-volume excitations of the 111
interface in the quantum XXZ model}}
\begin{center}
{\LARGE \bf Finite-volume excitations of the 111\\
interface in the quantum XXZ model\\[27pt]}
{\large \bf Oscar Bolina, Pierluigi Contucci, Bruno Nachtergaele 
and Shannon Starr\\[10pt]}
{\large  Department of Mathematics\\
University of California, Davis\\
Davis, CA 95616-8633, USA\\[15pt]}
{\normalsize bolina@math.ucdavis.edu, contucci@math.ucdavis.edu, 
bxn@math.ucdavis.edu, sstarr@math.ucdavis.edu}\\[30pt]
\end{center}

\noindent
{\bf Abstract}
We show that the ground states of the three-dimensional XXZ Heisenberg
ferromagnet with a 111 interface have excitations localized in a
subvolume of linear size $R$ with energies bounded by $O(1/R^2)$. As
part of the proof we show the equivalence of ensembles for the 111
interface states in the following sense: In the thermodynamic limit the
states with fixed magnetization yield the same expectation values for
gauge invariant local observables as a suitable grand canonical state
with fluctuating magnetization. Here, gauge invariant means commuting with
the total third component of the spin, which is a conserved quantity of
the Hamiltonian. As a corollary of equivalence of ensembles we also prove
the convergence of the thermodynamic limit of sequences of canonical
states (i.e., with fixed magnetization).
\vspace{8pt}
\newline 
{\small \bf Keywords:} Anisotropic Heisenberg ferromagnet, XXZ model,
rigidity
of interfaces, interface excitations, $111$ interface, equivalence of
ensembles.
\vskip .2 cm
\noindent
{\small \bf PACS numbers:} 05.30.Ch, 05.70.Nb, 05.50.+q 
\newline
{\small \bf MCS numbers:} 82B10, 82B24, 82D40 
\vfill
\hrule width2truein \smallskip {\baselineskip=10pt \noindent Copyright
\copyright\ 1999 by the authors. Reproduction of this article in its entirety, 
by any means, is permitted for non-commercial purposes.\par }}

\newpage

\Section{Introduction and main results}
\label{sec:intro}

A determining factor in the stability of the magnetic state of small 
ferromagnetic particles is the structure of the spectrum of their 
low-lying excitations. Stability against thermal (and quantum) fluctuations
is a major concern when one is interested in increasing the 
density of information stored on magnetic hard disks. Higher density of
information requires smaller magnetic particles to store the bits. The
smaller these particles get, the less stable their magnetic state tends to
be. It is also well-known that ferromagnets spontaneously form domains with 
different orientations of the magnetization. These two facts motivate us
to study the excitation spectrum of finite size ferromagnets with a domain
wall or {\it interface}. From examples, it is known that the presence of an 
interface, in general, has an effect on the low-lying excitation spectrum 
\cite{KN2,KN3}.

We consider the spin 1/2 XXZ Heisenberg model on the three-dimensional
lattice $\Ir^3$. For any finite volume $\Lambda\subset\Ir^3$, the 
Hamiltonian is given by 
\be
H_\Lambda = - \sum_{\substack{x,y\in\Lambda \\ \vert x-y\vert=1}}
\Delta^{-1} (S_x^{(1)} S_y^{(1)} + S_x^{(2)} S_y^{(2)}) 
+ S_x^{(3)} S_y^{(3)},
\ee
where $\Delta>1$ is the anisotropy. It will be convenient to work with
the usual parametrization $\Delta=(q+q^{-1})/2$, $0<q<1$. Note that
in the limit $\Delta\to\infty$ ($q\to 0$), one recovers the Ising model.
The case $\Delta=1$ ($q=1$) is the XXX Heisenberg model. 

It is well-known that this model has two ferromagnetically ordered
translation invariant ground states. What is less well-known is that
there are also ground states describing an interface between two domains
with opposite magnetization. The 100 interfaces are similar to the
Dobrushin interfaces found in the Ising model. They exist for
sufficiently small temperatures, as was recently proved in \cite{BCF}.
Unlike the Ising model, the XXZ model also possesses ground states with a
rigid 111 interface at zero temperature \cite{KN2}. Its stability at
positive temperatures is still an open problem. 

In this paper we are interested in estimating the low-lying excitation\~s
above the ground state with a 111 interface. It is easy to show that the
excitation spectrum above the translation invariant ground states has a 
non-vanishing gap. In \cite{KN2} it was proved that, in the corresponding
two-dimensional model, the excitations above the 11 interface are gapless.
By an extension of the methods in \cite{LPW}, Matsui \cite{Mat2} showed
that the excitation spectrum has to be gapless in all dimensions $\geq 2$.
Here, we are interested in the nature of the low-lying excitations for the
three-dimensional model, and in particular their dependence on size.
We prove the following bound for the energy of an excitation localized in 
a finite domain $\Lambda_R$ of linear size $R$. \\

{\bf Main Result:}
{\em Excitations localized in $\Lambda_R$ have a gap $\gamma_R$ bounded by
\be
\gamma_R\leq 100\frac{q^{2(1-\delta(q,\nu))}}{(1-q^2)}\frac{1}{R^2},      
\quad \textrm{for}\quad R>70.
\label{main}
\ee
where $\delta(q,\nu)$ is an exponent between $0$ and $1/2$ that depends on
the filling factor $\nu$ of the interface plane (see explanation below),
as well as the parameter $q$.}\\

The meaning of this bound is the following. We consider the model in a finite
volume $\Lambda$, with a fixed magnetization and boundary conditions that
induce an interface.  By perturbing the ground state in a cylindrical
subvolume $\Lambda_R$, with circular cross-section of radius $R$, we then 
construct an orthogonal  state with the same magnetization. The bound 
\eq{main} is an upper bound for the difference in
energy of this state with respect to the ground state  in the limit $\Lambda
\nearrow \Ir^3$. For finite volumes $\Lambda$, the same bound holds as long as
$\Lambda$ is substantially larger than $R$.  When $R$ and the finite volume are
comparable in size, a similar bound holds but with a larger constant factor and
additional error terms (see Section 4).

The dependence on $q$ of the bound \eq{main} has some interesting features,
which we explain next. First, in the limit $q\to 1$, the bound diverges.
This means that our Ansatz for the excitations of the 111 interface does not
work for the isotropic model. This is not surprising as the isotropic model
does not have a rigid 111 interface, although it does possess gapless 
excitations, as is well-known from spinwave theory. In the limit $q\to 0$,
the Ising limit, the bound vanishes. This is to be expected, as the 111
interface contours of the Ising model are highly degenerate. 

In order to explain the role of the exponent $\delta(q,\nu)$ in \eq{main}
we first need to discuss some properties of the interface states 
themselves. For $0<q<1$,
the model has a two-parameter family of pure ground states with an interface
in the 111 direction. One parameter is an angle, playing the same role as
the angles $\phi_x$ in the Ansatz \eq{ansatz} for the excitations. The
second parameter, which is relevant for the present discussion, corresponds
to the mean position of the interface in the lattice. 
If we think of spin up at any site as describing an empty site, and spin down 
as a site occupied by a particle, the third component of the spin becomes
equivalent to the number of particles. In Section 2,
\eq{grand_canonical}, we will introduce the chemical potential $\mu$ 
to control the expected number of particles, alias the third component of 
the total spin. In the limit $q\to 0$, the filling factor $\nu$ of the 
interface has a simple interpretation: $\nu=0$ means that interface separates 
a region entirely filled with particles from a region that is empty. A non-zero $\nu$ means that there is a partially
filled plane in between the filled and the empty region, with filling factor
$\nu$. It turns out that the exponent $\delta(q,\nu)$, can be considered as  
a function of $\mu$ alone. For each value of $\mu\in\Rl$, we get an interface 
state, and $\delta$ is the distance of $\mu$ to the integers, i.e.,
$\delta(\mu)=\min(\vert\mu-\floor{\mu}\vert, \vert 1-\mu+\floor{\mu}\vert)$,
where $\floor{\mu}$ is the integer part of $\mu$. In general, the relation
between $\mu$ and $\nu$ depends nontrivially on $q$. But for all $q$,
$0<q<1$, one has $\delta(q,1/2)=0$ and $\delta(q,0)=1/2$. For further
details on the interdependence of the parameters $q,\delta,\mu$, and $\nu$,
we refer to Section 6.1.

We believe that $O(1/R^2)$ is the true behavior of the low-lying excitations.
There are indications in the physics literature that this should indeed be
the case \cite{HN}. Our rigorous bounds are obtained using the variational 
principle: If $\psi_0$ is a ground state of $H_\Lambda$, and $\psi$ is any 
other state that is linearly independent of $\psi_0$, then
\be
\gamma:=E_1-E_0\leq \frac{ \op{\psi}{H^{(q)}_{\Lambda}}{\psi} }{\|\psi\|^2}
	\cdot \frac{1}{1 
	- \frac{| \ip{\psi_0}{\psi} |^2}{\|\psi_0\|^2\|\psi\|^2}} \; .
\label{vp}
\ee
The first factor in the RHS is the energy of the perturbed state $\psi$.
The second factor is necessary to correct for the non-orthogonality of
$\psi$ and the ground state. In general, one would need to consider the
orthogonal complement of $\psi$ to the entire ground state subspace of
$H_\Lambda$. In the present case however, we know that for each eigenvalue
of the third component of the total spin, $J^{(3)}$, there is exactly one
ground state.  As we will only consider perturbations that commute with
$J^{(3)}$, it is sufficient to take the orthogonal complement of $\psi$ to
$\psi_0$.

Our ansatz for $\psi$ is of the following form
\be
\psi=\prod_{x\in\Lambda_R}e^{i2\phi_x S_x^{(3)}}\psi_0\quad .
\label{ansatz}
\ee
The energy of such a state can be written as follows
\be
\frac{\langle \psi \mid H_{\Lambda} \mid \psi \rangle}{\Vert\psi\Vert^2}
=\sum_{\substack{x\in\Lambda_R,y\in \Lambda\\ \vert x-y\vert=1}}
P_{x,y}[1 - \cos(\phi_x - \phi_y)].
\label{ene_form}\ee
where the $P_{x,y}$ are probabilities determined by the interface ground
state. $P_{x,y}$ can be interpreted as the probability that the bond
$(x,y)$ belongs to ``the interface contour'', i.e., one of the sites is
occupied by an up spin and one by a down spin. These probabilities decay
exponentially fast as a function of the distance to the expected location
of the interface. In particular, this shows that the interface is rigid
and that the problem of calculating its excitation energies is quasi
two-dimensional. In fact, the next step in our proof makes this explicit. 
We consider excitations of the form \eq{ansatz} with
$$
\phi_x = \mathcal{S}\phi(\frac{x_\perp}{R}), \quad R\geq 1
$$
where $\mathcal{S}$ is a suitable scale factor, $\phi$ is a smooth function 
with compact support in $\Rl^2$, and $x_\perp$ is the component of 
$x\in\Ir^2$, orthogonal to the $111$ direction. It is shown that the energy 
$\gamma_R$ of  such excitations satisfies the bound
$$
\gamma_R\leq \frac{C(q)}{R^2}\frac{\Vert \nabla \phi\Vert_{L^2}^2}{\Vert\phi
\Vert^2_{L^2}}\quad .
$$
In principle, $\phi$ is a map from $\Rl^2$ to the circle, and as such could 
have nontrivial topology. As we will only be considering small perturbations,
this will be of no relevance here. It is, therefore, natural to take for $\phi$
an eigenfunction belonging  to the smallest eigenvalue of $-\Delta$ on a
circular domain with Dirichlet boundary conditions, which minimizes of the
Rayleigh quotient on the  RHS, i.e., the Bessel function $J_0$. This is
different from the so-called superinstanton Ansatz of Patrascioiu and Seiler in
\cite{PS}, where they use the fundamental solution of the Laplace equation,
instead of an eigenfunction.

All our results are for ground states that are eigenstates of the third
component of the total spin, which is a conserved quantity, and for 
thermodynamic limits of such states. We will call this {\em the canonical 
ensemble}. Our derivation, however, relies on an equivalence of ensembles 
result for the interface ground states of the XXZ model.  The state of the
``small'' volume $\Lambda_R$, immersed in the much larger  volume $\Lambda$, is
well approximated by a grand canonical state with  suitable chemical potential
(see Chapter 2 for the precise definitions), which does not have a fixed
magnetization. As expected, this equivalence of  ensembles holds only for
observables that commute with the third component of the total spin which are
analogous to the gauge invariant observables in  particle systems. This
equivalence of ensembles result is non-trivial. Although we only give the proof
in dimensions 3, it is straightforward to generalize the proof to all
dimensions $\geq 3$. Equivalence of ensembles (in the above sense) does not
hold for the one-dimensional model. This can be derived from the results in
\cite{GW}. In two dimensions, our method without modifications, yields the
equivalence of ensembles for volumes that  grow as $\sqrt{L}$ in the $11$
direction and as $L$ in the direction of the  interface. With additional work
one can obtain equivalence of ensembles result  for standard sequences of
increasing volumes.

As another application of equivalence of ensembles we prove the existence 
of the thermodynamic limit of sequences canonical ground states with a given
density, i.e., magnetization per site.

Concerning the gap above diagonal interface states in dimensions other than
three we can make the following comments. First of all, diagonal interface
states exist in all dimensions \cite{ASW}. In one dimension there is a spectral
gap above the ground states \cite{KN1}. In two dimensions an upper bound of
order $1/R$ was proved in \cite{KN2}. The method of this paper can be used to
obtain a bound of order $1/R^2$ also in two dimensions. In all dimensions
greater than three our method can be applied without change to obtain
equivalence of ensembles, the existence of the thermodynamic limit and an upper
bound of order $1/R^2$ for the  excitation energies.

The paper is organized as follows. Chapter 2 introduces the model and the
geometrical setting.  Chapter 3 deals with the equivalence of ensembles result
which is a main ingredient of our proofs. The bound on the excitation energy is
a product of two factors as in  \eq{vp}. A bound on the first factor, called
the {\em energy bound}, is  derived in Section 4. The second factor requires an
estimate for the inner  product of the ground state with the perturbed state,
which is derived in  Section 5. In Section 6 we prove a number of results
for the grand canonical ensemble in one dimension that we use 
in the paper. 

\Section{Interface states of the XXZ model}

\begin{figure}[t]
\begin{center}
\resizebox{!}{7truecm}{\includegraphics{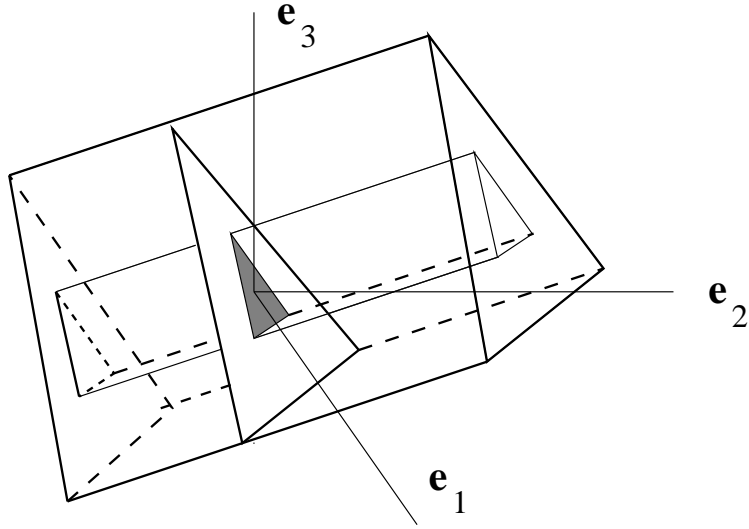}}
\parbox{14truecm}{\caption{\baselineskip=16 pt\small\label{fig:cylinder}
Example of a cylindrical $\Lambda$ embedded in $\Ir^3$.
A small cylindrical subvolume as used in the construction of the perturbed
states is also shown.}
}
\end{center}
\end{figure}

Our magnet occupies a volume $\Lambda$ which is a subset of $\Ir^3$.
Let $e_1,e_2,e_3$ denote the standard basis vectors in $\Ir^3$.
(See Figure \ref{fig:cylinder}.)
We let $l(x)$ denote the signed distance from the origin: 
$l(x) = x^1 + x^2 + x^3$, where $x = (x^1, x^2, x^3)\in\Ir^3$. 
Then
\be
B(\Lambda)=\{(x_0,x_1) :  | x_0 - x_1 | = 1, l(x_1)=l(x_0)+1\}
\label{set_bonds}\ee
describes the set of oriented bonds in $\Ir^3$. The infinite {\it stick} 
$\Sigma_0^\infty$ is, by definition, the set of vertices of the form
$$
\ldots -e_2-e_3,-e_3,0,e_1,e_1+e_2,e_1+e_2+e_3,e_1+e_2+e_3+e_1,\ldots
$$
For any even integer $L$, the finite stick $\Sigma_0$ of length $L+1$ is 
then given by
$$
\Sigma_0=\{x\in\Sigma_0^\infty\mid -L/2\leq l(x)\leq L/2\}\quad.
$$
We will take for $\Lambda$ is a cylindrical region whose axis points in the 
111 direction, where by {\it cylindrical\/} we mean that $\Lambda$ can be 
obtained from a subset $\Gamma$ of the $l(x)=0$ plane, which we will call the 
base, by adding to all vertices $x\in\Gamma$ the finite stick $\Sigma_0$: 
$$
\Lambda=\{x+y\mid x\in\Gamma, y\in\Sigma_0\}
$$
The equation $l(x)=c$, for any constant c, defines a 
cross-section of $\Lambda$, which contains exactly 
$A=\vert\Gamma\vert$ vertices. Hence, $\vert \Lambda\vert =(L+1)A$.
We refer to these cross-sections as planes.

\begin{figure}
\begin{center}
\resizebox{!}{7truecm}{\includegraphics{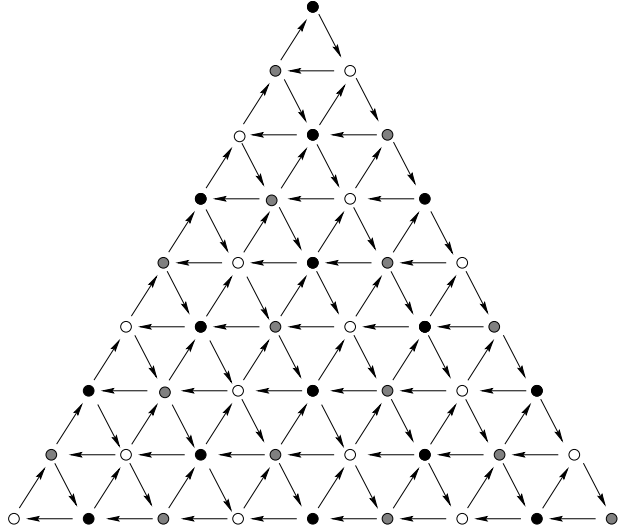}}
\parbox{14truecm}{\caption{\baselineskip=16 pt\small
\label{fig:trilat}
The projection onto the $111$ plane of a cylindrical volume $\Lambda$ with 
triangular base. The shading of the vertices depends on the value of $l(x)$ 
modulo 3. The orientation of the bonds is indicated by arrows. Observe
that each site has an equal number of incoming and outgoing bonds.}
}
\end{center}
\end{figure}

\begin{figure}
\begin{center}
\resizebox{!}{7truecm}{\includegraphics{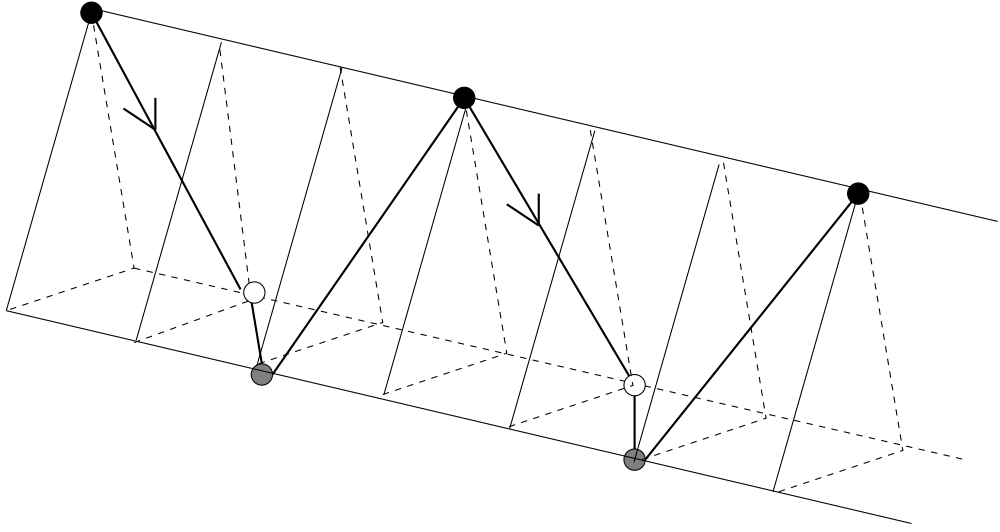}}
\parbox{14truecm}{\caption{\baselineskip=16 pt\small
\label{fig:stick}
The bonds connecting the vertices of a stick $\Sigma$ form a 
one-dimensional subsystem.}
}
\end{center}
\end{figure}

As an example, the projection onto the plane $l(x)=0$, of the vertices of 
$\Lambda$ with triangular base 
is shown in Figure \ref{fig:trilat}, with different shades
depending on the value of $l(x)$ modulo 3.
The orientation of the bonds is indicated by arrows, and one may
observe that each site on the interior of $\Lambda$ 
has an equal number of incoming and outgoing bonds.
By construction, $\Lambda$ can be decomposed into one-dimensional sticks 
running parallel to the cylindrical axis, which we will generically call
$\Sigma$. (See Figure \ref{fig:stick}.) One should observe that $\Sigma$ is
comprised entirely of nearest-neighbor pairs so that every site on $\Sigma$ is
connected to every other site by a sequence of bonds. This will allows us
to exploit the well-known properties of the one-dimensional Heisenberg XXZ 
model to describe $\Sigma$.
The Hamiltonian for the spin-$\frac{1}{2}$ ferromagnetic $XXZ$ Heisenberg
model is given by
\be\label{ham}
H_\Lambda = \sum_{(x_0,x_1)\in B(\Lambda)} h^{q}_{x_0,x_1},
\ee
where 
\be
h^{q}_{x_0,x_1} = - \Delta^{-1} (S_{x_0}^{(1)} S_{x_1}^{(1)}
+ S_{x_0}^{(2)} S_{x_1}^{(2)}) - S_{x_0}^{(3)} S_{x_1}^{(3)} 
+ \frac{1}{4} + \frac{1}{4} A(\Delta) (S_{x_1}^{(3)} - S_{x_0}^{(3)}).
\ee
and $\Delta \geq 1$ is the ``anisotropic coupling'', 
$A(\Delta) = \frac{1}{2} \sqrt{1 - 1/\Delta^2}$, and
$q$, $0<q<1$, is the solution of $\Delta = \frac{1}{2}(q + q^{-1})$
The matrices $S_x^{(\alpha)}$ ($\alpha = 1,2,3$) are the Pauli spin
matrices acting on the site $x$,
\be
S^{(1)} = \left[\begin{array}{cc} 0 & 1/2 \\
1/2 & 0 \end{array}\right],\quad
S^{(2)} = \left[\begin{array}{cc} 0 & -i/2 \\
i/2 & 0 \end{array}\right],\quad
S^{(3)} = \left[\begin{array}{cc} 1/2 & 0 \\
0 & -1/2 \end{array}\right].		
\ee
The terms containing $A(\Delta)$ cancel on all sites except at the 
top and bottom plane of the cylinder.
The usefulness of the nearest-neighbor Hamiltonian stems from the 
fact that its action on any bond is given by
\begin{eqnarray*}
h^q \dd = 0, && \quad
h^q \du = \frac{1}{q + q^{-1}}
\left( q \du - \ud \right) ,\\
h^q \uu = 0, && \quad
h^q \ud = - \frac{1}{q + q^{-1}} 
\left( \du - q^{-1} \ud \right).
\end{eqnarray*}
In other words, $h^q$ is the orthogonal projection on the unit vector
\be
\xi_q = \frac{1}{\sqrt{1 + q^2}} (q \du - \ud). \label{def:xi}
\ee
There is a $(|\Lambda|+1)$-fold degeneracy in the ground states
with a unique ground state for each value of total third component
of the spin $\sum_{x\in\Lambda} S^{(3)}_x$. The basis vectors of the
Hilbert space $(\Cx^2)^{\otimes \vert \Lambda\vert}$ can be labeled with
particle configurations $\alpha =\{\alpha(x)\}_{x \in \Lambda}$, 
where $\alpha(x)$ is 0 or 1, corresponding to $\ket{\uparrow}$ and
$\ket{\downarrow}$, respectively. We write $\Num$ for the operator
defined by
$$
\Num\ket{\alpha} = (\sum_{x \in \Lambda} \alpha(x))\ket{\alpha},
$$
and let $\A(\Lambda,n)$ denote the collection of all 
configurations with $\Num(\alpha) = n$.

Following \cite{ASW} the ground states are given by
\be
\label{def:gs}
\psi_0(\Lambda,n) = \sum_{\alphavec \in \A(\Lambda,n)} 
\bigotimes_{x \in \Lambda} q^{l(x) \alpha(x)} \ket{\alpha(x)} ,
\ee
Note that the weights of $\alpha$ are invariant under 
any permutation of the sites for which planes are invariant. 
These states describe an interface located, on the average, in the
plane determined by $(L/2+l_x)A=n$ \cite{KN2}.

We denote $\Vert \psi_0(\Lambda,n)\Vert^2$ by $Z(\Lambda,n)$.
This quantity is given by
\be\label{uuyy}
Z(\Lambda,n) =\sum_{\alphavec \in \A(\Lambda,n)} \ 
\prod_{x \in \Lambda} q^{2 l(x) \alpha(x)} 
\ee
We will treat $Z(\Lambda,n)$ as a canonical partition function.
It will be useful to consider, also, its grand canonical analogue:
\be
\ZGC{\Lambda}{\mu} = \sum_{n=0}^L Z(\Lambda,n) q^{-2\mu n}
= \prod_{x \in \Lambda} (1 + q^{2 (l(x)-\mu)}) .
\label{grand_canonical}\ee
Then it is easily seen that $\ZGC{\Lambda}{\mu}$ is the squared-norm
of the grand canonical vector defined by
\be\label{def:gc}
\psi^{GC}(\Lambda,\mu) =\sum_{n=0}^{\vert \Lambda\vert}
q^{-n\mu}\psi_0(\Lambda,n)= \bigotimes_{x \in \Lambda}
(\up + q^{l(x)-\mu} \down).
\ee
Due to the product structure, the thermodynamic limit is simply given by
\be
\avgGC{X}{\Ir^3,\mu}=\bigotimes_{x \in \Ir^3} \frac{\bra{\uparrow} 
  + q^{l(x) - \mu} \bra{\downarrow}}{\sqrt{1 + q^{2(l(x)-\mu)}}}
 \;\; X \;\bigotimes_{x \in \Ir^3} \frac{\ket{\uparrow} 
  + q^{l(x) - \mu} \ket{\downarrow}}{\sqrt{1 + q^{2(l(x)-\mu)}}}
\label{product}\ee
for all local observables $X$.

\Section{Equivalence of Ensembles}
\label{sect:eq:ens}

A key step in our argument is the development of an equivalence of ensembles.
Specifically, we will show that for a gauge-invariant local observable
the canonical expectation is close to the grand canonical expectation
for some suitably chosen chemical potential $\mu$.
Here $\mu$ only depends on the total spin of the canonical ensemble, not on
the form of the observable.
From this, naturally follows a thermodynamic limit for gauge-invariant
observables. We begin with activity bounds that show that the ratio of
two canonical partition functions with different particle numbers is
approximately exponential in the difference of the particle numbers,
i.e.,
$$
Z(\Lambda,n-k)\approx Z(\Lambda,n) q^{-2k\mu}
$$
for $\vert k\vert \ll n$. More precisely, we have the following lemma.

{
\lemma[Activity bounds] 
For every volume ${\Lambda}$, 
$|{\Lambda}|=(L+1)A$, the ratio of canonical partition functions
for different number of particles can be bounded from above and 
below by {\rm activity} bounds as follows. Let $A_0$ be any constant.
Suppose $n$, $0\leq n \leq A(L+1)$, and $\mu$ are 
such that
\be\label{activ:hyp1}
n - A\navg \leq \frac{1}{2}A_0 A^{1/2}.
\ee
Then, for every $k$ satisfying
\be\label{activ:hyp2}
\vert k\vert\leq \frac{1}{2}A_0 A^{1/2},
\ee
one has the bounds
\be
\frac{Z(\Lambda,n)}{Z(\Lambda,n-k)} \; \le \;
C(A_0,A)
q^{k[2\frac{n}{A}-2\navg+2\mu a\sigma^2-\frac{k}{A}]/(a\sigma^2)},
\label{rrat}
\ee
and
\be
\frac{Z(\Lambda,n)}{Z(\Lambda,n-k)} \; \ge \;
C(A_0,A)^{-1}
q^{k[2\frac{n}{A}-2\navg+2\mu a\sigma^2-\frac{k}{A}]/(a \sigma^2)},
\label{lrat}
\ee
where $a = 2 |\ln q|$,
$$
\sigma^2:=\sigma^2(\mu,L)=\frac{1}{4}\sum_{l=-L/2}^{L/2} 
\frac{1}{\cosh^2(\frac{a}{2}(l-\mu))},
$$
and
\be
C(A_0,A) = \frac{1 + \frac{A_0}{\sigma^2 A^{1/2}}}
{1 - \frac{A_0}{\sigma^2 A^{1/2}}}.
\ee
Moreover, if $\mu$ is the solution of
$ \frac{n}{A}-\navg=0 $,
then, also using the bounds for $\sigma^2$ given in \eq{lower_sigma}, we
obtain
\be
\label{special:rat1}
C(A_0/2,A)^{-1} q^{-\frac{k^2(1-q^2)}{2 a (1+q^2) A}} \leq
q^{-2k\mu} \frac{Z(\Lambda,n)}{Z(\Lambda,n-k)} \leq 
C(A_0/2,A) q^{-\frac{2 k^2(1-q^2)}{a q^2 A}}.
\ee
Alternatively, if $\mu$ solves $\frac{n-k}{A} - \navg = 0$, then we obtain
\be
\label{special:rat2}
C(A_0/2,A)^{-1} q^{\frac{k^2(1-q^2)}{2 a (1+q^2) A}} \leq
q^{-2k\mu} \frac{Z(\Lambda,n)}{Z(\Lambda,n-k)} \leq 
C(A_0/2,A) q^{\frac{2 k^2(1-q^2)}{a q^2 A}}.
\ee
}
\begin{proof}
This can be obtained as follows. Let consider the grand canonical probability 
\be
p(\mu,{\bf n})=q^{-2\mu |{\bf n}|}\frac{Z({\bf n})}{Z_{GC}(\mu)} \, ;
\label{gcprob}
\ee
with
\be
Z({\bf n})=\sum_{\alpha:\A(\Sigma_1,n_1)\otimes\cdots\otimes
\A(\Sigma_{A-A_0},n_{A-A_0})}q^{w(\alpha)}
\label{cprob}
\ee
where $\Sigma_i$ is the i-th one dimensional stick that we are decomposing
our volume in, and where $Z_{GC}(\mu)$ is the grand-canonical partition
function. Clearly, we have
\be
Z(n) \, = \, \sum_{{\bf n}: |{\bf n}|=n} Z({\bf n}) \; .
\ee
Define
\be
p(\mu,n) \, = \, \sum_{{\bf n}: |{\bf n}|=n} p(\mu,{\bf n}) \; ,
\ee
and we have
\be
\frac{Z(n)}{Z(n-k)}=\frac{p(\mu,n)}{p(\mu,n-k)}q^{2k\mu}
\label{ratio}
\ee
The idea now is to make use of the {\it local central limit theorem}
for the probability distribution of the occupation number in the i-th
stick (see \cite{Fel} Theorem XVI.4.3.).
Let $\xi_i=\sum_{x\in\Sigma_i} \alpha_x$. For any integer $N$, consider,  
the probability
\be
P_\mu(\xi_1=n_1,...,\xi_N = n_N) \, = \, p(\mu,{\bf n}) \, .
\label{defpro}
\ee
Due to the factorization property of $p(\mu,{\bf n})$, the $\xi$'s are
independent identically distributed random variables. 
For centered i.i.d. random variables $X_i$ with variance $\sigma^2$,
the local central limit 
theorem guarantees that the random variable
\be
S_N =\frac{1}{\sigma\sqrt{N}}\sum_{n=1}^N X_n\quad.
\label{lclt}
\ee
is close to a Gaussian in the sense that the quantity
\be
P_N(x):=\Prob(\sum_{n=1}^N X_n=x)
\label{cumu}
\ee
fulfills the bounds
\be
\frac{1}{\sigma\sqrt{2\pi N}}e^{-\frac{x^2}{2\sigma^2 N}}
\left(1-\frac{c}{\sqrt{N}}\right)
\leq P_N(x)\leq\frac{1}{\sigma\sqrt{2\pi
N}}e^{-\frac{x^2}{2\sigma^2 N}}\left(1+\frac{c}{\sqrt{N}}\right)
\label{prob_bounds1}
\ee
where $c$ is the constant
\be
c = \frac{\max(|x|,|x-k|)}{\sigma^2 \sqrt{N}} .
\label{ci}
\ee
By applying \eq{prob_bounds1} to the centered quantity 
$X_n=\xi_n - \langle{\xi_n}\rangle$, we obtain the following bounds on the 
ratio of probabilities:
\be
C(N)^{-1}e^{-k(2x-k)/2\sigma^2 N}\leq
\frac{P_N(x)}{P_N(x-k)}\leq C(N)e^{-k(2x-k)/2\sigma^2 N}
\label{ragio}
\ee
where
\be
C(N)=\frac{1+cN^{-1/2}}{1-cN^{-1/2}}\quad .
\label{cicci}
\ee
In terms of the non-centered variables $\xi_i$ we have
\be
p(\mu,n) \; = \; P_A\left(n-A \navg \right)
\label{ncclt}
\ee
where $\navg$ is the average number of particles of a 1D stick $\Sigma$,
in the grand canonical ensemble with chemical potential $\mu$. 	
From this and the hypotheses \eq{activ:hyp1}, \eq{activ:hyp2}, we obtain
\be
c = \frac{A_0}{\sigma^2}
\quad \textrm{ and } \quad
C(A_0,A) = \frac{1 + \frac{A_0}{\sigma^2 A^{1/2}}}
{1 - \frac{A_0}{\sigma^2 A^{1/2}}}.
\ee
Note that in case $\mu$ is chosen so that $\navg = n/A$
or $\navg = (n-k)/A$ then we can replace $c$ by $c/2$,
with the result that $C(A_0,A)$ may be replaced by
$$
C(A_0/2,A) = \frac{1 + \frac{A_0}{2 \sigma^2 A^{1/2}}}
{1 - \frac{A_0}{2 \sigma^2 A^{1/2}}},
$$
as well.
\newline
Also, from \eq{ncclt} and \eq{ragio}, we have
\be
C(A_0,A)^{-1}e^{-\frac{k(2 n - 2 A \navg - k)}{2 \sigma^2 A}}
\leq \frac{p(\mu,n)}{p(\mu,n-k)} 
\leq C(A_0,A) e^{-\frac{k(2 n - 2 A \navg - k)}{2\sigma^2 A}}.
\ee
Using \eq{ratio} (and observing that $q^{2\mu k} = e^{-a \mu}$), 
we have
\be
\frac{Z(n)}{Z(n-k)} \; \le \; 
  C(A_0,A) e^{-k[2\frac{n}{A}-2\navg+2a\sigma^2 \mu - \frac{k}{A}]/2 \sigma^2}\; ,
\label{rratio2}
\ee
and
\be
\frac{Z(n)}{Z(n-k)} \; \ge \; 
  C(A_0,A) e^{-k[2\frac{n}{A}-2\navg+2a\sigma^2 \mu - \frac{k}{A}]/2 \sigma^2}\; .
\label{lratio2}
\ee
Changing to base $q$ then leads to equations \eq{rrat} and \eq{lrat}
of the theorem.
By the derivation of Section \ref{sect:var_n}, we have the bounds
on the variance for the number of particles in a 1D stick:
\be
\label{var:bounds}
\frac{1}{4}\frac{q^2}{1-q^2}\leq \sigma^2(\mu) \leq \frac{1+q^2}{1-q^2}.
\ee
In conjunction with the remark about replacing $C(A_0,A)$ by 
$C(A_0/2,A)$, this gives equations \eq{special:rat1} and \eq{special:rat2}.
\end{proof}

As an application of this lemma, let us consider the case where
$n$ is replaced by $\rho |\Lambda| - n_0$, $k$ is replaced by 
$\rho |\Lambda_0| - n_0$ and $\Lambda$ is replaced by
$\Lambda_0^c := \Lambda \setminus \Lambda_0$.
This means that in the lemma $A$ is replaced by $A - A_0$,
and $(n-k)/A$ is replaced by 
$\rho (|\Lambda| - |\Lambda_0|)/(A-A_0) = \rho (L+1)$.
Then, direct substitution shows
\begin{eqnarray}
&&\frac{Z(\Lambda_0^c,\rho |\Lambda| - n_0)}{Z(\Lambda_0^c,\rho 
|\Lambda_0^c|)} \nonumber\\
&&\qquad \le C(A_0/2,A-A_0)\, q^{-2 k \mu}
e^{-k[2\rho(L+1) - 2 \navg + \frac{k}{A-A_0}]/2 \sigma^2} ,\\
&&\frac{Z(\Lambda_0^c,\rho |\Lambda| - n_0)}{Z(\Lambda_0^c,\rho 
|\Lambda_0^c|)}\; \nonumber\\
&&\qquad \ge C(A_0/2,A-A_0)^{-1}\, q^{-2 k \mu}
e^{-k[2\rho(L+1) -  2 \navg + \frac{k}{A-A_0}]/2 \sigma^2},
\end{eqnarray}
where we have retained $k$, for the moment.
If, further, we choose $\mu$ so that $\navg = \rho (L+1)$, which is
always possible (see Section \ref{sect:avgexp}), then, 
by equation \eq{special:rat2}, we have
\begin{eqnarray}
q^{2 \mu k} 
\frac{Z(\Lambda_0^c,\rho |\Lambda| - n_0)}
{Z(\Lambda_0^c,\rho |\Lambda_0^c|)}\; \le \; 
  C(A_0/2,A-A_0)\, e^{-\frac{k^2}{2 (A-A_0) \sigma^2}} ,\\
q^{2 \mu k} 
\frac{Z(\Lambda_0^c,\rho |\Lambda| - n_0)}
{Z(\Lambda_0^c,\rho |\Lambda_0^c|)}\; \ge \; 
  C(A_0/2,A-A_0)^{-1}\, e^{-\frac{k^2}{2 (A-A_0) \sigma^2}} .
\end{eqnarray}
Using our bounds for $\sigma^2$, we have
\begin{eqnarray}
\label{spef:rat1}
q^{2 \mu k} \frac{Z(\Lambda_0^c,\rho |\Lambda| - n_0)}
{Z(\Lambda_0^c,\rho |\Lambda_0^c|)}\; \le \; 
  C(A_0/2,A-A_0)\, e^{-\frac{(1-q^2) k^2}{2 (1+q^2) (A-A_0)}},\\
\label{spef:rat2}
q^{2 \mu k} \frac{Z(\Lambda_0^c,\rho |\Lambda| - n_0)}
{Z(\Lambda_0^c,\rho |\Lambda_0^c|)}\; \ge \; 
  C(A_0/2,A-A_0)^{-1}\, e^{-\frac{2(1-q^2)k^2}{2 q^2 (A-A_0)}} .
\end{eqnarray}
By our choice of $\mu$, conditions (\ref{activ:hyp1}) and (\ref{activ:hyp2})
are satisfied as long as the order of $L$ does not exceed the order of
$(A-A_0)^{1/2}$.
This estimate will be of use in the next theorem.

Let $\|X\|_{gs}$ denote the operator-norm of $X$ restricted to the 
subspace of ground states. For observables $X$, localized in $\Lambda$
and commuting with $J^{(3)}$, $\|X\|_{gs}$ is also given by
$$
\|X\|_{gs} = \sup_{0\leq n \leq |\Lambda|}
\vert\avg{X}{\Lambda,n}\vert .
$$ 

{\theorem[Equivalence of Ensembles]\label{thm:eqv}
Consider two cylindrical volumes $\Lambda$ and $\Lambda_0$, $\Lambda_0
\subset \Lambda$, of the type defined in Section 2 (in particular
$\vert\Lambda\vert=A (L+1)$, $\vert\Lambda_0\vert=A_0 (L+1)$), and fix a total
number of particles $n_\Lambda$. Define $\rho=n_\Lambda/\vert \Lambda\vert$.
Suppose $X$ is a local observable in the volume $\Lambda_0$, 
which commutes with $J^{(3)} := \sum_x S^{(3)}_x$. Then we have
\be
|\avg{X}{\Lambda,n}
  -  \avgGC{X}{\Lambda_0,\mu}| \; \le \; 
  \varepsilon \|X\|_{gs} \; ,
\label{equicgrc}
\ee
where 
\be
\label{vare}
\varepsilon
  = \frac{\ln^2(A-A_0) + 2 (1+a^2) A_0^2 + 4}{2(A-A_0)} 
  + \frac{4 A_0}{q^2 (A-A_0)^{1/2} - 2 A_0},
\ee
$a = 2 \vert\ln q\vert$,
and the chemical potential $\mu$ is determined by the equation
\be
 \navg = \rho (L+1).
\label{themu}\ee
In particular, for $\rho = 1/2$ the calculations of Section \ref{sect:navg}
will show that $\mu = 0$.
}

{\corollary[Existence of the Thermodynamic limit]\hfill\break
\label{ThermoLimitCor}
(i) Suppose we have a sequence of pairs $(\Lambda_k,n_k)$ 
with $\Lambda_k$ cylindrical volumes and $\Lambda_k \nearrow \Ir^3$ in such
a way that the length does not grow faster than the linear size of the 
base. Let $\mu_k$ solve $\avgGC{\Num}{\Lambda_k,\mu_k} = n_k$.
Then the convergence $\mu_k \to \mu$ guarantees the convergence,
of $\avg{.}{\Lambda_k,n_k}$ to $\avgGC{.}{\Ir^3,\mu}$, for all local 
observables $X$ commuting with $J^{(3)}$ :
\be \label{thermo:limit}
\avg{X}{\Lambda_k,n_k} \to \avgGC{X}{\Ir^3,\mu} 
\ee
\newline
(ii) Moreover, for any choice of $\mu$, we may find a sequence of pairs
$(\Lambda_k,n_k)$ such that 
\be \label{thermo:limit2}
\avg{X}{\Lambda_k,n_k} \to \avgGC{X}{\Ir^3,\mu} .
\ee
}

\begin{proof}
(Proof of Corollary)
It follows from the monotonicity of $\navg$ proved in Section \ref{sect:navg},
that the equation
\be
\avgGC{\Num}{\Lambda_k,\mu_k} = n_k
\label{mu_eq}\ee
always has a unique solution for $\mu_k$. Then, (i) follows immediately 
from the inequality \eq{equicgrc}, once we observe that $\epsilon \searrow 0$ 
as $\Lambda \nearrow \Ir^3$ in the sense prescribed in the corollary.
\newline
For (ii), take $\Lambda_k$, with base $A_k$, and $n_k$ such that
$$
n_k=\floor{A_k\navg}\quad .
$$
where $\floor{x}$ denotes the largest integer $\leq x$.
Then, $\mu_k$ solving \eq{mu_eq}, is easily seen to converge to $\mu$,
and \eq{thermo:limit2} follows from (i).
\end{proof}

The interpretation of the condition $\mu_k\to\mu$ in (i) of the Corollary
is that, not only does $n_k/\vert\Lambda_k\vert$ converge to $\rho=1/2$,
but, more precisely
$$
n_k=\rho\vert\Lambda_k\vert+\nu A_k + o(A_k)\quad.
$$
The term proportional to $\vert\Lambda_k\vert$ guarantees that the 
interface is in the center of the volume, the second term fixes its
filling factor.

\begin{proof}
(Proof of Theorem \ref{thm:eqv})
Let $\mu$ be determined by \eq{themu}, and define $\Xi$ as follows:
\be
\Xi=\frac{Z(\Lambda, n_\Lambda) q^{-2\mu \rho |\Lambda_0|}}{Z(\Lambda_0^c,
\rho \vert \Lambda_0^c\vert) Z^{GC}(\Lambda_0, \mu)}
\label{Xi}\ee
where $\Lambda_0^c:=\Lambda\setminus\Lambda_0$.
We will obtain the equivalence of ensembles by combining two facts.
The first is that $\Xi$ is approximately equal to $1$, and the second is
an estimate showing that 
$$
\vert \avg{X}{\Lambda,n_\Lambda}\Xi-\avgGC{X}{\Lambda_0,\mu}\vert 
  \leq \varepsilon\Vert X\Vert_{gs} 
$$
But first, let us recall the definitions of the 
expectation of an observable $X$:
\begin{eqnarray}
  \avg{X}{\Lambda,n} 
  &=& \frac{\op{\psi(\Lambda,n)}{X}{\psi(\Lambda,n)}}
  {\ip{\psi(\Lambda,n)}{\psi(\Lambda,n)}}, \\
  \avgGC{X}{\Lambda,\mu} 
  &=& \frac{\op{\psi^{GC}(\Lambda,\mu)}{X}{\psi^{GC}(\Lambda,\mu)}}
  {\ip{\psi^{GC}(\Lambda,\mu)}{\psi^{GC}(\Lambda,\mu)}}.
\end{eqnarray}
Since $X$ is an observable localized in $\Lambda_0$,
we note that $\avgGC{X}{\Lambda,\mu} = \avgGC{X}{\Lambda_0,\mu}$.
Moreover, we may decompose the grand canonical state into
a superposition of canonical states:
\be
\psi^{GC}(\Lambda_0,\mu) 
  = \sum_{n_0 = 0}^{|\Lambda_0|} q^{-\mu n_0} \psi(\Lambda_0,n_0).
\ee
Since $X$ commutes with $J^{(3)}$, it does not have off-diagonal
matrix elements between these canonical states with all different values
of the total spin. Therefore, 
\be
\avgGC{X}{\Lambda,\mu} 
  = \ZGC{\Lambda}{\mu}^{-1} \sum_{n_0 = 0}^{|\Lambda_0|} 
  q^{-2 \mu n_0} Z(\Lambda_0,n_0) \avg{X}{\Lambda_0,n_0}.
\ee
Note also, that since we have a decomposition
\be
\psi(\Lambda,n) = \sum_{n_0 = 0}^{|\Lambda_0|} 
  \psi(\Lambda \setminus \Lambda_0,n-n_0) \otimes 
  \psi(\Lambda_0,n_0),
\ee
and using the previously described properties, we have
\begin{eqnarray}
\avg{X}{\Lambda,n}
  &=& \sum_{n_0 = 0}^{|\Lambda_0|} 
  \frac{Z(\Lambda\setminus\Lambda_0,n-n_0)
  Z(\Lambda_0,n_0)}{Z(\Lambda,n)} \avg{X}{\Lambda_0,n_0} \\
  &=& Z^{GC}(\Lambda_0,\mu)^{-1} \sum_{n_0=0}^{|\Lambda_0|}
  q^{-2\mu n_0} Z(\Lambda_0,n_0)\avg{X}{\Lambda_0,n_0} \times \nonumber\\
  & & \quad \times \frac{Z(\Lambda_0^c,n-n_0) Z^{GC}(\Lambda_0,\mu)}
	{q^{-2\mu n_0} Z(\Lambda,n)}.
\end{eqnarray}
This differs from the definition of $\avgGC{X}{\Lambda_0,\mu}$ only
by the final factor, which is a ratio of partition functions hence amenable
to our activity bounds.
\newline
In fact, we have
\begin{eqnarray}
\avg{X}{\Lambda,n} \Xi - \avgGC{X}{\Lambda,\mu}
  &=& Z^{GC}(\Lambda_0,\mu)^{-1} \sum_{n_0=0}^{|\Lambda_0|}
    q^{-2\mu n_0} \avg{X}{\Lambda_0,n_0} Z(\Lambda_0,n_0) \times \nonumber\\
  & & \quad \times \left[q^{2\mu(n_0 - \avg{n_0}{})} \
    \frac{Z(\Lambda_0^c,n-n_0)}{Z(\Lambda_0^c,\floor{\rho |\Lambda_0|})} 
    - 1\right] 
\end{eqnarray}
where $\avg{n_0}{}=\avgGC{\Num}{\Lambda_0,\mu}$, which equals 
$\rho |\Lambda_0|$ for our choice of $\mu$. Thus we obtain
$|\avg{X}{\Lambda,n} \Xi - \avgGC{X}{\Lambda,\mu}|
  \leq \|X\|_{gs} \avgGC{|g|}{\Lambda_0,\mu},$
where
\be
  g = q^{2\mu(n_0 - \avg{n_0}{})} \
    \frac{Z(\Lambda_0^c,n-n_0)}{Z(\Lambda_0^c,\floor{\rho |\Lambda_0|})} 
    - 1 .
\ee
Now we use the activity bounds \eq{spef:rat1} and \eq{spef:rat2}, 
but replacing $k$ by its actual value, $\avg{n_0}{} - n_0$.
We arrive at the bounds
\begin{eqnarray}
  g &\leq& g_1 := 
  C(A_0/2,A-A_0) e^{-\frac{(1-q^2) (\avg{n_0}{} - n_0)^2}{2 (1+q^2) (A-A_0)}} - 1,\\
  g &\geq& g_2 := 
  C(A_0/2,A-A_0)^{-1} e^{-\frac{2(1-q^2)(\avg{n_0}{} - n_0)^2}{2 q^2 (A-A_0)}} - 1,
\end{eqnarray}
where
\be
C(A_0/2,A-A_0) = \frac{1 + \frac{A_0}{2 \sigma^2 (A-A_0)^{1/2}}}
{1 - \frac{A_0}{2 \sigma^2 (A-A_0)^{1/2}}}.
\ee
Therefore, $|g| \leq \max(|g_1|,|g_2|) \leq |g_1| + |g_2|$.
\newline
We now use the triangle inequality and the fact that the exponent is
negative to obtain:
\be
|g_1| \leq
  \left|1 - e^{-\frac{(1-q^2)(\avg{n_0}{} - n_0)^2}{2 (1+q^2) (A-A_0)}}\right| 
  + |1 - C(A_0/2,A-A_0)|,
\ee
so that
\be
\label{g1:bound1}
\avg{|g_1|}{\Lambda_0,\mu} \leq  
\avgGC{1 - e^{-\frac{(1-q^2) (\avg{n_0}{} - n_0)^2}{2 (1+q^2) (A-A_0)}}}
{\Lambda_0,\mu} + C(A_0/2,A-A_0) - 1.
\ee
Similarly,
\be
\avg{|g_2|}{\Lambda_0,\mu} \leq  
\avgGC{1 - e^{-\frac{2(1-q^2)(\avg{n_0}{} - n_0)^2}{2 q^2 (A-A_0)}}}
{\Lambda_0,\mu} + 1 - C(A_0/2,A-A_0)^{-1}.
\ee
\newline
We will use the Chebyshev inequality to control the expectation term in 
\eq{g1:bound1}. Specifically, for any $B>0$,
\begin{eqnarray*}
\avgGC{1 - e^{-\frac{(1-q^2) (\avg{n_0}{} - n_0)^2}{2 (1+q^2) (A-A_0)}}}
{\Lambda_0,\mu}
	&\leq& \Prob(2 |n_0 - \avg{n_0}{}| \geq 2 B) 
	+ 1 - e^{-\frac{(1-q^2)B^2}{2(1+q^2)(A-A_0)}} \\
	&\leq& q^{2 B} \avgGC{q^{-2 | n_0 - \avg{n_0}{} |}}{\Lambda_0,\mu}
	+ 1 - e^{-\frac{(1-q^2)B^2}{2(1+q^2)(A-A_0)}}.
\end{eqnarray*}
In Section \ref{sect:avgexp} we show that
$\avgGC{q^{-2 |n_0 - \avg{n_0}{}|}}{\Lambda_0,\mu} \leq 2 (2 q^{-2})^{A_0}$.
One choice for $B$ is $a^{-1} [\ln(A-A_0) + A_0 \ln(2 q^{-2})]$.
This gives the bound
\begin{eqnarray}
\avgGC{ 1 - q^{\frac{(n_0 - \avg{n_0}{})^2}{A-A_0}} }{\Lambda_0,\mu}
  &\leq& \frac{2 + \frac{1-q^2}{a^2 (1+q^2)} 
  \left[2 (1+a^2) A_0^2 + \ln^2(A-A_0)\right]}{A-A_0} \nonumber\\
  &\leq& \frac{2 + (1+a^2) A_0^2 + \frac{1}{2} \ln^2(A-A_0)}{A-A_0} \\
  &=:& C_1(A,A_0,q) \nonumber
\end{eqnarray}
The leading order term in the bound is  
$\frac{\ln^2(A-A_0)}{2(A-A_0)}$ for fixed $q$, strictly between 0 and 1.
Also, let 
\be
C_2(q,A,A_0) 
	= \frac{4 A_0}{q^2 (A-A_0)^{1/2} - 2 A_0},
\ee
which is greater than both $C(A_0/2,A-A_0)-1$ and $1 - C(A_0/2,A-A_0)^{-1}$.
Then 
$|\avg{f}{\Lambda,n} \Xi - \avgGC{f}{\Lambda,\mu}|
\leq (C_1 + C_2) \|X\|_{gs}$.
In particular,
$|\avg{\unity}{\Lambda,n} \Xi - \avgGC{\unity}{\Lambda,\mu}|
\leq (C_1 + C_2) \|\unity\|_{gs}$,
which is to say that
$|\Xi - 1| \leq C_1 + C_2$.
Then, using the triangle inequality, we have
\begin{eqnarray*}
|\avg{X}{\Lambda,n} - \avgGC{X}{\Lambda,\mu}|
  &\leq&  |1 - \Xi| \cdot |\avg{X}{\Lambda,n}|
	+ |\avg{X}{\Lambda,n} \Xi - \avgGC{X}{\Lambda,\mu}| \\
  &\leq&  2 (C_1 + C_2) \|X\|_{gs}.
\end{eqnarray*}
So, defining 
$\varepsilon = 2 C_1(q,\Lambda,\Lambda_0,n) + 2 C_2(q,\Lambda,\Lambda_0)$,
the theorem is proved. 
\end{proof}

Note that the restriction to observables $X$ that commute with the
third component of the total spin $J^{(3)}$ is necessary. E.g., the
expectation of $S^+_x$ obviously vanishes in any canonical state,
while it is easy to see, by direct computation, that it does not
vanish in the grand canonical states. This is entirely analogous to
the restriction to gauge invariant observables in particle systems.


\Section{Bound on the energy}

In this section we will estimate the energy of a class of perturbations
of the ground state $\psi_0$ given in \eq{def:gs}. Let $\Lambda$ and 
$\Lambda_R$ be two cylindrical volumes as described in Section 2, 
$\Lambda_R\subset\Lambda$. E.g., $\Lambda_R$ and $\Lambda$, may have triangular
cross-sections (see Figure \ref{fig:cylinder}). We will generally assume
that the radius $R$ of $\Lambda_R$ is much less than that of $\Lambda$. 
We consider $\psi$ of the form
\be
\label{ps}
\psi(\Lambda,n,\phi) = \sum_{\alphavec \in \A(\Lambda,n)} 
\bigotimes_{x \in \Lambda} e^{i \phi(x) \alpha(x)} 
q^{l(x) \alpha(x)} \ket{\alpha(x)} ,
\ee
where $\supp(\phi) \subset \Lambda_R$.

We will also suppose that 
\be
\phi = \frac{\Sc}{R} \tilde \phi(\tilde{y}_1,\tilde{y}_2)
\ee
where $\tilde\phi$ is a smooth functions of its variables
and $\Sc$ is a parameter, which we will eventually take to zero 
independent of $R$. The coordinates $\tilde{y}^1,tilde{y}^2$, are
defined by
\be
\tilde{y}^1 = \frac{2 x^1 - x^2 - x^3}{\sqrt{6} R}
\quad {\rm and} \quad \tilde{y}^2 = \frac{x^2 - x^3}{\sqrt{2} R},
\ee
and are  to be viewed as rescaled coordinates for $x$ along the plane 
perpendicular to the 111 axis.

There are two points to our assumptions on $\phi$: First, that $\phi$ is
independent of the 111 component of $x$. Second, that $\phi$ is 
associated to a scale-invariant phase $\tilde\phi$ by 
$\phi(x) = R^{-1} \tilde\phi(x/R)$. Ultimately, the constant $\Sc$ will vanish.
The leading term in our estimate of the gap  is independent of $\Sc$ 
as long as $\Sc \ll 1$. 

Let $\Gamma_R$ be the projection of $\Lambda_R$ onto the plane
$l(x)=0$, $A_R=\vert\Gamma_R\vert$, $\Omega_R$ be the convex hull of 
$\Gamma_R$, and $\tilde{\Omega} = \{x \in\Rl^2: R x \in \Omega_R\}$, 
the rescaled region, and let $m(\tilde{\Omega})$ be the area of 
$\tilde{\Omega}$ (for the standard Lebesgue measure on $\Rl^2$). 

We will also use the following notation: 
$\partial_{\tilde y}\tilde\phi$ and $\partial^2_{\tilde y}\tilde\phi$
are the first- and second-derivative tensors of $\tilde{\phi}$, and by the 
$L^\infty$ norm of a tensor we mean the maximum of the $L^\infty$ norms of 
the components.

Then we have the following theorem.


{\theorem[Bound on $\frac{\op{\psi}{H^{(q)}_{\Lambda}}{\psi}
}{\|\psi\|^2}$] \label{energy:bound}
Considering a perturbed state as in (\ref{ps}), the energy is bounded by
\begin{eqnarray}
\label{bound:en}
\frac{\langle \psi \mid H^{(q)}_{\Lambda} \mid \psi
\rangle}{\|\psi\|^2} \leq
2\frac{1+q^2}{1-q^2}\left(\frac{A_R \Sc^2}{R^4} 
\frac{\|\nabla_{\tilde{y}} \tilde\phi\|_{L^2(\tilde\Omega)}^2}{m(\tilde\Omega)}
+ \mathcal{E}_{\textrm{num}}\right)
\end{eqnarray} 
where
\be
\mathcal{E}_{\textrm{num}} = \frac{6 A_R \Sc^2}{R^5} 
\|\partial_{\tilde y}^2 \tilde\phi\|_{L^\infty}
\|\partial_{\tilde y} \tilde\phi\|_{L^\infty}  
\ee
is a correction to the main term which becomes negligible as $R \to \infty$.
}

\begin{proof}
We begin by calculating how a two-site hamiltonian $h_b^q$ 
acts on the perturbed state. 

We consider the decomposition of our lattice into the relevant bond
$b = (x_0,x_1)$ and everything else $\Lambda \setminus b$.
Thus
\be
  h_b^q = \unity_{\Lambda \setminus b} \otimes \ket{\xi_b} \bra{\xi_b},
\ee
where $\xi_b$ is the unit vector from (\ref{def:xi}) on the 
pair $b$, and
\be
\psi(\Lambda,n)=\sum_{n_b=0}^2 \psi(\Lambda \setminus b,n-n_b) \otimes
\psi(b,n_b).
\ee
Here $\psi(b,n_b)$ is as would be defined by (\ref{ps}),
but with $\Lambda$ replaced by $b$ and $n$ replaced by $n_b$.
For example $\psi(b,1) = q^{l(x_0)} e^{i \phi(x_0)} \du 
+ q^{l(x_1)} e^{i \phi(x_1)} \ud$. But $\xi_b$ is orthogonal to 
$\psi(b,0)$ and $\psi(b,1)$, since $\xi_b$ lies in the sector of total spin 1.
And
\be
\ip{\xi_b}{\psi(b,1)} = \frac{1}{\sqrt{1+q^2}} 
q^{l(x_0)+1} e^{i \phi(x_0)} (1 - e^{i [\phi(x_1) - \phi(x_0)]}).
\ee
Now it is straightforward to see
\begin{eqnarray}
&&\op{\psi(\Lambda,n)}{h_b^q}{\psi(\Lambda,n)}\\
&&\quad =\|\psi(\Lambda \setminus b,n-1)\|^2\, |\ip{\xi_b}{\psi(b,1)}|^2 
\nonumber \\
&&\quad =\frac{2}{(q+q^{-1})^2}Z(\Lambda,n)P^{q}(b) (1 - \cos[\phi(x_1) - \phi(x_0)]),
\end{eqnarray}
where we have defined 
\be
P^q(b) = \frac{Z(\Lambda \setminus b,n-1)Z(b,1)}{Z(\Lambda,n)}.
\ee
Then we may write
\be\label{ene}
\frac{\langle \psi \mid H^{(q)}_{\Lambda} \mid \psi \rangle}{Z(\Lambda,n)}
= \frac{2}{(q+q^{-1})^2}~ \sum_{b \in B(\Lambda)} \ P^q(b)
(1 - \cos[\phi(x_1) - \phi(x_0)]).
\ee
Actually, $P^q(b)$ depends on $b$ only through $l(x_0)$.
So from here on, we'll write it as $P^q(l(x_0))$, and observe the following:
\be\label{pb}
\frac{\langle \psi \mid H^{(q)}_{\Lambda} \mid \psi \rangle}{Z(\Lambda,n)}
= \frac{2}{(q+q^{-1})^2} \sum_{l=-L/2}^{L/2-1} P^q(l) 
  \sum_{x \in \Gamma^l_R} \sum_{j=1}^3 
  (1 - \cos[\phi(x+e_j) - \phi(x)]) ,
\ee
where $\Gamma^l_R = \{x \in \Lambda_R : l(x) = l\}$.

Let us estimate the term
$\sum_{x \in \Gamma^l_R} \sum_{j=1}^3 (1 - \cos[\phi(x+e_j) - \phi(x)])$.
We have an inequality
\be
\label{cos:ineq}
 1 - \cos[\phi(x+e_j) - \phi(x)] \leq \frac{1}{2} [\phi(x+e_j) - \phi(x)]^2
\ee
(which is actually an  equality in the limit $R \to \infty$ for our ansatz).
Also,
\be
\label{diff:approx}
\sum_{i=1}^3 [\phi(x+e_j) - \phi(x)]^2 \approx |\nabla_x \phi(x)|^2 
	= \frac{\Sc^2}{R^4} |\nabla_{\tilde y} \tilde \phi|^2
\ee
In fact, using the inequality
\be
|[\tilde\phi(\tilde{y}+v) - \tilde\phi(\tilde y)]^2 - 
[c\cdot\nabla_{\tilde y} \tilde\phi(\tilde{y})]^2|
  \leq \| \partial_{\tilde y}^2 \tilde\phi \|_{L^\infty}
  \| \partial_{\tilde y} \tilde\phi\|_{L^\infty}
  \|v\|_{l^1}^3
\ee
one may conclude that the error in (\ref{diff:approx}) is bounded by 
$\frac{3 \Sc^2}{R^5} \|\partial_{\tilde y}^2 \tilde\phi\|_{L^\infty}
\|\partial_{\tilde y} \tilde\phi\|_{L^\infty}$.

Incorporating this estimate into the inequality of (\ref{cos:ineq}),
we have
\begin{eqnarray}
&& \sum_{x \in \Gamma^l_R} \sum_{j=1}^3 (1 - \cos[\phi(x+e_j) - \phi(x)]) 
  \leq \qquad \qquad \nonumber\\
&& \qquad \qquad 
\frac{1}{2 R^2} \sum_{x \in \Gamma^l_R} 
  |\nabla_{\tilde{y}} \phi(x)|^2
  + \frac{3 \Sc^2 |\Gamma^l_R|}{2 R^5} 
  \|\partial_{\tilde y}^2 \tilde\phi\|_{L^\infty}
  \|\partial_{\tilde y} \tilde\phi\|_{L^\infty} 
  \label{Energ:Ineq1} 
\end{eqnarray}
Finally, as $R \to \infty$, the sum over each $\Gamma^l_R$ becomes increasingly
well-approximated by the integral over $\Omega_R$,
we is proved in Lemma \ref{lem:voronoi} immediately following this proof.
The lemma gives us a bound
\be
\sum_{x \in \Gamma^l_R} 
|\nabla_{\tilde{y}} \phi(x)|^2 
  \leq \frac{\Sc^2 |\Gamma^l_R|}{R^2} \left[\frac{1}{m(\tilde{\Omega})} 
  \int_{\tilde{\Omega}} |\nabla_{\tilde{y}} \tilde\phi|^2\, d^2 y 
  + \frac{\rho}{R} 
  \|\nabla^2_{\tilde{y}} \tilde\phi \nabla_{\tilde{y}} \tilde\phi
	\|_{L^\infty(\tilde{\Omega})}\right] ,
  \label{Energ:Ineq2}
\ee
where $\nabla^2$ is the Laplacian and $\rho = \sqrt{2/3}$ is the
maximum radius for the Voronoi domain.
(Note that by its definition, as the trace of the second-derivative tensor,
the Laplacian enjoys the bounds
\be
\|\nabla^2_{\tilde{y}} \tilde\phi \nabla_{\tilde{y}} \tilde\phi
  \|_{L^\infty(\tilde{\Omega})} 
  \leq 2 \|\partial_{\tilde y}^2 \tilde\phi\|_{L^\infty}
  \|\partial_{\tilde y} \tilde\phi\|_{L^\infty}, 
\label{lasteq}\ee  
which may be combined with error term in (\ref{Energ:Ineq1}).)
Combining \eq{Energ:Ineq2} and \eq{lasteq} gives us the theorem, modulo
the term $\sum_{l=-L/2}^{L/2-1} P^q(l)$, for which we derive the necessary
in Lemma \ref{lem:Pq}.
\end{proof}

\begin{lemma}\label{lem:voronoi}
Suppose $\Gamma$ is a region in a regular lattice.
For each $x \in \Gamma$, let $\Omega_x$ be the Voronoi domain of $x$ with
respect to the whole lattice, and
let $\Omega_\Gamma$ be the union of all the individual domains $\Omega_x$.
If $f$ is a smooth function on $\Omega_\Gamma$, then
\be
\left| \frac{1}{|\Gamma|} \sum_{x \in \Gamma} f(x)
	- \frac{1}{m(\Omega_\Gamma)} \int_{\Omega_\Gamma} f(y)\, dy \right|
	\leq \rho \|\nabla_y f\|_{L^\infty(\Omega_\Gamma)}
\ee
where $\rho$ is the maximum radius of a Voronoi domain.
\end{lemma}

\begin{proof}
For each $x \in \Gamma$,
\begin{eqnarray*}
f(x) - \frac{1}{m(\Omega_x)} \int_{\Omega_x} f(y)\, dy 
  &\leq& -\frac{1}{m(\Omega_x)} \int_{\Omega_x} [f(y) - f(x)]\, dy \\
  &=& -\frac{1}{m(\Omega_x)} \int_{\Omega_x} \int_0^1 
    \frac{d}{dt} f(x + t (y-x))\, dt\, dy \\
  &=& -\frac{1}{m(\Omega_x)} \int_{\Omega_x} \int_0^1 
    \nabla_y f(x + t(y-x)) \cdot (y-x)\, dt\, dy .
\end{eqnarray*}
This clearly leads to the bound
\be
\left| f(x) - \frac{1}{m(\Omega_x)} \int_{\Omega_x} f(y)\, dy \right|
	\leq \rho(\Omega_x) \|\nabla_y f\|_{L^\infty(\Omega_x)}.
\ee
From this, the lemma follows easily.
\end{proof}

Now, we will derive the necessary bound on
$$
\sum_{l=-L/2}^{L/2-1} P^q(l)\quad .
$$
We will rely on bounds for similar quantities in the 
one-dimensional model proved in \cite{BCN}.

\begin{lemma}[Bound on $\sum_{l=-L/2}^{L/2-1} P^q(l)$]\label{lem:Pq}
\be
\sum_{l=-L/2}^{L/2-1} P^q(l) \leq  2\frac{1+q^2}{1 - q^2} .
\ee
\end{lemma}

\begin{proof}
Recall
\be
P^q(l)=\frac{Z(\Lambda \setminus b,n-1) Z(b,1)}
{Z(\Lambda,n)}.
\ee
The ratio of partition functions in the equation above is clear:
It is the probability of finding one particle shared by the 
sites of $b$, and $n-1$ particles shared by the sites of 
$\Lambda \setminus b$, conditioned on finding $n$ total particles on 
$\Lambda$. We consider the operator
$$Y_b = \unity_{\Lambda \setminus b} \otimes 
  \left(\ket{\uparrow \downarrow}_b \bra{\uparrow \downarrow}_b
  + \ket{\downarrow \uparrow}_b \bra{\downarrow \uparrow}_b\right).$$
Then
\be
\frac{Z(\Lambda \setminus b,n-1) Z(b,1)}{Z(\Lambda,n)}
	= \avg{Y_b}{\Lambda,n},
\ee
and
\be
\sum_{l=-L/2}^{L/2-1} P^q(l)
  =\left\langle\sum_{l=-L/2}^{L/2-1} Y_{b(l)}
  \right\rangle_{\Lambda,n}.
\label{PtoY}\ee
where $b(l)= (x_0,x_1)$, where $l(x_0)=l$, and $(x_0,x_1)$ is a bond in the
stick containing the origin, which we denote by $\Sigma_0$.
The restriction of the state in $\Lambda$ with $n$ spins down is of the
form
$$
\langle X\rangle_{\Sigma_0}=\sum_{k=0}^{L+1} c_k \avg{X}{\Sigma_0,k}
$$
where $X$ is any observable commuting with $J^{(3)}=\sum_{x\in\Sigma_0}
S^{(3)}_x$, as is, e.g., $Y_{b(l)}$, and the $c_k$ are non-negative
numbers summing up to one. We will now derive an upper bound for 
$\avg{\sum_{l=-L/2}^{L/2-1} Y_l}{\Sigma_0}$, that is independent of the
coefficients $c_k$.
We start from
\be
\avg{Y_l}{\Sigma_0,k}
\leq \Prob_k( S^{(3)}_l=\uparrow, S^{(3)}_{l+1}=\downarrow)
+\Prob_k( S^{(3)}_l=\downarrow, S^{(3)}_{l+1}=\uparrow)
\label{oneDprob}\ee
where $\Prob_k$ denotes the probability in the ground state with $k$
spins down for a one-dimensional system on $[-L/2,L/2]$, the sites of which
we label by $l$. Each term in the RHS of \eq{oneDprob} can be estimate as
follows.
\be
\Prob_k( S^{(3)}_l=\uparrow, S^{(3)}_{l+1}=\downarrow)
\leq
\min\left(\Prob_k(S^{(3)}_l=\uparrow),\Prob_k(S^{(3)}_{l+1}=\downarrow)\right)
\ee
Theorem 7.1 of \cite{BCN} gives the following bounds
\beann
\Prob_k(S^{(3)}_{l+1}=\downarrow)\leq q^{2(l-(k+1-L/2)}
&\mbox{if}& l\geq k+1-L/2\\
\Prob_k(S^{(3)}_l=\uparrow)\leq q^{2(k+1-L/2-l)}
&\mbox{if}& l < k+1-L/2
\eeann
Combining these inequalities and summing over $l$ yields
\be
\sum_{l=-L/2}^{L/2-1} \avg{Y_l}{\Sigma_0,k}\leq 2\frac{1+q^2}{1-q^2}
\ee
for all $k=0,\ldots,L+1$. Together with \eq{PtoY} this concludes the proof.
\end{proof}

\Section{Bound for the denominator}

Note that $\psi(\Lambda,n)=T(\phi)\psi_0(\Lambda,n)$, where $T(\phi)$ is 
the unitary operator defined by,
\be
T(\phi) = \bigotimes_{x \in \Lambda} (\ket{\uparrow}\bra{\uparrow}
  + e^{i \phi(x)} \ket{\downarrow} \bra{\downarrow}).
\ee
In particular, $\Vert T(\phi)\psi_0(\Lambda,n)\Vert^2=
\Vert\psi(\Lambda,n)\Vert^2=Z(\Lambda,n)$. For convenience, we will
sometimes omit the arguments $\Lambda$ and $n$ from the notation.
In this section we will consider the half-filled system, i.e,
$\rho=n/\vert\Lambda\vert=1/2$. This corresponds to $\mu=0$. 

\begin{theorem}[Bound on $|\frac{<\psi_0|\psi>}{<\psi_0|\psi_0>}|$ ]
\label{den:bounds}
Considering a perturbed state in the volume $\Lambda_0$ defined by
(\ref{ps}) we have that canonical and grand-canonical expectations
of the perturbed state are arbitrarily close for large volumes $\Lambda$
in the sense:
\be
\label{gcbl}
\left\vert\frac{\ip{\psi}{\psi_0}}{\ip{\psi_0}{\psi_0}}
  - \avgGC{T(\phi)}{\Lambda,\mu} \right\vert
  \le \frac{\ln^2(A-A_0) + 2 (1+a^2) A_0^2 + 4}{2(A-A_0)} 
  + \frac{4 A_0}{q^2 A^{1/2} - 2 A_0}.
\ee
Moreover, with the ansatz defined by (\ref{ps}), the grand canonical 
expectation is bounded as
\begin{eqnarray}
\label{deb}
&&\ln \left|\avgGC{T(\phi)}{\Lambda,\mu}\right|^2
  \leq \qquad \qquad \\
&& \qquad \qquad \leq - q^{2 \delta(\mu)} \frac{A_R \Sc^2}{4R^2} 
  \Bigg[\frac{\|\tilde\phi\|_{L^2(\tilde\Omega)}^2}{m(\tilde\Omega)} 
  - \frac{\sqrt{6}}{R} \|\partial_{\tilde{y}} \tilde\phi \|_{L^\infty} 
    \|\tilde\phi\|_{L^\infty}
  - \frac{\Sc^2}{12 R^2} \|\tilde\phi\|_{L^\infty}^4 
\Bigg] \nonumber
\end{eqnarray}
where $\delta(\mu)$ is the distance of $\mu$ from its closest integer
neighbor.
(Recall that we have defined the $L^\infty$-norm of a tensor 
to be the $L^\infty$-norm of its maximum component.)
\end{theorem}

\begin{proof}
The proof of equation (\ref{gcbl}) is a direct consequence of the 
equivalence of ensembles because, since $T(\phi)$ is a unitary operator,
$\|T(\phi)\| = 1$. Let us now consider the proof of equation (\ref{deb}).

We wish to bound the denominator from below;
i.e.\ to demonstrate that $1 - |\avg{T(\phi)}{\Lambda,n}|^2$ is not too small.
This is tantamount to showing that $|\avg{T(\phi)}{\Lambda,n}|^2$
is not too close to 1.
Furthermore, we know this quantity 
lies between 0 and 1.
We estimate the actual canonical average with the grand canonical average,
and take the logarithm in order to exploit the factorization properties
of the grand canonical ensemble.
First, we note
\be\label{rat}
\left|\avgGC{T(\phi)}{\Lambda,\mu}\right|
  = \left| \prod_{x \in \Lambda_0} \frac{1+e^{i \phi(x)} q^{2 (l(x)-\mu)}} 
  {1+q^{2 (l(x)-\mu)}} \right|.
\ee
Recall the definition $a=-2 \ln{q}$. 
This allows us a more convenient form
in place of (\ref{rat})
\begin{eqnarray}
\label{uu}
&&\left| \prod_{x \in \Lambda_0} \frac{1+e^{i \phi(x)} q^{2 (l(x)-\mu)}}
  {1+q^{2 (l(x)-\mu)}} \right|^2 \nonumber\\
  &&\quad = \prod_{x \in \Lambda_0} 
  \frac{e^{2 a (l(x)-\mu)} + 2 \cos \phi(x) e^{a (l(x)-\mu)} + 1}
  {e^{2 a (l(x)-\mu)} + 2 e^{a (l(x)-\mu)} + 1} \\
  &&\quad = \prod_{x \in \Lambda_0} 
  \left(1 - \frac{1}{2}(1-\tanh^2[a (l(x)-\mu)/2])
	(1 - \cos \phi(x))\right).
\nonumber
\end{eqnarray}
We partition the product over planes and estimate the logarithm, thus:
\begin{eqnarray*}
\ln \left|\avgGC{T(\phi)}{\Lambda,\mu}\right|^2
  &=& \ln \left( \prod_{x\in \Lambda_0} 1 - \frac{1}{2}
  (1-\tanh^{2}[a (l(x)-\mu)/2])(1 - \cos \phi(x)) \right ) \\
  &\leq& - \frac{1}{2} \sum_{x \in \Lambda-0} (1 - \tanh^2[a (l(x)-\mu)/2]) 
	(1 - \cos \phi(x)) \\
  &=& - \frac{1}{2} \sum_{l=-L/2}^{L/2} (1 - \tanh^2[a (l-\mu)/2])
	\sum_{x \in \Gamma^l_R} (1 - \cos \phi(x)) .
\end{eqnarray*}
We may approximate $1 - \cos(\phi(x))$
by $\frac{1}{2} \phi(x)^2$, with an error no larger than  
$\frac{1}{24} \|\phi\|_{L^\infty}^4 $
which is the same as $ \frac{\Sc^4}{24 R^4} \|\tilde\phi\|_{L^\infty}^4$.
In this case
\be
\label{sumsum}
\ln \left|\frac{Z_{GC}(\Lambda_0,\mu,\phi)}{Z_{GC}(\Lambda_0,\mu,0)}\right|^2
\leq - \frac{1}{2} \sum_{l=-L/2}^{L/2} (1 - \tanh^2 [a (l-\mu)/2])
  \left[\sum_{x \in \Gamma^l_R} \frac{1}{2} \phi_x^2 
  - \frac{\Sc^4 |\Gamma^l_R|}{24 R^4} \|\tilde\phi\|_\infty^4 \right] .
\ee
We may approximate the sum over $\Gamma^l_R$ with an integral such that 
the error is bounded by 
$\frac{\rho \Sc^2 |\Gamma^l_R|}{R^3} 
\|\nabla_{\tilde{y}} \phi \|_{L^\infty} \|\tilde\phi\|_{L^\infty}$.
We may bound the sum 
$\sum_{l=-L/2}^{L/2} (1 - \tanh^2[a (l-\mu)/2])$
from below by its largest term (since all the terms are positive).
The largest term occurs for that integer $l$ which is closest to $\mu$.
Thus, defining 
$\delta(\mu) = \min(\mu - \lfloor{\mu}\rfloor,\lceil{\mu}\rceil - \mu)$, 
we see
\be
\sum_{l=-L/2}^{L/2} (1 - \tanh^2[a(l-\mu)/2])  
  \geq 1 - \tanh^2[a \delta(\mu)/2] 
  = \frac{4}{(q^{\delta(\mu)} + q^{-\delta(\mu)})^2} 
  \geq q^{2\delta(\mu)},
\ee
Using these bounds, we may continue the estimate of $\eq{sumsum}$.
We arrive at
\begin{eqnarray}
\label{den:exp:bound}
&&\ln \left|\avgGC{T(\phi)}{\Lambda,\mu}\right|^2
  \leq \qquad \qquad \\
&& \qquad \qquad \leq - q^{2 \delta(\mu)}
  \frac{\Sc^2 |\Gamma^l_R| }{4R^2} 
  \Bigg[\frac{\|\tilde\phi\|_{L^2(\tilde\Omega)}^2}{m(\tilde\Omega)} 
  - \frac{\rho}{R} \|\nabla_{\tilde{y}} \tilde\phi \|_{L^\infty} 
    \|\tilde\phi\|_{L^\infty}
  - \frac{\Sc^2}{12 R^2} \|\tilde\phi\|_{L^\infty}^4 
\Bigg] \nonumber
\end{eqnarray}
Since $|\nabla_{\tilde y}\tilde\phi| 
\leq 2 \|\partial_{\tilde y}\tilde\phi\|_{l^\infty}$
and since $\rho = \sqrt{3/2}$, we have equation \eq{deb}.
\end{proof}

\subsection{Bound on the Ratio}
We will now combine the results of the bound on the numerator and the bound 
on the denominator to get a true bound on the spectral gap.
We first allow $\Lambda \nearrow \Ir^3$ in the appropriate fashion so that
$\varepsilon \searrow 0$.
Then we consider the case that $S\to 0$, holding $R$ fixed.
This means that we consider a perturbation to the ground state which is 
very small.
But since the ground state has energy zero, the energy of the 
perturbed state is entirely due to the small perturbation.
In fact it is proportional to the size of the perturbation, and from this we 
obtain a linearized (with respect to amplitude of $\phi$) bound: 
In fact we have, combining \eq{vp}, \eq{bound:en}, and \eq{gcbl}
\be
\label{rb}
\gamma_1 \leq \frac{16 q^{2(1-\delta(\mu))}}{(1-q^2)R^2} \cdot
  \frac{\|\nabla_{\tilde{y}}\tilde\phi\|_{L^2(\tilde\Omega)}^2/m(\tilde\Omega) 
  + \frac{6}{R} \|\partial_{\tilde y}^2\tilde\phi\|_\infty 
  \|\partial_{\tilde y}\tilde\phi\|_\infty}
  {\|\tilde\phi\|_{L^2(\tilde\Omega)}^2/m(\tilde\Omega) - \frac{\sqrt{6}}{R} 
  \|\partial_{\tilde y}\tilde\phi\|_\infty
  \|\tilde\phi\|_\infty}
\ee
Note that this bound is homogeneous with respect to the amplitude of $\phi$,
which is the result of our linearization.
We observe that, whatever the form for $\tilde\phi$, as long as it is smooth
we have the same asymptotic behavior for the bound on the spectral gap.
Namely $\gamma_1 = O(1/R^2)$.
This said, it is certainly worthwhile to find a best bound, which we 
take up presently.


\subsection{The Bessel Function Ansatz}
Let us write the leading-order term in the bound for the spectral gap:
\be
E(\tilde\phi) 
  = \frac{\|\nabla_{\tilde{y}}\tilde\phi\|_2^2}{\|\tilde\phi\|_2^2}.
\ee
In order to minimize the bound on the spectral gap, we will minimize
the functional $E(\phi)$ amongst all functions $\phi$ which possess two
continuous derivatives and which vanish on the boundary of 
the rescaled perturbed region $\tilde\Omega$.
(In order that the ``small'' phase $\phi$ match the external phase 
of $0,\pm2 \pi,\dots$ on $\partial\Omega$, it must be zero there.
Thus $\tilde\phi \equiv 0$ on $\partial\tilde\Omega$.)
Therefore, we consider the first variation
\be
\lim_{\tau \to 0} \frac{1}{\tau}[E(\phi+\tau \phi') - E(\phi)] 
  = \frac{2 \int \nabla\phi \cdot \nabla\phi'}{\int \phi^2}
  - \frac{2 \int \phi \phi' \int |\nabla \phi|^2}{\int\phi^2 \int\phi^2}.
\ee
Setting the first variation to zero for all test functions 
$\phi'$ leads to the eigenvalue problem for Laplace's equation
\be
\label{pde:1}
\left\{ \begin{array}{ll}
  -\nabla^2 \tilde\phi = \lambda \tilde\phi 
  & \textrm{ in } \tilde\Omega, \\
  \tilde\phi = 0 & \textrm{ on } \partial\tilde\Omega,
  \end{array} \right.
\ee
where $\lambda = E(\phi)$.

We choose, for our domain, the unit disk.
We seek the solution to equation (\ref{pde:1}) which minimizes $\lambda$,
but with the restriction that $\phi$ must possess two continuous derivatives.
So the fundamental solution, which is the logarithm,  is disallowed
(and, in fact, has higher energy).
We seek the first eigenstate of the Laplacian above the ground state.
This is a classic problem, found in any elementary PDE text, 
with the Bessel Function for the solution:
$$\tilde\phi(\tilde y) = J_0(z_0 r) ,$$
where $r = |\tilde y|$, $J_0$ is the zeroth Bessel function,
and $z_0 \approx 2.406$ is its first zero.
Now, using this choice for $\phi$ and the bounds \eq{rb}, we obtain
\be
\gamma_1 \leq \frac{16 q^{2(1-\delta(\mu))}}{(1-q^2)R^2} \cdot
  \frac{1.56 + \frac{6}{R} (2.90)(1.40)}
  {0.27 - \frac{\sqrt{6}}{R} (1.40)(1)}.
\ee
Thus,
\be
\gamma_1 \leq \frac{100 q^{2(1-\delta(\mu))}}{(1-q^2)R^2} 
\quad \textrm{for}
\quad R>70.
\ee

\Section{Results from the 1D grand canonical ensemble}

\subsection{The mean number of particles in a stick}
\label{sect:navg}
Recall that $\Sigma$ is a 1D stick running parallel to the 111 axis.
So, it is actually a 1D spin chain.
We wish to estimate the mean number of particles in
$\Sigma$, for the grand canonical ensemble.
This is
\begin{eqnarray}
  \navg := \ZGC{\Sigma}{\mu}^{-1}
	\sum_{n=1}^{L+1} n q^{-2 \mu n} Z(\Sigma,n)\\ 
	= \ZGC{\Sigma}{\mu}^{-1}
	\sum_{n=1}^{L+1} n e^{a \mu n} Z(\Sigma,n) .\nonumber
\end{eqnarray}
where $\Sigma$ is the interval 
$\{-\frac{L}{2},-\frac{L}{2}+1,\dots,\frac{L}{2}\}$.
(Recall $a = - 2 \log q$.)
By a standard calculation, we have
\be
\navg = \frac{1}{a} \frac{\partial}{{\partial \mu}} 
\log \ZGC{\Sigma}{\mu} .
\ee
On the other hand, the grand canonical partition function factorizes, 
as we have seen, so that
\be 
  \navg = \sum_{l=-L/2}^{L/2} 
	\frac{e^{a(\mu - l)} {1 + e^{a(\mu - l)}}}
	= \sum_{l=-L/2}^{L/2} \frac{1}{2} 
	\left[1 - \tanh\left(\frac{a}{2} (l - \mu)\right)\right].
\ee
An examination of the graph of the function $x \mapsto 1 - \tanh(x)$
reveals an approximate heaviside function, with support on the negative axis.
We define the function 
\be
\eta(x) = \left\{ 
  \begin{array}{ll}
  1 & x < 0,\\
  1/2 & x = 0,\\
  0 & x > 0.\end{array} \right. 
\ee
Then, as long as $-L/2 \leq \mu \leq L/2$, we remark
\be
\navg = 
  \left\{ \begin{array}{ll}
  \floor{\mu} + \frac{L}{2} & \mu \not\in \Ir,\\
  \mu + \frac{L+1}{2} & \mu \in \Ir\end{array} \right\} 
  +\sum_{l=-L/2}^{L/2} \left(\frac{1}{2} 
  - \frac{1}{2} \tanh\left(\frac{a}{2}(l - \mu)\right) - \eta(l - \mu)\right).
\ee
We make the definition
\be
F_L(\mu) = \navg - \left(\mu + \frac{L+1}{2}\right) 
\ee
For $\mu$ in the range above one may determine (by combining the two tails
in the series and estimating upwards by an integral) that
\be 
|F_\infty(\mu) - F_L(\mu)| 
  \leq \frac{1}{a} \ln\left(\frac{1 + \exp(-\frac{a}{2}(\frac{L}{2} - \mu))}
  {1 + \exp(-\frac{a}{2}(\frac{L}{2} + \mu))}\right)
\ee
Notice that in case $\mu = 0$, there is no error at all in estimating 
$F_L$ by $F_\infty$,
and, furthermore, $F_\infty(0)=0$.
It is clear that $F_\infty(\mu)$ is periodic in $\mu$ with period 1, 
because it is a sum over the entire integer lattice, so it will suffice for 
us to consider $\mu$ in the range $]0,1[$.
A straightforward calculation then yields
\begin{eqnarray*}
F_\infty(\mu) 
  & = & -\mu + \frac{1}{2} - \frac{1}{1 + e^{a \mu}}
  + \sum_{l=1}^\infty \left[ \frac{1}{1 + e^{a(l - \mu)}} -
  \frac{1}{1 + e^{a(l + \mu)}} \right] \\
  & = & -\mu + \frac{1}{2}\tanh(a \mu)
  + \sum_{l=1}^\infty \frac{\sinh(a \mu)}
	{\cosh(a \mu) + \cosh(a l)}  
\end{eqnarray*}
Defining $\{\mu\} = \mu - \floor{\mu}$ we have
\be
F_\infty(\mu) = - \{\mu\} + \frac{1}{2}\tanh(a \{\mu\})
  + \sum_{l=1}^\infty \frac{\sinh(a \{\mu\})}
	{\cosh(a \{\mu\}) + \cosh(a l)}
\label{defF}\ee
for all values of $\mu$.

\begin{lemma}
\label{lem:avgn}
The function $F_\infty$ defined in \eq{defF} has the following properties:
i) $F_\infty$ is periodic with period $1$, i.e, 
$F_\infty(\mu+1)=F_\infty(\mu)$, for all
$\mu\in\Rl$.\newline
ii) $F_\infty$ is odd about $\mu=1/2$, i.e., 
$F_\infty(1-\mu)=-F_\infty(\mu)$, for
all $\mu\in\Rl$.\newline
iii) $-1\leq F_\infty(\mu)\leq 1$, for all $\mu\in\Rl$.\newline
iv) $F_\infty(\mu) = 0$ for $\mu \in \Z$ and $\mu \in \frac{1}{2} + \Z$.
I.e.\ the estimate $\navg = \mu + \frac{L+1}{2}$ is exact for 
half-integer and integer filling.
\end{lemma}

\begin{figure}[t]
\begin{center}
\resizebox{13cm}{!}{\includegraphics{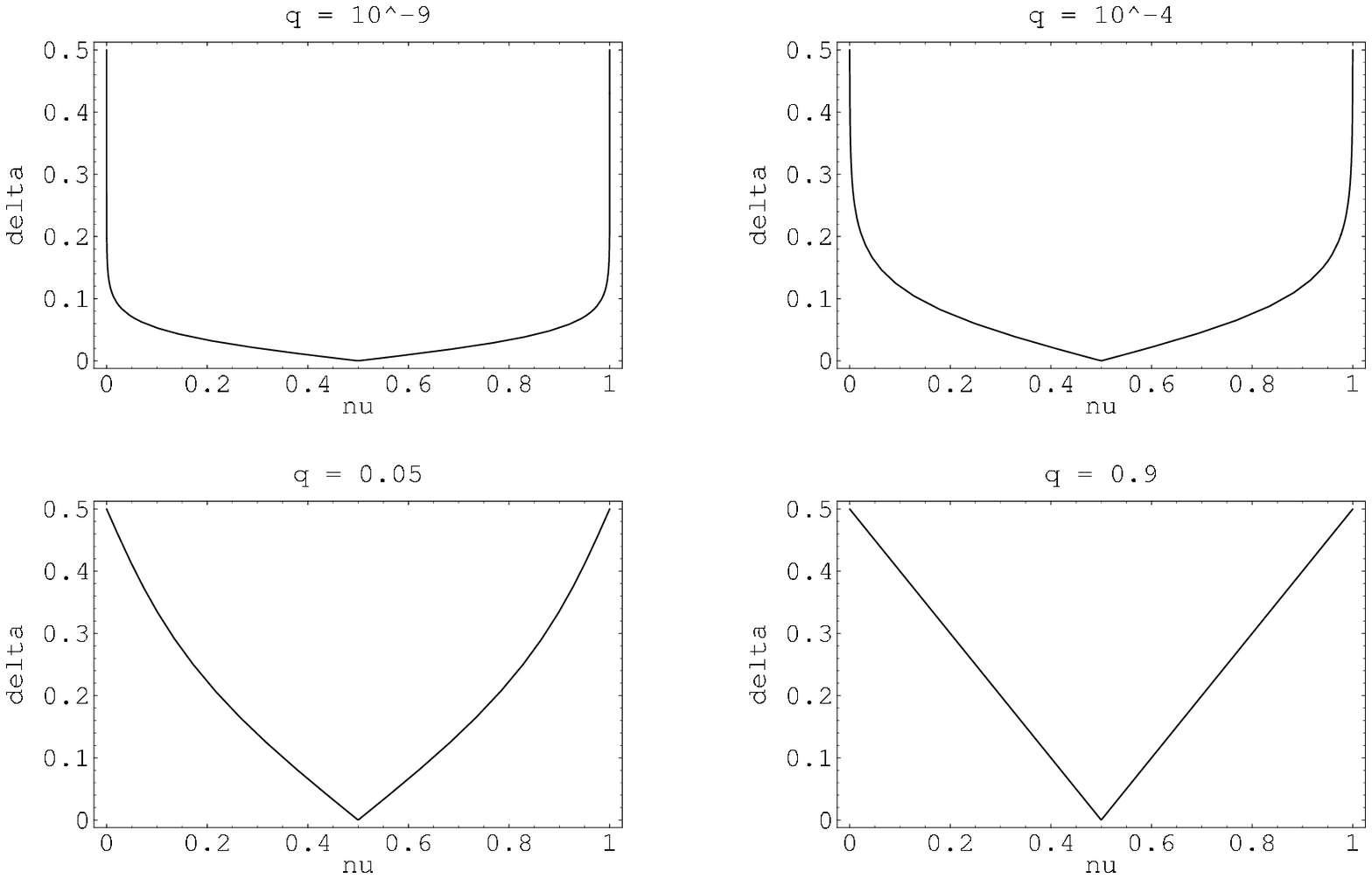}}
\parbox{14truecm}{\caption{\baselineskip=16 pt\small\label{fig:deltanu}
A plot of the function $\delta(\nu,q)$ for four different values of $q$.}
}
\end{center}
\end{figure}

\begin{proof}
The periodicity of $F_\infty$ follows directly from its definition. 
To prove (ii), define $F(\mu)$ for $0<\mu<1$ as
\be
F(\mu)
  = \sum_{k=1}^\infty
	\left[\frac{1}{1+e^{a(l-\mu)}}
	-\frac{1}{1+e^{a(l+\mu)}}\right]
    -\frac{1}{1+e^{a\mu}}
\label{otherF}\ee
Then,
\beann
F(1-\mu)&=&\sum_{l=1}^\infty\left[\frac{1}{1+e^{a(l-1+\mu)}}
-\frac{1}{1+e^{a(l+1-\mu)}}\right]-\frac{1}{1+e^{a(1-\mu)}}\\
&=&\sum_{l=1}^\infty\left[\frac{1}{1+e^{a(l+\mu)}}
-\frac{1}{1+e^{a(l-\mu)}}\right]\\
&&+\frac{1}{1+e^{a\mu}}+\frac{1}{1+e^{a(1-\mu)}}-\frac{1}{1+e^{a(1-\mu)}}\\
&=&-F(\mu)
\eeann
And clearly the remainder term
$$
\begin{cases}
\frac{1}{2}-\{\mu\},& \textrm{if $\mu \not\in \Ir$}\\
                   0 ,& \textrm{if $\mu \in \Ir$}
\end{cases}
$$
satisfies property (ii).
For the bounds, we first restrict ourselves to $\mu\in[0,1]$. 
For $\mu\geq 0$, we note that \eq{defF} implies
$$
F_\infty(\mu) \geq -\{\mu\}\geq -1.
$$
Then we use property ii) in combination with this bound
to also get the upper bound for $\mu\in[0,1]$.
$$
F_\infty(\mu)=-F_\infty(1-\mu)\leq 1
$$
Due to the peridicity property i), the upper and lower bound
are automatically extended to all real $\mu$.
The special values stated in iv) are straightforward from
\eq{defF} and \eq{otherF}.
\end{proof}
\begin{figure}[t]
\begin{center}
\resizebox{13cm}{!}{\includegraphics{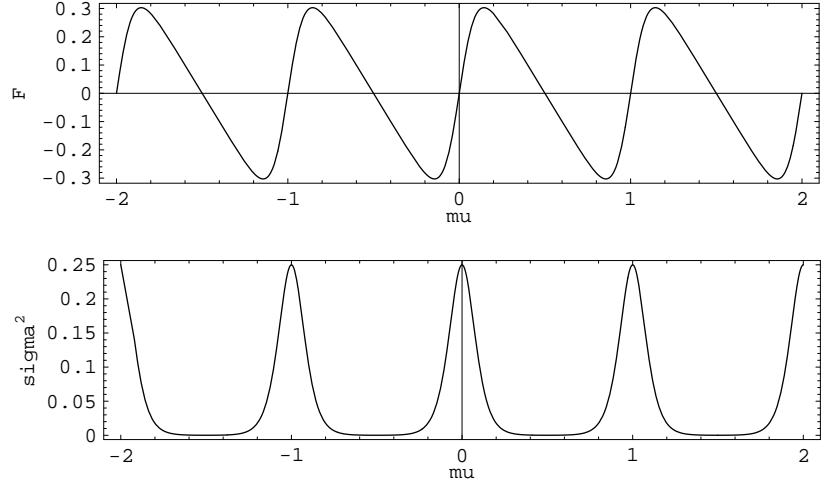}}
\parbox{14truecm}{\caption{\baselineskip=16 pt\small\label{fig:F_sigma}
A plot of the functions $F_\infty(\mu)$ and $\sigma^2(\mu)$, with $q
=e^{-10}$.}
}
\end{center}
\label{F-sigma figure}
\end{figure}
We can define the quantity $\delta(\mu)=\min(\vert\mu-\floor{\mu}\vert,
\vert 1-\mu+\floor{\mu}\vert)$, where $\floor{\mu}$ is the integer part of
$\mu$. In general, the relation between $\mu$ and $\nu$ depends
nontrivially on $q$ and the function $\delta$ can be thought as
$\delta(q,\nu)$. But for all $q$, $0<q<1$, one has $\delta(q,1/2)=0$ and
$\delta(q,0)=1/2$. See Figure \ref{fig:deltanu}.

\subsection{The variance of the number of particles in a stick}
\label{sect:var_n}

In the same way as was done above for the mean, we can compute
the variance of the number of particles in a stick in the 
grand canonical ensemble by using the standard formula
\be
\sigma^2(\mu,L) =\avgGC{\Num^2}{\Sigma,\mu} -(\avgGC{\Num}{\Sigma,\mu})^2
= \frac{1}{a^2} \frac{\partial^2}{{\partial \mu^2}} 
\log \ZGC{\Sigma}{\mu} ,
\ee 
which gives
\be
\sigma^2(\mu,L) = \frac{1}{4}\sum_{l=-L/2}^{L/2} \frac{1}{\cosh^2
(\frac{a}{2}
(l-\mu))}
\ee
Define 
\be
\sigma^2(\mu)=\lim_{L\to\infty}\sigma^2(\mu,L)
\ee
Then, the speed of convergence of this limit is bounded as follows:
\be
\vert \sigma^2(\mu) - \sigma^2(\mu,L)\vert
\leq 2\sum_{n=0}^{\infty}e^{-a(n-\mu + L/2)}=\frac{2q^{2(L/2-\mu)}}{1-q^2}
\ee
It is clear that $\sigma^2(\mu)$ is a periodic function of $\mu$ with 
period 1. It is not hard to see that $\sigma^2(\mu,L)$ is $C^\infty$ and 
attains its maximum in all integers and its minimum in the integers $+
1/2$.
It is easy to derive upper and lower bounds for $\sigma^2(\mu,L)$.
An upper bound is given by 
\be
\sigma^2(\mu,L)\leq \sum_{l=-L/2}^{L/2} e^{-\vert a(l-\mu)\vert}
\leq \sum_{l=-L/2}^{L/2} e^{-a\vert l\vert}\leq 1+
\frac{2e^{-a}}{1-e^{-a}}
\label{upper_sigma}\ee
and a lower bound  can be obtained using the crude bound
$2\cosh x\leq 2 e^{\vert x\vert }$:
\be
\sigma^2(\mu,L)\geq \frac{1}{4}\sum_{n=1}^{L} e^{-\vert an\vert}
\geq \frac{1}{4}\frac{e^{-a}-e^{-a (L+1)}}{1-e^{-a}}
\label{lower_sigma}\ee
{From} \eq{upper_sigma} and \eq{lower_sigma} we see that the limit 
$\sigma^2(\mu)$ satisfies the bounds
\be
\frac{1}{4}\frac{q^2}{1-q^2}\leq \sigma^2(\mu) \leq \frac{1+q^2}{1-q^2},
\ee
for all real $\mu$ and where we have again used the relation $e^{-a}=q^2$.

For the afficionados, one can also show that 
\be\
\lim_{q\downarrow 0}\sigma^2(\mu)=\begin{cases}0          & \textrm{if $\mu\not\in\Ir$}\\
                                         \frac{1}{4}& \textrm{if $\mu\in\Ir$}\end{cases}					 
\ee
The interpretation is simple. When $\mu\in\Ir$, the interface (kink) in
the one-dimensional system is located at a lattice site, which is occupied
by a particle with probability 1/2. Clearly, the variance of the particle
number is them $1/4$. However, for $\mu\not\in\Ir$, the kink is centered
at a position not belonging to the lattice and the state converges, as
$q\downarrow 0$, to a deterministic configuration with zero variance
for the particle number.

\subsection{Estimating $\avgGC{q^{2|\Num - \avg{\Num}{}|}}{\Sigma,\mu}$}
\label{sect:avgexp}
We begin with the obvious fact
\be
q^{2|\Num - \avg{\Num}{}|} 
  \leq q^{2\Num - 2\avg{\Num}{}} + q^{2\avg{\Num}{} - 2\Num}
\ee
from which it follows that
\be
\avgGC{q^{2|\Num - \avg{\Num}{}|}}{\Sigma,\mu}
  \leq q^{-2\avg{\Num}{}} \avgGC{q^{2\Num}}{\Sigma,\mu}
  + q^{2\avg{\Num}{}} \avgGC{q^{-2\Num}}{\Sigma,\mu}.
\ee
Now, we observe
\be
\label{qexp:avg}
\avg{q^{2\Num}}{\Sigma,\mu} 
  = \frac{\sum_{n=0}^{L+1} q^{2n} q^{-2 \mu n} Z(\Sigma,n)}
{Z^{GC}(\Sigma,\mu)}
  = \frac{Z^{GC}(\Sigma,\mu-1)}{Z^{GC}(\Sigma,\mu)} .
\ee
Since
\be
  Z^{GC}(\Sigma,\mu) = \prod_{l=-L/2}{L/2} (1 + q^{2(l - \mu)})
\ee
equation \eq{qexp:avg} leads us to conclude
\be
\avg{q^{2\Num}}{\Sigma,\mu}
  = \frac{1 + q^{2(L/2+1-\mu)}}{1+q^{-2(L/2+\mu)}} 
  \leq 2 q^{2(L/2+\mu)}.
\ee
Similarly,
\be
\avg{q^{-2\Num}}{\Sigma,\mu}
  = \frac{1 + q^{-2(L/2+1+\mu)}}{1 + q^{2(L/2 - \mu)}}
  \leq 2 q^{-2(L/2+1+\mu)}.  
\ee
Using the results of section \ref{sect:navg}, we then have
\be
\avgGC{q^{2|\Num-\avg{\Num}{}|}}{\Sigma,\mu}
	\leq 4 q^{-1-|F_L(\mu)|}
	\leq 4 q^{-2}.
\ee
If we wish to calculate $\avgGC{q^{2|\Num-\avg{\Num}{}|}}{\Lambda,\mu}$,
where $\Lambda$ is comprised of $A$ sticks, then nothing changes except 
that each estimate is raised to the power $A$.
Thus, $\avgGC{q^{2|\Num-\avg{\Num}{}|}}{\Lambda,\mu} \leq 2^{A+1} q^{-2A}$.

\section*{Acknowledgements}

O.B. was supported by Fapesp under grant 97/14430-2. B.N. was partially
supported by the National Science Foundation under grant \# DMS-9706599.


\pagebreak

\addcontentsline{toc}{chapter}{\textit{An Extension of Equivalence of Ensembles}}

\Section{Second Proof of the Activity Bounds}

We recall that the canonical ground state in a cylindrical region 
$\Lambda=\Lambda(L,A)$
with length $L$ and cross-sectional area $A$ is determined by
$$
\psi_0(\Lambda,n) = \sum_{\substack{\{\alpha(x)\} \in \{0,1\}^\Lambda\\
\sum_x \alpha(x) = n}} q^{\sum_x l(x) \alpha(x)} 
\bigotimes_{x \in \Lambda} \left|{\frac{1}{2}-\alpha(x)}\right\rangle_x\, .
$$
We prefer to index the state a little bit differently.
First of all, for the purposes of calculating the activity bounds, we may take
$$
\Lambda = \Lambda(L,n) 
  = \{-\frac{1}{2}(L-1),-\frac{1}{2}(L-3),\dots,\frac{1}{2}(L-1)\}
  \times \{1,2,\dots,n\}\, ,
$$ 
with points labelled by $x=(l,j)$, $l(x) = l$.
Second, we choose to multiply the state by a normalizing factor so that the limit
$L \to \infty$ can be taken more easily.
Thus we define a canonical ground state $\Psi_0(L,n,M)$ as a state with magnetization
$M$, by
$$
\Psi_0(L,n,M) = \sum_{\substack{\{m(l,j)\} \in \{\pm  1/2\}^{\Lambda(L,n)}\\
\sum_{(l,j)} m(l,j) = M}}
q^{\sum_{(l,j)} (\frac{1}{2} |l| - m(l,j) l)} \bigotimes_{(l,j) \in \Lambda(L,n)}
\ket{m(l,j)}_{(l,j)}\, .
$$
(Note, $M \in \{-\frac{1}{2} |\Lambda|,-\frac{1}{2}|\Lambda|+1,\dots,
\frac{1}{2}|\Lambda|\}$.)
This is related to the original definition by the formula
$$
\psi_0([1,L]\times [1,n],N) 
  = q^{\frac{1}{2} N (L+1) - \sum_{(l,r)} \frac{1}{2} |l|}\,
  \Psi_0(L,n,,\frac{1}{2}|\Lambda|-N)\, ,
$$
and one can easily translate the results back and forth between these two pictures.

We define
$$
Z(L,n,M) = \|\Psi_0(L,n,M)\|^2\, ,
$$
and we wish to find an asymptotic expression for the value of $Z(L,n,M)$ as
$n$ and $M$ approach infinity, such that $M/n$ converges, or even more genreally if $M/n$
is just $o(n)$.
At first, we consider the case that $L$ is finite;
then we may consider the case of an infinitely long cylinder as the limit of
finite cylinders, by taking the thermodynamic limit.
The normalization, $Z(L,n,M)$ is called the canonical partition function at 
zero temperature.
In order to calculate the asymptotics of $Z(L,n,M)$, it is useful to define the 
``grand canonical partition function'', which is simply the generating
function for the canonical partition functions:
\begin{equation}
\label{ForEOEProof}
\begin{split}
Z^{\textrm{GC}}(L,n,z) 
  &:= \sum_{M = - \frac{1}{2}|\Lambda|}^{\frac{1}{2}|\Lambda|} z^{2M} Z(L,n,M) \\
  &= \prod_{l \in \{-\frac{1}{2}(L-1),-\frac{1}{2}(L-3),\dots,\frac{1}{2}(L-1)\}} 
  (z q^{|l|-l} + z^{-1} q^{|l|+l})^n \\
  &= F_L(z)^n\, ,
\end{split}
\end{equation}
where
\begin{align*}
F_L(z) 
  &= \prod_{l \in \{-\frac{1}{2}(L-1),-\frac{1}{2}(L-3),\dots,\frac{1}{2}(L-1)\}} 
  (z q^{|l|-l} + z^{-1} q^{|l|+l}) \\
  &= \begin{cases} \displaystyle
  \prod_{l=1}^{L/2} (1 + q^{2l-1} z^2) (1 + q^{2l-1} z^{-2})\, ,
  & \textrm{$L$ even}\, ;\\
  \displaystyle (z + z^{-1})
  \prod_{l=1}^{(L-1)/2} (1 + q^{2l} z^2) (1 + q^{2l} z^{-2})\, ,
  & \textrm{$L$ odd}\, .
  \end{cases}
\end{align*}
Using the Cauchy integral formula, we then have, for $M \in \frac{1}{2} \Ir$,
and any $r \in \Rl^+$,
\begin{align*}
Z(L,n,M) 
  &= \oint_{C(0;r)} z^{-2M} F_L(z)^n \frac{dz}{2 \pi i z} \\
  &= r^{-2M} \int_{-1/2}^{1/2} e^{-4 \pi i M \theta} 
  F_L(r e^{2 \pi i \theta})^n\, d\theta\, .
\end{align*}
Since $Z(L,n,M)$ is definitely real, we can take the real part of the integrand.
Also, by consideration of the definition of $F_L(z)$, we know that the integral
is zero unless $M \in \{-\frac{1}{2} |\Lambda|,-\frac{1}{2}(|\Lambda|+1),\dots,
\frac{1}{2} |\Lambda|\}$, which in particular implies that $2M$ has the same 
parity as $|\Lambda|$.
For such $M$, it is apparent that the integrand of
$$
Z(L,n,M) = r^{-2M} \int_{-1/2}^{1/2} \real\left[e^{-4 \pi i M \theta} 
  F_L(r e^{2 \pi i \theta})^n\right]\, d\theta
$$
is periodic with period equal to $1/2$.
Hence, for such $M$,
$$
Z(L,n,M) = r^{-2M} \int_{-1/2}^{1/2} \real\left[e^{-2 \pi i M \theta} 
  F_L(r e^{\pi i \theta})^n\right]\, d\theta\, .
$$
Since $r \in \Rl^+$ we can define $t \in \Rl$ such that $r = e^{t/2}$.
We also define $M = m n$, where 
$m \in \left\{-\frac{L}{2},-\frac{L}{2}+\frac{1}{n},\dots,\frac{L}{2}\right\}$.
Then
\begin{equation}
\label{PreLemma}
\begin{split}
Z(L,n,mn)
  &= \left(e^{-mt} F_L(e^{t/2})\right)^n 
  \int_{-1/2}^{1/2} \exp\left[ \frac{n}{2} 
  \ln \left|\frac{F_L(e^{t/2} e^{\pi i \theta})}{F_L(e^{t/2})}\right|^2\right]\\
  &\hspace{125pt}
  \cos[n\{-2\pi m \theta + \arg(F_L(e^{t/2} e^{\pi i \theta}))\}]\, d\theta\, .
\end{split}
\end{equation}
This is the type of equation that can be solved by the method of steepest
descents.
All one needs to do is figure out the leading-order behavior of the logarithm
of $F_L(e^{t/2} e^{\pi i \theta})$.
We go to the trouble of proving the following lemma, instead of simply referring
to one of the many complex analysis texts in which one can find the method
of steepest descents, because we need control of the error terms.
(We need better control of the error terms then was available with the 
previous activity bounds.)

\begin{lemma}
\label{2ndACBoundsLemma}
For each $L \in \Nl_{\geq 2}$,
define a diffeomorphism $\phi_L:\left(-\frac{L}{2},\frac{L}{2}\right) \to \Rl$ 
by the inverse formula
$$
\phi_L^{-1}(t) = \frac{1}{2}
  \sum_{l \in \left\{-\frac{L-1}{2},-\frac{L-3}{2},\dots,
  \frac{L-1}{2}\right\}} \frac{\sinh(t)}{\cosh(t) + \cosh(\eta l)}\, ,
$$
where $q = e^{-\eta}$.
(See the remark at the end of the lemma for a description of the function $\phi_L^{-1}(t)$.)
Then one has the bounds, 
\begin{equation}
\label{2ndACBoundsLemmaEq1}
\begin{split}
Z(L,n,mn) &\leq q^{\frac{n}{2}\floor{\frac{1}{2} L^2}}
  \sqrt{\frac{\phi_L'(m)}{2 \pi n}}  
  \exp\left[n \int_m^L \phi_L(s)\, ds\right]
  (1 + \varepsilon_{\textrm{upper}})\, ,\\
Z(L,n,mn) &\geq q^{\frac{n}{2}\floor{\frac{1}{2} L^2}}
  \sqrt{\frac{\phi_L'(m)}{2 \pi n}}  
  \exp\left[n \int_m^L \phi_L(s)\, ds\right]
  (1 - \varepsilon_{\textrm{lower}})\, ,
\end{split}
\end{equation}
for 
$m \in \left\{-\frac{L}{2},-\frac{L}{2}+\frac{1}{n},\dots,\frac{L}{2}\right\}$,
where, for any $0<\epsilon<\frac{1}{4}$,
\begin{equation}
\label{2ndACBoundsLemmaEq2}
\begin{split}
\varepsilon_{\textrm{upper}} &= 
  \frac{n^{-1+2\epsilon}}{6 - n^{-1+2\epsilon}}
  + \sqrt{\frac{2 \pi n }{\phi_L'(m)}} 
  \exp\left[\frac{-2 n ^{2 \epsilon}(1 - \frac{1}{6} n^{-1+2\epsilon})^2}
  {\phi_L'(m)}\right]\, ,\\
\varepsilon_{\textrm{lower}} &=
  n^{-1+2\epsilon} 
  + \frac{4 n^{-1+4\epsilon} (m \phi_L'(m))^2}{(1-2 n^{-1+2\epsilon})^4}\\
  &\hspace{25pt} + \left(1  + \sqrt{\frac{2 \pi n}{\phi_L'(m)}}\right)
  \exp\left[\frac{-2 n^{2\epsilon} (1 - \frac{1}{6} n^{-1+2\epsilon})^2}
  {\phi_L'(m)}\right]\, .
\end{split}
\end{equation}
\end{lemma}

\textbf{Remark:}
For information about the functions $\phi_L^{-1}(t)$,
the reader should refer back to Sections \ref{sect:navg}.
Note the $F_L$ of that section has nothing to do with the $F_L$ of this section.
$F_L(\mu)$ and $\phi_L^{-1}(t)$ are connected by the identity
$$
\phi_L(t) = \frac{t}{\eta} + F_L\left(\frac{t}{\eta}\right)\, ,
$$
where $\eta$ of this section is identical to $a$ from before.
We apologize for the change of notation, some of this is to facilitate the proof,
and some is by accident.

\textbf{Erratum:} In Section \ref{sect:navg}, we wrote:
A straightforward calculation then yields
\begin{eqnarray*}
F_\infty(\mu) 
  & = & -\mu + \frac{1}{2} - \frac{1}{1 + e^{a \mu}}
  + \sum_{l=1}^\infty \left[ \frac{1}{1 + e^{a(l - \mu)}} -
  \frac{1}{1 + e^{a(l + \mu)}} \right] \\
  & = & -\mu + \frac{1}{2}\tanh(a \mu)
  + \sum_{l=1}^\infty \frac{\sinh(a \mu)}
	{\cosh(a \mu) + \cosh(a l)}  
\end{eqnarray*}
But this is wrong because
$$
\frac{1}{2} - \frac{1}{1+e^{a\mu}} = - \frac{1}{2} \tanh\left(\frac{a\mu}{2}\right)\, .
$$
Thus
\begin{eqnarray*}
F_\infty(\mu) 
  & = & -\mu + \frac{1}{2}\tanh\left(\frac{a \mu}{2}\right)
  + \sum_{l=1}^\infty \frac{\sinh(a \mu)}
	{\cosh(a \mu) + \cosh(a l)}  \\
  & = & -\{\mu\} + \frac{1}{2}\tanh\left(\frac{a \{\mu\}}{2}\right)
  + \sum_{l=1}^\infty \frac{\sinh(a \{\mu\})}
	{\cosh(a \{\mu\}) + \cosh(a l)}  \, ,
\end{eqnarray*}
instead. 
This does not change any of the results which followed the mistake, because we never used that particular
calculation.

\textbf{More Remarks:}
We will define in the course of the proof $\sigma_L^2(t) = (\phi_L^{-1})'(t)$.
This definition almost agrees with the definition from \ref{sect:var_n}, except that
$\sigma_L^2(t) = \sigma\left(L,\frac{t}{\eta}\right)$.
The most important feature of the function $\phi_L^{-1}(t)$ 
is that $\phi_L$ has two limits : one for $L$
even, and one for $L$ odd.
Choosing a sequence $L(i)$ to be of a definite parity
with $L(i) \to \infty$, the functions
$\phi_{L(i)}$ and $\phi_{L(i)}'$ converge uniformly on compact sets.
This means that if $m$ is bounded, the error bounds $\varepsilon_{\textrm{upper}}$
and $\varepsilon_{\textrm{lower}}$ can be made uniform in $L$.
Plots of the limiting functions $F_\infty(\mu)$ and $\sigma^2(\mu)$,
which are limits for $L$ odd, are given in Figure \ref{F-sigma figure}.
In case the reader wants to know something else about $\phi_L^{-1}(t)$ and $\sigma_L^2(t)$,
we note that one can easily derive the following sine-series
\begin{align*}
\lim_{\substack{L \to \infty \\ L\ \textrm{odd}}} \phi_L^{-1}(t) 
  &= \frac{1}{2} \sum_{l \in \Ir} \frac{\sinh(t)}{\cosh(t) + \cosh(\eta l)} \\
  &= \frac{t}{\eta} + \frac{\pi}{\eta} \sum_{n\in\Ir\setminus\{0\}} \frac{\sin(2 n \pi t /\eta)}
  {\sinh(2 n \pi^2/\eta)}\, ;\\
\Rightarrow
\lim_{\substack{L \to \infty \\ L\ \textrm{even}}} \phi_L^{-1}(t)
  &= \frac{t}{\eta} - \frac{\pi}{\eta} \sum_{n\in\Ir\setminus\{0\}} \frac{\sin(2 n \pi t /\eta)}
  {\sinh(2 n \pi^2/\eta)}\, .
\end{align*}
This obviously implies
$$
\lim_{\substack{L \to \infty \\ L\ \textrm{odd, even}}}
  = -\frac{1}{\eta^2} \mp \frac{1}{\eta^2} \sum_{n\in\Ir\setminus\{0\}} \frac{2 n \pi^2}{\eta} 
  \operatorname{csch} \left(\frac{2n\pi^2}{\eta}\right) \cos\left(\frac{2 n \pi t}{\eta}\right)\, .
$$

\begin{proof}
We observe that for $\theta \in \left(-\frac{1}{2},\frac{1}{2}\right)$,
$$
\frac{F_L(e^{t/2} e^{\pi i \theta})}{F_L(e^{t/2})}
  = \prod_{l \in \left\{-\frac{L-1}{2},\dots,\frac{L-1}{2}\right\}}
  \left(\frac{\cosh(2 l \eta) + \cosh t \cos 2 \pi \theta + i \sinh t \sin 2\pi \theta}
  {\cosh(2 l \eta) + \cosh t}\right)^{1/2}\, ,
$$
where for the square root we use the principle branch (see Figure \ref{BranchFig}).
This is well-defined, i.e.\ in a single connected, simply connected domain
not intersecting the branch cut, since 
$\theta \in \left(-\frac{1}{2},\frac{1}{2}\right)$.
\begin{figure}
\label{BranchFig}
\begin{center}
\epsfig{file=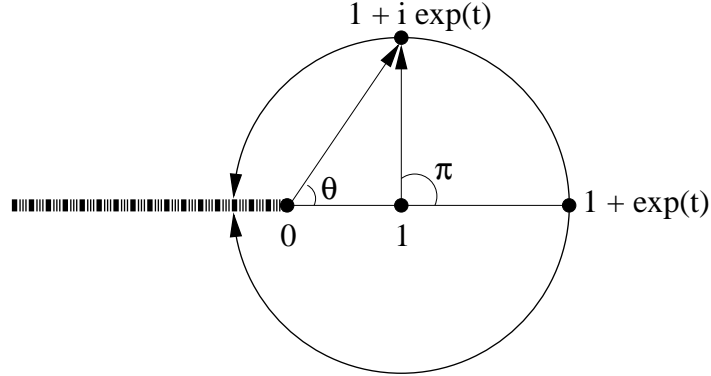,height=5truecm}
\end{center}
\caption{Principle branch for logarithm and square-root}
\end{figure}
From this it follows that
$$
\arg\left(\frac{F_L(e^{t/2} e^{\pi i \theta})}{F_L(e^{t/2})}\right)
  = \frac{1}{2} \sum_{l \in \left\{-\frac{L-1}{2},\dots,\frac{L-1}{2}\right\}}
  \arctan\left(\frac{\sinh t \sin 2\pi \theta}{\cosh t \cos 2\pi \theta
  + \cosh 2 l \eta}\right)\, .
$$
We also have
\begin{align*}
\left|\frac{F_L(e^{t/2} e^{\pi i \theta})}{F_L(e^{t/2})}\right|^2
  &= \prod_{l \in \left\{-\frac{L-1}{2},\dots,\frac{L-1}{2}\right\}}
  \frac{1 + 2 q^{2l} e^t \cos 2 \pi \theta + q^{4l} e^{2t}}
  {1 + 2 q^{2l} e^t + q^{4l} e^{2t}} \\
  &= \prod_{l \in \left\{-\frac{L-1}{2},\dots,\frac{L-1}{2}\right\}}
  \left(1 - \sech^2\left(\frac{t}{2} - l \eta\right) \sin^2 \pi \theta\right)\, .
\end{align*}
So, for $L \in \Nl_{\geq 2}$, and
$m \in \left\{-\frac{L}{2},-\frac{L}{2}+\frac{1}{n},\dots,\frac{L}{2}\right\}$,
and any $t \in \Rl$,
\begin{equation}
\label{StartingPt}
\begin{split}
&Z(L,n,mn) 
  = \left(e^{-mt} F_L(e^{t/2})\right)^n 
  \int_{-1/2}^{1/2} \exp[n \mathcal{F}_L(t,\theta)]
  \cos[n \mathcal{G}_L(t,\theta)]\, d\theta\, ,\\
&\mathcal{F}_L(t,\theta) 
 = \frac{1}{2} \sum_{l \in \{-\frac{L-1}{2},\dots,\frac{L-1}{2}\}}
  \ln\left(1 - \sech^2\left(\frac{t + 2 l \eta}{2}\right) 
  \sin^2\pi \theta\right)\, ,\\
&\mathcal{G}_L(t,\theta) 
  = - 2 \pi m \theta + \frac{1}{2}
  \sum_{l \in \left\{-\frac{L-1}{2},\dots,\frac{L-1}{2}\right\}}
  \arctan\left(\frac{\sinh t \sin 2\pi\theta}{\cosh t \cos 2\pi\theta
  + \cosh 2 l \eta}\right)\, .
\end{split}
\end{equation}
So, to estimate $Z(L,n,mn)$, we have to estimate two things: the integral
$$
\int_{-1/2}^{1/2} \exp[n \mathcal{F}_L(t,\theta)] \cos[n \mathcal{G}_L(t,\theta)]\,
  d\theta\, ,
$$
and the prefactor $\left(e^{-mt} F_L(e^{t/2})\right)^{n}$.
We estimate the integral first, since this will lead us to a particular choice
of $t$, from which it will be easy to extract $e^{-mt} F_L(e^{t/2})$.

The point to estimating the integral is to show that the largest constribution
comes from a very small neighborhood of $\theta = 0$, roughly of length $n^{-1/2}$.
For this reason, for any $\epsilon$ such that $0<\epsilon<\frac{1}{4}$,
we break the integral up as follows
\begin{align*}
\int\limits_{|\theta|<\frac{1}{2}} \exp[n\mathcal{F}_L(t,\theta)] 
\cos[n \mathcal{G}_L(t,\theta)]\, d\theta
  &= \int\limits_{\frac{1}{2}>|\theta|>\pi^{-1} n^{-\frac{1}{2}+\epsilon}} 
  \exp[n\mathcal{F}_L(t,\theta)] \cos[n \mathcal{G}_L(t,\theta)]\, d\theta\\
  &+ \int\limits_{|\theta|<\pi^{-1} n^{-\frac{1}{2} + \epsilon}} 
  \exp[n\mathcal{F}_L(t,\theta)] \cos[n \mathcal{G}_L(t,\theta)]\, d\theta\, .
\end{align*}

We estimate the large $|\theta|$-integral first.
For $\frac{1}{2} > |\theta| > \pi^{-1} n^{-\frac{1}{2} + \epsilon}$,
$$
\sin^2 \pi \theta > \sin^2 (n^{-\frac{1}{2} + \epsilon}) > 
  n^{-1+2\epsilon} (1 - \frac{1}{6} n^{-1+2\epsilon})^2\, .
$$
Hence
\begin{align*}
\ln\left(1 - \sech^2\left(\frac{t + 2 l \eta}{2}\right) \sin^2\pi \theta\right) 
  &< - \sech^2\left(\frac{t + 2 l \eta}{2}\right) \sin^2\pi \theta \\
  &< - n^{-1+2\epsilon} (1 - \frac{1}{6} n^{-1+2\epsilon})^2
  \sech^2\left(\frac{t + 2 l \eta}{2}\right)\, .
\end{align*}
Hence,
$$
\mathcal{F}_L(t,\theta) \leq - \frac{1}{2} n^{-1+2\epsilon} 
  (1 - \frac{1}{6} n^{-1+2\epsilon})^2 
  \sum_{l \in \{-\frac{L-1}{2},\dots,\frac{L-1}{2}\}}
  \sech^2\left(\frac{t + 2 l \eta}{2}\right)\, .
$$
We define
$$
\sigma_L^2(t) = \frac{1}{4} \sum_{l \in \{-\frac{L-1}{2},\dots,\frac{L-1}{2}\}}
  \sech^2\left(\frac{t + 2 l \eta}{2}\right)
  = (\phi_L^{-1})'(t)\, .
$$
Then
$$
\mathcal{F}_L(t,\theta) 
  \leq - 2 n^{-1+2\epsilon} \sigma_L^2(t) (1 - \frac{1}{6} n^{-1+2\epsilon})^2\, .
$$
Hence
\begin{equation}
\label{BernoulliLemmaEq2}
\begin{split}
\left|\,\int\limits_{\frac{1}{2}>|\theta|>\pi^{-1} n^{-\frac{1}{2}+\epsilon}} 
  \exp[n \mathcal{F}_L(t,\theta)] \cos[n \mathcal{G}_L(t,\theta)]\, d\theta\right|
  &\leq \int\limits_{\frac{1}{2}>|\theta|>\pi^{-1} n^{-\frac{1}{2}+\epsilon}} 
  \exp[n \mathcal{F}_L(t,\theta)]\, d\theta \\
  &\leq \exp\left[- 2 n^{2\epsilon} \sigma_L^2(t) 
  (1 - \frac{1}{6} n^{-1+2\epsilon})^2\right]\, .
\end{split}
\end{equation}
This means that an exponentially small portion of the integral comes from
the large $|\theta|$ range.
However, we must choose $\epsilon > 0$, otherwise our estimate is order 1.

Next, we consider small $|\theta|$.
First we estimate $\mathcal{G}_L(t,\theta)$ for 
$|\theta| < \pi^{-1} n^{-\frac{1}{2} + \epsilon}$.
Suppose, first, $m>0$, $t>0$ and $\theta>0$.
Then
\begin{align*}
\frac{\sinh t \sin 2\pi\theta}{\cosh t \cos 2\pi\theta + \cosh 2 l \eta}
  &\leq \frac{2 \pi \theta \sinh t}{(\cosh t + \cosh 2 l \eta)
  (1 - 2 \pi^2 \theta^2)}\\
  &\leq 2 \pi \theta (1 - 2 n^{-1+2\epsilon})^{-1} \frac{\sinh t}
  {\cosh t + \cosh 2 l \eta}\, ,
\end{align*}
and
\begin{align*}
\frac{\sinh t \sin 2\pi\theta}{\cosh t \cos 2\pi\theta + \cosh 2 l \eta}
  &\geq 2 \pi \theta (1 - \frac{2}{3} \pi^2 \theta^2) \frac{\sinh t}
  {\cosh t + \cosh 2 l \eta}\\
  &\geq 2 \pi \theta (1 - \frac{2}{3} n^{-1+2\epsilon}) \frac{\sinh t}
  {\cosh t + \cosh 2 l \eta}\, .
\end{align*}
Of course, for $x\geq 0$,
$$
\arctan x = \int_0^x \frac{dy}{1+y^2}\, ,
$$
so
$$
x(1 - \frac{1}{3} x^2) \leq \arctan x \leq x\, .
$$
Thus,
\begin{align*}
\arctan\left(\frac{\sinh t \sin 2\pi\theta}{\cosh t \cos 2\pi\theta
  + \cosh 2 l} \eta\right)
  &\leq 2 \pi \theta (1 - 2 n^{-1+2\epsilon})^{-1} \frac{\sinh t}
  {\cosh t + \cosh 2 l \eta}\, ,\\
\arctan\left(\frac{\sinh t \sin 2\pi\theta}{\cosh t \cos 2\pi\theta
  + \cosh 2 l} \eta\right)
  &\geq 2 \pi \theta 
  \left(1 - \frac{2 n^{-1+2\epsilon}}{(1 - 2n^{-1+2\epsilon})^2}\right) 
  \frac{\sinh t}{\cosh t + \cosh 2 l \eta}\, .\\
\end{align*}
So
\begin{align*}
\mathcal{G}_L(t,\theta)
  &\leq - 2 \pi m \theta + 
  \pi \theta (1 - 2 n^{-1+2\epsilon})^{-1} 
  \sum_{l \in \left\{-\frac{L-1}{2},\dots,\frac{L-1}{2}\right\}}
  \frac{\sinh t}{\cosh t + \cosh 2 l \eta} \\
  &= 2 \pi \theta \left(-m + (1 - 2 n^{-1+2\epsilon})^{-1} \phi_L^{-1}(t)\right)\, ,\\
\mathcal{G}_L(t,\theta)
  &\geq - 2 \pi m \theta + 
  \pi \theta \left(1 - \frac{2 n^{-1+2\epsilon}}{(1 - 2 n^{-1+2\epsilon})^2}\right) 
  \sum_{l \in \left\{-\frac{L-1}{2},\dots,\frac{L-1}{2}\right\}}
  \frac{\sinh t}{\cosh t + \cosh 2 l \eta} \\
  &= 2 \pi \theta \left(-m + \left(1 - \frac{2 n^{-1+2\epsilon}}
  {(1 - 2 n^{-1+2\epsilon})^2}\right) \phi_L^{-1}(t)\right)\, .
\end{align*}
We will see that for the exponential, 
the proper scaling of $\theta$ comes from the definition
$$
  \theta = \frac{\Theta}{\sqrt{4 \pi^2 n \sigma_L^2(t)}}\, .
$$
Then the arument of cosine will be divergent as $n \to \infty$, unless
$\phi_L^{-1}(t) = m$, i.e.\ $t = \phi_L(m)$.
This is the condition for a stationary phase, because it is the condition that
makes the phase, i.e.\ the argument of cosine, become $o(1)$ as $n \to \infty$.
With this choice of $t$,
$$
- \frac{4 m n^{-1+2\epsilon} \pi \theta}{(1-2 n^{-1+2\epsilon})^2}
  \leq \mathcal{G}_L(t,\theta) 
  \leq \frac{4 m n^{-1+2\epsilon} \pi \theta}{1-2 n^{-1+2\epsilon}}\, .
$$
So
$$
|\mathcal{G}_L(t,\theta)|
  \leq \frac{4 |m| n^{-1+2\epsilon} \pi |\theta|}{(1 - 2n^{-1+2\epsilon})^2}\, .
$$
By symmetry, this same condition is true for any signs of $m$ and $\theta$,
as long as $t = \phi_L(m)$ and $|\theta| < \pi^{-1} n^{-\frac{1}{2} + \epsilon}$.
Implementing the scaling we mentioned before, 
$$
|n \mathcal{G}_L(t,\theta)|
  \leq \frac{2 n^{-\frac{1}{2}+2\epsilon} |\Theta|}{(1 - 2n^{-1+2\epsilon})^2} 
  |m \phi_L'(m)|\, .
$$
Then, for $|\theta| < \pi^{-1} n^{-\frac{1}{2} + \epsilon}$,
\begin{equation}
\label{BernoulliLemmaEq3}
1 \geq \cos[n \mathcal{G}_L(t,\theta)]
  \geq 1 - \frac{2 n^{-1+4\epsilon} \Theta^2}{(1 - 2 n^{-1+2\epsilon})^4}
  (m \phi_L'(m))^2\, .
\end{equation}
This equation will be useful to us, shortly.

Next we derive upper bounds for the exponential term.
For $|\theta| \leq \pi^{-1} n^{-\frac{1}{2}+\epsilon}$,
$$
\sin^2 \pi\theta \geq \pi^2 \theta^2 (1 - \frac{1}{3} \pi^2 \theta^2)
  \geq \pi^2 \theta^2 (1 - \frac{1}{6} n^{-1+2\epsilon})^2\, .
$$
Thus
\begin{align*}
\ln\left(1 - \sech^2\left(\frac{t + 2 l \eta}{2}\right) \sin^2\pi \theta\right) 
  &\leq - \sech^2\left(\frac{t + 2 l \eta}{2}\right) \sin^2\pi \theta \\
  &\leq - \pi^2 \theta^2 (1 - \frac{1}{6} n^{-1+2\epsilon})^2
  \sech^2\left(\frac{t + 2 l \eta}{2}\right)\, .
\end{align*}
This means
$$
\mathcal{F}_L(t,\theta)
  \leq - 2 \pi^2 \theta^2 (1 - \frac{1}{6} n^{-1+2\epsilon})^2 \sigma_L^2(t)\, ,
$$
and
$$
n \mathcal{F}_L(t,\theta)
  \leq -\frac{1}{2} \Theta^2 (1 - \frac{1}{6} n^{-1+2\epsilon})^2\, .
$$
Hence
\begin{align*}
\int\limits_{|\theta|<\pi^{-1} n^{-\frac{1}{2}+\epsilon}}
  \exp[n \mathcal{F}_L(t,\theta)] \cos[n \mathcal{G}_L(t,\theta)]\, d\theta
  &\leq \int\limits_{|\theta|<\pi^{-1} n^{-\frac{1}{2}+\epsilon}}
  \exp[n \mathcal{F}_L(t,\theta)]\, d\theta \\
  &\hspace{-100pt}
  \leq \int_{-2 n^\epsilon \sigma_L(t)}^{2 n^\epsilon \sigma_L(t)}
  \exp\left[-\frac{1}{2} \Theta^2 (1 - \frac{1}{6} n^{-1+2\epsilon})^2\right]
  \frac{d\Theta}{\sqrt{4 \pi^2 n \sigma_L^2(t)}} \\
  &\hspace{-100pt}
  \leq \int_{-\infty}^{\infty}
  \exp\left[-\frac{1}{2} \Theta^2 (1 - \frac{1}{6} n^{-1+2\epsilon})^2\right]
  \frac{d\Theta}{\sqrt{4 \pi^2 n \sigma_L^2(t)}} \\
  &\hspace{-100pt}
  = \frac{1}{\sqrt{2 \pi n \sigma_L^2(t) (1 - \frac{1}{6} n^{-1+2\epsilon})^2}}\, .
\end{align*}
Then, incorporating equation \eq{BernoulliLemmaEq2},
\begin{equation}
\label{BernoulliLemmaEq4}
\begin{split}
\int_{-1/2}^{1/2} \exp[n \mathcal{F}_L(t,\theta)] \cos[n \mathcal{G}_L(t,\theta)]\,
  d\theta
  &\leq \frac{1}{\sqrt{2 \pi n \sigma_L^2(t) (1 - \frac{1}{6} n^{-1+2\epsilon})^2}}\\
  &\hspace{50pt} + \exp\left[-2 n^{2\epsilon} \sigma_L^2(t) 
  (1 - \frac{1}{6} n^{-1+2\epsilon})^2\right]\, .
\end{split}
\end{equation}

Now we come to lower bounds for the exponential term.
For $|\theta|< \pi^{-1} n^{-\frac{1}{2} + \epsilon}$,
$$
\sin^2 \pi \theta \leq \pi^2 \theta^2 \leq n^{-1 + 2 \epsilon}\, .
$$
Since
$$
\ln(1 - x) = - \int_0^{x} \frac{dy}{1-y} \geq - \frac{x}{1-x}\, ,
$$
we have
$$
\ln\left(1 - \sech^2\left(\frac{t + 2 l \eta}{2}\right) \sin^2\pi \theta\right) 
  \geq - \pi^2 \theta^2 \sech^2\left(\frac{t + 2 l \eta}{2}\right) 
  (1 - n^{-1+2\epsilon})^{-1}\, .
$$
Thus,
$$
\mathcal{F}_L(t,\theta)
  \geq - 2 \pi^2 \theta^2 \sigma_L^2(t) (1 - n^{-1+2\epsilon})^{-1}\, ,
$$
which implies, using the scaling for $\theta$,
$$
n \mathcal{F}_L(t,\theta)
  \geq -\frac{1}{2} \Theta^2 (1 - n^{-1+2\epsilon})^{-1}\, .
$$
Then, employing equation \eq{BernoulliLemmaEq3},
\begin{align*}
&\int\limits_{|\theta|\leq \pi^{-1} n^{-\frac{1}{2}+\epsilon}}
  \exp[n\mathcal{F}_L(t,\theta)] \cos[n \mathcal{G}_L(t,\theta)]\, d\theta\\
  &\hspace{100pt}
  \geq \int_{-2 n^{\epsilon} \sigma_L(t)}^{2 n^{\epsilon} \sigma_L(t)}
  e^{-\frac{1}{2} \Theta^2(1-n^{-1+2\epsilon})} (1 - C \Theta^2) 
  \frac{d\Theta}{\sqrt{4 \pi^2 n \sigma_L^2(t)}}\\
  &\hspace{100pt}
  \geq \int_{-\infty}^{\infty}
  e^{-\frac{1}{2} \Theta^2(1-n^{-1+2\epsilon})} 
  (1 - e^{-2n^{2\epsilon}\sigma_L^2(t)} - C \Theta^2) 
  \frac{d\Theta}{\sqrt{4 \pi^2 n \sigma_L^2(t)}}\\
  &\hspace{100pt}
  \geq \frac{1}{\sqrt{2\pi n \sigma_L^2(t)}}
  \left(1 - n^{-1+2\epsilon} - \exp[-2n^{2\epsilon}\sigma_L^2(t)]
  - 2 C\right)\, ,
\end{align*}
where
$$
C = \frac{2 n^{-1+4\epsilon} (m \phi_L'(m))^2}{(1 - 2 n^{-1+2\epsilon})^4}\, .
$$
Combining this with \eq{BernoulliLemmaEq2}, gives
\begin{equation}
\label{BernoulliLemmaEq5}
\begin{split}
&\int_{-1/2}^{1/2} \exp[n\mathcal{F}_L(t,\theta)] 
  \cos[n \mathcal{G}_L(t,\theta)]\, d\theta\\
  &\hspace{50pt}
  \geq \frac{1}{\sqrt{2\pi n \sigma_L^2(t)}}
  \left(1 - n^{-1+2\epsilon} - \exp[-2n^{2\epsilon}\sigma_L^2(t)]
  - \frac{4 n^{-1+4\epsilon} (m \phi_L'(m))^2}{(1 - 2 n^{-1+2\epsilon})^4}\right)\\
  &\hspace{75pt}
  - \exp\left[-2n^{2\epsilon} \sigma_L^2(t) 
  (1 - \frac{1}{6} n^{-1+2\epsilon})^2\right]\, .
\end{split}
\end{equation}
 
Finally, we come to the prefactor $(e^{-mt} F_L(e^{t/2}))^n$.
We observe that
\begin{align*}
\ln(e^{-mt} F_L(e^{t/2}))
  &= - m t 
  + \frac{1}{2} \sum_{l \in \left\{-\frac{L-1}{2},\dots,\frac{L-1}{2}\right\}}
  \ln\left(q^{2|l|} (2 \cosh t + 2 \cosh 2 \eta l)\right) \\
  &= - \phi_L^{-1}(t) t
  + \frac{1}{2} \sum_{l \in \left\{-\frac{L-1}{2},\dots,\frac{L-1}{2}\right\}}
  \ln\left(q^{2|l|} (2 \cosh t + 2 \cosh 2 \eta l)\right) \\
  &=: \Phi_L(t)\, .
\end{align*}
Then we see
\begin{align*}
\Phi'_L(t) &= - t (\phi_L^{-1})'(t) - \phi_L^{-1}(t)
  + \frac{1}{2} \sum_{l \in \left\{-\frac{L-1}{2},\dots,\frac{L-1}{2}\right\}}
  \frac{\sinh t}{\cosh t + \cosh 2 l \eta} \\
  &= - t (\phi_L^{-1})'(t)\, .
\end{align*}
On the other hand,
\begin{align*}
\lim_{t \to \infty} \Phi_L(t)
  &= \lim_{t \to \infty} \Bigg[\ln q^{\sum_l |l|} + 
  \left(\frac{L}{2} - \phi_L^{-1}\right) t \\
  &\hspace{50pt}
  + \frac{1}{2} \sum_{l \in \left\{-\frac{L-1}{2},\dots,\frac{L-1}{2}\right\}}
  \ln\left(1 + e^{-2t} + 2 e^{-t} \cosh 2 \eta l\right) \Bigg] \\
  &= \ln q^{\frac{1}{2} \floor{\frac{1}{2} L^2}}\, .
\end{align*}
Thus,
\begin{equation}
\label{BernoulliLemmaEq6}
\begin{split}
e^{-mt} F_L(e^{t/2}) 
  &= q^{\frac{1}{2} \floor{\frac{1}{2} L^2}}
  \exp\left[\int_{t}^\infty s (\phi_L^{-1})'(s)\, ds\right]\\
  &= q^{\frac{1}{2} \floor{\frac{1}{2} L^2}}
  \exp\left[\int_{m}^{L/2} \phi_L(s)\, ds\right]\, .
\end{split}
\end{equation}
Puting equations \eq{BernoulliLemmaEq4}, \eq{BernoulliLemmaEq5} 
and \eq{BernoulliLemmaEq6} together yields 
\eq{2ndACBoundsLemmaEq1} and \eq{2ndACBoundsLemmaEq2}.
\end{proof}

\Section{Equivalence of ensembles in two dimensions}
\label{2d Generalization}
For $n_0 < n$, and
$X \in \Obs_{\Lambda(L,n_0)} \subset \Obs_{\Lambda(L,n_0)}$, 
(the inclusion is the natural inclusion obtained by taking
$X \otimes \unity_{\Lambda(L,n) \setminus \Lambda(L,n_0)}$,)
\begin{gather*}
\langle{X}\rangle^{\textrm{can}}_{(L,n,mn)} = 
  \frac{\ip{\Psi_0(L,n,mn)}{X\otimes\unity_{\Lambda(L,n)\setminus\Lambda(L,n_0)}
  \cdot \Psi_0(L,n,mn)}}
  {\|\Psi_0(L,n,mn)\|^2} \\
\langle{X}\rangle^{\textrm{GC}}_{(L,n_0,r)} = 
  \frac{\ip{\Psi^{\textrm{GC}}_0(L,n_0,r)}{X\Psi^{\textrm{GC}}_0(L,n_0,r)}}
  {\|\Psi^{\textrm{GC}}_0(L,n_0,r)\|^2} 
\end{gather*}
where
\begin{align*}
\Psi^{\textrm{GC}}_0(L,n_0,r) 
  &= \sum_{M_0} \Psi_0(L,n_0,M) r^{2M_0}\\
  &= \bigotimes_{(l,j) \in \Lambda(L,n_0)} 
  (r q^{(|l|-l)/2} \ket{\uparrow}_{(l,j)} 
  + r^{-1} q^{(|l|+l)/2} \ket{\downarrow}_{(l,j)})\, .
\end{align*}
The following theorem shows that these two expectations are close.
\begin{theorem}
\textrm{\bf(Equivalence of Ensembles)}
\label{EOE2}
Suppose $X \in \Obs_{\Lambda(L,n_0)}$, $n_0<n$ and 
$m \in \left\{-\frac{L-1}{n},\dots,\frac{L-1}{n}\right\}$.
Then
$$
\left|\langle{X}\rangle^{\textrm{can}}_{(L,n,mn)}
  - \langle{X}\rangle^{\textrm{GC}}_{(L,n_0,e^{t/2})}\right|
  \leq 2 (\mathcal{E}_1 + \mathcal{E}_2) \|X\|\, ,
$$
where $t = \phi_L(m)$ and for any $\epsilon_2$ satisfying $0<\epsilon_2<\frac{1}{2}$,
$$
\mathcal{E}_1 := \frac{4}{1 - e^{|t|} q^{2n^{\epsilon_2}}}
\left[\frac{e^{|t n^{\epsilon_2}|} q^{n^{2\epsilon_2}} \sum_{k \in \Ir} q^{k^2}}
  {F_L(e^{t/2}) (q^2;q^2)_\infty}\right]^{n_0}
$$
and, defining 
$\mathcal{D} = B\left(\frac{mn}{n-n_0};\frac{n_0 n^{\epsilon_2}}{n-n_0}\right)$,
\begin{align*}
\mathcal{E}_2 &:= \max\Bigg(\left[1 + \frac{n_0}{n-n_0}\right]^{1/2} 
  \left[1 + \frac{\|\phi_L''\|_{L^\infty(\mathcal{D})}}{\phi'_L(m)}
  \frac{(|m|+n^{\epsilon_2})n_0}{n-n_0}\right]^{1/2}\\
  &\qquad \exp\left[\frac{1}{2} \|\phi_L'\|_{L^\infty(\mathcal{D})}
  \frac{(|m|+n^{\epsilon_2})^2 n_0^2}{n-n_0}\right]
  \frac{1 + \|\varepsilon_{\textrm{upper}}(L,n-n_0,\cdot)\|_{L^\infty(\mathcal{D})}}
  {1 - \varepsilon_{\textrm{lower}}(L,n,m)} - 1\, ,\\
  &\qquad 1 - \left[1 + \frac{n_0}{n-n_0}\right]^{1/2} 
  \left[1 - \frac{\|\phi_L''\|_{L^\infty(\mathcal{D})}}{\phi'_L(m)}
  \frac{(|m|+n^{\epsilon_2})n_0}{n-n_0}\right]^{1/2}\\
  &\qquad \exp\left[-\frac{1}{2} \|\phi_L'\|_{L^\infty(\mathcal{D})}
  \frac{(|m|+n^{\epsilon_2})^2 n_0^2}{n-n_0}\right]
  \frac{1 - \|\varepsilon_{\textrm{lower}}(L,n-n_0,\cdot)\|_{L^\infty(\mathcal{D})}}
  {1 + \varepsilon_{\textrm{upper}}(L,n,m)} \Bigg)\, .
\end{align*}
\end{theorem}
Before proving the theorem, we will need a simple lemma.
\begin{lemma}
\label{EOELemma}
We have the following estimate, which is useful for large $|M_0|$,
$$
Z(L,n_0,M_0) 
  \leq \left[\frac{q^{\frac{M_0}{n_0}\left(\frac{M_0}{n_0}+1\right)} 
  \sum_{k \in \Ir} q^{k^2}}
  {(q^2;q^2)_\infty}\right]^2\, .
$$
\end{lemma}
\begin{proof} \textbf{(of Lemma). }
\begin{align*}
Z(L,n_0,M_0)
  &= \sum_{\substack{\{m(j)\} \in (\Ir(+\frac{1}{2}))^{n_0} \\
  \sum_j m(j) = M_0}} \prod_{j=1}^{n_0} Z(L,m_j) \\
  &\leq \sum_{\substack{\{m(j)\} \in (\Ir(+\frac{1}{2}))^{n_0} \\
  \sum_j m(j) = M_0}} \prod_{j=1}^{n_0} \lim_{\substack{L_\alpha \to \infty\\
  L_\alpha \equiv L (\operatorname{mod} 2)}} Z(L_\alpha,m_j) \\
  &\leq \sum_{\substack{\{m(j)\} \in (\Ir(+\frac{1}{2}))^{n_0} \\
  \sum_j m(j) = M_0}} \prod_{j=1}^{n_0}
  \frac{q^{m(j)^2 + m(j)}}{(q^2;q^2)_\infty}\, ,
\end{align*}
the last equation following by equation \eq{PartitionFunctionIdentity}.
Defining $\widetilde{m}(j) = m(j) - \frac{M_0}{n_0}$, we have
\begin{align*}
Z(L,n_0,M_0)
  &\leq 
  \sum_{\substack{\{\widetilde{m}(j)\} \in 
  (\Ir(+\frac{1}{2})-\frac{M_0}{n_0})^{n_0} \\
  \sum_j \widetilde{m}(j) = 0}}
  \frac{q^{\sum_{j=1}^{n_0} \left(\frac{M_0}{n_0} + \widetilde{m}(j)\right)
  \left(\frac{M_0}{n_0} + \widetilde{m}(j) + 1\right)}}{(q^2;q^2)_\infty^{n_0}}\\
  &\leq \frac{q^{M_0\left(\frac{M_0}{n_0} + 1\right)}}{(q^2;q^2)_\infty^{n_0}}
  \sum_{\substack{\{\widetilde{m}(j)\} \in 
  (\Ir(+\frac{1}{2})-\frac{M_0}{n_0})^{n_0} \\
  \sum_j \widetilde{m}(j) = 0}}
  q^{\sum_j \widetilde{m}(j)^2}\\
  &\leq\left(\frac{q^{\frac{M_0}{n_0}\left(\frac{M_0}{n_0}+1\right)}}
  {(q^2;q^2)_\infty}
  \sum_{\widetilde{m} \in \Ir} q^{\widetilde{m}^2}\right)^{n_0}\, .
\end{align*}
\end{proof}
Now we can prove the theorem.

\begin{proof}
We observe that 
\begin{align*}
\Psi_0(L,n,m n)
  &= \sum_{M_0 \in \left\{-\frac{1}{2}|\Lambda(L,n_0)|,\dots,
  \frac{1}{2}|\Lambda(L,n_0)|\right\}}
  \Psi_0(L,n_0,M_0) \otimes \\
  &\hspace{150pt} j(\Psi_0(L,n-n_0,m n - M_0))\, .
\end{align*}
where $j:\Hil_{\Lambda(L,n-n_0)} \to \Hil_{\Lambda(L,n)\setminus \Lambda(L,n_0)}$
is the obvious isomorphism, induced by the shift
$\Lambda(L,n-n_0) \equiv \Lambda(L,n)\setminus \Lambda(L,n_0)$,
$(l,j) \mapsto (l,j+n_0)$.
This means
\begin{equation}
\label{EOE2sum1}
\langle{X}\rangle^{\textrm{can}}_{(L,n,mn)} 
  = \sum_{M_0} \langle{X}\rangle^{\textrm{can}}_{(L,n_0,M_0)} 
  \frac{Z(L,n_0,M_0) Z(L,n-n_0,m n - M_0)}{Z(L,n,m n)}\, .
\end{equation}
Also, by its definition,
\begin{equation}
\label{EOE2sum2}
\langle{X}\rangle^{\textrm{GC}}_{(L,n_0,e^{t/2})}
  = \sum_{M_0} \langle{X}\rangle^{\textrm{can}}_{(L,n_0,M_0)}
  \frac{e^{t M_0} Z(L,n_0,M_0)}{Z^{\textrm{GC}}(L,n_0,e^{t/2})}\, .
\end{equation}
We view these two sums as expectations on a discrete probability space.
Specifically, define two probability measures $P^{\textrm{can}}$,
$P^{\textrm{GC}}$ on the discrete space 
$$
\Omega = \left\{-\frac{1}{2}|\Lambda(L,n_0)|,-\frac{1}{2}|\Lambda(L,n_0)|+1,\dots,
  \frac{1}{2}|\Lambda(L,n_0)|\right\}\, ,
$$
given by
\begin{align*}
P^{\textrm{can}}(\{M_0\}) 
  &= \frac{Z(L,n_0,M_0) Z(L,n-n_0,m n - M_0)}{Z(L,n,m n)}\, ,\\
P^{\textrm{GC}}(\{M_0\}) 
  &= \frac{e^{t M_0} Z(L,n_0,M_0)}{Z^{\textrm{GC}}(L,n_0,e^{t/2})}\, .
\end{align*}
For any random variable $\mathcal{X}$ on $\Omega$, any subset $\Omega_1 \subset \Omega$,
and any finite (signed) measure $Q$ on $\Omega$, let
$E[\mathcal{X},\Omega_1,Q]$ be the expectation value,
i.e.\
$$
E[\mathcal{X},\Omega_1,Q] = \sum_{x \in \Omega_1} \mathcal{X}(x) Q(\{x\})\, .
$$
Then
\begin{align*}
\langle{X}\rangle^{\textrm{can}}_{(L,n,mn)} 
  - \langle{X}\rangle^{\textrm{GC}}_{(L,n_0,e^{t/2})}
  &= \sum_{M_0 \in \Omega} \langle{X}\rangle^{\textrm{can}}_{(L,n_0,M_0)}\,
      (P^{\textrm{can}}(\{M_0\}) - P^{\textrm{GC}}(\{M_0\}))\\
  &= E[\langle{X}\rangle^{\textrm{can}}_{(L,n_0,\cdot)},\Omega,
    P^{\textrm{can}}-P^{\textrm{GC}}]\, .
\end{align*}

For a fixed $\epsilon_2 \in \left(0,\frac{1}{2}\right)$, we decompose
$\Omega$ into two subsets 
\begin{gather*}
\Omega_1 = \{M_0 \in \Omega : |M_0| < n_0 n^{\epsilon_2}\}\, ,
\Omega_2 = \Omega \setminus \Omega_1 
  = \{M_0 \in \Omega : |M_0| \geq n_0 n^{\epsilon_2}\}\, .
\end{gather*}
From Lemma \ref{EOELemma}, we know that $Z(L,n_0,M_0)$ is small for large $M_0$,
and in particular it is decreasing like $q$ to the quadratic function
$M_0\left(\frac{M_0}{n_0}+1\right)$.
This is a faster decay rate than the exponential factor in the 
grand canonical expectation.
Thus, if $n$ is large enough then $P^{\textrm{GC}}(\Omega_2)$ is quite small.
Specifically,
\begin{equation}
\label{GCLargeMBounds}
\begin{aligned}
P^{\textrm{GC}}(\Omega_2) &=
\sum_{|M_0| \geq n_0 n^{\epsilon_2}} \frac{e^{tM_0} Z(L,n_0,M_0)}
  {Z^{\textrm{GC}}(L,n_0,e^{t/2})} \\
  &\leq \left[\frac{\sum_{k \in \Ir} q^{k^2}}{(q^2;q^2)_\infty 
  F_L(e^{t/2})}\right]^{n_0}
  \sum_{|M_0|\geq n_0 n^{\epsilon_2}} e^{tM_0} q^{M_0^2/n_0} \\
  &\leq \frac{4}{1 - e^{|t|} q^{2n^{\epsilon_2}}}
  \left[\frac{e^{|tn^{\epsilon_2}|} q^{n^{2\epsilon_2}}
  \sum_{k \in \Ir} q^{k^2}}{(q^2;q^2)_\infty F_L(e^{t/2})}\right]^{n_0}\, .
\end{aligned}
\end{equation}
We define $\mathcal{E}_1 = P^{\textrm{GC}}(\Omega_2)$.

Next, we observe that for any bounded random variable $x$ on $\Omega$,
$$
\left|E[x,\Omega,P^{\textrm{can}}] - E[x,\Omega,P^{\textrm{GC}}]\right|
  \leq \|x\|_{\sup,\Omega} E[1,\Omega,|P^{\textrm{can}}-P^{\textrm{GC}}|]\, ,
$$
where for any signed measure, $Q$, the absolute value of $Q$ is the nonnegative measure
$$
|Q|(\Omega_1) = \sum_{x \in \Omega_1} |Q(\{x\})|\, .
$$
Then by the triangle inequality $|Q_1 + Q_2| \leq |Q|_1 + |Q|_2$.
So,
\begin{align*}
E[1,\Omega,|P^{\textrm{can}}-P^{\textrm{GC}}|]
  &= E[1,\Omega_1,|P^{\textrm{can}}-P^{\textrm{GC}}|] + 
     E[1,\Omega_2,|P^{\textrm{can}}-P^{\textrm{GC}}|] \\
  &\leq E[1,\Omega_1,|P^{\textrm{can}}-P^{\textrm{GC}}|] \\ 
  &\qquad + E[1,\Omega_2,P^{\textrm{can}}] + E[1,\Omega_2,P^{\textrm{GC}}]\\
  &\leq \left\|\frac{dP^{\textrm{can}}}{dP^{\textrm{GC}}} 
  - 1\right\|_{\sup,\Omega_1}  P^{\textrm{GC}}(\Omega_1)
  + P^{\textrm{can}}(\Omega_2) + P^{\textrm{GC}}(\Omega_2)\\
  &\leq \left\|\frac{dP^{\textrm{can}}}{dP^{\textrm{GC}}} 
  - 1\right\|_{\sup,\Omega_1}
  + P^{\textrm{can}}(\Omega_2) + P^{\textrm{GC}}(\Omega_2)\, ,
\end{align*}
where as usual $dQ_1/dQ_2$ is defined when $Q_1 \ll Q_2$ by
$$
\frac{dQ_1}{dQ_2}(x) = \frac{Q_1(\{x\})}{Q_2(\{x\})} \quad \forall x \in \supp(Q_2)\, .
$$
By the definition of $P^{\textrm{GC}}$, $\supp P^{\textrm{GC}} = \Omega$, and therefore
$P^{\textrm{can}} \ll P^{\textrm{GC}}$.
We define
$$
\mathcal{E}_2 = \left\|\frac{dP^{\textrm{can}}}{dP^{\textrm{GC}}} 
  - 1\right\|_{\sup,\Omega_1}\, .
$$
On the other hand, we also know $P^{\textrm{can}}(\Omega) = P^{\textrm{GC}}(\Omega)=1$.
So
\begin{align*}
P^{\textrm{can}}(\Omega_2)
  &= P^{\textrm{GC}}(\Omega_2) + P^{\textrm{GC}}(\Omega_1) - P^{\textrm{can}}(\Omega_1)\\
  &\leq P^{\textrm{GC}}(\Omega_2) + E[1,\Omega_1,|P^{\textrm{can}}-P^{\textrm{GC}}|] \\
  &\leq \mathcal{E}_1 + \mathcal{E}_2\, .
\end{align*}
Thus, we obtain
$$
E[1,\Omega,|P^{\textrm{can}}-P^{\textrm{GC}}|] 
  \leq 2 (\mathcal{E}_1 + \mathcal{E}_2)\, ,
$$
so that
$$
\left|E[x,\Omega,P^{\textrm{can}}] - E[x,\Omega,P^{\textrm{GC}}]\right|
  \leq 2 \|x\|_{\sup,\Omega} (\mathcal{E}_1 + \mathcal{E}_2)\, .
$$
In particular, this implies
$$
\langle{X}\rangle^{\textrm{can}}_{(L,n,mn)} 
  - \langle{X}\rangle^{\textrm{GC}}_{(L,n_0,e^{t/2})}
  \leq 2 \|X\| (\mathcal{E}_1 + \mathcal{E}_2)\, ,
$$
since for any $M_0 \in \Omega$,
$$
\left|\langle{X}\rangle^{\textrm{can}}_{(L,n_0,M_0)}\right|
  \leq \|X\|\, .
$$

We already have a bound for $\mathcal{E}_1$, all we need to do now is bound
$\mathcal{E}_2$.
This is where the activity bounds come in.
We observe
$$
\frac{dP^{\textrm{can}}}{dP^{\textrm{GC}}}(M_0)
  = \frac{Z(L,n-n_0,mn-M_0) Z^{\textrm{GC}}(L,n_0,e^{t/2})}
  {Z(L,n,mn) e^{M_0 t}}\, .
$$
Now, by Lemma \ref{2ndACBoundsLemma}, 
$$
Z(L,n,mn) = q^{\frac{n}{2} \floor{\frac{L^2}{2}}}
  \sqrt{\frac{\phi_L'\left( m \right)}{2 \pi n}} 
  \exp\left[n\int_{m}^{L/2} \phi_L(s)\, ds\right] C_1\, ,
$$
where
$$
1 - \varepsilon_{\textrm{lower}}(L,n,m) \leq C_1 
  \leq 1 + \varepsilon_{\textrm{upper}}(L,n,m)\, .
$$
Similarly,
\begin{align*}
Z(L,n-n_0,mn-M_0) 
  &= q^{\frac{n-n_0}{2} \floor{\frac{L^2}{2}}}
  \sqrt{\frac{\phi_L'\left( m + \frac{mn_0 -M_0}{n-n_0}\right)}
{2 \pi (n-n_0)}} \\
  &\qquad
  \exp\left[(n-n_0)\int_{m+\frac{mn_0 -M_0}{n-n_0}}^{L/2} \phi_L(s)\, ds\right] C_2\, ,
\end{align*}
where
\begin{align*}
C_2 &\geq 1 - 
  \varepsilon_{\textrm{lower}}\left(L,n-n_0,m+\frac{mn_0 -M_0}{n-n_0}\right)\, ,\\ 
C_2 &\leq 1 + 
  \varepsilon_{\textrm{upper}}\left(L,n-n_0,m+\frac{mn_0 -M_0}{n-n_0}\right)\, .
\end{align*}
Also, from \eq{BernoulliLemmaEq6} and \eq{ForEOEProof}, we have
$$
e^{-m n_0 t} Z^{\textrm{GC}}(L,n_0,e^{t/2})
  = q^{\frac{n_0}{2} \floor{\frac{L^2}{2}}}
  \exp\left[n_0 \int_m^{L/2} \phi_L(s)\, ds\right]\, ,
$$
while simply, from the definition of $t = \phi_L(m)$, we have
\begin{align*}
e^{(m n_0 - M_0) t}
  &= \exp\left[(m n_0 - M_0) \phi_L(m)\right] \\
  &= \exp\left[(n-n_0)\int_{m}^{m+\frac{mn_0 -M_0}{n-n_0}} \phi_L(m)\, ds\right]\, .
\end{align*}
Puting it all together, we have
\begin{align*}
\frac{dP^{\textrm{can}}}{dP^{\textrm{GC}}}(M_0)
  &= \frac{C_2}{C_1} \left[\frac{n}{n-n_0}\right]^{1/2}
\left[\frac{\phi_L'\left( m + \frac{mn_0 -M_0}{n-n_0}\right)}
  {\phi_l'(m)}\right]^{1/2}\\
  &\qquad \exp\left[-(n-n_0)
  \int_{m}^{m+\frac{mn_0 -M_0}{n-n_0}} [\phi_L(s) - \phi_L(m)]\, ds\right]\\
  &= \frac{C_2}{C_1} \left[1 + \frac{n_0}{n-n_0}\right]^{1/2}
  \left[1 + \frac{mn_0 - M_0}{n-n_0} \int_0^1 
  \frac{\phi_L''\left(m + \frac{mn_0 -M_0}{n-n_0}t\right)}
  {\phi'_L(m)}\, dt\right]^{1/2}\\
  &\qquad
  \exp\left[-\frac{(mn_0 -M_0)^2}{n-n_0} 
  \int_0^1 (1 - t)
  \phi_L'\left(m + \frac{mn_0 -M_0}{n-n_0} t\right)\right]\, dt\, .
\end{align*}
Taking the supremum and infemum over $\Omega_1$, yields the stated bound
for $\mathcal{E}_2$.
\end{proof}
From this theorem, one can derive the existence of the thermodynamic limit,
as in Corollary \ref{ThermoLimitCor}, but for two dimensions. 
One can extend the results of Sections 5.6--5.11, i.e.\ the upper-bound on
the spectral gap to two dimensions.
The only difference is that in place of radial Bessel functions, which were used to 
minimize the two-dimensional Laplacian (since the interface plane in three-dimensions is
two-dimensional) with Dirichlet boundary conditions, 
one uses trigonometric functions since one is now solving the one-dimensional Laplacian.


\nocite{*}
\newpage
\pagestyle{myheadings} 
\markright{  \rm \normalsize BIBLIOGRAPHY. \hspace{0.5cm}
Some spectral properties ...}
\thispagestyle{myheadings}
\addcontentsline{toc}{chapter}{\bf Bibliography}
\bibliography{xxz}
\end{document}